\documentclass[economy]{dukedissertation}

\usepackage{amsmath, amssymb, amsfonts, amsthm}
\usepackage{graphicx}
\usepackage{natbib}
\usepackage{color}
\usepackage{bm}
\usepackage{subfigure}
\usepackage{mathabx}
\usepackage{multirow}
\usepackage{setspace}
\usepackage{newlfont}
\usepackage{xy}
\xyoption{all}
\usepackage{color}
\usepackage{epsfig}
\usepackage{fancyhdr}
\usepackage{float}
\usepackage{lscape}
\usepackage{pdflscape}

\newtheorem{theorem}{Theorem}[section]
\newtheorem{prop}[theorem]{Proposition}
\newtheorem{cor}[theorem]{Corollary}
\newtheorem{lem}[theorem]{Lemma}
\newtheorem{conc}[theorem]{Conclusion}
\newtheorem{rej}[theorem]{Rejection Criteria}
\newenvironment{nprop}[2][Proposition]{\begin{trivlist}
\item[\hskip \labelsep {\bfseries #1}\hskip \labelsep {\bfseries #2}]}{\end{trivlist}}

\newcommand{\RR}{\mathbb{R}}

\newcommand{\C}{\mathbb{C}}

\newcommand{\e}{\mathrm{e}}

\newcommand{\p}{\partial}

\renewcommand{\S}{\mathcal{S}}

\newcommand{\F}{\mathcal{F}}

\newcommand{\hV}{\hat{V}}
\newcommand{\tV}{\tilde{V}}

\newcommand{\hF}{\hat{F}}
\newcommand{\x}{\mathbf{x}}

\newcommand{\Hv}{\vec{\mathrm{H}}}
\newcommand{\cf}{\bar{f}}
\newcommand{\cp}{\bar{p}}
\newcommand{\cq}{\bar{q}}

\newcommand{\al}{\alpha}
\newcommand{\be}{\beta}
\newcommand{\vep}{\varepsilon}
\newcommand{\la}{\lambda}
\newcommand{\La}{\Lambda}
\newcommand{\ga}{\gamma}
\newcommand{\del}{\delta}
\newcommand{\Del}{\Delta}
\newcommand{\Si}{\Sigma}
\newcommand{\si}{\sigma}
\newcommand{\Om}{\Omega}
\newcommand{\om}{\omega}
\newcommand{\Ga}{\Gamma}
\newcommand{\Up}{\Upsilon}
\newcommand{\ka}{\kappa}

\newcommand{\cov}{\nabla}

\newcommand{\pa}[2]{\dfrac{\partial #1}{\partial #2}}
\newcommand{\da}[2]{\dfrac{d #1}{d #2}}
\newcommand{\ppa}[2]{\dfrac{\partial^{2} #1}{\partial #2^{2}}}

\newcommand{\glap}[1]{\Box_{#1}}
\renewcommand{\div}{\nabla \cdot}
\newcommand{\divx}[1]{\nabla_{#1} \cdot}
\newcommand{\lap}{\Delta}
\newcommand{\dtg}{\dot{g}}
\newcommand{\dtf}{\dot{f}}

\newcommand{\tr}{\operatorname{tr}}

\newcommand{\Quad}{\operatorname{Quad}}
\newcommand{\Schrodinger}{Schr\"{o}dinger}
\newcommand{\LHopital}{L'H\^{o}pital}
\newcommand{\Ric}{\operatorname{Ric}}
\newcommand{\R}{\mathbf{R}}
\renewcommand{\H}{\mathbf{H}}
\newcommand{\II}{\operatorname{\text{I\hspace{-1 pt}I}}}

\renewcommand{\Re}{\operatorname{Re}}
\renewcommand{\Im}{\operatorname{Im}}
\newcommand{\Taylor}{\operatorname{Taylor}}

\newcommand{\conj}[1]{\bar{#1}}
\newcommand{\conjL}[1]{\overline{#1}}

\newcommand{\pnth}[1]{\left( #1 \right)}
\newcommand{\norm}[1]{\left\| #1 \right\|}
\newcommand{\abs}[1]{\left| #1 \right|}

\newcommand{\brce}[1]{\left\{ #1 \right\}}
\newcommand{\brkt}[1]{\left[ #1 \right]}
\newcommand{\ip}[1]{\left\langle #1 \right\rangle}

\author{Alan R. Parry}
\title{Wave Dark Matter and Dwarf Spheroidal Galaxies}
\supervisor{Hubert L.\ Bray}
\department{Mathematics} 

\date{2013} 

\copyrighttext{All rights reserved except the rights granted by the\\
   \href{http://creativecommons.org/licenses/by-nc/3.0/us/}
        {Creative Commons Attribution-Noncommercial Licence}
}

\member{Paul S.\ Aspinwall}
\member{Arlie O.\ Petters}
\member{Thomas P.\ Witelski}

\usepackage{hyperref}

\begin{document}

\maketitle

\abstract

We explore a model of dark matter called wave dark matter (also known as scalar field dark matter and boson stars) which has recently been motivated by a new geometric perspective by Bray (\cite{Bray10}).  Wave dark matter describes dark matter as a scalar field which satisfies the Einstein-Klein-Gordon equations.  These equations rely on a fundamental constant $\Up$ (also known as the ``mass term'' of the Klein-Gordon equation).  Specifically, in this dissertation, we study spherically symmetric wave dark matter and compare these results with observations of dwarf spheroidal galaxies as a first attempt to compare the implications of the theory of wave dark matter with actual observations of dark matter.  This includes finding a first estimate of the fundamental constant $\Up$.

In the introductory Chapter \ref{ch1}, we present some preliminary background material to define and motivate the study of wave dark matter and describe some of the properties of dwarf spheroidal galaxies.

In Chapter \ref{ch2}, we present several different ways of describing a spherically symmetric spacetime and the resulting metrics.  We then focus our discussion on an especially useful form of the metric of a spherically symmetric spacetime in polar-areal coordinates and its properties.  In particular, we show how the metric component functions chosen are extremely compatible with notions in Newtonian mechanics.  We also show the monotonicity of the Hawking mass in these coordinates.  Finally, we discuss how these coordinates and the metric can be used to solve the spherically symmetric Einstein-Klein-Gordon equations.

In Chapter \ref{ch3}, we explore spherically symmetric solutions to the Einstein-Klein-Gordon equations, the defining equations of wave dark matter, where the scalar field is of the form $f(t,r) = \e^{i\om t}F(r)$ for some constant $\om \in \RR$ and complex-valued function $F(r)$.  We show that, under appropriate conditions, the corresponding metric is static if and only if $F(r) = h(r)\e^{i a}$ for some constant $a \in \RR$ and real-valued function $h(r)$.  We describe the behavior of the resulting solutions, which are called spherically symmetric static states of wave dark matter.  We also describe how, in the low field limit, the parameters defining these static states are related and show that these relationships imply important properties of the static states.

In Chapter \ref{ch4}, we compare the wave dark matter model to observations to obtain a working value of $\Up$.  Specifically, we compare the mass profiles of spherically symmetric static states of wave dark matter to the Burkert mass profiles that have been shown by Salucci et al.\ (\cite{Salucci11}) to predict well the velocity dispersion profiles of the eight classical dwarf spheroidal galaxies.  We show that a reasonable working value for the fundamental constant in the wave dark matter model is $\Up = 50 \text{ yr}^{-1}$.  We also show that under precise assumptions the value of $\Up$ can be bounded above by $1000 \text{ yr}^{-1}$.

In order to study non-static solutions of the spherically symmetric Einstein-Klein-Gordon equations, we need to be able to evolve these equations through time numerically.  Chapter \ref{ch5} is concerned with presenting the numerical scheme we will use to solve the spherically symmetric Einstein-Klein-Gordon equations in our future work.  We will discuss how to appropriately implement the boundary conditions into the scheme as well as some artificial dissipation.  We will also discuss the accuracy and stability of the scheme.  Finally, we will present some examples that show the scheme in action.

In Chapter \ref{ch6}, we summarize our results.  Finally, Appendix \ref{app1} contains a derivation of the Einstein-Klein-Gordon equations from its corresponding action. 

The majority of this thesis has also been presented by the author in three separate shorter papers (\cite{Parry12-1,Parry12-3,Parry12-2}); note that \cite{Parry12-2}, and hence part of Chapter \ref{ch4}, represents joint work with Hubert Bray. 

\dedication{This dissertation is dedicated to my wife and children.}

\tableofcontents 
\listoftables	
\listoffigures	
\abbreviations

\section*{Symbols}

This is a list of commonly used symbols in this dissertation.  Unless otherwise noted in the text, each symbol is defined as follows.

\begin{symbollist}
  \item[$\RR$] The set of real numbers.
  \item[$\C$] The set of complex numbers.
  \item[$\e$] Euler's number, $\e = 2.71828...$.
  \item[$i$] The imaginary unit.
  \item[$\p_{x}$] The coordinate vector field $\pa{}{x}$ corresponding to the coordinate function $x$.
  \item[$g$] The spacetime metric.
  \item[$\ga$] The induced metric of a submanifold of the spacetime, usually that of a spacelike hypersurface or a metric sphere.
  \item[$\cov$] Used to denote a connection or covariant derivative.  It will be clear from the context what it is.
  \item[$\Ga$] The Christoffel symbols of a connection.
  \item[$\R$] The Riemann curvature tensor.
  \item[$\Ric$] The Ricci curvature tensor.
  \item[$R$] The scalar curvature.
  \item[$G$] The Einstein curvature tensor, $G = \Ric - \dfrac{1}{2}Rg$.
  \item[$T$] The stress energy tensor.
  \item[$\Si$] Usually a two-dimensional surface in a manifold.
  \item[$\Si_{t,r}$] The metric sphere of radius $r$ and time $t$.
  \item[$m_{H}(\Si)$] The Hawking mass of the 2-dimensional surface $\Si$.
  \item[$\II$] The second fundamental form.
  \item[$\Hv$] The mean curvature vector field.
  \item[$H$] The Mean Curvature in Chapter \ref{ch2} and the derivative of the function $F(r)$ otherwise.  
  \item[$\H^{f}$] The Hessian of the function $f$.
  \item[$\glap{g}$] The Laplacian operator on the spacetime with respect to the metric $g$.
  \item[$t,r,\theta,\varphi$] The names of the coordinates of a spherically symmetric spacetime.
  \item[$f$] The scalar field of the Einstein-Klein-Gordon equations.
  \item[$p$] A multiple of the time derivative of $f$, defined specifically by equation (\ref{p-Def}).
  \item[$V$] One of two of the metric functions in the spherically symmetric metric in equation (\ref{metric}).  This function is analogous to the Newtonian potential in the low-field limit.
  \item[$M$] One of two of the metric functions in the spherically symmetric metric in equation (\ref{metric}).  This function can be interpreted as the mass inside a metric sphere of the spacetime.
  \item[$\om$] Usually denotes the angular frequency of the scalar field $f$ when of the form $f(t,r) = \e^{i\om t}F(r)$.
  \item[$F$] The function $F$ as defined immediately above.
  \item[$\Up$] The fundamental constant of the Einstein-Klein-Gordon equations.
  \item[$\mu_{0}$] The parameter in the Einstein-Klein-Gordon equations that controls the magnitude of the energy density.
  \item[$m$] Usually used as the total mass of a spherically symmetric spacetime.
  \item[$r_{h}$] The half mass radius, or the radius which contains a mass equal to $m/2$, where $m$ is the total mass as above.
\end{symbollist}

\section*{Abbreviations}

These are common abbreviations relevant to this dissertation.  Not all of these may appear in this dissertation, but they are useful nonetheless.

\begin{symbollist}
  \item[NFW] Navarro, Frenk, and White.  This acronym is generally used to denote the dark matter energy density profile described by them.
  \item[ADM] Arnowitt, Deser, and Misner.  This acronym is generally used to denote either the ADM mass or ADM formulation of general relativity first described by these three physicists.
  \item[EKG] Einstein-Klein-Gordon.  Usually used to describe the Einstein-Klein-Gordon equations for a scalar field.  These equations consist of the Einstein equation coupled to the Klein-Gordon equation.
  \item[KG] Klein-Gordon.  Used to describe just the Klein-Gordon equation.
  \item[PS] Poisson-\Schrodinger.  Used to describe the coupled Poisson and \Schrodinger\ equations.
  \item[WKB] Wentzel-Kramers-Brillouin.  The WKB approximation generally refers to the leading asymptotic terms of a solution to an differential equation.
  \item[dSph] Dwarf Spheroidal Galaxy (or Galaxies).
  \item[DM] Dark Matter.
  \item[WDM] Wave Dark Matter.
  \item[SR] Special Relativity.
  \item[GR] General Relativity.
\end{symbollist}


\acknowledgements

I would like to thank my advisor, Hubert Bray, whose instruction has not only educated me in mathematics but also in how best to work in the world of academia.  He looks out for his graduate students on not just a professional level but on a personal one as well and wants them all to succeed in every aspect of life.  An advisor and mentor like him is a rare find.

I would also like to thank the remaining members of my dissertation committee, Paul Aspinwall, Arlie Petters, and Thomas Witelski for being willing to advise me over the years and for being willing to read this lengthy dissertation.  Your time and willingness to help me complete my degree has been invaluable.

Next, I wish to thank Justin Corvino and Pengzi Miao for selecting me as a TA during the MSRI summer school they co-organized.  I also thank Justin Corvino and the mathematics and physics departments at Lafayette college as well as Pengzi Miao and the mathematics department at the University of Miami for inviting me to give a talk or two at their respective universities.  All of these trips helped me meet new people and build experience that I can take into the professional world.

I also thank Thomas Curtright and the other organizers at the Miami 2012 topical physics conference for inviting me to speak there.  It was a great opportunity for me to present my physics related research to the world of physicists and get feedback from and network with them.

I offer my thanks to Fernando Schwartz and the University of Tennessee at Knoxville, for paying my way to visit their university twice for two separate conferences.

I also gratefully acknowledge the financial support of National Science Foundation Grant \# DMS-1007063.

Most of all, I thank my beautiful wife, Alesha, and my three wonderful children, Madison, Ethan, and Lindsey, without whose support, love, and encouragement at home, I surely would not have completed this dissertation and degree.  I love all of you!  It is to them that I dedicate this work.

\begin{flushright}
  Alan R.\ Parry
\end{flushright} 

%
%
%
\chapter{Introduction}\label{ch1}

General relativity is Einstein's theory of gravity and currently the dominant theory to explain the large scale structure of the universe making it an exciting field full of big questions about some of the most fascinating objects in the cosmos.  Furthermore, general relativity is described using the language of semi-Riemannian geometry, a beautiful subset of differential geometry that naturally extends the ideas of calculus to manifolds that are not necessarily flat and succinctly describes their curvature.  Thus general relativity lies at one of the most innate and appealing intersections between mathematics and physics.

The topic of this dissertation concerns the possibility of describing dark matter, which makes up nearly one-fourth of the energy density of our universe, from a geometrical point of view using a scalar field.  This idea has been around for roughly twenty years under the name of scalar field dark matter or boson stars (see \cite{Lee09, Matos09, MSBS, Seidel90, Seidel98, Bernal08, Sin94, Mielke03, Sharma08, Ji94, Lee92, Lee96, Guzman01, Guzman00} for a few references).  However, this theory of dark matter has been recently championed by Hubert Bray (see \cite{Bray10,Bray12}) due to its role as a consequence of a very simple set of geometric axioms.

To set up the appropriate background material to the research presented in this dissertation, a brief description about the theory of general relativity and the concept of dark matter as well as some recent work by mathematicians and physicists alike concerning dark matter and galaxies is required.  We devote the next three sections to discussing such a background.  Section \ref{WDMsec} will describe the problem this dissertation addresses.  Throughout this dissertation, we will assume a general understanding of the main concepts in semi-Riemannian geometry (for an excellent reference on this material see \cite{Oneill83}).

\section{Basic Ideas of General Relativity}

\begin{figure}

  \begin{center}
    \includegraphics[height = 2.75 in]{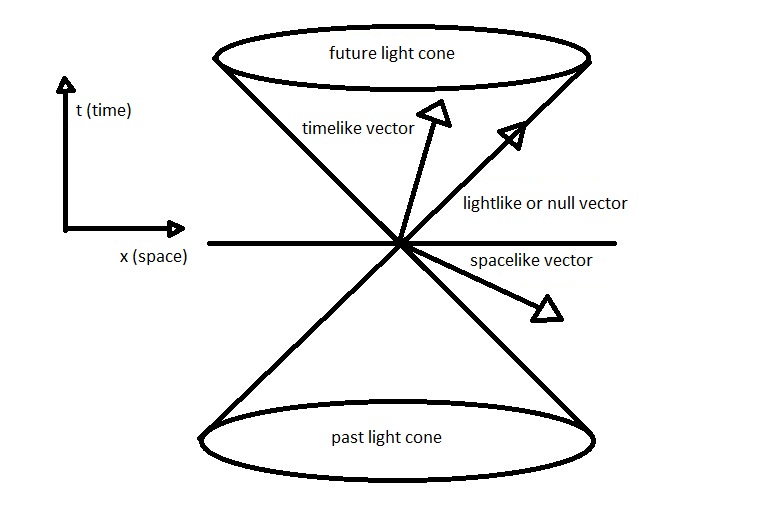}
  \end{center}

  \caption{The light cone in Minkowski space projected into the $(t,x)$ plane.  A timelike, spacelike, and lightlike or null vector is depicted in the figure.}

\label{timecone}

\end{figure}

As a precursor to general relativity, Einstein published his theory of special relativity in 1905 (\cite{Einstein05}), which overturned many of the natural assumptions about the universe that Newton had made in his theory of gravitation and relativity.  Chief among these differences was that special relativity asserted that time and space were two parts of the same thing and that the speed of light was constant, two ideas completely foreign to the Newtonian model.  This theory was given an elegant mathematical setting by Hermann Minkowski in 1908 (\cite{Minkowski08}), who described the special relativity spacetime as a differentiable manifold, $N$, of three spatial dimensions and one time dimension, coupled with a semi-Riemannian metric, $g$, whose line element is of the form
\begin{equation}\label{Mink}
  ds^{2} = -dt^{2} + dx^{2} + dy^{2} + dz^{2}
\end{equation}
where we have used geometrized units to set the speed of light equal to $1$.  This is called the Minkowski spacetime.  In Minkowski spacetime, the change in time of two events is not invariant in all inertial frames as it was assumed to be by Newton, but in fact, the invariant interval is the spacetime interval $ds$ defined by the line element above.  From this fact, that $ds$ is invariant in all inertial frames, one can obtain, among other things, that the speed of light is invariant in all frames of reference.  This metric also splits the set of vectors in the tangent space to $N$ at $p$, $T_{p}N$, into three sets, timelike, spacelike, or null vectors depending on the sign of the dot product of the vector with itself with respect to the inner product induced on each tangent space by the metric in (\ref{Mink}).  This is illustrated in Figure \ref{timecone}.  Finally, this mathematical description of special relativity gave an elegant interpretation of inertial observers in the spacetime as being those who follow the geodesics, or non-accelerating curves, whose velocity vectors are timelike.

\begin{figure}

  \begin{center}
    \includegraphics[height = 2 in]{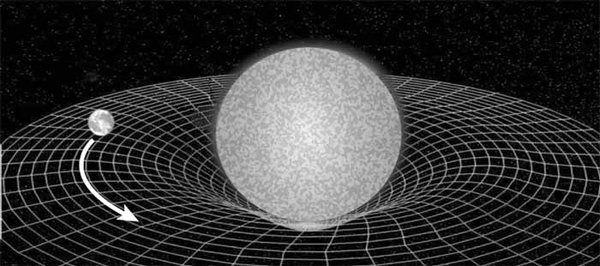}
  \end{center}

  \caption{The earth traveling along a geodesic in the curved spacetime generated by the sun.  Image credit: http://einstein.stanford.edu/.}

  \label{curvgeo}

\end{figure}

These ideas were carried over into Einstein's theory of general relativity, first presented in 1915 and published in 1916 (\cite{Einstein16, Hilbert15}), which generalizes special relativity to apply to non-inertial reference frames as well.  In this theory, Einstein removed the requirement that the metric be the Minkowski one in (\ref{Mink}).  By removing this requirement, the set of allowable spacetimes and metrics dramatically increases to include spacetimes which are intrinsically curved.  This intrinsic curvature was interpreted by Einstein as the presence of energy density, whether it be made up of matter, radiation, etc.  This is done via a beautiful equation which identifies a purely mathematical object describing the curvature of the spacetime with a physical object that describes where the energy density lies in the spacetime.  This equation is called the Einstein equation and is given by
\begin{equation}{\label{Eineq}}
  G = 8\pi T
\end{equation}
where $G = \Ric - (R/2)g$ is the Einstein curvature tensor, consisting of a formula involving the Ricci curvature tensor, $\Ric$, the scalar curvature, $R$, and the metric, $g$, and $T$ is the classical stress energy tensor from physics.  Note that $G$, $\Ric$, and $R$ are all objects that contain information about the intrinsic curvature of the spacetime.  This equating of energy density and curvature is completed by the concept that test particles which follow timelike geodesics are in free fall, that is, they are only acting under the influence of gravity.  This takes advantage of the natural covariant derivative induced by the Levi-Civita connection, whose corresponding non-accelerating curves (i.e. geodesics) are not necessarily straight lines.  From a physical point of view, this fundamentally changed the way we understood gravity.  Gravity was no longer a force.  Instead, gravity is the phenomenon that free falling objects follow the curved geodesics of a curved spacetime.  For example, the earth orbits the sun not due to some imaginary cord tethering it to the sun, but instead orbits because it is following a geodesic of the curved spacetime created by the sun, that is, the earth is trying to follow a straight line, but since the spacetime around it is curved it gets stuck in the dimple caused by the sun and orbits.  Figure \ref{curvgeo} illustrates this.

This notion of equating energy density with spacetime curvature has been incredibly successful at predicting and explaining observed phenomena, including many which were inconsistent with a Newtonian view of gravity.  We have already mentioned that it is consistent with the observation that the speed of light is constant, but general relativity also explains the observations of (1) time dilation and length contraction for moving reference frames and observers near massive bodies, a concept used in practice today to sync the clock of a GPS on the ground with that of the clock on a GPS satellite, (2) relativistic precession, which explained the discrepancy in the observed precession of the perihelion of Mercury with the prediction of Newtonian gravity, (3) black holes, (4) the big bang, and most recently (5) dark energy, which is responsible for the accelerating expansion of the universe and can be described by a simple modification of the Einstein equation with a cosmological constant (\cite{Perlmutter99, Riess98}).

This physical interpretation of the geometry of spacetime has also led to many connections with geometric analysis.  The positive mass theorem, first proved in certain dimensions by Schoen and Yau (\cite{Schoen79,Schoen81}), and the Riemannian Penrose inequality, proved in the case of a single black hole by Huisken and Ilmanen (\cite{Huisken01}) and then in the case of any number of black holes by Bray (\cite{Bray01}), are examples of such connections.

The next topic we present is a problem in astrophysics that, while it has been known for several decades, it is still not very well understood.  That is the problem of dark matter.

\section{Dark Matter}

While general relativity has been extremely successful at explaining a lot of physical phenomena in the universe, there does exist a small number of observations which, at first glance, are in apparent contradiction with the predictions of general relativity.  One of these is concerning how fast stars, gas, and dust are orbiting in spiral galaxies and galaxy clusters.  This was first noticed in the Milky Way by Oort (\cite{Oort32}) and in clusters of galaxies by Zwicky (\cite{Zwicky33}).

There are two methods of determining the rotational speeds of these objects in a spiral galaxy, if the galaxy is at the appropriate angle that objects at different radii can be resolved but also tipped enough that objects on the left and right of the center of the galaxy from our vantage point are either moving toward us or away from us.  Under these conditions, red and blue shift can be used to directly determine the rotational velocities of objects at different radii.  Moreover, due to the fact that the luminosity at all wavelengths of the galaxy is proportional to the amount of regular mass in a galaxy, one can obtain an approximate mass profile of the regular mass and use Newtonian mechanics, which is what general relativity reduces to on a galactic scale, to compute the rotational speed at each radii.  We call a plot of the rotational speed at each radii the rotation curve.

\begin{figure}

  \begin{center}
    \includegraphics[width = 4 in]{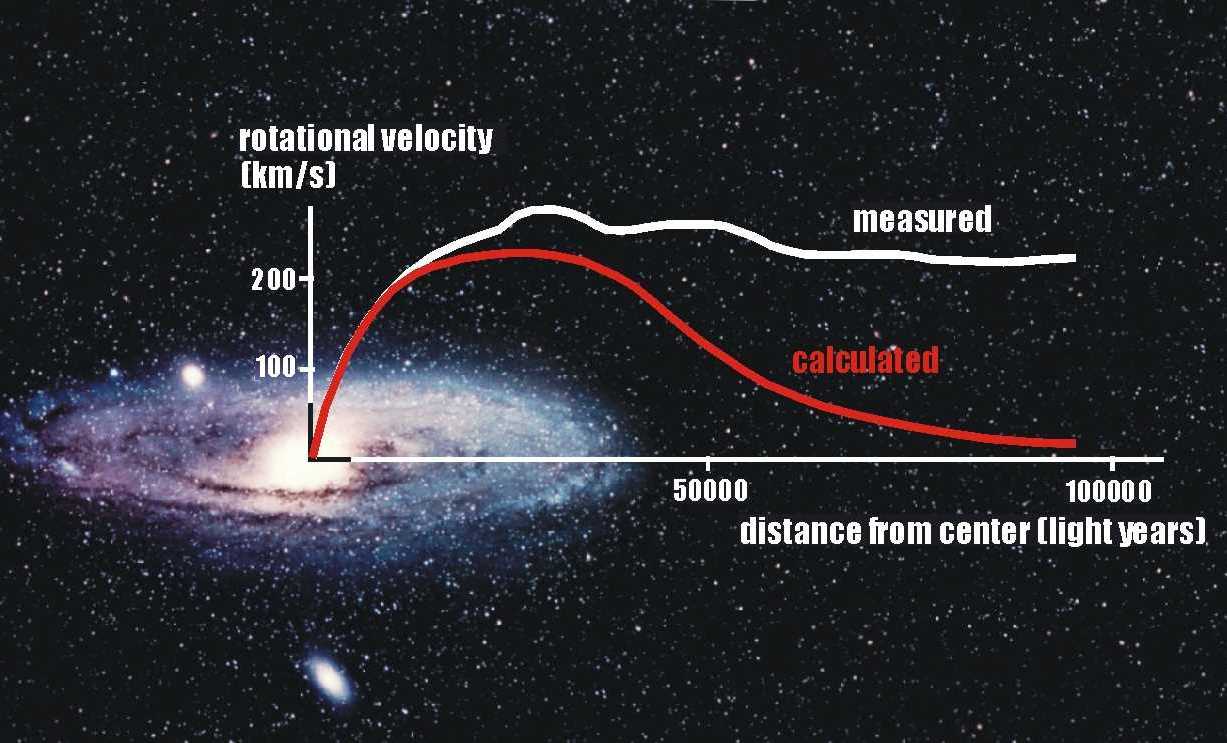}
  \end{center}

  \caption{The rotation curves, both observed and calculated, for the Andromeda galaxy.  Credit: Queens University.}

  \label{DMobs1}

\end{figure}

These two methods have been performed on many spiral galaxies, see \cite{Begeman89} and \cite{Bosma81} amongst other references.  What has been found in every spiral galaxy is that the stars, gas, and dust at distant radii are moving much faster than predicted and that instead of the rotation curve dropping off quickly at large radii, it tends to remain flat.  Figure \ref{DMobs1} shows the two rotation curves for the Andromeda galaxy overlayed on a picture of the galaxy itself.

There are only two possible explanations for such a discrepancy between computation and correct data.  Either the law of gravity used is incorrect on at least the galactic scale and requires an overhaul similar to how general relativity overhauled Newtonian mechanics, or there is more matter present in the galaxy than can be accounted for by the luminosity alone.  This kind of matter is called dark matter since it interacts gravitationally but does not give off any kind of light or interact in any other observed way (e.g. it is collisionless and hence frictionless).  Both approaches to this problem have been and are still being considered, but due to another observation, most astrophysicists favor the solution of dark matter.

\begin{figure}

  \begin{center}
    \includegraphics[width = 4 in]{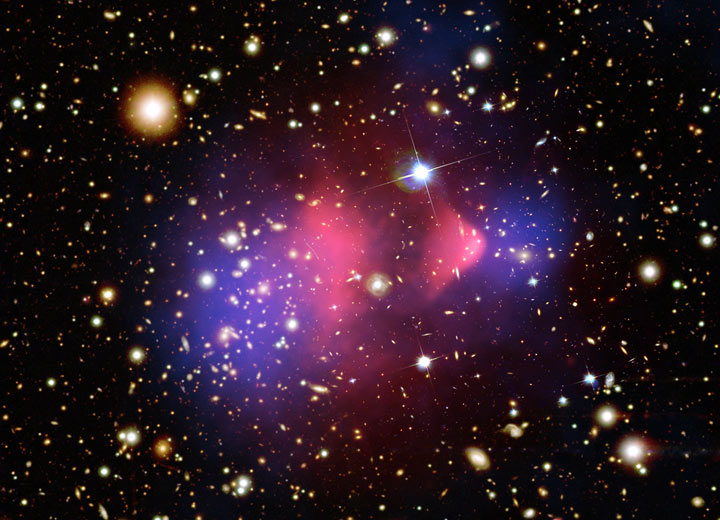}
  \end{center}

  \caption{The Bullet Cluster.  The pink clouds are where the visible matter from the galaxy clusters is located, while the blue represents the dark matter. Credit: NASA/CXC/CfA/M.\  Markevitch et al.; Optical: NASA/STScI; Magellan/U.Arizona/D.Clowe et al.; Lensing Map: NASA/STScI; ESO WFI; Magellan/U.Arizona/D.Clowe et al.}

  \label{DMobs2}

\end{figure}

This observation is of the bullet cluster, an image of which is presented in Figure \ref{DMobs2}.  The bullet cluster is actually two clusters of galaxies (a cluster of galaxies being a group of many galaxies gravitationally bound together) that have recently collided with one another.  In this collision, the vast majority of the regular luminous mass, that part which is comprised of gas and dust, was slowed down due to friction in the colliding gas clouds.  However, the stars and planets, which are too far apart to collide, as well as the dark matter passed through the collision without slowing down.  The two separate components have been resolved using gravitational lensing.  The pink cloud is the regular matter, while the blue is the dark matter.  Thus the bullet cluster represents an event where dark matter has literally been stripped away from its host galaxy clusters.  This is not something that would be expected if the rotational curves could be explained by simply correcting the law of gravity on the galactic scale.

There are additional observational results that support the existence of dark matter including the velocity dispersion profiles of dwarf spheroidal galaxies (\cite{Walker07,Salucci11,Walker09,Walker10}), and gravitational lensing (\cite{Dahle07}).  These and other observations support the idea that most of the matter in the universe is not baryonic, but is, in fact, some form of exotic dark matter and that almost all astronomical objects from the galactic scale and up contain a significant amount of this dark matter.  Specifically, the energy density of the universe seems to be currently made up of three constituents, dark energy, dark matter, and regular or baryonic matter.  Dark energy accounts for about $72\%$ of the energy density of the universe, while dark matter accounts for $23\%$ and baryonic matter accounts for only about $4.6\%$ (\cite{Hinshaw09}).

Baryonic matter is described extremely well in the realm of quantum mechanics, while dark energy, on the other hand, only seems to be described adequately using general relativity.  However, describing dark matter is currently one of the biggest open problems in astrophysics (\cite{DMAW, Hooper08, Bertone05, Ostriker93, Trimble87, Binney98, Binney08}).  Since both quantum mechanics and general relativity have been successful at describing the other components of the energy density of the universe, there is active research to describe dark matter from both the quantum mechanical and general relativistic perspectives.

Quantum mechanics considers dark matter a particle and explores the consequences of this idea, see \cite{Hooper08} and \cite{Primack88} for review articles. Part of this research that is particularly important to this dissertation is the attempts to obtain an energy density profile for the dark matter halo around a galaxy that agrees well with observation.  Two profiles, one from Navarro, Frenk, and White (\cite{NFW96}) and another from Burkert (\cite{Burkert95}), are two such profiles that have been well studied.

Another avenue of approach to finding a way to describe dark matter may lie in the field of general relativity.  In the next section, we introduce some important ideas presented by Hubert Bray in a recent paper on the axioms of general relativity (\cite{Bray10}) that suggest such an approach.

\section{Axioms of General Relativity}

Hubert Bray recently developed a pair of axioms that construct general relativity upon two principles (\cite{Bray10}).  The first, called Axiom 0, is the statement that the fundamental mathematical objects of the universe are described by geometry.  The second, Axiom 1, states that the fundamental laws of the universe are described by analysis.  We reprint these axioms here because they are particularly relevant to our work.


\begin{description}
  \item[Axiom 0] Let $N$ be a smooth spacetime manifold with a smooth metric $g$ of signature $(-+++)$ and a smooth connection $\cov$.  In addition, given a fixed coordinate chart, let $\{\p_{i}\}$, for \mbox{$0 \leq i \leq 3$,} be the standard basis vector fields in this coordinate chart and define $g_{ij} = g(\p_{i},\p_{j})$ and \mbox{$\Ga_{ijk} = g(\cov_{\p_{i}}\p_{j},\p_{k})$.}  Moreover, let
      \begin{equation}
        M = \{g_{ij}\} \quad \text{and} \quad C = \{\Ga_{ijk}\} \quad \text{and} \quad M' = \{g_{ij,k}\} \quad \text{and} \quad C' = \{\Ga_{ijk,\ell}\}
      \end{equation}
      be the components of the metric and connection in the coordinate chart and all of their first derivatives.
  \item[Axiom 1] For all coordinate charts $\Phi:\Om \subset N \to \RR^{4}$ and open sets $U$ whose closure is compact and in the interior of $\Om$, $(g,\cov)$ is a critical point of the functional
      \begin{equation}\label{func}
        \F_{\Phi,U}(g,\cov) = \int_{\Phi(U)} \Quad_{M}(M' \cup M \cup C' \cup C)\, dV_{\RR^{4}}
      \end{equation}
      with respect to smooth variations of the metric and connection compactly supported in $U$, for some fixed quadratic function $\Quad_{M}$ with coefficients in $M$.  Note that a quadratic function, $\Quad_{Y}(\{x_{\al}\})$, has the form
      \begin{equation}
        \Quad_{Y}(\{x_{\al}\}) = \sum_{\al,\be} F^{\al\be}(Y)x_{\al}x_{\be}
      \end{equation}
      for some functions $\{F^{\al\be}\}$ of $Y$.
\end{description}

Though, at first glance, these axioms may appear daunting to understand, they are actually remarkably simple in their construction of general relativity.  Axiom 0 suggests that the fundamental objects of the universe are the spacetime, metric, and connection.  All three of these geometric objects are required to describe the geometry of a semi-Riemannian manifold.  Axiom 1 imposes a simple analytical requirement: that via variational calculus, the metric and connection are critical points of an action whose integrand is an appropriate quadratic functional.  This is perhaps the simplest type of action that will still produce the vacuum Einstein equation.  In addition, it produces even more laws, some of which have already been shown to be physically relevant.  We paraphrase these constructions below.

By fixing the connection as the Levi-Civita connection, vacuum general relativity, which yields important solutions like the Schwarzschild metric describing stars and black holes, is recovered from these axioms.  This is because the Einstein-Hilbert action (\cite{Hilbert15})
\begin{equation}
  \int R\, dV,
\end{equation}
where $R$ is again the scalar curvature of the spacetime, satisfies the requirements of the axioms and because metrics which are critical points of this action, with respect to the aforementioned variations, satisfy
\begin{equation}
  G = 0.
\end{equation}
Additionally, vacuum general relativity with a cosmological constant, which describes dark energy (\cite{Perlmutter99, Riess98}), is recovered in the same way from the action
\begin{equation}
  \int R - 2\La\, dV,
\end{equation}
which also satisfies the requirements of the axioms.  That is, metrics which are critical points of this action satisfy
\begin{equation}
  G + \La g = 0.
\end{equation}

Thus both vacuum general relativity and dark energy can be obtained from these axioms without removing the usual assumption that the connection is the Levi-Civita one.  And we could stop there, if desired.  However, many major advances in the theory of gravity have been made by removing assumptions.  From Newtonian gravity to special relativity, the assumptions that time intervals were invariant among inertial reference frames and that there could be no maximum velocity were removed.  From special relativity to general relativity, the assumption that the metric was flat was removed.  The metric is the most fundamental object of the spacetime and so was a natural choice to allow to be general.  Arguably, the second most fundamental object is the connection, but it is often overlooked because the Levi-Civita connection, which is induced by the metric, is normally assumed.  However, since the connection is arguably the second most fundamental object, it is natural to be curious of what results if the assumption that the connection is the Levi-Civita connection is removed.

This is exactly one of the questions treated in Bray's paper, in particular detail in one of the appendices (\cite{Bray10}).  A connection that is not the Levi-Civita connection is not necessarily metric compatible or torsion free.  While we leave the details to Bray's paper, ultimately the fully antisymmetric part of the torsion tensor can be related to a scalar field $f:N \to \RR$.  In order for the connection and metric to be critical points of the corresponding functional of the type in \ref{func}, the scalar field and metric must satisfy the following set of equations
\begin{align}
  \label{EKG1a} G + \La g &= 8\pi \mu_{0}\pnth{ \frac{2}{\Up^{2}} df \otimes df - \pnth{\frac{\abs{df}^{2}}{\Up^{2}} + f^{2}}g} \\
  \label{EKG1b} \glap{g} f &= \Up^{2} f
\end{align}
where $\glap{g}$ is the Laplacian operator induced by the metric $g$, $\mu_{0}$ is some constant that controls the magnitude of the energy density (though not to be confused with the value of the energy density $\mu(t,0)$ at $r=0$) which can be absorbed into $f$ if desired, and $\Up$ is a fundamental constant of the system.  This system is called the Einstein-Klein-Gordon system of equations and can also be directly obtained from an action, though not strictly speaking an action of the type in \ref{func}.  Specifically this action is
\begin{equation}
  \label{EKGaction} \F_{\Phi,U}(g,f) = \int_{\Phi(U)} R - 2\La - 16\pi\mu_{0}\pnth{\frac{\abs{df}^{2}}{\Up^{2}} + \abs{f}^{2}}\, dV.
\end{equation}
That is, if $g$ and $f$ are critical points of this functional under compactly supported variations, then $g$ and $f$ satisfy the Einstein-Klein-Gordon equations, (\ref{EKG1a}) and (\ref{EKG1b}).  In our current research, we consider a complex scalar field, $f:N \to \C$ and this same action.  In this case, the Einstein-Klein-Gordon equations are
\begin{align}
  \label{EKG2a} G + \La g &= 8\pi \mu_{0}\pnth{ \frac{df \otimes d\cf + d\cf \otimes df}{\Up^{2}} - \pnth{\frac{\abs{df}^{2}}{\Up^{2}} + \abs{f}^{2}}g} \\
  \label{EKG2b} \glap{g} f &= \Up^{2}f
\end{align}
which reduces to (\ref{EKG1a}) and (\ref{EKG1b}) if $f$ is real-valued.  In Appendix \ref{app1}, we derive equations (\ref{EKG2a}) and (\ref{EKG2b}) from equation (\ref{EKGaction}) via a variational argument.

We find this axiomatic treatment of the construction of general relativity philosophically appealing because of its simplicity and structure.  Instead of constructing theories from arbitrary actions and arbitrarily included matter fields, these axioms yield a set of parameters that govern allowable actions and show how the different actions, and hence theories, are related.  The requirements of these axioms attempt to be as minimalistic as possible, but still general enough to recover the most important results in general relativity.  Moreover, the entire scope of solutions to this simple pair of axioms has not yet been fully developed and considered as a model for the universe.  In particular, the solution involving a scalar field has no verified physical analogue yet.  Thus not only does this simple pair of axioms encompass the theory we already have, it also includes candidates to describe other physical phenomena.

To be more precise, vacuum general relativity and vacuum general relativity with a cosmological constant are arguably the most natural general relativity theories that lie within the scope of the axioms.  These two theories describe two of the most important cases of the universe, that of vacuum and of dark energy.  Perhaps the next simplest theory consistent with the axioms is that of a single scalar field satisfying the Einstein-Klein-Gordon equations.  So it is natural to ask if this could describe something physical as well.  In particular, dark matter is the next largest portion of the energy density of the universe after dark energy.  Thus the big question we ask is could this scalar field theory describe dark matter and its effects?  From here on, we will call this theory of dark matter, {\bf wave dark matter}, due to equation (\ref{EKG2b}) being a wave-like equation.

\section{Wave Dark Matter}\label{WDMsec}

Wave dark matter has already been shown to be consistent with many cosmological observations (\cite{Bernal08,Matos09,Matos00,Matos01}).  This dissertation makes contributions towards determining if wave dark matter is also consistent at the galactic level.  In \cite{Bray10} and \cite{Bray12}, Bray presented preliminary evidence that wave dark matter could account for previously unexplained wave-like behavior in spiral and elliptical galaxies.  The main goal of this dissertation is to consider dwarf spheroidal galaxies and use observations of these galaxies to constrain the fundamental constant of the Einstein-Klein-Gordon equations, $\Up$.  We will also present a numerical scheme that can be used to evolve the spherically symmetric Einstein-Klein-Gordon equations in time; this scheme will be useful in our later work on this subject.

To achieve this goal of constraining $\Up$, we utilize some of the simplest solutions to the Einstein-Klein-Gordon equations, namely, spherically symmetric static states.  We construct these static states in detail before making our comparisons to dwarf spheroidal galaxies.  Specifically, in Chapter \ref{ch2}, we will survey spherically symmetric spacetimes in order to present some results about a metric with convenient properties.  In Chapter \ref{ch3}, we discuss the static states and their properties.  In Chapter \ref{ch4}, we compare these static states to dwarf spheroidal galaxies to constrain $\Up$.  Finally, in Chapter \ref{ch5}, we present the numerical scheme mentioned above to solve the time dependent Einstein-Klein-Gordon equations in spherical symmetry and also a few test simulations.

\section{Dwarf Spheroidal Galaxies}

\begin{figure}

  \begin{center}
    \includegraphics[width = 4 in]{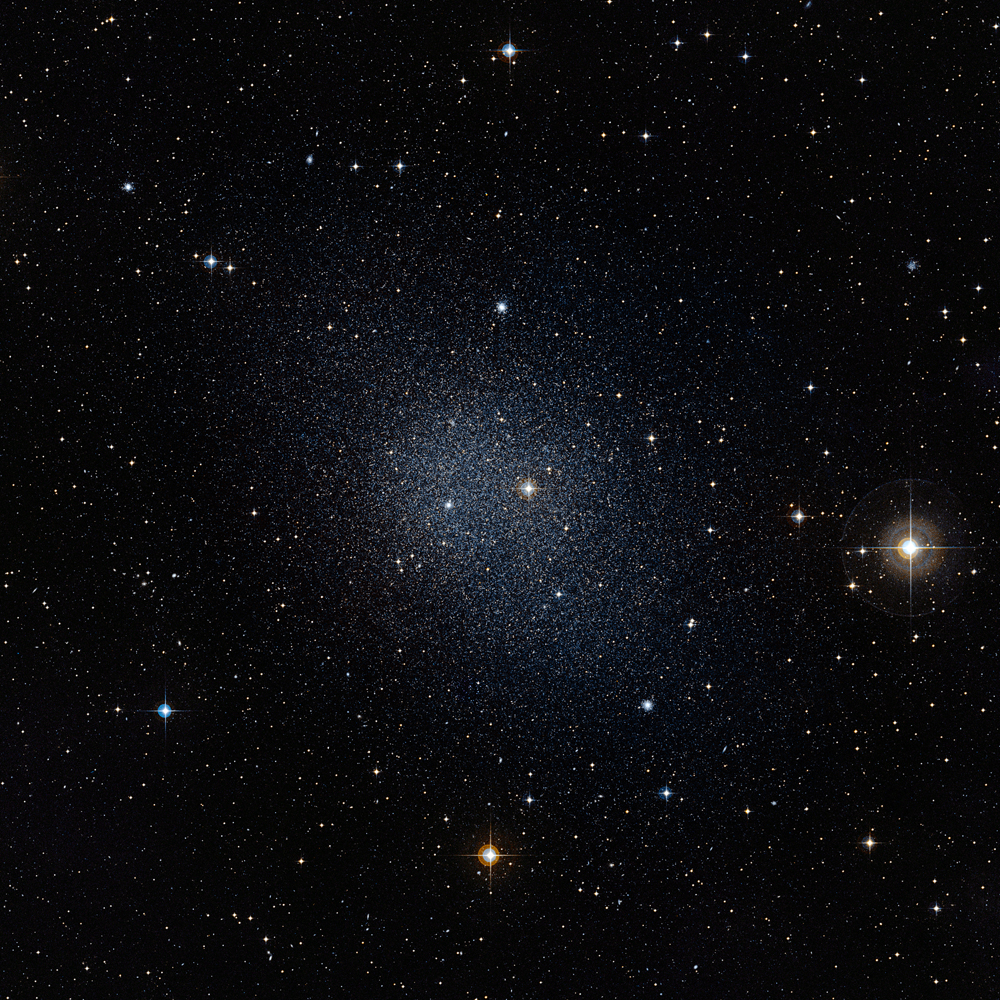}
  \end{center}

  \caption{Fornax Dwarf Spheroidal Galaxy. Photo Credit: ESO/Digital Sky Survey 2}

  \label{fornax}

\end{figure}

In this final section, we discuss briefly the subject of dwarf spheroidal galaxies and why they are of interest and use to the wave dark matter model. Dwarf spheroidal galaxies, like the Fornax galaxy in Figure \ref{fornax}, are the smallest cosmological objects known to contain a significant amount of dark matter and, in fact, this has been used as a factor to distinguish dwarf spheroidal galaxies from globular clusters, in which there is evidence against the presence of dark matter (\cite{Bergh08,Mateo98,Conroy11}).

The evidence for dark matter in these dwarf spheroidal galaxies comes at least in part from their observed velocity dispersion profiles.  The velocity dispersion at a particular radii in a dwarf spheroidal galaxy is effectively the standard deviation of rotational stellar velocities near that radii.  If the mass in a dwarf spheroidal galaxy followed the light as in the King model (\cite{King62}), the velocity dispersion profile should rapidly decrease at large radii.  However, observed velocity dispersion profiles are generally flat out to large radii indicating the presence of dark matter (\cite{Walker07}).  In fact, dwarf spheroidal galaxies appear to be the most dark matter dominated galaxies known (\cite{Mash06, Walker07, Mateo98, Kleyna02, Strigari08}).

Moreover, dwarf spheroidal galaxies are approximately spherically symmetric (\cite{Mash06, Walker07, Mateo98}) and, even though they are almost always satellite galaxies and at first glance could possibly be subject to tidal forces from their host galaxies, they appear to be in dynamical equilibrium (\cite{Salucci11, Cote99}).

These observations make dwarf spheroidal galaxies some of the best places to study dark matter.  Being extremely dark matter dominated and assuming dynamical equilibrium, their internal kinematics are almost entirely controlled by their dark matter component, making their individual stars valuable tracers of the gravitational effects of dark matter.  Furthermore, the fact that they are approximately spherically symmetric allows for considerable simplification in the mathematics required to reliably model them.  For this reason, we have directed this work towards the goal of comparing the wave dark matter model to observations of dwarf spheroidal galaxies. 
\chapter{A Survey of Spherically Symmetric Spacetimes}\label{ch2}

Spherically symmetric spacetimes are an important case in the study of general relativity for a number of reasons.  Foremost among them is that it is often a good starting point in the study of a problem in general relativity.  For example, one of the first projects undergone in general relativity was to compute nontrivial spherically symmetric spacetimes that are exact solutions of the Einstein equation.  This resulted in the discovery of the Schwarzschild spacetime, which is by far the most important spherically symmetric solution to date, and later Birkhoff's theorem about Ricci flat or vacuum spherically symmetric spacetimes (\cite{Wald84, Jebsen21, Birkhoff23}).  Spherically symmetric spacetimes also create a situation where the dynamics of the system are less complicated by effectively reducing a 4-dimensional solution to a 2-dimensional one.  This accessibility makes using spherically symmetric spacetimes all the more attractive as a starting point.  Finally, while Birkhoff's theorem classifies all vacuum spherically symmetric spacetimes, there are still some nonvacuum spherically symmetric spacetimes that are interesting as well, both from a physical and mathematical standpoint, such as dwarf spheroidal galaxies, which are dominated by their dark matter halos and also closely approximated by spherical symmetry (\cite{Mash06}), and spherically symmetric scalar fields.

Since spherically symmetric spacetimes are still of interest, it would be useful to find an efficient way to describe and model these spacetimes in a general sense.  In particular, it would be useful to have a form of a general spherically symmetric metric that is well-suited to numerical evolutions of the Einstein equation.  There are many possible forms of the metric of a general spherically symmetric spacetime to use and the purpose of this chapter is first to collect these metrics, discuss the advantages and disadvantages of using each one, and then present a metric that is not only extremely well-suited to numerical evolutions, but also is described in terms that are very natural analogues to the low field or Newtonian limit.  Most of these metrics can be described well within the established framework of numerical relativity and as such, we will use that framework in the following.

\section{Metrics for a Spherically Symmetric spacetime}

The study of numerical relativity is devoted to devising ways of evolving the Einstein equation in order to solve for the components of the spacetime metric in different situations of interest as well as actually conducting such numerical experiments and comparing them to real data.  This evolution usually takes place in a spacetime which is foliated by $t=constant$ spacelike hypersurfaces.  The common method in numerical relativity is to decouple the time component from the space components into what is commonly called the (3+1)-formalism of general relativity (\cite{Chop98,Gour12,Alcu08}).

\begin{figure}

\begin{center}
  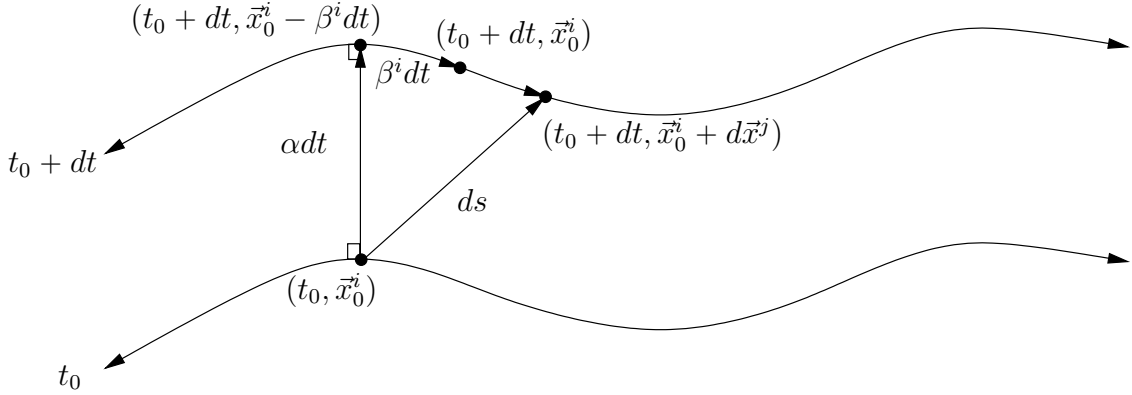
\end{center}

\caption{Infinitesimal distance in a $t=constant$ hypersurface foliated spacetime.}

\label{3plus1_fig}

\end{figure}

The framework for this formalism is, as stated before, a spacetime, $N$, foliated by $t=constant$ spacelike hypersurfaces described by a Riemannian 3-metric $\ga$ which may change with time.  Note that such a foliation is possible for any globally hyperbolic spacetime (\cite{Gour12,Wald84}).  Consider a coordinate chart on $U \subseteq N$, $\{t,x^{1},x^{2},x^{3}\}$, where $\p_{t}$ is timelike and the $\p_{x^{j}}$ are all spacelike.  Now consider an observer starting on the $t=t_{0}$ hypersurface at the coordinate $(t_{0},\vec{x}^{j}_{0})$.  This observer then travels to another infinitesimally close hypersurface $t = t_{0}+dt$ to the coordinate $(t_{0} + dt, \vec{x}^{j}_{0} + d\vec{x}^{j})$ as in Figure \ref{3plus1_fig}.

The observer has now traveled an infinitesimal distance $ds$.  We can measure the ``square'' of this infinitesimal distance, $ds^{2}$, using the analogue of Pythagorean's theorem and this will give us the line element form of the metric.  If another observer travels normal to the hypersurface from $(t_{0},\vec{x}^{j}_{0})$ to the hypersurface $t=t_{0}+dt$, since the normal direction is not necessarily the same direction as the $t$ coordinate direction, it will arrive at the coordinate $(t_{0} + dt, \vec{x}^{j}_{0} - \beta^{j}dt)$.  We call the 3-vector field, $\beta$, the shift vector because it measures the spacelike shift of the coordinates while traveling normally.  Note that the components of $\beta$ can vary with all the coordinates.  This normal observer, having traveled in a timelike direction, has experienced some proper time $d\tau$, which is some multiple of the change in time coordinate, that is,
\begin{equation}
  d\tau = \al dt.
\end{equation}
Hence the length of its normal movement from one surface to the other is $\al dt$.  The value of $\al$ can vary with all of the coordinates making it a function on the manifold.  This function is called the lapse function since it measures the lapse in proper time compared to coordinate time.  To get the length in the spatial direction between where the normal observer ended up and where the original observer did, we need only to use the metric on the hypersurfaces and the difference of the two space coordinates.  This difference, for each $i$, takes the form
\begin{equation}
  (x^{j}_{0} + dx^{j}) - (x^{j}_{0} - \beta^{j}dt) = dx^{j} + \beta^{j}dt
\end{equation}
and so the length squared of the spatial movement will be
\begin{equation}
  \ga_{jk}(dx^{j} + \beta^{j}dt)(dx^{k} + \beta^{k}dt)
\end{equation}
where we have implemented the Einstein summation convention.  Then using the generalized version of Pythagorean's theorem and recalling that $t$ is a timelike direction we get that
\begin{equation}\label{3plus1_metric}
  ds^{2} = -\al^{2}\, dt^{2} + \ga_{jk}(dx^{j} + \beta^{j} dt)(dx^{k} + \beta^{k}dt),
\end{equation}
which is the line element of the metric.  This is the most general form of the metric for any foliated spacetime, that is, all metrics of a foliated spacetime can be written in this form.  Note that we will often write $g$ instead of $ds^{2}$ to refer interchangeably to the metric and the line element.

From this metric, and a choice of slicing condition, a complete system of partial differential equations that evolve the Einstein equation can be constructed.  This system is often referred to as the ADM formulation of general relativity in reference to the authors of the paper in which it was first introduced (\cite{ADM08}).  It involves evolution equations of both the metric and the extrinsic curvature of the $t=constant$ hypersurfaces (\cite{ADM08,Chop98,Bona02}).  These equations are very commonly used in numerical relativity, but we find they overcomplicate the situation in spherical symmetry, which is why we have elected not to use this formulation of general relativity directly.

If we know more about the spacetime in question, we will be able to determine more of the components of the metric.  In a spherically symmetric spacetime, with coordinates, $r,\theta,\varphi$, chosen so that $\theta$ and $\varphi$ are the polar-angular coordinates on the hypersurface, the shift vector must be completely radial so that the metric remains invariant under rotations.  In this case, we will denote the radial component of the shift vector by simply $\beta$.  Moreover, the $3$-metric can be written as
\begin{equation}
  \ga = \ga_{rr}\, dr^2 + \ga_{\theta\theta}\, d\si^{2}.
\end{equation}
where $d\si^{2} = d\theta^{2} + \sin^{2}\theta\, d\varphi^{2}$ is the standard metric on the unit sphere.  This implies that the most general spherically symmetric metric can be written in the form
\begin{align}\label{3plus1_ssmetric_a}
  \notag g &= -\al^{2}\, dt^{2} + \ga_{rr}(dr + \beta\, dt)(dr + \beta\, dt) + \ga_{\theta\theta}\, d\si^{2} \\
  &= -\pnth{\al^{2} - \ga_{rr}\beta^{2}}\, dt^{2} + \ga_{rr}\beta(dr\, dt + dt\, dr) + \ga_{rr}\, dr^{2} + \ga_{\theta\theta}\, d\si^{2}.
\end{align}
Note that all the metric component functions can only depend on $t$ and $r$ due to spherical symmetry.  For convenience, we will define two positive functions $a(t,r)$ and $q(t,r)$ so that
\begin{equation}
  a(t,r)^{2} = \ga_{rr} \qquad \text{and} \qquad q(t,r)^{2} = \ga_{\theta\theta}.
\end{equation}
Then we can rewrite (\ref{3plus1_ssmetric_a}) as
\begin{equation}\label{3plus1_ssmetric_b}
  g = -\pnth{\al^{2} - a^{2}\beta^2}\, dt^{2} + a^2\beta(dr\, dt + dt\, dr) + a^2\, dr^{2} + q^2\, d\si^{2},
\end{equation}
where $\al,a,\beta$, and $q$ are functions of only $t$ and $r$.  It is important to note here that since we have not defined the coordinates $t$ and $r$ geometrically yet, there remain two degrees of freedom left in this metric. There are several different choices that can be made in this regard.  We mention the most common here, but there is a very useful and more extensive list of several choices that can be made in a more general setting in Gourgoulhon's recent book (\cite{Gour12}).

There are three rather common slicing conditions that are often used in many settings, all of which place a condition on the lapse function $\al$.  These conditions are the maximal slicing, harmonic slicing, and geodesic slicing conditions.

Under the maximal slicing condition, one requires that each $t=constant$ hypersurface be a maximal hypersurface, that is, it has zero mean curvature,
\begin{equation}
  H=0.
\end{equation}
The metric stays of the form in equation (\ref{3plus1_ssmetric_a}), but since the evolution equations in the ADM formulation evolve the components of the second fundamental form and $H$ is the trace of the second fundamental form, this places a constraint on some of the evolution variables, which can be used to simplify the evolution equations (\cite{Cord11,Gour12,Baum10}).  Note here that after making this choice, there remains one more degree of freedom which can be used to constrain the $r$ coordinate.

Harmonic slicing requires that the coordinate function $t$ be a harmonic function under the metric $g$.  That is, $\glap{g}t = 0$, where $\Box_{g}$ is the d'Alembertian or Laplacian operator with respect to the metric $g$.  This is often accompanied with the condition that the hypersurfaces remain orthogonal to the time direction, which uses the remaining degree of freedom, and indeed some refer to both of these choices together as harmonic slicing.  In the case that both conditions are satisfied, this would yield a metric of the form
\begin{equation}
  g = -\al^{2}\, dt^{2} + a^{2}\, dr^{2} + q^{2}\, d\si^{2}
\end{equation}
with the added condition $\glap{g}t = 0$, which can be used to compute an evolution equation for the lapse function $\al$ (\cite{Bona92}).

Geodesic slicing requires that movement along the curve $\xi = (t,0,0,0)$, which is given in our coordinates, be geodesic, that is, coordinate observer worldlines are geodesics.  This requirement is satisfied by choosing
\begin{equation}
  \al = constant \qquad \text{and} \qquad \be=0.
\end{equation}
However, the most reasonable choice for the constant is 1, since the condition $\al=1$ on the lapse function has the added implication that normal observers proper time is the same as coordinate time and in fact, since $\be =0$, normal observers are coordinate observers (\cite{Gour12,Baum10,Alcu00}).  This results in a metric of the form
\begin{equation}
  g = -dt^{2} + a^2\, dr^{2} + q^2\, d\si^{2}.
\end{equation}
While this choice seems very attractive at first, since it either eliminates or greatly simplifies the evolution equations, it does have a tendency to develop coordinate singularities when evolved in time and hence must be used with caution (\cite{Baum10}).  Note also that this requirement is sometimes reduced to simply $\al = 1$ without necessarily requiring the shift parameter or vector to vanish.

We can alternatively use the degrees of freedom to make choices concerning the $r$-coordinate.  We will mention three such choices here that are standard in the study of spherically symmetric spacetimes.

The first is the choice of normal slicing.  That is, choose the $t$ coordinate so that the $\p_{t}$ vector field is always normal to the hypersurfaces.  This choice makes all normal observers coordinate observers as well.  This is equivalent to choosing the shift parameter or vector to be identically 0 and results in a metric of the form
\begin{equation}
  g = -\al^{2}\, dt^{2} + a^{2}\, dr^{2} + q^{2}\, d\si^{2}.
\end{equation}
This choice has already been mentioned above as it is often coupled with harmonic or geodesic slicing conditions (\cite{Gour12}).  Since the coordinate vector fields on the hypersurfaces were already orthogonal, this results, as seen above, in a diagonal metric.

The next choice is to require that the metric on the hypersurfaces to be conformal to the flat metric.  This amounts to choosing the function $q$ in (\ref{3plus1_ssmetric_b}) to satisfy $q = ra$, which would make the metric become
\begin{equation}
  g = -\pnth{\al^{2} - a^{2}\beta^2}\, dt^{2} + a^2\beta(dr\, dt + dt\, dr) + a^2 \pnth{dr^{2} + r^2\, d\si^{2}}.
\end{equation}
This choice is referred to as isotropic coordinates.  It is often coupled with the normal slicing choice above, which uses both of the degrees of freedom and results in a metric of the form
\begin{equation}
  g = -\al^{2}\, dt^{2} + a^2 \pnth{dr^{2} + r^2\, d\si^{2}}.
\end{equation}
A common example of the normal-isotropic case is the Schwarzschild metric in isotropic coordinates (\cite{Wald84}).

The last choice we mention in the (3+1) framework is to give the coordinate $r$ geometric significance by choosing it to be the areal coordinate.  That is, choose the coordinate $r$ so that the area of each metric $2$-sphere on the hypersurface is exactly $4\pi r^{2}$.  This choice requires that $q = r$.  Additionally, it is almost always accompanied with the normal slicing choice above, again using both degrees of freedom.  This results in the polar-areal coordinates on a spherically symmetric spacetime and yields a metric of the form
\begin{equation}
  g = -\al^{2}\, dt^{2} + a^{2}\, dr^{2} + r^{2}\, d\si^{2}.
\end{equation}
This chart is probably the most familiar form of a general spherically symmetric metric, and is the chart most commonly used when introducing the Schwarzschild spacetime.  And rightly so, as it has some very clear advantages.  For one, the $r$-coordinate's role is analogous to its role in a flat spacetime.  Additionally, these coordinates give the Einstein curvature tensor a very simple form.  However, they may not be well suited to dealing with high gravitational fields as we would likely run into the same limitations that the polar areal metric has in describing the Schwarzschild spacetime inside the Schwarzschild radius.  This is not much of a problem for us, since we are mostly interested in describing objects on the galactic scale which are in the low field limit.

We present one more useful coordinate system and metric for a general spherically symmetric spacetime that does not fit into the (3+1)-formalism framework, nor does it depend on the ability to foliate the spacetime, but is useful from a theoretical standpoint nonetheless.  In these coordinates, we still choose to use the polar-angular coordinates $\theta$ and $\varphi$ to describe the rotations, but instead of separating the time and radial coordinates, we choose coordinates $u$ and $v$ such that $\p_{u}$ and $\p_{v}$ are nonparallel future pointing null vectors.  This coordinate system is descriptively called null coordinates.  In this case, the metric on the spacetime takes the form
\begin{equation}
  g = A\, (du \, dv + dv\, du) + Q^{2}\, d\si^{2}
\end{equation}
where $A$ and $Q$ are functions of only $u$ and $v$.  Note that there are no extra degrees of freedom with this metric as all coordinates have been well-defined.  While these coordinates are not very well suited to numerical evolutions, they can be very useful for theoretical discussions about the spherically symmetric spacetime. For example, in these coordinates, it is very straightforward to prove the monotonicity of the Hawking mass given the dominant energy condition.  Additionally, it seems to perform well when in the presence of high gravitational fields.  In the Schwarzschild case, these coordinates are known as the Kruskal coordinate system and is the system generally used to describe the region inside the Schwarzschild radius (\cite{Wald84}).

We make one final note here about static spacetimes.  If the spherically symmetric spacetime is also known to be static, meaning that there exists a timelike killing vector field with orthogonal spacelike hypersurfaces, then we can automatically eliminate the cross term in the general metric (\ref{3plus1_ssmetric_b}) by selecting the time coordinate to be in the direction of the timelike killing vector field and choosing the remaining coordinates to be the general spherical coordinates on the orthogonal spacelike hypersurfaces.  This effectively sets $\be = 0$.  A survey article on static spherically symmetric spacetimes can be found in \cite{Deser05}.

All of these different coordinate choices and different forms of the metric on a spherically symmetric spacetime have advantages to them.  However, for the problem of numerically evolving a spacetime metric in a low gravitational field, we find that the polar-areal coordinates are extremely well suited due to the very simple system one gets from the Einstein equation.  As such, polar-areal coordinates is the coordinate system that we will use throughout this dissertation, but, in addition, we will introduce new variables that will give the metric a different form.  This new form of the metric will result in the added advantages that the new metric functions have very clear analogues in the Newtonian or low-field limit and the Einstein curvature tensor will become even more simplified.

\section{A Newtonian-Compatible Metric}

Let $N$ be a spherically symmetric, (3+1)-dimensional spacetime, which can be foliated by $t=constant$ spacelike hypersurfaces, with a Lorentzian metric $g$.  Choose the polar-areal coordinate system discussed above so that the $t$-coordinate direction is always normal to the hypersurfaces, the radial coordinate $r$ is the areal coordinate (that is, so that the metric sphere of metric radius $r$ always has a surface area of $4\pi r^2$), and $\theta$ and $\varphi$ are the usual polar-angular coordinates on the hypersurface.  In these coordinates, the metric $g$ has the line element form
\begin{equation} \label{metric-1}
  g = -\al(t,r)^{2}\, dt^{2} + a(t,r)^{2}\, dr^{2} + r^{2}\, d\si^{2}
\end{equation}
Now we define the functions $V(t,r)$ and $M(t,r)$ as follows,
\begin{align}
  V &= \ln \al & M &= \frac{r}{2}\pnth{\frac{a^{2}-1}{a^{2}}}
\end{align}
and rewrite the metric (\ref{metric-1}) in terms of these functions to obtain,
\begin{equation} \label{metric}
  g = - \e^{2V}\, dt^{2} + \pnth{1 - \frac{2M}{r}}^{-1}\, dr^{2} + r^{2}\, d\si^{2}
\end{equation}
Note that we will always be interested in the low-field limit, where $M \ll r$, which necessarily requires that $M(t,0)=0$ for all $t$.

\subsection{Compatibility with Newtonian Physics}

Here, we compute several important properties about this metric and the physical interpretations of both $V$ and $M$.  To physically interpret this metric, we will introduce the Einstein equation, but first, we present a property that doesn't require the Einstein equation, the proof of which can be found in Section \ref{Proofs}.

\begin{prop} \label{mH-Prop}
  The function $M(t,r)$ is the spacetime Hawking mass of the metric sphere, $\Si_{t,r}$, for any given $t$ and $r$.
\end{prop}

This suggests that we should interpret $M$ as the mass of the system.  However, there is an even stronger reason to do so, which we will investigate later.  In order to prepare for that discussion, we need some preliminary results first about the relationship between the metric and its stress-energy tensor.  To begin, consider this metric as a solution to the Einstein equation
\begin{equation} \label{EinEq-1}
  G = 8\pi T
\end{equation}
for some stress-energy tensor $T$.  To facilitate our discussion, we will compute the Einstein curvature tensor of this metric in certain directions.  To that end, define the following unit vector fields.
\begin{align} \label{norvec}
  \nu_{t} &=  \e^{-V}\, \p_{t} & \nu_{r} &= \sqrt{1 - \frac{2M}{r}} \, \p_{r} & \nu_{\theta} &= \frac{1}{r}\, \p_{\theta} & \nu_{\varphi} &= \frac{1}{r\sin \theta} \, \p_{\varphi}
\end{align}
Note that at every point, $p \in N$, except the coordinate singularities $r=0$ and $\theta=\pm \pi$ (that is, all points where all the vector fields above are well defined), these vector fields form an orthonormal basis of $T_{p}N$ and hence are a frame field.  The Einstein curvature tensor is defined as
\begin{equation}\label{ECT}
  G = \Ric - \frac{1}{2}Rg
\end{equation}
where $\Ric$ and $R$ are the Ricci curvature tensor and the scalar curvature of the spacetime respectively.  Since both the Ricci curvature tensor and its trace $R$ are present in this equation, we will need to know a few results about the Ricci curvature in these coordinates.  We have the following lemma and subsequent corollary.  While the proof of the corollary is short and is presented here, the proof of the lemma can be found in Section \ref{Proofs}.

\begin{lem} \label{RicLem-2}
  The only nonzero components of the Ricci curvature tensor in the $\nu_{\eta}$ basis and the scalar curvature are as follows.
    \begin{enumerate}
      \item\label{RicLem-2a} $\displaystyle \Ric(\nu_{t},\nu_{t}) = \pnth{V_{rr} + V_{r}^{2} + \frac{2V_{r}}{r}}\pnth{1-\frac{2M}{r}} + \frac{V_{r}}{r}\pnth{\frac{M}{r} - M_{r}} \\ \phantom{.} \hspace{7 eM} + \frac{V_{t}M_{t} - M_{tt}}{r \e^{2V}}\pnth{1-\frac{2M}{r}}^{-1} - \frac{3M_{t}^{2}}{r^{2} \e^{2V}}\pnth{1-\frac{2M}{r}}^{-2}$
      \item\label{RicLem-2b} $\displaystyle \Ric(\nu_{t},\nu_{r}) = \frac{2M_{t}}{r^{2} \e^{V}}\pnth{1 - \frac{2M}{r}}^{-1/2}$
      \item\label{RicLem-2c} $\displaystyle \Ric(\nu_{r},\nu_{r}) = -(V_{rr} + V_{r}^2)\pnth{1 - \frac{2M}{r}} - \pnth{\frac{2}{r^{2}} + \frac{V_{r}}{r}}\pnth{\frac{M}{r} - M_{r}} \\ \phantom{.} \hspace{7 eM} - \frac{V_{t}M_{t} - M_{tt}}{r \e^{2V}}\pnth{1-\frac{2M}{r}}^{-1} + \frac{3M_{t}^{2}}{r^2 \e^{2V}}\pnth{1-\frac{2M}{r}}^{-2}$
      \item\label{RicLem-2d} $\displaystyle \Ric(\nu_{\theta},\nu_{\theta}) = \Ric(\nu_{\varphi},\nu_{\varphi}) = \frac{1}{r^{2}}\pnth{\frac{M}{r} + M_{r}} - \frac{V_{r}}{r}\pnth{1 - \frac{2M}{r}}$
      \item\label{RicLem-2e} $\displaystyle R = -2\pnth{V_{rr} + V_{r}^{2} + \frac{2V_{r}}{r}}\pnth{1-\frac{2M}{r}} - \frac{2V_{r}}{r}\pnth{\frac{M}{r} - M_{r}} + \frac{4M_{r}}{r^2} \\ \phantom{.} \hspace{7 eM} - \frac{2(V_{t}M_{t} - M_{tt})}{r \e^{2V}}\pnth{1-\frac{2M}{r}}^{-1} + \frac{6M_{t}^{2}}{r^{2} \e^{2V}}\pnth{1-\frac{2M}{r}}^{-2}$
    \end{enumerate}
\end{lem}

\begin{cor} \label{GCor-1}
  The only nonzero components of the Einstein curvature tensor in the $\nu_{\eta}$ basis are as follows.
    \begin{enumerate}
      \item\label{GCor-1a} $\displaystyle G(\nu_{t},\nu_{t}) = \frac{2 M_{r}}{r^{2}}$
      \item\label{GCor-1b} $\displaystyle G(\nu_{t},\nu_{r}) = \frac{2M_{t}}{r^{2} \e^{V}}\pnth{1 - \frac{2M}{r}}^{-1/2}$
      \item\label{GCor-1c} $\displaystyle G(\nu_{r},\nu_{r}) = -\frac{2M}{r^{3}} + \frac{2V_{r}}{r}\pnth{1 - \frac{2M}{r}}$
      \item\label{GCor-1d} $\displaystyle G(\nu_{\theta},\nu_{\theta}) = G(\nu_{\varphi},\nu_{\varphi}) = \pnth{V_{rr} + V_{r}^{2} + \frac{V_{r}}{r}}\pnth{1-\frac{2M}{r}} \\ \phantom{.} \hspace{12 eM} + \pnth{\frac{M}{r} - M_{r}}\pnth{\frac{1}{r^{2}} + \frac{V_{r}}{r}} \\ \phantom{.} \hspace{12 eM} + \frac{V_{t}M_{t} - M_{tt}}{r \e^{2V}}\pnth{1 - \frac{2M}{r}}^{-1} - \frac{3M_{t}^{2}}{r^{2} \e^{2V}}\pnth{1-\frac{2M}{r}}^{-2}$
    \end{enumerate}
\end{cor}

\begin{proof} Since $\{\nu_{t},\nu_{r},\nu_{\theta},\nu_{\varphi}\}$ form an orthonormal basis everywhere.  We have the following.
  \begin{align}
    G(\nu_{t},\nu_{t}) &= \Ric(\nu_{t},\nu_{t}) - \frac{R}{2}g(\nu_{t},\nu_{t}) = \Ric(\nu_{t},\nu_{t}) + \frac{R}{2} \\
    G(\nu_{t},\nu_{r}) &= \Ric(\nu_{t},\nu_{r}) - \frac{R}{2}g(\nu_{t},\nu_{r}) = \Ric(\nu_{t},\nu_{r}) \displaybreak[0]\\
    G(\nu_{r},\nu_{r}) &= \Ric(\nu_{r},\nu_{r}) - \frac{R}{2}g(\nu_{r},\nu_{r}) = \Ric(\nu_{r},\nu_{r}) - \frac{R}{2} \displaybreak[0]\\
    G(\nu_{\theta},\nu_{\theta}) &= \Ric(\nu_{\theta},\nu_{\theta}) - \frac{R}{2}g(\nu_{\theta},\nu_{\theta}) = \Ric(\nu_{\theta},\nu_{\theta}) - \frac{R}{2} \displaybreak[0]\\
    \notag G(\nu_{\varphi},\nu_{\varphi}) &= \Ric(\nu_{\varphi},\nu_{\varphi}) - \frac{R}{2}g(\nu_{\varphi},\nu_{\varphi}) = \Ric(\nu_{\varphi},\nu_{\varphi}) - \frac{R}{2} \\
    &= \Ric(\nu_{\theta},\nu_{\theta}) - \frac{R}{2} = G(\nu_{\theta},\nu_{\theta})
  \end{align}
We then use Lemma \ref{RicLem-2} to substitute in the values for $\Ric$ and $R$.  The rest is algebra.  These are the only nonzero components by Lemma \ref{RicLem-2}, equation (\ref{ECT}), and the fact that the $\nu_{\eta}$ basis is orthonormal.
\end{proof}

Next, we define the function $\mu(t,r)$ to be the energy density of an observer at $p=(t,r)$ moving through the slices with 4-velocity $\nu_{t}$.  That is,
\begin{equation} \label{ED-def}
  \mu = T(\nu_{t}, \nu_{t}).
\end{equation}
We now have enough information to prove the following proposition, which contains the promised stronger reason for interpreting $M$ as the mass inside each metric sphere.

\begin{prop} \label{ED-Prop-1}
  For a fixed $t$ and $r$, $M(t,r)$ is the flat integral of the energy density, $\mu(t,r)$, over the ball $E_{t,r}$ of radius $r$ at time $t$.
\end{prop}

\begin{proof}
  By the Einstein equation (\ref{EinEq-1}), equation (\ref{ED-def}), and Corollary \ref{GCor-1}, we have that
  \begin{align} \label{EDEq-1}
    \notag G(\nu_{t},\nu_{t}) &= 8\pi T(\nu_{t},\nu_{t}) \\
    \notag \frac{2 M_{r}}{r^{2}} &= 8\pi \mu \\
    M_{r} &= 4\pi r^{2}\mu
  \end{align}
  Since $\int_{\Si_{t,r}} dA = 4\pi r^{2}$, where $\Si_{t,r}$ is the sphere of radius $r$ at time $t$, we have then that for a fixed $t$ and $r$,
  \begin{align}
    M(t,r) &= \int_{0}^{r} 4\pi s^{2} \mu(t,s)\, ds =\int_{0}^{r} \int_{\Si_{t,s}} \mu(t,s) \, dA \, ds \\
    &= \int_{0}^{r}\int_{0}^{2\pi} \int_{0}^{\pi} \mu(t,s) s^{2}\sin\theta \, d\theta\, d\varphi\, ds \\
    &= \int_{E_{t,r}} \mu(t,s) \, dV_{0},
  \end{align}
  where $dV_{0} = s^{2}\sin\theta \, d\theta\, d\varphi\, ds$ is the flat volume form on the ball of radius $s$, and we have introduced $s$ as a dummy integrating variable in the ``$r$'' position.  Thus $M(t,r)$ is the flat volume integral of the energy density over the ball of radius $r$.
\end{proof}

Note that $M$ is not the integral of the energy density with respect to the metric's volume form, $dV = (1 - 2M/s)^{-1/2}\, s^{2}\sin\theta \, ds \, d\theta \, d\varphi$, but rather the following is true.
  \begin{equation}
    M(t,r) = \int_{E_{t,r}}\mu(t,s)\, \sqrt{1 - \frac{2M(t,s)}{s}} \, dV.
  \end{equation}
However, in the Newtonian limit, $M\ll r$, the above integral becomes approximately the integral of the energy density with respect to the metric's volume form over the ball $E_{t,r}$ of radius $r$.  Thus referring to $M(t,r)$ as the mass inside the metric sphere of radius $r$ at time $t$ not only makes sense from a geometrical point of view given the Hawking mass, but also from a physical point of view.

Furthermore, since the energy density is spherically symmetric and smooth at the origin, we must have $\mu_{r}(t,0) = 0$ for all $t$.  Thus for small $r$, $\mu$ is approximately constant and nonnegative.  Then the above integral yields for small $r$ that
\begin{equation}
    M(t,r) = \int_{0}^{r} 4\pi s^{2}\mu(t,s)\, dr \approx \int_{0}^{r} 4\pi s^{2}\mu(t,0)\, dr = \frac{4\pi \mu(t,0)}{3}r^{3}.
\end{equation}
Thus the initial behavior of $M$ near $r=0$ is that of a cubic power function (or in the case when $\mu(t,0)=0$, the zero function).  In particular, this implies that for all $t$
\begin{equation}
    M(t,0) = 0, \qquad M_{r}(t,0) = 0, \qquad \text{and} \qquad M_{rr}(t,0) = 0.
\end{equation}
This fact will be useful in a proof in Section \ref{Proofs}.

Next we define the function $P$ to be the pressure in the stress-energy tensor for an observer at $(t,r)$, that is, let
\begin{equation} \label{PR-def}
  P = T(\nu_{r},\nu_{r}).
\end{equation}
Then we can use Corollary \ref{GCor-1} and the Einstein equation (\ref{EinEq-1}) to prove the following proposition.

\begin{prop} \label{PR-Prop-1}
  In the Newtonian limit, where $P=0$ and $M\ll r$, we have that $\Del V = 4\pi \mu$, where $\Del$ is the flat Laplacian on $\RR^{3}$.
\end{prop}

\begin{proof}
  By the Einstein equation (\ref{EinEq-1}), equation (\ref{PR-def}), and Corollary \ref{GCor-1}, we have that
  \begin{align}\label{PREq-1}
    \notag G(\nu_{r},\nu_{r}) &= 8\pi T(\nu_{r},\nu_{r}) \\
    \notag -\frac{2M}{r^{3}} + \frac{2V_{r}}{r}\pnth{1 - \frac{2M}{r}} &= 8\pi P \displaybreak[0]\\
    \notag \frac{2V_{r}}{r} &= \pnth{1 - \frac{2M}{r}}^{-1}\pnth{\frac{2M}{r^{3}} + 8\pi P} \\
    r^{2}V_{r} &= \pnth{1 - \frac{2M}{r}}^{-1}\pnth{M + 4\pi r^{3}P}.
  \end{align}
  Note that in the Newtonian limit, where $P=0$ and $M\ll r$, this equation is approximated by
  \begin{equation} \label{PotEq-1}
    r^{2}V_{r} = M.
  \end{equation}
  Also, the metric on the hypersurface, under these assumptions, is approximately the polar-areal metric on $\RR^{3}$.  Note that the Laplacian on $\RR^{3}$ of a spherically symmetric function $f$ is given by
  \begin{equation}
    \Del f= \ppa{f}{r} + \frac{2}{r}\pa{f}{r} = \frac{1}{r^{2}}\pa{}{r}\pnth{r^{2}\pa{f}{r}}
  \end{equation}
  Then applying the operator $\dfrac{1}{r^{2}}\pa{}{r}$ to (\ref{PotEq-1}) and using equation (\ref{EDEq-1}) yields
  \begin{align}\label{PotEq-2}
    \notag \frac{1}{r^{2}}\pa{}{r}(r^{2}V_{r}) &= \frac{1}{r^{2}}\pa{}{r}(M) \\
    \notag \Del V &= \frac{1}{r^{2}}M_{r} \\
    \Del V &= 4\pi \mu
  \end{align}
\end{proof}

Equation (\ref{PotEq-2}) is Poisson's equation and is the defining equation of the gravitational potential in Newtonian mechanics.  Moreover, equation (\ref{PotEq-1}) reduces to $V_{r} = M/r^{2}$ which yields the inverse square law for Newtonian gravity.  Then we can interpret $V$ as the analogue in our scenario of the Newtonian potential.  The interpretation of $M$ and $V$ via Propositions \ref{mH-Prop}, \ref{ED-Prop-1}, and \ref{PR-Prop-1} are what we mean by saying that the metric (\ref{metric}) is Newtonian compatible.

\subsection{Other Useful Properties of this Metric}

In this section, we produce two additional useful results which are readily obtainable with these coordinates and metric functions.  The first is the well-known result of the monotonicity of the Hawking mass in spherical symmetry, which is made particularly straightforward using this form of the metric.  It follows almost immediately from Corollary \ref{GCor-1}.

\begin{prop}\label{mH-Prop2}
  If the spacetime satisfies the dominant energy condition that $G(X,Y) \geq 0$ for all future-pointing causal vector fields, $X$ and $Y$, then,  whenever $2M(t,r) \leq r$, the Hawking Mass, $M(t,r)$, is monotonically increasing in any non-timelike direction for which the radial coordinate increases.
\end{prop}

\begin{proof}
  The vector field $\nu_{t}$ is the future-pointing timelike unit vector field in the $t$ direction and the vector fields $\nu_{r} + \nu_{t}$ (future-pointing) and $\nu_{r} - \nu_{t}$ (past-pointing) are null vector fields in the null directions where the $r$ coordinate increases.  By Corollary \ref{GCor-1}, we have that
  \begin{align}
    \notag (\nu_{r} + \nu_{t})(M) &= M_{r}\pnth{1 - \frac{2M}{r}}^{1/2} + \frac{M_{t}}{ \e^{V}} \\
    \notag &= \frac{r^{2}}{2}\pnth{1 - \frac{2M}{r}}^{1/2}\pnth{\frac{2M_{r}}{r^{2}} + \frac{2M_{t}}{r^{2} \e^{V}}\pnth{1 - \frac{2M}{r}}^{-1/2}} \displaybreak[0]\\
    \notag &= \frac{r^{2}}{2}\pnth{1 - \frac{2M}{r}}^{1/2}\pnth{G(\nu_{t},\nu_{t}) + G(\nu_{t},\nu_{r})} \\
    &= \frac{r^{2}}{2}\pnth{1 - \frac{2M}{r}}^{1/2} G(\nu_{t},\nu_{t}+\nu_{r})
  \end{align}
  The last factor here is positive by the dominant energy condition since both $\nu_{t}$ and $\nu_{t} + \nu_{r}$ are future-pointing causal vector fields.  Since $r^{2}\geq 0$ and $2M \leq r$ everywhere, it must be that $(\nu_{r} + \nu_{t})(M) \geq 0$ everywhere as well.  By the same corollary, we also have
  \begin{align}
    \notag (\nu_{r} - \nu_{t})(M) &= M_{r}\pnth{1 - \frac{2M}{r}}^{1/2} - \frac{M_{t}}{ \e^{V}} \\
    \notag &= \frac{r^{2}}{2}\pnth{1 - \frac{2M}{r}}^{1/2}\pnth{\frac{2M_{r}}{r^{2}} - \frac{2M_{t}}{r^{2} \e^{V}}\pnth{1 - \frac{2M}{r}}^{-1/2}} \displaybreak[0]\\
    \notag &= \frac{r^{2}}{2}\pnth{1 - \frac{2M}{r}}^{1/2}\pnth{G(\nu_{t},\nu_{t}) - G(\nu_{t},\nu_{r})} \\
    &= \frac{r^{2}}{2}\pnth{1 - \frac{2M}{r}}^{1/2} G(\nu_{t},\nu_{t}-\nu_{r})
  \end{align}
  The last factor here is positive by the dominant energy condition since $\nu_{t} - \nu_{r}$ is also future-pointing causal.  Then, as before, it must be that $(\nu_{r} - \nu_{t})(M) \geq 0$ everywhere.  Since both $(\nu_{r} + \nu_{t})(M)$ and $(\nu_{r} - \nu_{t})(M)$ are both nonnegative everywhere, any positive linear combination of the two is also nonnegative, which is the desired result.
\end{proof}

This next property will be useful later when we consider the Einstein-Klein-Gordon system.  By Corollary \ref{GCor-1} and the Einstein equation, any stress-energy tensor $T$ for this spacetime could have no other nonzero components than the corresponding components of the Einstein curvature tensor presented in Corollary \ref{GCor-1}.  Thus we define the following functions, two of which have already been introduced in equations (\ref{ED-def}) and (\ref{PR-def}),
\begin{subequations} \label{SETFun-def}
\begin{align}
  \mu(t,r) &= T(\nu_{t}, \nu_{t}) & \rho(t,r) &= T(\nu_{t},\nu_{r}) \\
  P(t,r) &= T(\nu_{r},\nu_{r}) & Q(t,r) &= T(\nu_{\theta},\nu_{\theta}) = T(\nu_{\varphi},\nu_{\varphi})
\end{align}
\end{subequations}
and these functions account for all the nonzero components of $T$ in terms of the orthonormal frame $\{\nu_{t},\nu_{r},\nu_{\theta},\nu_{\varphi}\}$.  Then we have the following result which follows directly from the required property of all stress-energy tensors that $\divx{g} T = 0$, where $\divx{g} T$ denotes the divergence of the tensor $T$ with respect to the metric $g$.
We will leave the lengthy proof of this proposition to Section \ref{Proofs}.


\begin{prop}\label{EinEq-prop1}
Suppose that the metric is spherically symmetric and of the form in equation (\ref{metric}) and that $T$ is a suitable spherically symmetric tensor of the form in equation (\ref{SETFun-def}).  Then solving the equation
\begin{equation}\label{divfreeT}
  \divx{g} T = 0
\end{equation}
along with solving the ODEs obtained from (\ref{EDEq-1}) and (\ref{PREq-1}), namely
  \begin{align}
    \label{EDEq-1-thm} M_{r} &= 4\pi r^{2}\mu \\
    \label{PREq-1-thm} V_{r} &= \pnth{1 - \frac{2M}{r}}^{-1}\pnth{\frac{M}{r^{2}} + 4\pi r P},
  \end{align}
  for each value of $t$ is equivalent to solving the entire Einstein equation.
\end{prop}

In this proposition, the requirement that $T$ is ``suitable'' refers to its having to come from some kind of physical notion or some set of mathematical conditions that yield a physical-like stress-energy tensor or, at least, a $T$ for which the Einstein equation is solvable.

\section{The Einstein-Klein-Gordon Equations}

A good example of the usefulness of this choice of metric is in the evolution in spherical symmetry of the Einstein-Klein-Gordon equations described in Chapter \ref{ch1} and given in equations (\ref{EKG2a}) and (\ref{EKG2b}).  As stated before, in our current research and this dissertation, we will work specifically with a complex valued scalar field.  Before we derive the spherically symmetric Einstein-Klein-Gordon equations in this metric, we should note that several other references including, but certainly not limited to, \cite{Hawley00, Seidel90, Seidel98, Bernal08, Gleiser89, MSBS, Matos09, Lee09, Sharma08, Lai07, Hawley03, Lee96} have written the Einstein-Klein-Gordon equations in spherical symmetry using either the metric presented here or another form of a general spherically symmetric metric.

Consider a spherically symmetric spacetime $N$ with Lorentzian metric $g$ as described in the previous section.  Let $f:N\to\C$ be a spherically symmetric complex valued scalar field.  We work on the galactic scale where the effects of the cosmological constant are negligible and so we will set $\La = 0$.  Then the Einstein-Klein Gordon equations in this case are those found in (\ref{EKG2a}) and (\ref{EKG2b}) with $\La = 0$.  For reference, we reprint here those equations without $\La$.
\begin{subequations}\label{EKG}
\begin{align}
  \label{EKG-1} G &= 8\pi \mu_{0}\pnth{\frac{df \otimes d\cf + d\cf \otimes df}{\Up^{2}} - \pnth{\frac{\abs{df}^{2}}{\Up^{2}} + \abs{f}^{2}}g} \\
  \label{EKG-2} \glap{g}f &= \Up^2 f.
\end{align}
\end{subequations}
Equations (\ref{EinEq-1}) and (\ref{EKG-1}) imply that the stress-energy tensor corresponding to a complex scalar field is given by
\begin{equation} \label{SET-Def}
   T = \mu_{0}\pnth{\frac{df \otimes d\cf + d\cf \otimes df}{\Up^{2}} - \pnth{\frac{\abs{df}^{2}}{\Up^{2}} + \abs{f}^{2}}g}.
\end{equation}

With these equations, we will construct a system of partial differential equations which can be used to numerically evolve the scalar field $f$ and the metric from consistent initial data.  Since in the previous section we have already computed the components of the Einstein curvature tensor, in order to construct such a system of PDEs, we need to compute the components of the stress energy tensor and the Laplacian in the metric $g$.  Before we do, we will define the function, $p(t,r)$, by the equation
\begin{equation}\label{p-Def}
  p = f_{t}\e^{-V}\pnth{1 - \frac{2M}{r}}^{-1/2}.
\end{equation}
This is done to make the resulting system of PDEs first order in time and results in a more convenient choice than choosing only $p=f_{t}$.  We now have the following lemmas.
\begin{lem}\label{SET-Lem-1}
    For the stress-energy tensor given in equation (\ref{SET-Def}), the following are true.
    \begin{enumerate}
      \item $\displaystyle T(\nu_{t},\nu_{t}) = \mu_{0}\pnth{\abs{f}^{2} + \pnth{1 - \frac{2M}{r}}\frac{\abs{f_{r}}^{2} + \abs{p}^{2}}{\Up^{2}}}$
      \item $\displaystyle T(\nu_{t},\nu_{r}) = \frac{2\mu_{0}}{\Up^2}\pnth{1 - \frac{2M}{r}}\Re(f_{r}\cp)$
      \item $\displaystyle T(\nu_{r},\nu_{r}) = \mu_{0}\pnth{-\abs{f}^{2} + \pnth{1 - \frac{2M}{r}}\frac{\abs{f_{r}}^{2} + \abs{p}^{2}}{\Up^{2}}}$
      \item $\displaystyle T(\nu_{\theta},\nu_{\theta}) = T(\nu_{\varphi},\nu_{\varphi}) = -\mu_{0}\pnth{\abs{f}^{2} + \pnth{1 - \frac{2M}{r}}\frac{\abs{f_{r}}^{2} - \abs{p}^{2}}{\Up^{2}}}$
    \end{enumerate}
\end{lem}

\noindent The proof of this lemma is left to Section \ref{Proofs}.

\begin{lem}\label{EinEq-Lem-1}
    The components of the Einstein equation in the $\nu_{\eta}$ directions result in the following PDEs.
    \begin{align}
      \label{NCpde1} M_{r} &= 4\pi r^{2}\mu_{0}\pnth{\abs{f}^{2} + \pnth{1 - \frac{2M}{r}}\frac{\abs{f_{r}}^{2} + \abs{p}^{2}}{\Up^{2}}} \\
      \label{NCceq1} M_{t} &= \frac{8\pi r^{2}\mu_{0}\e^{V}}{\Up^{2}}\pnth{1 - \frac{2M}{r}}^{3/2}\Re(f_{r}\cp) \displaybreak[0]\\
      \label{NCpde2} V_{r} &= \pnth{1 - \frac{2M}{r}}^{-1}\pnth{\frac{M}{r^{2}} - 4\pi r\mu_{0}\pnth{\abs{f}^{2} - \pnth{1 - \frac{2M}{r}}\frac{\abs{f_{r}}^{2} + \abs{p}^{2}}{\Up^{2}}}} \displaybreak[0] \\
      \notag V_{t}M_{t} &= -r\e^{2V}\Bigg[\pnth{V_{rr} + V_{r}^{2} + \frac{V_{r}}{r}}\pnth{1 - \frac{2M}{r}}^{2} \\
      \notag &\qquad + \pnth{\frac{M}{r} - M_{r}}\pnth{\frac{1}{r^{2}} + \frac{V_{r}}{r}}\pnth{1 - \frac{2M}{r}} - \frac{M_{tt}}{r\e^{2V}} - \frac{3M_{t}^{2}}{r^{2}\e^{2V}}\pnth{1 - \frac{2M}{r}}^{-1} \\
      \label{NCceq2} &\qquad + 8\pi \mu_{0}\pnth{1 - \frac{2M}{r}}\pnth{\abs{f}^{2} + \pnth{1 - \frac{2M}{r}}\frac{\abs{f_{r}}^{2} - \abs{p}^{2}}{\Up^{2}}}\Bigg]
    \end{align}
\end{lem}

\begin{proof}  This follows directly from the Einstein equation (\ref{EinEq-1}), Corollary \ref{GCor-1}, and Lemma \ref{SET-Lem-1} and the algebra necessary to solve for each of the quantities above.  Equation (\ref{NCpde1}) follows from the $\nu_{t},\nu_{t}$ component of the Einstein equation, (\ref{NCceq1}) follows from the $\nu_{t},\nu_{r}$ component, (\ref{NCpde2}) from the $\nu_{r},\nu_{r}$ component, and (\ref{NCceq2}) from the $\nu_{\theta},\nu_{\theta}$ or $\nu_{\varphi},\nu_{\varphi}$ component since they're identical. \end{proof}

The Klein-Gordon equation and the fact that
\begin{equation}\label{glap-def}
  \glap{g}f = \frac{1}{\sqrt{\abs{g}}}\p_{\eta}\pnth{\sqrt{\abs{g}}g^{\eta\om}\p_{\om}f}.
\end{equation}
yield the following lemma.
  \begin{lem}\label{KG-Lem-1}
    The Klein-Gordon equation in this metric yields the following PDE.
    \begin{equation}\label{NCpde4}
      p_{t} = \e^{V}\pnth{-\Up^{2}f\pnth{1 - \frac{2M}{r}}^{-1/2} + \frac{2f_{r}}{r}\sqrt{1 - \frac{2M}{r}}} + \p_{r}\pnth{\e^{V}f_{r}\sqrt{1 - \frac{2M}{r}}}
    \end{equation}
  \end{lem}

\begin{proof}
  To compute the Laplacian, we will first need the quantity $\sqrt{\abs{g}}$.  We have
  \begin{equation}
    \sqrt{\abs{g}} = \sqrt{\e^{2V}\pnth{1 - \frac{2M}{r}}^{-1}r^{4}\sin^{2}\theta} = \e^{V}\pnth{1-\frac{2M}{r}}^{-1/2}r^{2}\sin\theta
  \end{equation}
  Then since $f = f(t,r)$ and $g$ is diagonal, we have by equation (\ref{glap-def})
  \begin{align}
    \notag \glap{g}f &= \frac{1}{\e^{V}r^{2}\sin\theta}\sqrt{1 - \frac{2M}{r}}\Bigg(\p_{t}\pnth{\e^{V}\pnth{1 - \frac{2M}{r}}^{-1/2}r^{2}\sin\theta\, g^{tt}f_{t}} \\
    \notag &\hspace{2.5 in} + \p_{r}\pnth{\e^{V}\pnth{1 - \frac{2M}{r}}^{-1/2}r^{2}\sin\theta\, g^{rr}f_{r}}\Bigg) \displaybreak[0]\\
    \notag &= \frac{1}{\e^{V}r^{2}\sin\theta}\sqrt{1 - \frac{2M}{r}}\Bigg(\p_{t}\pnth{-\e^{-V}\pnth{1 - \frac{2M}{r}}^{-1/2}r^{2}\sin\theta\, f_{t}} \\
    \notag &\hspace{2.5 in} + \p_{r}\pnth{\e^{V}\sqrt{1 - \frac{2M}{r}}r^{2}\sin\theta\, f_{r}}\Bigg) \displaybreak[0]\\
    \notag &= \frac{1}{\e^{V}r^{2}\sin\theta}\sqrt{1 - \frac{2M}{r}}\Bigg(\p_{t}\pnth{-p r^{2}\sin\theta} \\
    \notag &\hspace{1.5 in} + 2r\e^{V}\sqrt{1 - \frac{2M}{r}}\sin\theta\, f_{r} + r^{2}\sin\theta\, \p_{r}\pnth{\e^{V}f_{r}\sqrt{1 - \frac{2M}{r}}}\Bigg) \\
    &= -p_{t}\e^{-V}\sqrt{1 - \frac{2M}{r}} + \frac{2f_{r}}{r}\pnth{1 - \frac{2M}{r}} + \e^{-V}\sqrt{1 - \frac{2M}{r}}\, \p_{r}\pnth{\e^{V}f_{r}\sqrt{1 - \frac{2M}{r}}}
  \end{align}
  Then the Klein-Gordon equation (\ref{EKG-2}) becomes
  \begin{equation}
    -p_{t}\e^{-V}\sqrt{1 - \frac{2M}{r}} + \frac{2f_{r}}{r}\pnth{1 - \frac{2M}{r}} + \e^{-V}\sqrt{1 - \frac{2M}{r}}\, \p_{r}\pnth{\e^{V}f_{r}\sqrt{1 - \frac{2M}{r}}} = \Up^{2}f
  \end{equation}
  Solving for $p_{t}$ yields the desired result.
\end{proof}

Then rewrite (\ref{p-Def}) as
\begin{equation} \label{NCpde3}
  f_{t} = p\e^{V}\sqrt{1 - \frac{2M}{r}}
\end{equation}
to yield an evolution equation for $f$.

The next lemma shows that $T$ given in (\ref{SET-Def}) is divergence free.
\begin{lem}\label{divTEKG-Lem}
   Equation (\ref{EKG-2}) implies that $\divx{g}T = 0$, where $T$ is given by (\ref{SET-Def}).
\end{lem}
\noindent This lemma's proof is also included in Section \ref{Proofs}.  Now we can prove the following.
\begin{lem}\label{NCpde-Lem-1}
  If $f,p,M$, and $V$ solve (\ref{NCpde1}), (\ref{NCpde2}), (\ref{NCpde4}), and (\ref{NCpde3}), then they also solve (\ref{NCceq1}) and (\ref{NCceq2}).
\end{lem}

\begin{proof}
  This follows from equation (\ref{SETFun-def}), Lemmas \ref{SET-Lem-1}, \ref{EinEq-Lem-1}, \ref{divTEKG-Lem}, and Proposition \ref{EinEq-prop1}.
\end{proof}

Lemmas \ref{EinEq-Lem-1}, \ref{KG-Lem-1}, and \ref{NCpde-Lem-1} and equation (\ref{NCpde3}) prove the following proposition.
\begin{prop}\label{NCpdeprop}
  Solving the following PDEs for $f(t,r)$, $p(t,r)$, $V(t,r)$, and $M(t,r)$ with consistent initial data is necessary and sufficient to solve the Einstein-Klein-Gordon equations (\ref{EKG-1}) and (\ref{EKG-2}) with a metric of the form (\ref{metric}) on the same initial data.
  \begin{subequations} \label{NCpde}
  \begin{align}
    \label{NCpde1a} M_{r} &= 4\pi r^{2}\mu_{0}\pnth{\abs{f}^{2} + \pnth{1 - \frac{2M}{r}}\frac{\abs{f_{r}}^{2} + \abs{p}^{2}}{\Up^{2}}} \\
    \label{NCpde2a} V_{r} &= \pnth{1 - \frac{2M}{r}}^{-1}\pnth{\frac{M}{r^{2}} - 4\pi r\mu_{0}\pnth{\abs{f}^{2} - \pnth{1 - \frac{2M}{r}}\frac{\abs{f_{r}}^{2} + \abs{p}^{2}}{\Up^{2}}}} \displaybreak[0]\\
    \label{NCpde3a} f_{t} &= p\e^{V}\sqrt{1 - \frac{2M}{r}} \\
    \label{NCpde4a} p_{t} &= \e^{V}\pnth{-\Up^{2}f\pnth{1 - \frac{2M}{r}}^{-1/2} + \frac{2f_{r}}{r}\sqrt{1 - \frac{2M}{r}}} + \p_{r}\pnth{\e^{V}f_{r}\sqrt{1 - \frac{2M}{r}}}
  \end{align}
  \end{subequations}
\end{prop}

Thus these equations are effectively the spherically symmetric Einstein-Klein-Gordon equations (specifically though only in the metric (\ref{metric})).  Throughout the remainder of this dissertation, when solving the Einstein-Klein-Gordon equations in spherical symmetry, this is the system we will solve.

\section{Proofs}\label{Proofs}

In this section, we present the proofs which were omitted in the previous sections.  Every proof of a proposition or lemma from above will refer back to the number of the statement it is proving.  We also introduce a new lemma here which is useful for proving some of the above statements.  We will present all of the proofs here in the order with which their corresponding statements are given above.

\begin{proof}[Proof of Proposition \ref{mH-Prop}]
The Hawking mass of a metric sphere is defined as
\begin{equation}
  m_{H}(\Si_{t,r}) = \sqrt{\frac{\abs{\Si_{t,r}}}{16\pi}}\pnth{1 - \frac{1}{16\pi}\int_{\Si_{t,r}} g\pnth{\Hv,\Hv}\, dA}
\end{equation}
where $\abs{\Si_{t,r}}$ is the surface area of the metric sphere, $\Hv$ is the mean curvature vector of the sphere in the spacetime, and $dA$ is the volume element on the sphere (\cite{Hawk68,Bray07}).  By the definition of our coordinate $r$, we have that $\abs{\Si_{t,r}}=4\pi r^{2}$, but it is also easily computed in this metric since
\begin{equation}\label{dA}
  dA = \sqrt{\abs{d\si^{2}}}\, d\theta \, d\varphi = \sqrt{r^{4} \sin^{2}\theta}\, d\theta \, d\varphi = r^{2}\sin \theta\, d\theta \, d\varphi,
\end{equation}
which yields that
\begin{equation}\label{Area}
  \abs{\Si_{t,r}} = \int_{\Si_{t,r}} dA = \int_{0}^{2\pi}\int_{0}^{\pi} r^{2}\sin \theta\, d\theta \, d\varphi = \int_{0}^{2\pi} 2r^{2}\, d\varphi = 4\pi r^{2}.
\end{equation}
To compute the mean curvature vector, $\Hv$, we have that
\begin{equation} \label{MCV-1}
  \Hv = \ga^{jk}\II(\p_{j},\p_{k})
\end{equation}
where $j,k \in \{\theta,\varphi\}$, $\ga$ is the metric on $\Si_{t,r}$ (note that we have reused the variable $\ga$ here to refer to the metric on the sphere and not to the metric on the $t=constant$ hypersurfaces as we did in the section describing the framework of numerical relativity), and $\II$ is the second fundamental form tensor, which sends a pair of vectors tangent to the sphere to a vector normal to the sphere.  It is defined as follows for all $X,Y \in T\Si_{t,r}$
\begin{equation}
  \II(X,Y) = \cov_{X}Y -\, ^{2}\cov_{X}Y
\end{equation}
where $\cov$ is the covariant derivative operator on $N$ and $^{2}\cov$ is the induced covariant derivative operator on $\Si_{t,r}$.  Since $\ga$ is diagonal, we only need to concern ourselves with the diagonal components of this tensor in order to compute $\Hv$.  To perform these computations, first note that the Christoffel symbols on the sphere in these coordinates have the following property (note that a superscript of 2 will always denote that that quantity corresponds to the sphere, also for the following computations, roman letters will denote subscripts in $\{\theta,\varphi\}$, while Greek letters will denote subscripts ranging over all four coordinates).  Since both the radial and time directions are normal to the sphere, we have that
\begin{align}
  \notag \Ga_{jk}^{\ \ \ell} &= \frac{1}{2}g^{\ell\eta}(g_{j\eta,k} + g_{k\eta,j} - g_{jk,\eta}) \\
  \notag &= \frac{1}{2}g^{\ell m}(g_{jm,k} + g_{km,j} - g_{jk,m}) \displaybreak[0] \\
  \notag &= \frac{1}{2}\ga^{\ell m}(g_{jm,k} + g_{km,j} - g_{jk,m}) \\
  &= \,^{2}\Ga_{jk}^{\ \ \ell}.
\end{align}
Thus we need not distinguish between the Christoffel symbols for the metric on the sphere and those on the entire manifold.  We have then that
\begin{align}
  \II(\p_{\theta},\p_{\theta}) &= \cov_{\theta}\p_{\theta} -\, ^{2}\cov_{\theta}\p_{\theta} = \Ga_{\theta\theta}^{\ \ \ \eta}\, \p_{\eta} -\, ^{2}\Ga_{\theta\theta}^{\ \ \ k}\, \p_{k} = \Ga_{\theta\theta}^{\ \ \ t}\, \p_{t} + \Ga_{\theta\theta}^{\ \ \ r}\, \p_{r}
\intertext{and similarly}
  \II(\p_{\varphi},\p_{\varphi}) &= \cov_{\varphi}\p_{\varphi} -\, ^{2}\cov_{\varphi}\p_{\varphi} = \Ga_{\varphi\varphi}^{\ \ \ \eta}\, \p_{\eta} -\, ^{2}\Ga_{\varphi\varphi}^{\ \ \ k}\, \p_{k} = \Ga_{\varphi\varphi}^{\ \ \ t}\, \p_{t} + \Ga_{\varphi\varphi}^{\ \ \ r}\, \p_{r}
\end{align}
Then we need to compute the above Christoffel symbols.  We have that
\begin{subequations} \label{Chistoffel-1}
\begin{align}
  \Ga_{\theta\theta}^{\ \ \ t} &= \frac{1}{2}g^{t\eta}(g_{\theta \eta,\theta } + g_{\theta \eta,\theta } - g_{\theta \theta ,\eta}) = \frac{1}{2}g^{tt}(- g_{\theta \theta ,t}) = \frac{1}{2 \e^{2V}}(-\p_{t}(r^{2})) = 0 \displaybreak[0]\\
  \notag \Ga_{\theta\theta}^{\ \ \ r} &= \frac{1}{2}g^{r\eta}(g_{\theta \eta,\theta } + g_{\theta \eta,\theta } - g_{\theta \theta ,\eta}) = \frac{1}{2}g^{rr}(- g_{\theta \theta ,r}) \\
  &= \frac{1}{2}\pnth{1-\frac{2M}{r}}(-\p_{r}(r^{2})) = -r\pnth{1 - \frac{2M}{r}} \displaybreak[0]\\
  \Ga_{\varphi\varphi}^{\ \ \ t} &= \frac{1}{2}g^{t\eta}(g_{\varphi \eta,\varphi } + g_{\varphi \eta,\varphi } - g_{\varphi \varphi ,\eta}) = \frac{1}{2}g^{tt}(- g_{\varphi \varphi ,t}) = \frac{1}{2 \e^{2V}}(-\p_{t}(r^{2}\sin^{2}\theta)) = 0 \displaybreak[0]\\
  \notag \Ga_{\varphi\varphi}^{\ \ \ r} &= \frac{1}{2}g^{r\eta}(g_{\varphi \eta,\varphi } + g_{\varphi \eta,\varphi } - g_{\varphi \varphi ,\eta}) = \frac{1}{2}g^{rr}(- g_{\varphi \varphi ,r}) \\
  &= \frac{1}{2}\pnth{1-\frac{2M}{r}}(-\p_{r}(r^{2}\sin^{2}\theta)) = -r\pnth{1 - \frac{2M}{r}}\sin^{2}\theta.
\end{align}
\end{subequations}
Then we have that
\begin{align}
  \II(\p_{\theta},\p_{\theta}) &= \Ga_{\theta\theta}^{\ \ \ t}\, \p_{t} + \Ga_{\theta\theta}^{\ \ \ r}\, \p_{r} = -r\pnth{1 - \frac{2M}{r}}\, \p_{r} \\
  \II(\p_{\varphi},\p_{\varphi}) &= \Ga_{\varphi\varphi}^{\ \ \ t}\, \p_{t} + \Ga_{\varphi\varphi}^{\ \ \ r}\, \p_{r} = -r\pnth{1 - \frac{2M}{r}}\sin^{2}\theta\, \p_{r}
\end{align}
This makes (\ref{MCV-1}) become
\begin{align}
  \notag \Hv &= \ga^{\theta\theta}\II(\p_{\theta},\p_{\theta}) + \ga^{\varphi\varphi}\II(\p_{\varphi}, \p_{\varphi}) \\
  \notag &= -\frac{1}{r^{2}}\cdot r\pnth{1 - \frac{2M}{r}}\, \p_{r} - \frac{1}{r^{2}\sin^{2}\theta}\cdot r\pnth{1 - \frac{2M}{r}}\sin^{2}\theta\, \p_{r} \\
  \label{MCV-2} &= -\frac{2}{r}\pnth{1 - \frac{2M}{r}}\, \p_{r}
\end{align}
Finally, we can compute the Hawking mass as follows.
\begin{align}
  \notag m_{H}(\Si_{t,r}) &= \sqrt{\frac{\abs{\Si_{t,r}}}{16\pi}}\pnth{1 - \frac{1}{16\pi}\int_{\Si_{t,r}} g\pnth{\Hv,\Hv}\, dA} \\
  \notag &= \sqrt{\frac{4\pi r^{2}}{16\pi}}\pnth{1 - \frac{1}{16\pi}\int_{\Si_{t,r}} g\pnth{-\frac{2}{r}\pnth{1 - \frac{2M}{r}}\, \p_{r},\, -\frac{2}{r}\pnth{1 - \frac{2M}{r}}\, \p_{r}}\, dA} \displaybreak[0]\\
  \notag &= \frac{r}{2}\pnth{1 - \frac{1}{16\pi}\int_{0}^{2\pi}\int_{0}^{\pi} \frac{4}{r^{2}}\pnth{1 - \frac{2M}{r}}^{2}g(\p_{r},\p_{r}) r^{2}\sin\theta \, d\theta\, d\varphi} \displaybreak[0]\\
  \notag &= \frac{r}{2}\pnth{1 - \frac{1}{16\pi}\int_{0}^{2\pi}\int_{0}^{\pi} 4\pnth{1 - \frac{2M}{r}} \sin\theta \, d\theta\, d\varphi} \displaybreak[0]\\
  \notag &= \frac{r}{2}\pnth{1 - \frac{1}{4\pi} \pnth{1 - \frac{2M}{r}}\int_{0}^{2\pi}\int_{0}^{\pi} \sin\theta \, d\theta\, d\varphi} \displaybreak[0]\\
  \notag &= \frac{r}{2}\pnth{1 - \frac{1}{4\pi}\pnth{1 - \frac{2M}{r}}(4\pi)} \displaybreak[0]\\
  \notag &= \frac{r}{2}\pnth{\frac{2M}{r}} \\
  \label{mH-1} m_{H}(\Si_{t,r}) &= M
\end{align}
\end{proof}

To prove Lemma \ref{RicLem-2}, we will utilize the following additional lemma.

\begin{lem} \label{RicLem-1}
  For all $\eta \in \{t,r,\theta,\varphi\}$, let $\nu_{\eta}$ be as defined by equation (\ref{norvec}).  Then the following are true.
    \begin{enumerate}
      \item\label{RicLem-1a} For all $\eta \in \{t,r,\theta,\varphi\}$, $\Ric(\nu_{\eta},\nu_{\eta}) = g(\nu_{\eta},\nu_{\eta})g^{\eta\eta}\Ric_{\eta\eta}$.
      \item\label{RicLem-1b} $\displaystyle R = \sum_{\eta} g(\nu_{\eta},\nu_{\eta})\Ric(\nu_{\eta},\nu_{\eta})$
    \end{enumerate}
\end{lem}

\begin{proof}
Since $\{\nu_{\eta}\}$ is a frame field, we have that
\begin{equation}
  \Ric(\nu_{t},\nu_{t}) =  \e^{-2V}\Ric(\p_{t},\p_{t}) = -g^{tt}\Ric_{tt} = g(\nu_{t},\nu_{t})g^{tt}\Ric_{tt}.
\end{equation}
By similar arguments, we have that for all $\eta \in \{t,r,\theta,\varphi\}$,
\begin{equation}
  \Ric(\nu_{\eta},\nu_{\eta}) = g(\nu_{\eta},\nu_{\eta})g^{\eta\eta}\Ric_{\eta\eta}
\end{equation}
which proves \ref{RicLem-1a}.  To prove \ref{RicLem-1b}, we use the first result and find that since $g$ is diagonal and $g(\nu_{\eta},\nu_{\eta})=\pm 1$, we have that
\begin{equation}
  R = \sum_{\eta\om} g^{\eta\om}\Ric_{\eta\om} = \sum_{\eta} g^{\eta\eta}\Ric_{\eta\eta} = \sum_{\eta} \frac{\Ric(\nu_{\eta},\nu_{\eta})}{g(\nu_{\eta},\nu_{\eta})} = \sum_{\eta} g(\nu_{\eta},\nu_{\eta})\Ric(\nu_{\eta},\nu_{\eta}).
\end{equation}
Fact \ref{RicLem-1b} also follows from a well understood more general property of frame fields and traces of tensors.
\end{proof}

\noindent With this, we prove Lemma \ref{RicLem-2}.

\begin{proof}[Proof of Lemma \ref{RicLem-2}]
In this proof, all indices range over the coordinates $\{t,r,\theta,\varphi\}$.  By Lemma \ref{RicLem-1} and the well known formula for the components of the Riemann curvature tensor, $\R$, in terms of the Christoffel symbols
\begin{equation}
  \R_{jk\ell}^{\ \ \ \, m} = \Ga_{j\ell\ \ ,k}^{\ \ m} - \Ga_{k\ell\ \ ,j}^{\ \ m} + \Ga_{j\ell}^{\ \ s}\Ga_{ks}^{\ \ m} - \Ga_{k\ell}^{\ \ s}\Ga_{js}^{\ \ m}.
\end{equation}
Since $\Ric_{jk} = \R_{j\ell k}^{\ \ \ \, \ell}$, we have that
\begin{subequations} \label{RicEq-1}
  \begin{align}
    \notag \Ric(\nu_{t},\nu_{t}) &= g(\nu_{t},\nu_{t})g^{tt}\Ric_{tt} \\
    &=  \e^{-2V}\pnth{\Ga_{tt\ ,k}^{\ \ k} - \Ga_{kt\ ,t}^{\ \ k} + \Ga_{tt}^{\ \ s}\Ga_{ks}^{\ \ k} - \Ga_{kt}^{\ \ s}\Ga_{ts}^{\ \ k}} \displaybreak[0]\\
    \notag \Ric(\nu_{t},\nu_{r}) &=  \e^{-V}\sqrt{1 - \frac{2M}{r}}\, \Ric_{tr} \\
    &=  \e^{-V}\sqrt{1 - \frac{2M}{r}} \pnth{\Ga_{tr\ ,k}^{\ \ k} - \Ga_{kr\ ,t}^{\ \ k} + \Ga_{tr}^{\ \ s}\Ga_{ks}^{\ \ k} - \Ga_{kr}^{\ \ s}\Ga_{ts}^{\ \ k}} \displaybreak[0]\\
    \notag \Ric(\nu_{r},\nu_{r}) &= g(\nu_{r},\nu_{r})g^{rr}\Ric_{rr} = \pnth{1 - \frac{2M}{r}}\Ric_{rr} \\
    &= \pnth{1 - \frac{2M}{r}} \pnth{\Ga_{rr\ ,k}^{\ \ k} - \Ga_{kr\ ,r}^{\ \ k} + \Ga_{rr}^{\ \ s}\Ga_{ks}^{\ \ k} - \Ga_{kr}^{\ \ s}\Ga_{rs}^{\ \ k}} \displaybreak[0]\\
    \notag \Ric(\nu_{\theta},\nu_{\theta}) &= g(\nu_{\theta},\nu_{\theta})g^{\theta\theta}\Ric_{\theta\theta} = \frac{1}{r^{2}}\Ric_{\theta\theta} \\
    &= \frac{1}{r^{2}} \pnth{\Ga_{\theta \theta \ ,k}^{\ \ k} - \Ga_{k\theta \ ,\theta }^{\ \ k} + \Ga_{\theta \theta }^{\ \ s}\Ga_{ks}^{\ \ k} - \Ga_{k\theta }^{\ \ s}\Ga_{\theta s}^{\ \ k}} \displaybreak[0]\\
    \notag \Ric(\nu_{\varphi},\nu_{\varphi}) &= g(\nu_{\varphi},\nu_{\varphi})g^{\varphi\varphi}\Ric_{\varphi\varphi} = \frac{1}{r^{2}\sin^{2} \theta}\Ric_{\varphi\varphi} \\
    &= \frac{1}{r^{2}\sin^{2}\theta} \pnth{\Ga_{\varphi \varphi \ ,k}^{\ \ \ k} - \Ga_{k\varphi \ ,\varphi }^{\ \ \ k} + \Ga_{\varphi \varphi }^{\ \ \ s}\Ga_{ks}^{\ \ k} - \Ga_{k\varphi }^{\ \ \ s}\Ga_{\varphi s}^{\ \ \ k}}.
  \end{align}
\end{subequations}
Then to compute these, we need the Christoffel symbols.  For convenience, we will write the Christoffel symbols in four matrices, $\Ga^{t}$, $\Ga^{r}$, $\Ga^{\theta}$, and $\Ga^{\varphi}$, which are the $4\times4$ matrices of Christoffel symbols corresponding to a given fixed upper index.  To do so, recall, that the formula for the Christoffel symbols in terms of the metric components is the following
\begin{equation} \label{CS-Def}
  \Ga_{jk}^{\ \ \ell} = \frac{1}{2}g^{\ell m} (g_{jm,k} + g_{km,j} - g_{jk,m})
\end{equation}
Using (\ref{CS-Def}) and the fact that $g$ is diagonal, we have that
\begin{subequations} \label{CS-1}
\begin{align}
  \notag \Ga_{jk}^{\ \ t} &= \frac{1}{2}g^{t m}(g_{jm,k} + g_{km,j} - g_{jk,m}) = \frac{1}{2}g^{tt}(g_{jt,k} + g_{kt,j} - g_{jk,t}) \\
  &= -\frac{1}{2 \e^{2V}}(g_{jt,k} + g_{kt,j} - g_{jk,t}) \displaybreak[0]\\
  \notag \Ga_{jk}^{\ \ r} &= \frac{1}{2}g^{r m}(g_{jm,k} + g_{km,j} - g_{jk,m}) = \frac{1}{2}g^{rr}(g_{jr,k} + g_{kr,j} - g_{jk,r}) \\
  &= \frac{1}{2}\pnth{1 - \frac{2M}{r}} (g_{jr,k} + g_{kr,j} - g_{jk,r}) \displaybreak[0]\\
  \notag \Ga_{jk}^{\ \ \theta} &=  \frac{1}{2}g^{\theta m}(g_{jm,k} + g_{km,j} - g_{jk,m}) = \frac{1}{2}g^{\theta\theta}(g_{j\theta,k} + g_{k\theta,j} - g_{jk,\theta}) \\
  &= \frac{1}{2r^{2}} (g_{j\theta,k} + g_{k\theta,j} - g_{jk,\theta}) \displaybreak[0]\\
  \notag \Ga_{jk}^{\ \ \varphi} &= \frac{1}{2}g^{\varphi m}(g_{jm,k} + g_{km,j} - g_{jk,m}) = \frac{1}{2}g^{\varphi\varphi}(g_{j\varphi,k} + g_{k\varphi,j} - g_{jk,\varphi}) \\
  &= \frac{1}{2r^{2}\sin^{2}\theta} (g_{j\varphi,k} + g_{k\varphi,j} - g_{jk,\varphi}) = \frac{1}{2r^{2}\sin^{2}\theta} (g_{j\varphi,k} + g_{k\varphi,j}).
\end{align}
\end{subequations}
The last line is due to the fact that none of the metric components depend on $\varphi$.  To help compute these quantities, it would be useful to note the following.
\begin{subequations} \label{MDer}
\begin{align}
  \notag g_{tt,t} &= \p_{t}(- \e^{2V}) = -2V_{t} \e^{2V} & g_{tt,\theta} &= \p_{\theta}(- \e^{2V}) = 0 \\
  g_{tt,r} &= \p_{r}(- \e^{2V}) = -2V_{r} \e^{2V} & g_{tt,\varphi} &= \p_{\varphi}(- \e^{2V}) = 0 \displaybreak[0] \\
  \notag g_{rr,t} &= \p_{t}\pnth{1 - \frac{2M}{r}}^{-1}  &  g_{rr,\theta} &= \p_{\theta}\pnth{1 - \frac{2M}{r}}^{-1} = 0 \\
  \notag &= \pnth{1 - \frac{2M}{r}}^{-2}\pnth{\frac{2M_{t}}{r}} \displaybreak[0]\\
  \notag g_{rr,r} &= \p_{r}\pnth{1 - \frac{2M}{r}}^{-1} & g_{rr,\varphi} &= \p_{\varphi}\pnth{1 - \frac{2M}{r}}^{-1} = 0 \\
  &= \pnth{1 - \frac{2M}{r}}^{-2}\pnth{\frac{2M_{r}}{r} - \frac{2M}{r^{2}}} \displaybreak[0]\\
  \notag g_{\theta\theta,t} &= \p_{t}(r^{2}) = 0 & g_{\theta\theta,\theta} &= \p_{\theta}(r^{2}) = 0 \displaybreak[0]\\
  g_{\theta\theta,r} &= \p_{r}(r^{2}) = 2r & g_{\theta\theta,\varphi} &= \p_{\varphi}(r^{2}) = 0 \displaybreak[0]\\
  \notag g_{\varphi\varphi,t} &= \p_{t}(r^{2}\sin^{2}\theta) = 0 & g_{\varphi\varphi,\theta} &= \p_{\theta}(r^{2}\sin^{2}\theta) = 2r^{2}\sin\theta \cos\theta \\
  g_{\varphi\varphi,r} &= \p_{r}(r^{2}\sin^{2}\theta) = 2r\sin^{2}\theta & g_{\varphi\varphi,\varphi} &= \p_{\varphi}{r^{2}\sin^{2}\theta} = 0.
\end{align}
\end{subequations}
All the other metric components are 0.  Next, if
\begin{equation}
  \Phi = \pnth{1 - \frac{2M}{r}},
\end{equation}
then, using (\ref{CS-1}) and (\ref{MDer}), it is a series of straightforward algebra computations to obtain the following.
\begin{subequations} \label{CS-2}
\begin{align}
  \Ga^{t} &= \begin{pmatrix} V_{t} & V_{r} & 0 & 0 \\ V_{r} & \dfrac{M_{t}}{r \e^{2V}}\Phi^{-2} & 0 & 0 \\ 0 & 0 & 0 & 0 \\ 0 & 0 & 0 & 0 \end{pmatrix} \displaybreak[0] \\
  \Ga^{r} &= \begin{pmatrix} V_{r} \e^{2V}\Phi & \dfrac{M_{t}}{r}\Phi^{-1} & 0 & 0 \\ \dfrac{M_{t}}{r}\Phi^{-1} & \Phi^{-1}\pnth{\dfrac{M_{r}}{r} - \dfrac{M}{r^{2}}} & 0 & 0 \\ 0 & 0 & -r\Phi & 0 \\ 0 & 0 & 0 & -r \Phi\sin^{2}\theta \end{pmatrix} \displaybreak[0]\\
  \Ga^{\theta} &= \begin{pmatrix} 0 & 0 & 0 & 0 \\ 0 & 0 & \dfrac{1}{r} & 0 \\ 0 & \dfrac{1}{r} & 0 & 0 \\ 0 & 0 & 0 & -\sin\theta\cos\theta \end{pmatrix} \displaybreak[0]\\
  \Ga^{\varphi} &= \begin{pmatrix} 0 & 0 & 0 & 0 \\ 0 & 0 & 0 & \dfrac{1}{r} \\ 0 & 0 & 0 & \dfrac{\cos\theta}{\sin\theta} \\ 0 & \dfrac{1}{r} & \dfrac{\cos\theta}{\sin\theta} & 0 \end{pmatrix}
\end{align}
\end{subequations}
Using (\ref{CS-2}), we can now continue the computations started in (\ref{RicEq-1}).
\begin{align}
  \notag \Ric(\nu_{t},\nu_{t}) &=  \e^{-2V}\pnth{\Ga_{tt\ ,k}^{\ \ k} - \Ga_{kt\ ,t}^{\ \ k} + \Ga_{tt}^{\ \ s}\Ga_{ks}^{\ \ k} - \Ga_{kt}^{\ \ s}\Ga_{ts}^{\ \ k}} \displaybreak[0]\\
  \notag &=  \e^{-2V}\Bigg\{\p_{t}V_{t} + \p_{r}\pnth{V_{r} \e^{2V}\pnth{1 - \frac{2M}{r}}} - \p_{t}V_{t} \\
  \notag &\qquad - \p_{t}\pnth{\frac{M_{t}}{r}\pnth{1 - \frac{2M}{r}}^{-1}} + V_{t}\pnth{V_{t} + \frac{M_{t}}{r}\pnth{1 - \frac{2M}{r}}^{-1}} \displaybreak[0] \\
  \notag &\qquad + V_{r} \e^{2V}\pnth{1 - \frac{2M}{r}}\pnth{V_{r} + \pnth{1 - \frac{2M}{r}}^{-1}\pnth{\frac{M_{r}}{r} - \frac{M}{r^{2}}} + \frac{2}{r}} \\
  \notag &\qquad - V_{t}^{2} - 2 V_{r}^{2} \e^{2V}\pnth{1 - \frac{2M}{r}} - \frac{M_{t}^{2}}{r^{2}}\pnth{1 - \frac{2M}{r}}^{-2} \Bigg\} \displaybreak[0]\\
  \notag &=  \e^{-2V}\Bigg\{V_{rr} \e^{2V}\pnth{1 - \frac{2M}{r}} + 2V_{r}^{2} \e^{2V}\pnth{1 - \frac{2M}{r}} + V_{r} \e^{2V}\pnth{\frac{2M}{r^{2}} - \frac{2M_{r}}{r}} \\
  \notag &\qquad - \frac{M_{tt}}{r}\pnth{1 - \frac{2M}{r}}^{-1} - \frac{2M_{t}^{2}}{r^{2}}\pnth{1 - \frac{2M}{r}}^{-2} + \frac{V_{t}M_{t}}{r}\pnth{1 - \frac{2M}{r}}^{-1} \displaybreak[0]\\
  \notag &\qquad + V_{r}^{2} \e^{2V}\pnth{1 - \frac{2M}{r}} + V_{r} \e^{2V}\pnth{\frac{M_{r}}{r} - \frac{M}{r^{2}}} + \frac{2V_{r} \e^{2V}}{r}\pnth{1 - \frac{2M}{r}} \\
  \notag &\qquad - 2 V_{r}^{2} \e^{2V}\pnth{1 - \frac{2M}{r}} - \frac{M_{t}^{2}}{r^{2}}\pnth{1 - \frac{2M}{r}}^{-2} \Bigg\} \displaybreak[0]\\
  \notag \Ric(\nu_{t},\nu_{t}) &= \pnth{V_{rr} + V_{r}^{2} + \frac{2V_{r}}{r}}\pnth{1-\frac{2M}{r}} + \frac{V_{r}}{r}\pnth{\frac{M}{r} - M_{r}} \\
  &\qquad + \frac{V_{t}M_{t} - M_{tt}}{r \e^{2V}}\pnth{1-\frac{2M}{r}}^{-1} - \frac{3M_{t}^{2}}{r^{2} \e^{2V}}\pnth{1-\frac{2M}{r}}^{-2}
\end{align}
\begin{align}
  \notag \Ric(\nu_{t},\nu_{r}) &=  \e^{-V}\sqrt{1 - \frac{2M}{r}} \pnth{\Ga_{tr\ ,k}^{\ \ k} - \Ga_{kr\ ,t}^{\ \ k} + \Ga_{tr}^{\ \ s}\Ga_{ks}^{\ \ k} - \Ga_{kr}^{\ \ s}\Ga_{ts}^{\ \ k}} \displaybreak[0]\\
  \notag &=  \e^{-V}\sqrt{1 - \frac{2M}{r}} \Bigg\{\p_{t}(V_{r}) + \p_{r}\pnth{\frac{M_{t}}{r}\pnth{1 - \frac{2M}{r}}^{-1}} - \p_{t}(V_{r}) \\
  \notag &\qquad - \p_{t}\pnth{\pnth{1 - \frac{2M}{r}}^{-1}\pnth{\frac{M_{r}}{r} - \frac{M}{r^{2}}}} - 2\p_{t}\pnth{\frac{1}{r}} \\
  \notag &\qquad + V_{r}\pnth{V_{t} + \frac{M_{t}}{r}\pnth{1 - \frac{2M}{r}}^{-1}} \displaybreak[0] \\
  \notag &\qquad + \frac{M_{t}}{r}\pnth{1 - \frac{2M}{r}}^{-1}\pnth{V_{r} + \pnth{1 - \frac{2M}{r}}^{-1}\pnth{\frac{M_{r}}{r} - \frac{M}{r^{2}}} + \frac{2}{r}} \\
  \notag &\qquad - V_{r}V_{t} - \frac{2M_{t}V_{r}}{r}\pnth{1 - \frac{2M}{r}}^{-1} - \frac{M_{t}}{r}\pnth{1 - \frac{2M}{r}}^{-2}\pnth{\frac{M_{r}}{r} - \frac{M}{r^{2}}} \Bigg\} \displaybreak[0]\\
  \notag &=  \e^{-V}\sqrt{1 - \frac{2M}{r}} \Bigg\{\p_{r}\p_{t}\pnth{-\frac{1}{2}\ln\pnth{1 - \frac{2M}{r}}} - \p_{t}\p_{r}\pnth{-\frac{1}{2}\ln\pnth{1 - \frac{2M}{r}}} \\
  \notag &\qquad + \frac{2M_{t}}{r^{2}}\pnth{1 - \frac{2M}{r}}^{-1} \Bigg\} \\
  \Ric(\nu_{t},\nu_{r}) &= \frac{2M_{t}}{r^{2} \e^{V}}\pnth{1 - \frac{2M}{r}}^{-1/2}
\end{align}
\begin{align}
  \notag \Ric(\nu_{r},\nu_{r}) &= \pnth{1 - \frac{2M}{r}} \pnth{\Ga_{rr\ ,k}^{\ \ k} - \Ga_{kr\ ,r}^{\ \ k} + \Ga_{rr}^{\ \ s}\Ga_{ks}^{\ \ k} - \Ga_{kr}^{\ \ s}\Ga_{rs}^{\ \ k}} \displaybreak[0]\\
  \notag &= \pnth{1 - \frac{2M}{r}} \Bigg\{\p_{t}\pnth{\frac{M_{t}}{r \e^{2V}}\pnth{1 - \frac{2M}{r}}^{-2}} \\
  \notag & \qquad + \p_{r}\pnth{\pnth{1 - \frac{2M}{r}}^{-1}\pnth{\frac{M_{r}}{r} - \frac{M}{r^{2}}}} - \p_{r}V_{r} - 2\p_{r}\pnth{\frac{1}{r}} \\
  \notag &\qquad -\p_{r}\pnth{\pnth{1 - \frac{2M}{r}}^{-1}\pnth{\frac{M_{r}}{r} - \frac{M}{r^{2}}}} + \frac{V_{t}M_{t}}{r \e^{2V}}\pnth{1 - \frac{2M}{r}}^{-2} \displaybreak[0]\\
  \notag &\qquad + \frac{M_{t}^{2}}{r^{2} \e^{2V}}\pnth{1 - \frac{2M}{r}}^{-3} + \pnth{V_{r} + \frac{2}{r}}\pnth{1 - \frac{2M}{r}}^{-1}\pnth{\frac{M_{r}}{r} - \frac{M}{r^{2}}} \\
  \notag &\qquad + \pnth{1 - \frac{2M}{r}}^{-2}\pnth{\frac{M_{r}}{r} - \frac{M}{r^{2}}}^{2} - V_{r}^{2} - \frac{2M_{t}^{2}}{r^{2} \e^{2V}}\pnth{1 - \frac{2M}{r}}^{-3} - \frac{2}{r^{2}} \\
  \notag &\qquad - \pnth{1 - \frac{2M}{r}}^{-2}\pnth{\frac{M_{r}}{r} - \frac{M}{r^{2}}}^{2} \Bigg\} \displaybreak[0] \\
  \notag &= \pnth{1 - \frac{2M}{r}} \Bigg\{\frac{M_{tt}}{r \e^{2V}}\pnth{1 - \frac{2M}{r}}^{-2} - \frac{2V_{t}M_{t}}{r \e^{2V}}\pnth{1 - \frac{2M}{r}}^{-2} \\
  \notag &\qquad - V_{rr} + \frac{2}{r^{2}} + \frac{V_{t}M_{t}}{r \e^{2V}}\pnth{1 - \frac{2M}{r}}^{-2} - \frac{M_{t}^{2}}{r^{2} \e^{2V}}\pnth{1 - \frac{2M}{r}}^{-3} - V_{r}^{2} - \frac{2}{r^{2}} \\
  \notag &\qquad  + \frac{4M_{t}^{2}}{r^{2} \e^{2V}}\pnth{1 - \frac{2M}{r}}^{-3} + \pnth{V_{r} + \frac{2}{r}}\pnth{1 - \frac{2M}{r}}^{-1}\pnth{\frac{M_{r}}{r} - \frac{M}{r^{2}}} \Bigg\} \displaybreak[0] \\
  \notag \Ric(\nu_{r},\nu_{r}) &= -(V_{rr} + V_{r}^2)\pnth{1 - \frac{2M}{r}} + \frac{3M_{t}^{2}}{r^2 \e^{2V}}\pnth{1-\frac{2M}{r}}^{-2} \\
  &\qquad - \pnth{\frac{2}{r^{2}} + \frac{V_{r}}{r}}\pnth{\frac{M}{r} - M_{r}} - \frac{V_{t}M_{t} - M_{tt}}{r \e^{2V}}\pnth{1-\frac{2M}{r}}^{-1}
\end{align}
\begin{align}
  \notag \Ric(\nu_{\theta},\nu_{\theta}) &= \frac{1}{r^{2}} \pnth{\Ga_{\theta \theta \ ,k}^{\ \ k} - \Ga_{k\theta \ ,\theta }^{\ \ k} + \Ga_{\theta \theta }^{\ \ s}\Ga_{ks}^{\ \ k} - \Ga_{k\theta }^{\ \ s}\Ga_{\theta s}^{\ \ k}} \displaybreak[0]\\
  \notag &= \frac{1}{r^{2}}\Bigg\{\p_{r}\pnth{-r\pnth{1 - \frac{2M}{r}}} - \p_{\theta}\pnth{\frac{\cos \theta}{\sin\theta}} - r\pnth{1 - \frac{2M}{r}}\pnth{V_{r} + \frac{2}{r}} \\
  \notag &\qquad - r\pnth{\frac{M_{r}}{r} - \frac{M}{r^{2}}} + 2\pnth{1 - \frac{2M}{r}} - \frac{\cos^{2}\theta}{\sin^{2}\theta}\Bigg\} \displaybreak[0]\\
  \notag &= \frac{1}{r^{2}}\Bigg\{-\pnth{1 - \frac{2M}{r}} - r\pnth{\frac{2M}{r^{2}} - \frac{2M_{r}}{r}} + \frac{1}{\sin^{2}\theta}  \\
  \notag &\qquad - r\pnth{1 - \frac{2M}{r}}\pnth{V_{r} + \frac{2}{r}} + r\pnth{\frac{M}{r^{2}} - \frac{M_{r}}{r}} \\
  \notag &\qquad + 2\pnth{1 - \frac{2M}{r}} - \frac{\cos^{2}\theta}{\sin^{2}\theta}\Bigg\} \displaybreak[0] \\
  \notag &= \frac{1}{r^{2}}\Bigg\{-\frac{\sin^{2}\theta}{\sin^{2}\theta} + \frac{2M}{r} - \frac{M}{r} + M_{r} + \frac{1}{\sin^{2}\theta} - rV_{r}\pnth{1 - \frac{2M}{r}} \\
  \notag &\qquad - 2\pnth{1 - \frac{2M}{r}} + 2 \pnth{1 - \frac{2M}{r}} - \frac{\cos^{2}\theta}{\sin^{2}\theta} \Bigg\} \displaybreak[0] \\
  \Ric(\nu_{\theta},\nu_{\theta}) &= \frac{1}{r^{2}}\pnth{\frac{M}{r} + M_{r}} - \frac{V_{r}}{r}\pnth{1 - \frac{2M}{r}}
\end{align}
\begin{align}
  \notag \Ric(\nu_{\varphi},\nu_{\varphi}) &= \frac{1}{r^{2}\sin^{2}\theta} \pnth{\Ga_{\varphi \varphi \ ,k}^{\ \ \ k} - \Ga_{k\varphi \ ,\varphi }^{\ \ \ k} + \Ga_{\varphi \varphi }^{\ \ \ s}\Ga_{ks}^{\ \ k} - \Ga_{k\varphi }^{\ \ \ s}\Ga_{\varphi s}^{\ \ \ k}} \displaybreak[0] \\
  \notag &= \frac{1}{r^{2}\sin^{2}\theta} \Bigg\{\p_{r}\pnth{-r\sin^{2}\theta\pnth{1 - \frac{2M}{r}}} + \p_{\theta}\pnth{-\sin\theta\cos\theta} \\
  \notag &\qquad - r\sin^{2}\theta \pnth{1 - \frac{2M}{r}}\pnth{V_{r} + \frac{2}{r}} - r\sin^{2}\theta\pnth{\frac{M_{r}}{r} - \frac{M}{r^{2}}}\\
  \notag &\qquad  - \cos^{2}\theta + 2\sin^{2}\theta \pnth{1 - \frac{2M}{r}} + 2\cos^{2}\theta \Bigg\} \displaybreak[0] \\
  \notag &= \frac{1}{r^{2}\sin^{2}\theta} \Bigg\{-\sin^{2}\theta\pnth{1 - \frac{2M}{r}} - r\sin^{2}\theta\pnth{\frac{2M}{r^{2}} - \frac{2M_{r}}{r}} \\
  \notag &\qquad - r\sin^{2}\theta \pnth{1 - \frac{2M}{r}}\pnth{V_{r} + \frac{2}{r}} + r\sin^{2}\theta\pnth{\frac{M}{r^{2}} - \frac{M_{r}}{r}} \\
  \notag &\qquad  - \cos^{2}\theta - \cos^{2}\theta + \sin^{2}\theta + 2\sin^{2}\theta \pnth{1 - \frac{2M}{r}} + 2\cos^{2}\theta \Bigg\} \displaybreak[0] \\
  \notag &= \frac{1}{r^{2}}\Bigg(\frac{2M}{r} - \frac{M}{r} + M_{r} - rV_{r}\pnth{1 - \frac{2M}{r}} \\
  \notag &\qquad - 2\pnth{1 - \frac{2M}{r}} + 2\pnth{1 - \frac{2M}{r}}\Bigg) \displaybreak[0]\\
  \label{Ricthth} \Ric(\nu_{\varphi},\nu_{\varphi}) &= \frac{1}{r^{2}}\pnth{\frac{M}{r} + M_{r}} - \frac{V_{r}}{r}\pnth{1 - \frac{2M}{r}} = \Ric(\nu_{\theta},\nu_{\theta}).
\end{align}
These are the only nonzero components of the Ricci curvature tensor since
\begin{align}
  \notag \Ric(\nu_{t},\nu_{\theta}) &= \frac{1}{r\e^{V}}\Ric_{t\theta} = \frac{1}{r\e^{V}}\R_{tk\theta}^{\ \ \ \, k} \\
  \notag &= \frac{1}{r\e^{V}}\pnth{\Ga_{t\theta\ ,k}^{\ \ \, k} - \Ga_{k\theta\ ,t}^{\ \ \, k} + \Ga_{t\theta}^{\ \ \, s}\Ga_{ks}^{\ \ k} - \Ga_{k\theta}^{\ \ \, s}\Ga_{ts}^{\ \ k}} \displaybreak[0]\\
  \notag &= \frac{1}{r\e^{V}}\pnth{- \Ga_{r\theta\ ,t}^{\ \ \, r} - \Ga_{\theta\theta\ ,t}^{\ \ \ \theta} - \Ga_{\varphi\theta\ ,t}^{\ \ \ \varphi} - \Ga_{k\theta}^{\ \ \, r}\Ga_{tr}^{\ \ k} - \Ga_{k\theta}^{\ \ \, \theta}\Ga_{t\theta}^{\ \ \, k} - \Ga_{k\theta}^{\ \ \, \varphi}\Ga_{t\varphi}^{\ \ \, k}} \displaybreak[0]\\
  \notag &= \frac{1}{r\e^{V}}\pnth{- \Ga_{\theta\theta}^{\ \ \ r}\Ga_{tr}^{\ \ \theta} - \Ga_{r\theta}^{\ \ \, \theta}\Ga_{t\theta}^{\ \ \, r} - \Ga_{\varphi\theta}^{\ \ \ \varphi}\Ga_{t\varphi}^{\ \ \, \varphi}} \\
  &= 0
\end{align}
\begin{align}
  \notag \Ric(\nu_{t},\nu_{\varphi}) &= \frac{1}{r\sin \theta \e^{V}}\Ric_{t\varphi} = \frac{1}{r\sin \theta\e^{V}}\R_{tk\varphi}^{\ \ \ \, k} \\
  \notag &= \frac{1}{r\sin \theta\e^{V}}\pnth{\Ga_{t\varphi\ ,k}^{\ \ \, k} - \Ga_{k\varphi\ ,t}^{\ \ \, k} + \Ga_{t\varphi}^{\ \ \, s}\Ga_{ks}^{\ \ k} - \Ga_{k\varphi}^{\ \ \, s}\Ga_{ts}^{\ \ k}} \displaybreak[0]\\
  \notag &= \frac{1}{r\sin \theta\e^{V}}\pnth{- \Ga_{r\varphi\ ,t}^{\ \ \, r} - \Ga_{\theta\varphi\ ,t}^{\ \ \ \theta} - \Ga_{\varphi\varphi\ ,t}^{\ \ \ \varphi} - \Ga_{k\varphi}^{\ \ \, r}\Ga_{tr}^{\ \ k} - \Ga_{k\varphi}^{\ \ \, \theta}\Ga_{t\theta}^{\ \ \, k} - \Ga_{k\varphi}^{\ \ \, \varphi}\Ga_{t\varphi}^{\ \ \, k}} \displaybreak[0]\\
  \notag &= \frac{1}{r\sin \theta\e^{V}}\pnth{- \Ga_{\varphi\varphi}^{\ \ \ r}\Ga_{tr}^{\ \ \varphi} - \Ga_{\varphi\varphi}^{\ \ \ \theta}\Ga_{t\theta}^{\ \ \, \varphi} - \Ga_{r\varphi}^{\ \ \, \varphi}\Ga_{t\varphi}^{\ \ \, r} - \Ga_{\theta\varphi}^{\ \ \ \varphi}\Ga_{t\varphi}^{\ \ \, \theta}} \\
  &= 0
\end{align}
\begin{align}
  \notag \Ric(\nu_{r},\nu_{\theta}) &= \frac{1}{r}\sqrt{1 - \frac{2M}{r}}\Ric_{r\theta} = \frac{1}{r}\sqrt{1 - \frac{2M}{r}}\R_{rk\theta}^{\ \ \ \, k} \\
  \notag &= \frac{1}{r}\sqrt{1 - \frac{2M}{r}}\pnth{\Ga_{r\theta\ ,k}^{\ \ \, k} - \Ga_{k\theta\ ,r}^{\ \ \, k} + \Ga_{r\theta}^{\ \ \, s}\Ga_{ks}^{\ \ k} - \Ga_{k\theta}^{\ \ \, s}\Ga_{rs}^{\ \ k}} \displaybreak[0] \\
  \notag &= \frac{1}{r}\sqrt{1 - \frac{2M}{r}}\pnth{\Ga_{r\theta\ ,\theta}^{\ \ \, \theta} - \Ga_{k\theta\ ,r}^{\ \ \, k} + \Ga_{r\theta}^{\ \ \, \theta}\Ga_{k\theta}^{\ \ k} - \Ga_{k\theta}^{\ \ \, s}\Ga_{rs}^{\ \ k}} \displaybreak[0]\\
  \notag &= \frac{1}{r}\sqrt{1 - \frac{2M}{r}}(- \Ga_{r\theta\ ,r}^{\ \ \, r} - \Ga_{\theta\theta\ ,r}^{\ \ \ \theta} - \Ga_{\varphi\theta\ ,r}^{\ \ \ \varphi} + \Ga_{r\theta}^{\ \ \, \theta}\Ga_{r\theta}^{\ \ \, r} + \Ga_{r\theta}^{\ \ \, \theta}\Ga_{\theta\theta}^{\ \ \ \theta} \\
  \notag &\qquad + \Ga_{r\theta}^{\ \ \, \theta}\Ga_{\varphi\theta}^{\ \ \varphi} - \Ga_{r\theta}^{\ \ \, s}\Ga_{rs}^{\ \ r} - \Ga_{\theta\theta}^{\ \ \ s}\Ga_{rs}^{\ \ \theta} - \Ga_{\varphi\theta}^{\ \ \ s}\Ga_{rs}^{\ \ \varphi}) \displaybreak[0] \\
  \notag &= \frac{1}{r}\sqrt{1 - \frac{2M}{r}}\pnth{\frac{\cos \theta}{r\sin \theta} - \Ga_{r\theta}^{\ \ \, \theta}\Ga_{r\theta}^{\ \ \, r} - \Ga_{\theta\theta}^{\ \ \ r}\Ga_{rr}^{\ \ \theta} - \Ga_{\varphi\theta}^{\ \ \ \varphi}\Ga_{r\varphi}^{\ \ \, \varphi}} \displaybreak[0] \\
  \notag &= \frac{1}{r}\sqrt{1 - \frac{2M}{r}}\pnth{\frac{\cos \theta}{r\sin \theta} - \frac{\cos \theta}{r\sin\theta}} \\
  &= 0
\end{align}
\begin{align}
  \notag \Ric(\nu_{r},\nu_{\varphi}) &= \frac{1}{r\sin \theta}\sqrt{1 - \frac{2M}{r}}\Ric_{r\varphi} = \frac{1}{r\sin \theta}\sqrt{1 - \frac{2M}{r}}\R_{rk\varphi}^{\ \ \ \, k} \\
  \notag &= \frac{1}{r\sin\theta}\sqrt{1 - \frac{2M}{r}}\pnth{\Ga_{r\varphi\ ,k}^{\ \ \, k} - \Ga_{k\varphi\ ,r}^{\ \ \, k} + \Ga_{r\varphi}^{\ \ \, s}\Ga_{ks}^{\ \ k} - \Ga_{k\varphi}^{\ \ \, s}\Ga_{rs}^{\ \ k}} \displaybreak[0] \\
  \notag &= \frac{1}{r\sin\theta}\sqrt{1 - \frac{2M}{r}}\pnth{\Ga_{r\varphi\ ,\varphi}^{\ \ \, \varphi} - \Ga_{k\varphi\ ,r}^{\ \ \, k} + \Ga_{r\varphi}^{\ \ \, \varphi}\Ga_{k\varphi}^{\ \ k} - \Ga_{k\varphi}^{\ \ \, s}\Ga_{rs}^{\ \ k}} \displaybreak[0] \\
  \notag &= \frac{1}{r\sin\theta}\sqrt{1 - \frac{2M}{r}}(- \Ga_{r\varphi\ ,r}^{\ \ \, r} - \Ga_{\theta\varphi\ ,r}^{\ \ \ \theta} - \Ga_{\varphi\varphi\ ,r}^{\ \ \ \varphi} + \Ga_{r\varphi}^{\ \ \, \varphi}\Ga_{r\varphi}^{\ \ r} + \Ga_{r\varphi}^{\ \ \, \varphi}\Ga_{\theta\varphi}^{\ \ \ \theta} \\
  \notag &\qquad + \Ga_{r\varphi}^{\ \ \, \varphi}\Ga_{\varphi\varphi}^{\ \ \ \varphi} - \Ga_{r\varphi}^{\ \ \, s}\Ga_{rs}^{\ \ r} - \Ga_{\theta\varphi}^{\ \ \ s}\Ga_{rs}^{\ \ \theta} - \Ga_{\varphi\varphi}^{\ \ \ s}\Ga_{rs}^{\ \ \varphi}) \displaybreak[0] \\
  \notag &= \frac{1}{r\sin\theta}\sqrt{1 - \frac{2M}{r}}\pnth{- \Ga_{r\varphi}^{\ \ \, \varphi}\Ga_{r\varphi}^{\ \ \, r} - \Ga_{\theta\varphi}^{\ \ \ \varphi}\Ga_{r\varphi}^{\ \ \, \theta} - \Ga_{\varphi\varphi}^{\ \ \ r}\Ga_{rr}^{\ \ \varphi} - \Ga_{\varphi\varphi}^{\ \ \ \theta}\Ga_{r\theta}^{\ \ \, \varphi}} \\
  &= 0
\end{align}
\begin{align}
  \notag \Ric(\nu_{\theta},\nu_{\varphi}) &= \frac{1}{r^{2}\sin\theta}\Ric_{\theta\varphi} = \frac{1}{r^{2}\sin \theta}\R_{\theta k \varphi}^{\ \ \ \ k} \\
  \notag &= \frac{1}{r^{2}\sin\theta}\pnth{\Ga_{\theta\varphi\ ,k}^{\ \ \ k} - \Ga_{k\varphi\ ,\theta}^{\ \ \, k} + \Ga_{\theta\varphi}^{\ \ \ s}\Ga_{ks}^{\ \ k} - \Ga_{k\varphi}^{\ \ \, s}\Ga_{\theta s}^{\ \ \, k}} \displaybreak[0] \\
  \notag &= \frac{1}{r^{2}\sin\theta}\pnth{\Ga_{\theta\varphi\ ,\varphi}^{\ \ \ \varphi} - \Ga_{k\varphi\ ,\theta}^{\ \ \, k} + \Ga_{\theta\varphi}^{\ \ \ \varphi}\Ga_{k\varphi}^{\ \ k} - \Ga_{k\varphi}^{\ \ \, s}\Ga_{\theta s}^{\ \ \, k}} \displaybreak[0] \\
  \notag &= \frac{1}{r^{2}\sin\theta}(- \Ga_{r\varphi\ ,\theta}^{\ \ \, r} - \Ga_{\theta\varphi\ ,\theta}^{\ \ \ \theta} - \Ga_{\varphi\varphi\ ,\theta}^{\ \ \ \varphi} + \Ga_{\theta\varphi}^{\ \ \ \varphi}\Ga_{r\varphi}^{\ \ \, r} + \Ga_{\theta\varphi}^{\ \ \ \varphi}\Ga_{\theta\varphi}^{\ \ \ \theta} \\
  \notag &\qquad + \Ga_{\theta\varphi}^{\ \ \ \varphi}\Ga_{\varphi\varphi}^{\ \ \ \varphi} - \Ga_{k\varphi}^{\ \ \, r}\Ga_{\theta r}^{\ \ \, k} - \Ga_{k\varphi}^{\ \ \, \theta}\Ga_{\theta \theta}^{\ \ \ k} - \Ga_{k\varphi}^{\ \ \, \varphi}\Ga_{\theta \varphi}^{\ \ \ k}) \displaybreak[0] \\
  \notag &= \frac{1}{r^{2}\sin\theta}\pnth{- \Ga_{\theta\varphi}^{\ \ \ r}\Ga_{\theta r}^{\ \ \, \theta} - \Ga_{r\varphi}^{\ \ \, \theta}\Ga_{\theta \theta}^{\ \ \ r} - \Ga_{\varphi\varphi}^{\ \ \ \varphi}\Ga_{\theta \varphi}^{\ \ \ \varphi}} \\
  &= 0
\end{align}
The other components are 0 by the symmetry of the Ricci curvature tensor.  To get the last statement, we note that by equation (\ref{Ricthth}) and Lemma \ref{RicLem-1}-\ref{RicLem-1b}, we have that
\begin{equation}
  R = \sum_{\eta} g(\nu_{\eta},\nu_{\eta})\Ric(\nu_{\eta},\nu_{\eta}) = -\Ric(\nu_{t},\nu_{t}) + \Ric(\nu_{r},\nu_{r}) + 2\Ric(\nu_{\theta},\nu_{\theta}).
\end{equation}
From here, it is a matter of algebra and the statements above to get the desired equation for $R$.
\end{proof}

Next, we prove Proposition \ref{EinEq-prop1}.

\begin{proof}[Proof of Proposition \ref{EinEq-prop1}]

In solving equations (\ref{divfreeT}), (\ref{EDEq-1-thm}), and (\ref{PREq-1-thm}), we already solve the equations resulting from two of the components of the Einstein equation, namely (\ref{EDEq-1-thm}) and (\ref{PREq-1-thm}), which we reprint here for convenience.
  \begin{align}
    \label{EDEq-1-rp} M_{r} &= 4\pi r^{2}\mu \\
    \label{PREq-1-rp} V_{r} &= \pnth{1 - \frac{2M}{r}}^{-1}\pnth{\frac{M}{r^{2}} + 4\pi r P}
  \end{align}
In order to show that solving these two equations coupled with $\divx{g} T = 0$ is sufficient to solve the entire Einstein equation, we must show that if these equations hold, so do the other components of the Einstein equation.  To do so, we first need to write down the other components of the Einstein equation.  Recall that the above equations come from the $\nu_{t},\nu_{t}$ and $\nu_{r},\nu_{r}$ components of the Einstein equation.  Thus we only have the $\nu_{t},\nu_{r}$ and $\nu_{\theta},\nu_{\theta}$ components left to compute (the $\nu_{\varphi},\nu_{\varphi}$ component is identical to $\nu_{\theta},\nu_{\theta}$ component).  Then, by Corollary \ref{GCor-1} and equation (\ref{SETFun-def}), we have the following.
\begin{align}
  \notag G(\nu_{t},\nu_{r}) &= 8\pi T(\nu_{t},\nu_{r}) \\
  \notag \frac{2M_{t}}{r^{2} \e^{V}}\pnth{1 - \frac{2M}{r}}^{-1/2} &= 8\pi \rho \\
  \label{rhoEq-1} M_{t} &= 4\pi r^{2} \e^{V}\rho\sqrt{1 - \frac{2M}{r}}
\end{align}
and
\begin{align}
  \notag G(\nu_{\theta},\nu_{\theta}) &= 8\pi T(\nu_{\theta},\nu_{\theta}) \\
  \notag 8\pi Q &= \pnth{V_{rr} + V_{r}^{2} + \frac{V_{r}}{r}}\pnth{1-\frac{2M}{r}} + \pnth{\frac{M}{r} - M_{r}}\pnth{\frac{1}{r^{2}} + \frac{V_{r}}{r}} \\
  \label{QEq-1} &\qquad + \frac{V_{t}M_{t} - M_{tt}}{r \e^{2V}}\pnth{1 - \frac{2M}{r}}^{-1} - \frac{3M_{t}^{2}}{r^{2} \e^{2V}}\pnth{1-\frac{2M}{r}}^{-2}
\end{align}

Next 
we compute what $\divx{g} T = 0$
means in terms of the functions $\mu$, $P$, $\rho$, and $Q$.  In what follows, the summing indices run through the set $\{t,r,\theta,\varphi\}$ and we will use the Einstein summation convention wherever it applies, while specifically denoting any other summations of a different form.  First, we define
\begin{equation}
  \vep_{k} = g(\nu_{k},\nu_{k})
\end{equation}
 for all $k \in \{t,r,\theta,\varphi\}$.  Then we can write the divergence of $T$ as follows.
\begin{align}
  \notag \divx{g} T &= \sum_{k} \vep_{k}(\cov_{\nu_{k}}T)(\nu_{k},\p_{j})\, dx^{j} \\
  \label{divT1} &= -(\cov_{\nu_{t}}T)(\nu_{t},\p_{j})\, dx^{j} + (\cov_{\nu_{r}}T)(\nu_{r},\p_{j})\, dx^{j} + (\cov_{\nu_{\theta}}T)(\nu_{\theta},\p_{j})\, dx^{j} + (\cov_{\nu_{\varphi}}T)(\nu_{\varphi},\p_{j})\, dx^{j}
\end{align}
We will simplify each of these four terms individually.  By equations (\ref{CS-2}), (\ref{norvec}), and (\ref{SETFun-def}), we have that
\begin{align}
  \notag (\cov_{\nu_{t}}T)(\nu_{t},\p_{k})\, dx^{k} &=  \e^{-2V}(\cov_{t}T)(\p_{t},\p_{k})\, dx^{k} \\
  \notag &=  \e^{-2V}\Big[\p_{t}(T(\p_{t},\p_{k})) - \Ga_{tt}^{\ \ m}T(\p_{m},\p_{k}) - \Ga_{tk}^{\ \ m}T(\p_{t},\p_{m})\Big]\, dx^{k} \displaybreak[0]\\
  \notag &=  \e^{-2V}\Big[\p_{t}(T(\p_{t},\p_{t}))\, dt + \p_{t}(T(\p_{t},\p_{r}))\, dr - 2\Ga_{tt}^{\ \ t}T(\p_{t},\p_{t})\, dt - 2\Ga_{tt}^{\ \ r}T(\p_{t},\p_{r})\, dt \\
  \notag &\qquad - (\Ga_{tt}^{\ \ t} + \Ga_{tr}^{\ \ r})T(\p_{t},\p_{r})\, dr - \Ga_{tr}^{\ \ t}T(\p_{t},\p_{t})\, dr - \Ga_{tt}^{\ \ r}T(\p_{r},\p_{r})\, dr\Big] \displaybreak[0]\\
  \notag &=  \e^{-2V}\Bigg[\p_{t}\pnth{ \e^{2V}\mu}\, dt + \p_{t}\pnth{ \e^{V}\rho\pnth{1 - \frac{2M}{r}}^{-1/2}}\, dr - 2V_{t} \e^{2V}\mu\, dt  \\
  \notag &\qquad - 2\rho V_{r} \e^{3V}\sqrt{1 - \frac{2M}{r}}\, dt -  \e^{V}\rho\pnth{1 - \frac{2M}{r}}^{-1/2} \pnth{V_{t} + \frac{M_{t}}{r}\pnth{1 - \frac{2M}{r}}^{-1}}\, dr \\
  \notag &\qquad - V_{r} \e^{2V}\mu\, dr - V_{r} \e^{2V}P\, dr \Bigg] \displaybreak[0]\\
  \notag &=  \e^{-2V}\Bigg[\pnth{2V_{t} \e^{2V}\mu +  \e^{2V}\mu_{t} - 2V_{t} \e^{2V}\mu - 2\rho V_{r} \e^{3V}\sqrt{1 - \frac{2M}{r}}}\, dt \\
  \notag &\qquad + \left(V_{t} \e^{V}\rho\pnth{1 - \frac{2M}{r}}^{-1/2} +  \e^{V}\rho_{t}\pnth{1 - \frac{2M}{r}}^{-1/2} + \frac{ \e^{V}\rho M_{t}}{r}\pnth{1 - \frac{2M}{r}}^{-3/2} \right. \\
  \notag &\qquad \left. - \pnth{V_{t} + \frac{M_{t}}{r}\pnth{1 - \frac{2M}{r}}^{-1}} \e^{V}\rho\pnth{1 - \frac{2M}{r}}^{-1/2} - V_{r} \e^{2V}\mu - V_{r} \e^{2V}P\right)\, dr \Bigg] \displaybreak[0] \\
  \label{C_tT} (\cov_{\nu_{t}}T)(\nu_{t},\p_{k})\, dx^{k} &= \pnth{\mu_{t} - 2\rho V_{r} \e^{V}\sqrt{1 - \frac{2M}{r}}}\, dt  + \pnth{\rho_{t} \e^{-V}\pnth{1 - \frac{2M}{r}}^{-1/2} - V_{r}(\mu + P)}\, dr
\end{align}
Next, we have
\begin{align}
  \notag (\cov_{\nu_{r}}T)(\nu_{r},\p_{k})\, dx^{k} &= \pnth{1 - \frac{2M}{r}}(\cov_{r}T)(\p_{r},\p_{k})\, dx^{k} \\
  \notag &= \pnth{1 - \frac{2M}{r}}\Big[\p_{r}(T(\p_{r},\p_{k})) - \Ga_{rr}^{\ \ m}T(\p_{m},\p_{k}) - \Ga_{rk}^{\ \ m}T(\p_{r},\p_{m})\Big]\, dx^{k} \displaybreak[0] \\
  \notag &= \pnth{1 - \frac{2M}{r}}\Big[\p_{r}(T(\p_{r},\p_{t}))\, dt + \p_{r}(T(\p_{r},\p_{r}))\, dr - \pnth{\Ga_{rr}^{\ \ r} + \Ga_{rt}^{\ \ t}}T(\p_{t},\p_{r})\, dt \\
  \notag &\qquad - 2\Ga_{rr}^{\ \ t}T(\p_{t},\p_{r})\, dr - 2\Ga_{rr}^{\ \ r}T(\p_{r},\p_{r})\, dr - \Ga_{rr}^{\ \ t}T(\p_{t},\p_{t})\, dt - \Ga_{rt}^{\ \ r}T(\p_{r},\p_{r})\, dt \Big] \displaybreak[0] \\
  \notag &= \pnth{1 - \frac{2M}{r}}\Bigg[\p_{r}\pnth{\rho  \e^{V}\pnth{1 - \frac{2M}{r}}^{-1/2}}\, dt + \p_{r}\pnth{P\pnth{1 - \frac{2M}{r}}^{-1}}\, dr \\
  \notag &\qquad -\rho  \e^{V}\pnth{1 - \frac{2M}{r}}^{-1/2}\brkt{\pnth{1 - \frac{2M}{r}}^{-1}\pnth{\frac{M_{r}}{r} - \frac{M}{r^{2}}} + V_{r}}\, dt \\
  \notag &\qquad - \frac{2\rho M_{t}}{r  \e^{V}}\pnth{1 - \frac{2M}{r}}^{-5/2}\, dr - 2P\pnth{1 - \frac{2M}{r}}^{-2}\pnth{\frac{M_{r}}{r} - \frac{M}{r^{2}}}\, dr \\
  \notag &\qquad - \frac{\mu M_{t}}{r}\pnth{1 - \frac{2M}{r}}^{-2}\, dt - \frac{P M_{t}}{r}\pnth{1 - \frac{2M}{r}}^{-2}\, dt \Bigg] \displaybreak[0] \\
  \notag &= \pnth{1 - \frac{2M}{r}}\Bigg[\Bigg(\rho  \e^{V}\pnth{1 - \frac{2M}{r}}^{-3/2}\pnth{\frac{M_{r}}{r} - \frac{M}{r^{2}}} + \rho V_{r} \e^{V}\pnth{1 - \frac{2M}{r}}^{-1/2} \\
  \notag &\qquad + \rho_{r} \e^{V}\pnth{1 - \frac{2M}{r}}^{-1/2} - \frac{\mu M_{t}}{r}\pnth{1 - \frac{2M}{r}}^{-2} - \frac{P M_{t}}{r}\pnth{1 - \frac{2M}{r}}^{-2} \\
  \notag &\qquad -\rho  \e^{V}\pnth{1 - \frac{2M}{r}}^{-1/2}\brkt{\pnth{1 - \frac{2M}{r}}^{-1}\pnth{\frac{M_{r}}{r} - \frac{M}{r^{2}}} + V_{r}}\Bigg)\, dt \\
  \notag &\qquad + \Bigg(P_{r}\pnth{1 - \frac{2M}{r}}^{-1} + 2P\pnth{1 - \frac{2M}{r}}^{-2}\pnth{\frac{M_{r}}{r} - \frac{M}{r^{2}}} \\
  \notag &\qquad - \frac{2\rho M_{t}}{r  \e^{V}}\pnth{1 - \frac{2M}{r}}^{-5/2} - 2P\pnth{1 - \frac{2M}{r}}^{-2}\pnth{\frac{M_{r}}{r} - \frac{M}{r^{2}}} \Bigg)\, dr \Bigg] \displaybreak[0] \\
  \notag (\cov_{\nu_{r}}T)(\nu_{r},\p_{k})\, dx^{k} &= \pnth{\rho_{r} \e^{V}\sqrt{1 - \frac{2M}{r}} - \frac{M_{t}}{r}\pnth{1 - \frac{2M}{r}}^{-1}(\mu + P)}\, dt \\
  \label{C_rT} &\qquad + \pnth{P_{r} - \frac{2\rho M_{t}}{r  \e^{V}}\pnth{1 - \frac{2M}{r}}^{-3/2}}\, dr
\end{align}
Thirdly,
\begin{align}
  \notag (\cov_{\nu_{\theta}}T)(\nu_{\theta},\p_{k})\, dx^{k} &= \frac{1}{r^{2}}(\cov_{\theta}T)(\p_{\theta},\p_{k})\, dx^{k} \\
  \notag &= \frac{1}{r^{2}} \Big[\p_{\theta}(T(\p_{\theta},\p_{k})) - \Ga_{\theta\theta}^{\ \ m}T(\p_{m},\p_{k}) - \Ga_{\theta k}^{\ \ m}T(\p_{\theta},\p_{m})\Big]\, dx^{k} \displaybreak[0]\\
  \notag &= \frac{1}{r^{2}}\Big[-\Ga_{\theta\theta}^{\ \ r}T(\p_{r},\p_{t})\, dt - \Ga_{\theta\theta}^{\ \ r}T(\p_{r},\p_{r})\, dr - \Ga_{\theta r}^{\ \ \theta}T(\p_{\theta},\p_{\theta})\, dr \Big] \displaybreak[0]\\
  \notag &= \frac{1}{r^{2}}\brkt{r\rho  \e^{V}\sqrt{1 - \frac{2M}{r}}\, dt + r P \, dr - r Q\, dr} \\
  \label{C_thT} (\cov_{\nu_{\theta}}T)(\nu_{\theta},\p_{k})\, dx^{k} &= \frac{\rho  \e^{V}}{r}\sqrt{1 - \frac{2M}{r}}\, dt + \frac{P - Q}{r}\, dr
\end{align}
Lastly, we have that
\begin{align}
  \notag (\cov_{\nu_{\varphi}}T)(\nu_{\varphi},\p_{k})\, dx^{k} &= \frac{1}{r^{2}\sin^{2}\theta}(\cov_{\varphi}T)(\p_{\varphi},\p_{k})\, dx^{k} \\
  \notag &= \frac{1}{r^{2}\sin^{2}\theta}\Big[ \p_{\varphi}(T(\p_{\varphi},\p_{k})) - \Ga_{\varphi\varphi}^{\ \ \, m}T(\p_{m},\p_{k}) - \Ga_{\varphi k}^{\ \ m}T(\p_{\varphi},\p_{m}) \Big]\, dx^{k} \displaybreak[0]\\
  \notag &= \frac{1}{r^{2}\sin^{2}\theta} \Big[ - \Ga_{\varphi\varphi}^{\ \ \, r}T(\p_{r},\p_{t})\, dt - \Ga_{\varphi\varphi}^{\ \ \, r}T(\p_{r},\p_{r})\, dr - \Ga_{\varphi\varphi}^{\ \ \, \theta}T(\p_{\theta},\p_{\theta})\, d\theta \\
  \notag &\qquad - \Ga_{\varphi r}^{\ \ \, \varphi}T(\p_{\varphi},\p_{\varphi})\, dr - \Ga_{\varphi\theta}^{\ \ \, \varphi}T(\p_{\varphi},\p_{\varphi})\, d\theta \Big] \displaybreak[0]\\
  \notag &= \frac{1}{r^{2}\sin^{2}\theta} \Big[ r\rho  \e^{V} \sin^{2}\theta \sqrt{1 - \frac{2M}{r}}\, dt + r \sin^{2}\theta P\, dr \\
  \notag &\qquad + r^{2}Q\sin\theta \cos\theta \, d\theta - r Q \sin^{2}\theta \, dr - r^{2}Q \cos \theta \sin \theta\, d\theta \Big] \displaybreak[0]\\
  \notag &= \frac{\rho  \e^{V}}{r}\sqrt{1 - \frac{2M}{r}}\, dt + \frac{P - Q}{r}\, dr \\
  \label{C_phT} (\cov_{\nu_{\varphi}}T)(\nu_{\varphi},\p_{k})\, dx^{k} &= (\cov_{\nu_{\theta}}T)(\nu_{\theta},\p_{k})\, dx^{k}
\end{align}
Then equation (\ref{divT1}) becomes
\begin{align}
  \notag \divx{g} T &= -(\cov_{\nu_{t}}T)(\nu_{t},\p_{j})\, dx^{j} + (\cov_{\nu_{r}}T)(\nu_{r},\p_{j})\, dx^{j} + (\cov_{\nu_{\theta}}T)(\nu_{\theta},\p_{j})\, dx^{j} + (\cov_{\nu_{\varphi}}T)(\nu_{\varphi},\p_{j})\, dx^{j} \\
  \notag &= -\pnth{\mu_{t} - 2\rho V_{r} \e^{V}\sqrt{1 - \frac{2M}{r}}}\, dt  - \pnth{\rho_{t} \e^{-V}\pnth{1 - \frac{2M}{r}}^{-1/2} - V_{r}(\mu + P)}\, dr \\
  \notag &\qquad + \pnth{\rho_{r} \e^{V}\sqrt{1 - \frac{2M}{r}} - \frac{M_{t}}{r}\pnth{1 - \frac{2M}{r}}^{-1}(\mu + P)}\, dt \\
  \notag &\qquad + \pnth{P_{r} - \frac{2\rho M_{t}}{r  \e^{V}}\pnth{1 - \frac{2M}{r}}^{-3/2}}\, dr + \frac{2\rho  \e^{V}}{r}\sqrt{1 - \frac{2M}{r}}\, dt + \frac{2(P - Q)}{r}\, dr \\
  \notag &= \pnth{-\mu_{t} + 2\rho  \e^{V}\sqrt{1 - \frac{2M}{r}}\pnth{V_{r} + \frac{1}{r}} + \rho_{r} \e^{V}\sqrt{1 - \frac{2M}{r}} - \frac{M_{t}}{r}\pnth{1 - \frac{2M}{r}}^{-1}(\mu + P)}\, dt \\
  \label{divT2} &\qquad + \pnth{-\rho_{t} \e^{-V}\pnth{1 - \frac{2M}{r}}^{-1/2} + V_{r}(\mu + P) + P_{r} - \frac{2\rho M_{t}}{r  \e^{V}}\pnth{1 - \frac{2M}{r}}^{-3/2} + \frac{2(P - Q)}{r}}\, dr
\end{align}
Then $\divx{g} T = 0$ yields the following two equations
\begin{subequations} \label{divT3}
  \begin{align}
    \label{divT3a} \mu_{t} &= 2\rho  \e^{V}\sqrt{1 - \frac{2M}{r}}\pnth{V_{r} + \frac{1}{r}} + \rho_{r} \e^{V}\sqrt{1 - \frac{2M}{r}} - \frac{M_{t}}{r}\pnth{1 - \frac{2M}{r}}^{-1}(\mu + P) \displaybreak[0]\\
    \label{divT3b} \rho_{t} &=  \e^{V}\sqrt{1 - \frac{2M}{r}}\pnth{V_{r}(\mu + P) + P_{r} + \frac{2(P - Q)}{r}} - \frac{2\rho M_{t}}{r}\pnth{1 - \frac{2M}{r}}^{-1}
  \end{align}
\end{subequations}

We are now ready to show that solving $\divx{g} T = 0$ 
along with solving equations (\ref{EDEq-1-rp}) and (\ref{PREq-1-rp}) on each $t$=constant slice 
also solves the remaining two unique nonzero components of the Einstein equation, (\ref{rhoEq-1}) and (\ref{QEq-1}).  We will show first that (\ref{EDEq-1-rp}), (\ref{PREq-1-rp}), and (\ref{divT3}) implies that (\ref{rhoEq-1}) holds as well.  To do this, we first need to show that (\ref{EDEq-1-rp}) and (\ref{rhoEq-1}) are compatible, that is, given these equations, $M_{rt} = M_{tr}$.  Differentiating (\ref{EDEq-1-rp}) with respect to $t$ yields,
\begin{align}
  \notag \p_{t}M_{r} &= \p_{t}\pnth{4\pi r^{2}\mu} \\
  \label{EDEq-1_t} M_{rt} &= 4\pi r^{2}\mu_{t}
\end{align}
while differentiating (\ref{rhoEq-1}) with respect to $r$ yields,
\begin{align}
  \notag \p_{r}M_{t} &= \p_{r}\pnth{4\pi r^{2} \e^{V}\rho\sqrt{1 - \frac{2M}{r}}} \\
  \notag M_{tr} &= 4\pi\Bigg[ 2r  \e^{V} \rho \sqrt{1 - \frac{2M}{r}} + r^{2}V_{r} \e^{V}\rho\sqrt{1 - \frac{2M}{r}} \\
  &\qquad + r^{2} \e^{V}\rho_{r}\sqrt{1 - \frac{2M}{r}} - r^{2} \e^{V}\rho\pnth{1 - \frac{2M}{r}}^{-1/2}\pnth{\frac{M_{r}}{r} - \frac{M}{r^{2}}} \Bigg] \\
\intertext{To make the simplification process easier, to the right hand side above, we will add in and subtract out a copy of the second term in the sum above, namely, the term $4\pi r^{2} V_{r} \e^{V}\rho\sqrt{1 - 2M/r}$.}
  \notag M_{tr} &= 4\pi\Bigg[ 2r  \e^{V} \rho \sqrt{1 - \frac{2M}{r}} + 2r^{2}V_{r} \e^{V}\rho\sqrt{1 - \frac{2M}{r}} - r^{2}V_{r} \e^{V}\rho\sqrt{1 - \frac{2M}{r}} \\
  \notag &\qquad + r^{2} \e^{V}\rho_{r}\sqrt{1 - \frac{2M}{r}} - r^{2} \e^{V}\rho\pnth{1 - \frac{2M}{r}}^{-1/2}\pnth{\frac{M_{r}}{r} - \frac{M}{r^{2}}} \Bigg] \displaybreak[0]\\
  \notag &= 4\pi r^{2} \Bigg[ 2\rho  \e^{V}\sqrt{1 - \frac{2M}{r}}\pnth{V_{r} +\frac{1}{r}} + \rho_{r} \e^{V}\sqrt{1 - \frac{2M}{r}} \\
  \notag &\qquad - V_{r} \e^{V}\rho\sqrt{1 - \frac{2M}{r}} -  \e^{V}\rho\pnth{1 - \frac{2M}{r}}^{-1/2}\pnth{\frac{M_{r}}{r} - \frac{M}{r^{2}}} \Bigg] \displaybreak[0]\\
  \notag &= 4\pi r^{2}\Bigg[\mu_{t} + \frac{M_{t}}{r}\pnth{1 - \frac{2M}{r}}^{-1}(\mu + P) \\
  \notag &\qquad -  \e^{V}\rho\pnth{1 - \frac{2M}{r}}^{-1/2}\pnth{\frac{M}{r^{2}} + 4\pi r P} -  \e^{V}\rho\pnth{1 - \frac{2M}{r}}^{-1/2}\pnth{4\pi r \mu - \frac{M}{r^{2}}} \Bigg] \displaybreak[0]\\
  \notag &= 4\pi r^{2}\Bigg[\mu_{t} + \frac{M_{t}}{r}\pnth{1 - \frac{2M}{r}}^{-1}(\mu + P) - 4\pi r \rho  \e^{V}\pnth{1 - \frac{2M}{r}}^{-1/2}(\mu + P) \Bigg] \\
  \label{rhoEq-1_r} &= 4\pi r^{2}\mu_{t}
\end{align}
where in the next to last line we made substitutions using equations (\ref{EDEq-1-rp}), (\ref{PREq-1-rp}), and (\ref{divT3a}) and the last line used (\ref{rhoEq-1}).  Thus if $\divx{g}T=0$, (\ref{EDEq-1-rp}) and (\ref{rhoEq-1}) are compatible, that is, $M_{tr} = M_{rt}$ everywhere the equations are defined, namely, wherever $r \neq 0$.  However, by using \LHopital's rule on the equations above coupled with the fact that at $r = 0$, $M = M_{r} = M_{rr} = 0$ for all $t$, the equation $M_{tr} = M_{rt}$ still holds at the central value $r=0$.  This implies that there exists a function $M(t,r)$ which satisfies both (\ref{EDEq-1-rp}) and (\ref{PREq-1-rp}) everywhere, which of course is the metric function we seek.

As per our hypothesis, if for all values of $t$, we have solved (\ref{EDEq-1-rp}) with the initial condition $M = 0$ at $r=0$, we will obtain some function $M^{\ast}(t,r)$ which satisfies (\ref{EDEq-1-rp}) everywhere.  Then we have that 
  \begin{equation}
    M_{r}^{\ast} = M_{r}
  \end{equation}
everywhere, where $M$ is the function that satisfies both (\ref{EDEq-1-rp}) and (\ref{PREq-1-rp}).  This implies that 
  \begin{equation}
    M^{\ast}(t,r) = M(t,r) + f(t)
  \end{equation}
for some smooth function $f(t)$.  However, we also have that $M^{\ast}(t,0) = M(t,0) = 0$ for all $t$, which implies that $f(t) \equiv 0$.  Hence $M^{\ast}(t,r) = M(t,r)$ and the function we obtain by integrating (\ref{EDEq-1-rp}) with compatible initial conditions necessarily also satisfies (\ref{rhoEq-1}).

Finally, we use equations (\ref{EDEq-1-rp}), (\ref{PREq-1-rp}), (\ref{divT3}), and (\ref{rhoEq-1}) (since the first three imply equation (\ref{rhoEq-1})) to show that (\ref{QEq-1}) is automatically satisfied.  In order to do this, we will have to compute $V_{rr}$ and $M_{tt}$.  They are
\begin{align}
  \notag \p_{t} M_{t} &= \p_{t}\pnth{4\pi r^{2} \e^{V}\rho\sqrt{1 - \frac{2M}{r}}} \\
  \notag M_{tt} &= 4\pi r^{2}\brkt{V_{t} \e^{V}\rho \sqrt{1 - \frac{2M}{r}} +  \e^{V}\rho_{t}\sqrt{1 - \frac{2M}{r}} - \frac{ \e^{V} \rho M_{t}}{r}\pnth{1 - \frac{2M}{r}}^{-1/2}} \\
  \label{rhoEq-1_t} &= V_{t}M_{t} + 4\pi r^{2}\e^{V}\brkt{\rho_{t}\sqrt{1 - \frac{2M}{r}} - \frac{\rho M_{t}}{r}\pnth{1 - \frac{2M}{r}}^{-1/2}}
\end{align}
and
\begin{align}
  \notag \p_{r}V_{r} &= \p_{r}\brkt{\pnth{1 - \frac{2M}{r}}^{-1}\pnth{\frac{M}{r^{2}} + 4\pi r P}} \\
  \notag V_{rr} &= 2\pnth{1 - \frac{2M}{r}}^{-2}\pnth{\frac{M_{r}}{r} - \frac{M}{r^{2}}}\pnth{\frac{M}{r^{2}} + 4\pi r P} + \pnth{1 - \frac{2M}{r}}^{-1}\pnth{\frac{M_{r}}{r^{2}} - \frac{2M}{r^{3}} + 4\pi P + 4\pi r P_{r}} \\
  \notag &= 2V_{r}\pnth{1 - \frac{2M}{r}}^{-1}\pnth{\frac{M_{r}}{r} - \frac{M}{r^{2}}} + \pnth{1 - \frac{2M}{r}}^{-1}\pnth{\frac{M_{r}}{r^{2}} - \frac{M}{r^{3}} - \frac{M}{r^{3}} + 4\pi P + 4\pi r P_{r}} \\
  \notag &= 2V_{r}\pnth{1 - \frac{2M}{r}}^{-1}\pnth{\frac{M_{r}}{r} - \frac{M}{r^{2}}} - \frac{V_{r}}{r} + \pnth{1 - \frac{2M}{r}}^{-1}\pnth{\frac{M_{r}}{r^{2}} - \frac{M}{r^{3}} + 8\pi P + 4\pi r P_{r}} \\
  \label{PREq-1_r} V_{rr} &= \pnth{2\frac{V_{r}}{r} + \frac{1}{r^{2}}}\pnth{1 - \frac{2M}{r}}^{-1}\pnth{M_{r} - \frac{M}{r}} - \frac{V_{r}}{r} + 4\pi \pnth{1 - \frac{2M}{r}}^{-1}(2P + rP_{r})
\end{align}
Making these substitutions into (\ref{QEq-1}) yields
\begin{align}
  \notag 8\pi Q &= \pnth{V_{rr} + V_{r}^{2} + \frac{V_{r}}{r}}\pnth{1-\frac{2M}{r}} + \pnth{\frac{M}{r} - M_{r}}\pnth{\frac{1}{r^{2}} + \frac{V_{r}}{r}} \\
  \notag &\qquad + \frac{V_{t}M_{t} - M_{tt}}{r \e^{2V}}\pnth{1 - \frac{2M}{r}}^{-1} - \frac{3M_{t}^{2}}{r^{2} \e^{2V}}\pnth{1-\frac{2M}{r}}^{-2} \displaybreak[0] \\
  \notag &= 4\pi(2P + rP_{r}) + V_{r}^{2}\pnth{1 - \frac{2M}{r}} + \frac{V_{r}}{r}\pnth{M_{r} - \frac{M}{r}}  \\
  \notag &\qquad -\frac{4\pi r}{\e^{V}}\brkt{\rho_{t}\pnth{1 - \frac{2M}{r}}^{-1/2} - \frac{\rho M_{t}}{r}\pnth{1 - \frac{2M}{r}}^{-3/2}} - \frac{3M_{t}^{2}}{r^{2} \e^{2V}}\pnth{1-\frac{2M}{r}}^{-2} \displaybreak[0] \\
  \notag &= 4\pi(2P + rP_{r}) + V_{r}\pnth{\frac{M}{r^{2}} + 4\pi r P} + \frac{V_{r}}{r}\pnth{M_{r} - \frac{M}{r}}  \\
  \notag &\qquad -\frac{4\pi r}{\e^{V}}\brkt{\rho_{t}\pnth{1 - \frac{2M}{r}}^{-1/2} - \frac{\rho M_{t}}{r}\pnth{1 - \frac{2M}{r}}^{-3/2}} - \frac{12\pi \rho M_{t}}{\e^{V}}\pnth{1-\frac{2M}{r}}^{-3/2} \displaybreak[0] \\
  \notag &= 4\pi(2P + rP_{r}) + 4\pi r V_{r}P + \frac{V_{r}M_{r}}{r} -\frac{4\pi r \rho_{t}}{\e^{V}}\pnth{1 - \frac{2M}{r}}^{-1/2} - \frac{8\pi \rho M_{t}}{\e^{V}}\pnth{1-\frac{2M}{r}}^{-3/2} \displaybreak[0] \\
  \notag &= 4\pi(2P + rP_{r}) + 4\pi r V_{r}(P + \mu) - \frac{4\pi r \rho_{t}}{\e^{V}}\pnth{1 - \frac{2M}{r}}^{-1/2} - \frac{8\pi \rho M_{t}}{\e^{V}}\pnth{1-\frac{2M}{r}}^{-3/2} \displaybreak[0] \\
  \notag &= 4\pi(2P + rP_{r}) + 4\pi r V_{r}(P + \mu) - \frac{8\pi \rho M_{t}}{\e^{V}}\pnth{1-\frac{2M}{r}}^{-3/2} \\
  \notag &\qquad - 4\pi r \pnth{V_{r}(\mu + P) + P_{r} + \frac{2(P - Q)}{r}} + \frac{8\pi\rho M_{t}}{\e^{V}}\pnth{1 - \frac{2M}{r}}^{-3/2} \\
  \label{QEq-2} &= 8\pi Q
\end{align}
where the last two lines follow from using (\ref{divT3b}).  Thus so long as equation (\ref{divT3}) holds, equations (\ref{EDEq-1-rp}), (\ref{PREq-1-rp}), and (\ref{rhoEq-1}) imply equation (\ref{QEq-1}).  Since equations (\ref{divT3}), (\ref{EDEq-1-rp}), and (\ref{PREq-1-rp}) imply equation (\ref{rhoEq-1}) and equation (\ref{divT3}) follows from $\divx{g}T = 0$, we have that solving $\divx{g}T = 0$ 
and solving equations (\ref{EDEq-1-rp}) and (\ref{PREq-1-rp}) at each time $t$ also solves the entire Einstein equation, which was the desired result.
\end{proof}

The proof of Lemma \ref{SET-Lem-1} follows.

\begin{proof}[Proof of Lemma \ref{SET-Lem-1}]
  Recall equation (\ref{SET-Def}).  Then to compute the components of $T$, we first have that
    \begin{align}
      df &= f_{t}\, dt + f_{r}\, dr = p\e^{V}\sqrt{1 - \frac{2M}{r}}\, dt + f_{r}\, dr \\
      d\cf &= \cf_{t}\, dt + \cf_{r}\, dr = \cp\e^{V}\sqrt{1 - \frac{2M}{r}}\, dt + \cf_{r}\, dr.
    \end{align}
    Then, since $g$ is diagonal, this implies that
    \begin{align}\label{df2}
      \notag \abs{df}^{2} &= g(df,d\cf) = g\pnth{p\e^{V}\sqrt{1 - \frac{2M}{r}}\, dt + f_{r}\, dr,\, \cp\e^{V}\sqrt{1 - \frac{2M}{r}}\, dt + \cf_{r}\, dr} \\
      \notag &= \e^{2V}\pnth{1 - \frac{2M}{r}}p\cp\, g(dt,dt) + f_{r} \cf_{r}\, g(dr,dr) \\
      &= \pnth{1 - \frac{2M}{r}}\pnth{\abs{f_{r}}^{2} - \abs{p}^{2}}.
    \end{align}
    With these facts, we now compute the quantities in question and do so in the same order as they were presented.  Thus we have the following.
    \begin{align}
      \notag T(\nu_{t},\nu_{t}) &= \mu_{0}\pnth{\frac{2}{\Up^{2}} df(\nu_{t})d\cf(\nu_{t}) - \brkt{\pnth{1 - \frac{2M}{r}}\frac{\abs{f_{r}}^{2} - \abs{p}^{2}}{\Up^2} + \abs{f}^{2}}g(\nu_{t},\nu_{t})} \\
      \notag &= \mu_{0}\pnth{\frac{2}{\Up^{2}}\pnth{1 - \frac{2M}{r}}p\cp + \abs{f}^{2} + \pnth{1 - \frac{2M}{r}}\frac{\abs{f_{r}}^{2} - \abs{p}^{2}}{\Up^{2}}} \\
      &= \mu_{0}\pnth{\abs{f}^{2} + \pnth{1 - \frac{2M}{r}}\frac{\abs{f_{r}}^{2} + \abs{p}^{2}}{\Up^{2}}}.
    \end{align}
    Next, because $\nu_{\eta}$ is an everywhere orthonormal basis,
    \begin{align}
      \notag T(\nu_{t},\nu_{r}) &= \mu_{0}\Bigg(\frac{1}{\Up^{2}} \brkt{df(\nu_{t})d\cf(\nu_{r}) + d\cf(\nu_{t})df(\nu_{r})} \\
      \notag &\qquad - \brkt{\pnth{1 - \frac{2M}{r}}\frac{\abs{f_{r}}^{2} - \abs{p}^{2}}{\Up^2} + \abs{f}^{2}}g(\nu_{t},\nu_{r})\Bigg) \displaybreak[0] \\
      \notag &= \mu_{0}\pnth{\frac{1}{\Up^{2}}\pnth{1 - \frac{2M}{r}}\pnth{p\cf_{r} + f_{r}\cp}} \\
      &= \frac{2\mu_{0}}{\Up^{2}}\pnth{1 - \frac{2M}{r}}\Re(f_{r}\cp).
    \end{align}
    \begin{align}
      \notag T(\nu_{r},\nu_{r}) &= \mu_{0}\pnth{\frac{2}{\Up^{2}} df(\nu_{r})d\cf(\nu_{r}) - \brkt{\pnth{1 - \frac{2M}{r}}\frac{\abs{f_{r}}^{2} - \abs{p}^{2}}{\Up^2} + \abs{f}^{2}}g(\nu_{r},\nu_{r})} \\
      \notag &= \mu_{0}\pnth{\frac{2}{\Up^{2}} \pnth{1 - \frac{2M}{r}}f_{r}\cf_{r} - \brkt{\pnth{1 - \frac{2M}{r}}\frac{\abs{f_{r}}^{2} - \abs{p}^{2}}{\Up^2} + \abs{f}^{2}}} \\
      \notag &= \mu_{0}\pnth{-\abs{f}^{2} + \pnth{1 - \frac{2M}{r}}\frac{\abs{f_{r}}^{2} + \abs{p}^{2}}{\Up^{2}}}.
    \end{align}
    \begin{align}
      \notag T(\nu_{\theta},\nu_{\theta}) &= \mu_{0}\pnth{\frac{2}{\Up^{2}} df(\nu_{\theta})d\cf(\nu_{\theta}) - \brkt{\pnth{1 - \frac{2M}{r}}\frac{\abs{f_{r}}^{2} - \abs{p}^{2}}{\Up^2} + \abs{f}^{2}}g(\nu_{\theta},\nu_{\theta})} \\
      &= -\mu_{0}\pnth{\abs{f}^{2} + \pnth{1 - \frac{2M}{r}}\frac{\abs{f_{r}}^{2} - \abs{p}^{2}}{\Up^{2}}}.
    \end{align}
    Finally,
    \begin{align}
      \notag T(\nu_{\varphi},\nu_{\varphi}) &= \mu_{0}\pnth{\frac{2}{\Up^{2}} df(\nu_{\varphi})d\cf(\nu_{\varphi}) - \brkt{\pnth{1 - \frac{2M}{r}}\frac{\abs{f_{r}}^{2} - \abs{p}^{2}}{\Up^2} + \abs{f}^{2}}g(\nu_{\varphi},\nu_{\varphi})} \\
      &= -\mu_{0}\pnth{\abs{f}^{2} + \pnth{1 - \frac{2M}{r}}\frac{\abs{f_{r}}^{2} - \abs{p}^{2}}{\Up^{2}}}
    \end{align}
    so that $T(\nu_{\varphi},\nu_{\varphi}) = T(\nu_{\theta},\nu_{\theta})$.
\end{proof}

Finally, we prove Lemma \ref{divTEKG-Lem}.

\begin{proof}[Proof of Lemma \ref{divTEKG-Lem}]
  By equation (\ref{SETFun-def}) and Lemma \ref{SET-Lem-1}, we have that
  \begin{subequations}\label{SETEKGFun}
    \begin{align}
      \label{SETEKGFun-1} \mu &= \mu_{0}\pnth{\abs{f}^{2} + \pnth{1 - \frac{2M}{r}}\frac{\abs{f_{r}}^{2} + \abs{p}^{2}}{\Up^{2}}} \\
      \label{SETEKGFun-2} \rho &= \frac{2\mu_{0}}{\Up^2}\pnth{1 - \frac{2M}{r}}\Re(f_{r}\cp) \\
      \label{SETEKGFun-3} P &= \mu_{0}\pnth{-\abs{f}^{2} + \pnth{1 - \frac{2M}{r}}\frac{\abs{f_{r}}^{2} + \abs{p}^{2}}{\Up^{2}}} \\
      \label{SETEKGFun-4} Q &= -\mu_{0}\pnth{\abs{f}^{2} + \pnth{1 - \frac{2M}{r}}\frac{\abs{f_{r}}^{2} - \abs{p}^{2}}{\Up^{2}}}
    \end{align}
  \end{subequations}
  We know from a previous proof that $\divx{g}T=0$ is equivalent to equation (\ref{divT3}).  Moreover, by (\ref{p-Def}) and Lemma \ref{KG-Lem-1}, we have that (\ref{EKG-2}) is equivalent to (\ref{NCpde4}) and (\ref{NCpde3}).  Thus it suffices to show that given (\ref{SETEKGFun}), equations (\ref{NCpde4}) and (\ref{NCpde3}) imply equation (\ref{divT3}).

  We will start with (\ref{divT3a}).  On one hand, the left hand side is equal to differentiating (\ref{SETEKGFun-1}) with respect to $t$.  \begin{subequations}\label{divTEKG2-a}
    \begin{align}
      \notag \mu_{t} &= \p_{t}\brkt{\mu_{0}\pnth{\abs{f}^{2} + \pnth{1 - \frac{2M}{r}}\frac{\abs{f_{r}}^{2} + \abs{p}^{2}}{\Up^{2}}}} \\
      \notag &= \mu_{0}\brkt{f_{t}\cf + f \cf_{t} - \frac{2M_{t}}{r}\pnth{\frac{\abs{f_{r}}^{2} + \abs{p}^{2}}{\Up^{2}}} + \pnth{1 - \frac{2M}{r}}\frac{f_{rt}\cf_{r} + f_{r}\cf_{rt} + p_{t}\cp + p\cp_{t}}{\Up^{2}}} \displaybreak[0]\\
      \notag &= \mu_{0}\Bigg[\e^{V}\sqrt{1 - \frac{2M}{r}}\pnth{p\cf + f\cp} - \frac{2M_{t}}{r}\pnth{\frac{\abs{f_{r}}^{2} + \abs{p}^{2}}{\Up^{2}}} \\ \notag &\qquad + \pnth{1 - \frac{2M}{r}}\frac{f_{rt}\cf_{r} + f_{r}\cf_{rt} + p_{t}\cp + p\cp_{t}}{\Up^{2}}\Bigg] \displaybreak[0]\\
      \notag \mu_{t} &= \mu_{0}\Bigg[\cp \pnth{f \e^{V}\sqrt{1 - \frac{2M}{r}} + \frac{p_{t}}{\Up^{2}}\pnth{1 - \frac{2M}{r}}} \\
      \notag &\qquad + p \pnth{\cf \e^{V}\sqrt{1 - \frac{2M}{r}} + \frac{\cp_{t}}{\Up^{2}}\pnth{1 - \frac{2M}{r}}} - \frac{2M_{t}}{r}\pnth{\frac{\abs{f_{r}}^{2} + \abs{p}^{2}}{\Up^{2}}} \\
      \label{divTEKG2-aLHS} &\qquad + \pnth{1 - \frac{2M}{r}}\frac{\cf_{r}\p_{r}\pnth{p \e^{V}\sqrt{1 - \frac{2M}{r}}} + f_{r}\p_{r}\pnth{\cp \e^{V}\sqrt{1 - \frac{2M}{r}}}}{\Up^{2}} \Bigg]
    \end{align}
    On the other hand, we can substitute (\ref{SETEKGFun}) into (\ref{divT3a}).
    \begin{align}
      \notag \mu_{t} &= 2\rho \e^{V}\sqrt{1 - \frac{2M}{r}}\pnth{V_{r} + \frac{1}{r}} + \rho_{r}\e^{V}\sqrt{1 - \frac{2M}{r}} - \frac{M_{t}}{r}\pnth{1 - \frac{2M}{r}}^{-1}(\mu + P) \\
      \notag &= 2\e^{V}\sqrt{1 - \frac{2M}{r}}\pnth{\frac{2\mu_{0}}{\Up^2}\pnth{1 - \frac{2M}{r}}\Re(f_{r}\cp)}\pnth{V_{r} + \frac{1}{r}} \\
      \notag &\qquad + \p_{r}\brkt{\frac{2\mu_{0}}{\Up^2}\pnth{1 - \frac{2M}{r}}\Re(f_{r}\cp)}\e^{V}\sqrt{1 - \frac{2M}{r}} - \frac{2M_{t}\mu_{0}}{r}\pnth{\frac{\abs{f_{r}}^{2} + \abs{p}^{2}}{\Up^{2}}} \displaybreak[0] \\
      \notag &= \frac{2\mu_{0}}{\Up^{2}}\e^{V}\pnth{1 - \frac{2M}{r}}^{3/2}(f_{r}\cp + p \cf_{r})\pnth{V_{r} + \frac{1}{r}} \\
      \notag &\qquad + \frac{\mu_{0}}{\Up^2}\p_{r}\brkt{\pnth{1 - \frac{2M}{r}}(f_{r}\cp + p \cf_{r})}\e^{V}\sqrt{1 - \frac{2M}{r}}  - \frac{2M_{t}\mu_{0}}{r}\pnth{\frac{\abs{f_{r}}^{2} + \abs{p}^{2}}{\Up^{2}}} \displaybreak[0] \\
      \notag &= \cp \frac{\mu_{0}}{\Up^{2}}\pnth{1 - \frac{2M}{r}} \brkt{f_{r}V_{r}\e^{V}\sqrt{1 - \frac{2M}{r}} + \frac{2f_{r}\e^{V}}{r}\sqrt{1 - \frac{2M}{r}}} \\
      \notag & \qquad + p \frac{\mu_{0}}{\Up^{2}}\pnth{1 - \frac{2M}{r}} \brkt{\cf_{r}V_{r}\e^{V}\sqrt{1 - \frac{2M}{r}} + \frac{2\cf_{r}\e^{V}}{r}\sqrt{1 - \frac{2M}{r}}} \displaybreak[0]\\
      \notag & \qquad + \frac{\mu_{0}V_{r}\e^{V}}{\Up^{2}}\pnth{1 - \frac{2M}{r}}^{3/2}\pnth{f_{r}\cp + p \cf_{r}} - \frac{2M_{t}\mu_{0}}{r}\pnth{\frac{\abs{f_{r}}^{2} + \abs{p}^{2}}{\Up^{2}}} \displaybreak[0]\\
      \notag & \qquad + \frac{\mu_{0}\e^{V}}{\Up^2}\sqrt{1 - \frac{2M}{r}}\Bigg[(f_{r}\cp + p \cf_{r})\, \p_{r}\pnth{1 - \frac{2M}{r}} \\
      \notag & \qquad + \pnth{1 - \frac{2M}{r}}(f_{rr}\cp + p \cf_{rr} + f_{r}\cp_{r} + p_{r} \cf_{r})\Bigg] \displaybreak[0] \\
      \notag &= \cp \frac{\mu_{0}}{\Up^{2}}\pnth{1 - \frac{2M}{r}} \brkt{f_{r}V_{r}\e^{V}\sqrt{1 - \frac{2M}{r}} + f_{rr}\e^{V}\sqrt{1 - \frac{2M}{r}} + \frac{2f_{r}\e^{V}}{r}\sqrt{1 - \frac{2M}{r}}} \\
      \notag & \qquad + p \frac{\mu_{0}}{\Up^{2}}\pnth{1 - \frac{2M}{r}} \brkt{\cf_{r}V_{r}\e^{V}\sqrt{1 - \frac{2M}{r}} + \cf_{rr}\e^{V}\sqrt{1 - \frac{2M}{r}} + \frac{2\cf_{r}\e^{V}}{r}\sqrt{1 - \frac{2M}{r}}} \displaybreak[0]\\
      \notag & \qquad + \frac{\mu_{0}}{\Up^{2}}\pnth{1 - \frac{2M}{r}}\brkt{V_{r}\e^{V}\sqrt{1 - \frac{2M}{r}}\pnth{f_{r}\cp + p \cf_{r}}} - \frac{2M_{t}\mu_{0}}{r}\pnth{\frac{\abs{f_{r}}^{2} + \abs{p}^{2}}{\Up^{2}}} \displaybreak[0]\\
      \notag & \qquad + \frac{\mu_{0}}{\Up^2}\pnth{1 - \frac{2M}{r}}\Bigg[\e^{V}(f_{r}\cp + p \cf_{r})\pnth{1 - \frac{2M}{r}}^{-1/2}\p_{r}\pnth{1 - \frac{2M}{r}} \\
      \notag & \qquad + \e^{V}\sqrt{1 - \frac{2M}{r}}(f_{r}\cp_{r} + p_{r} \cf_{r})\Bigg] \displaybreak[0] \\
      \notag &= \cp \frac{\mu_{0}}{\Up^{2}}\pnth{1 - \frac{2M}{r}} \brkt{f_{r}V_{r}\e^{V}\sqrt{1 - \frac{2M}{r}} + f_{rr}\e^{V}\sqrt{1 - \frac{2M}{r}} + \frac{2f_{r}\e^{V}}{r}\sqrt{1 - \frac{2M}{r}}} \\
      \notag & \qquad + p \frac{\mu_{0}}{\Up^{2}}\pnth{1 - \frac{2M}{r}} \brkt{\cf_{r}V_{r}\e^{V}\sqrt{1 - \frac{2M}{r}} + \cf_{rr}\e^{V}\sqrt{1 - \frac{2M}{r}} + \frac{2\cf_{r}\e^{V}}{r}\sqrt{1 - \frac{2M}{r}}} \displaybreak[0]\\
      \notag & \qquad + \frac{\mu_{0}}{\Up^{2}}\pnth{1 - \frac{2M}{r}}\brkt{V_{r}\e^{V}\sqrt{1 - \frac{2M}{r}}\pnth{f_{r}\cp + p \cf_{r}}} - \frac{2M_{t}\mu_{0}}{r}\pnth{\frac{\abs{f_{r}}^{2} + \abs{p}^{2}}{\Up^{2}}} \displaybreak[0]\\
      \notag & \qquad + \frac{\mu_{0}}{\Up^2}\pnth{1 - \frac{2M}{r}}\Bigg[2\e^{V}(f_{r}\cp + p \cf_{r})\, \p_{r}\pnth{\sqrt{1 - \frac{2M}{r}}} \\
      \notag &\qquad + \e^{V}\sqrt{1 - \frac{2M}{r}}(f_{r}\cp_{r} + p_{r} \cf_{r})\Bigg] \displaybreak[0] \\
      \notag \mu_{t} &= \cp \frac{\mu_{0}}{\Up^{2}}\pnth{1 - \frac{2M}{r}} \brkt{\p_{r}\pnth{f_{r}\e^{V}\sqrt{1 - \frac{2M}{r}}} + \frac{2f_{r}\e^{V}}{r}\sqrt{1 - \frac{2M}{r}}} \\
      \notag & \qquad + p \frac{\mu_{0}}{\Up^{2}}\pnth{1 - \frac{2M}{r}} \brkt{\p_{r}\pnth{\cf_{r}\e^{V}\sqrt{1 - \frac{2M}{r}}} + \frac{2\cf_{r}\e^{V}}{r}\sqrt{1 - \frac{2M}{r}}} \displaybreak[0]\\
      \notag & \qquad + \frac{\mu_{0}}{\Up^2}\pnth{1 - \frac{2M}{r}}\brkt{\cf_{r}\p_{r}\pnth{p\e^{V}\sqrt{1 - \frac{2M}{r}}} + f_{r}\p_{r}\pnth{\cp\e^{V}\sqrt{1 - \frac{2M}{r}}}} \\
      \label{divTEKG2-aRHS} & \qquad - \frac{2M_{t}\mu_{0}}{r}\pnth{\frac{\abs{f_{r}}^{2} + \abs{p}^{2}}{\Up^{2}}}
    \end{align}
    Setting (\ref{divTEKG2-aLHS}) equal to (\ref{divTEKG2-aRHS}), we obtain
    \begin{align}
      \notag \mu_{0}\Bigg[&\cp \pnth{f \e^{V}\sqrt{1 - \frac{2M}{r}} + \frac{p_{t}}{\Up^{2}}\pnth{1 - \frac{2M}{r}}} + p \pnth{\cf \e^{V}\sqrt{1 - \frac{2M}{r}} + \frac{\cp_{t}}{\Up^{2}}\pnth{1 - \frac{2M}{r}}} \\
      \notag  - &\frac{2M_{t}}{r}\pnth{\frac{\abs{f_{r}}^{2} + \abs{p}^{2}}{\Up^{2}}} + \pnth{1 - \frac{2M}{r}}\frac{\cf_{r}\p_{r}\pnth{p \e^{V}\sqrt{1 - \frac{2M}{r}}} + f_{r}\p_{r}\pnth{\cp \e^{V}\sqrt{1 - \frac{2M}{r}}}}{\Up^{2}} \Bigg] \displaybreak[0]\\
      \notag &= \cp \frac{\mu_{0}}{\Up^{2}}\pnth{1 - \frac{2M}{r}} \brkt{\p_{r}\pnth{f_{r}\e^{V}\sqrt{1 - \frac{2M}{r}}} + \frac{2f_{r}\e^{V}}{r}\sqrt{1 - \frac{2M}{r}}} \displaybreak[0]\\
      \notag & \qquad + p \frac{\mu_{0}}{\Up^{2}}\pnth{1 - \frac{2M}{r}} \brkt{\p_{r}\pnth{\cf_{r}\e^{V}\sqrt{1 - \frac{2M}{r}}} + \frac{2\cf_{r}\e^{V}}{r}\sqrt{1 - \frac{2M}{r}}} \displaybreak[0] \\
      \notag & \qquad + \frac{\mu_{0}}{\Up^2}\pnth{1 - \frac{2M}{r}}\brkt{\cf_{r}\p_{r}\pnth{p\e^{V}\sqrt{1 - \frac{2M}{r}}} + f_{r}\p_{r}\pnth{\cp\e^{V}\sqrt{1 - \frac{2M}{r}}}} \\
      \notag &\qquad - \frac{2M_{t}\mu_{0}}{r}\pnth{\frac{\abs{f_{r}}^{2} + \abs{p}^{2}}{\Up^{2}}} \displaybreak[0]\\
      \notag \cp &\pnth{\Up^{2} f \e^{V}\pnth{1 - \frac{2M}{r}}^{-1/2} + p_{t}} + p \pnth{\Up^{2}\cf \e^{V}\pnth{1 - \frac{2M}{r}}^{-1/2} + \cp_{t}} \\
      \notag &= \cp \brkt{\p_{r}\pnth{f_{r}\e^{V}\sqrt{1 - \frac{2M}{r}}} + \frac{2f_{r}\e^{V}}{r}\sqrt{1 - \frac{2M}{r}}} \\
      \notag &\qquad + p \brkt{\p_{r}\pnth{\cf_{r}\e^{V}\sqrt{1 - \frac{2M}{r}}} + \frac{2\cf_{r}\e^{V}}{r}\sqrt{1 - \frac{2M}{r}}} \displaybreak[0]\\
      \notag 0 &= \cp\Bigg[- p_{t} + \p_{r}\pnth{f_{r}\e^{V}\sqrt{1 - \frac{2M}{r}}} \\
      \notag &\qquad + \e^{V}\pnth{-\Up^{2} f \pnth{1 - \frac{2M}{r}}^{-1/2} + \frac{2f_{r}}{r}\sqrt{1 - \frac{2M}{r}}} \Bigg] \\
      \notag &\qquad + p\Bigg[- \cp_{t} + \p_{r}\pnth{\cf_{r}\e^{V}\sqrt{1 - \frac{2M}{r}}} \\
      \label{divTEKG2-afin} &\qquad + \e^{V}\pnth{-\Up^{2} \cf \pnth{1 - \frac{2M}{r}}^{-1/2} + \frac{2\cf_{r}}{r}\sqrt{1 - \frac{2M}{r}}}\Bigg]
    \end{align}
  \end{subequations}
  Note that, since $V$ and $M$ are real valued, the second term above is the complex conjugate of the first term.  Then by equation (\ref{NCpde4}), equation (\ref{divTEKG2-afin}) holds.

  Next, we consider (\ref{divT3b}).  We first differentiate (\ref{SETEKGFun-2}) with respect to $t$ and then compare it to (\ref{divT3b}) after substituting (\ref{SETEKGFun}), (\ref{NCceq1}), and (\ref{NCpde2}).  This yields
  \begin{subequations}\label{divTEKG2-b}
    \begin{align}
      \notag \rho_{t} &= \p_{t}\brkt{\frac{2\mu_{0}}{\Up^2}\pnth{1 - \frac{2M}{r}}\Re(f_{r}\cp)} \\
      \notag &= \frac{\mu_{0}}{\Up^{2}}\brkt{-\frac{4M_{t}}{r}\Re(f_{r}\cp) + \pnth{1 - \frac{2M}{r}}\pnth{f_{rt}\cp + \cf_{rt}p + f_{r}\cp_{t} + \cf_{r}p_{t}}} \displaybreak[0]\\
      \notag &= \frac{\mu_{0}}{\Up^{2}}\Bigg\{-\frac{4M_{t}}{r}\Re(f_{r}\cp) + \pnth{1 - \frac{2M}{r}}\Bigg[\cp\, \p_{r}\pnth{p\e^{V}\sqrt{1 - \frac{2M}{r}}} \\
      \notag &\qquad + p\, \p_{r}\pnth{\cp\e^{V}\sqrt{1 - \frac{2M}{r}}} + f_{r}\cp_{t} + \cf_{r}p_{t}\Bigg]\Bigg\} \displaybreak[0]\\
      \notag &=\frac{\mu_{0}}{\Up^{2}}\Bigg\{-\frac{4M_{t}}{r}\Re(f_{r}\cp) + \pnth{1 - \frac{2M}{r}}\Bigg[\cp\, \p_{r}\pnth{p\e^{V}\sqrt{1 - \frac{2M}{r}}} \\
      \notag &\qquad + \cp_{r}p\e^{V}\sqrt{1-\frac{2M}{r}} + \abs{p}^{2} \p_{r}\pnth{\e^{V}\sqrt{1 - \frac{2M}{r}}} + f_{r}\cp_{t} + \cf_{r}p_{t}\Bigg]\Bigg\} \\
      \notag &= \frac{\mu_{0}}{\Up^{2}}\Bigg\{-\frac{4M_{t}}{r}\Re(f_{r}\cp) + \pnth{1 - \frac{2M}{r}}\Bigg[\p_{r}\pnth{\abs{p}^{2}\e^{V}\sqrt{1 - \frac{2M}{r}}} \\
      \label{divTEKG2-bLHS} &\qquad + \abs{p}^{2}\p_{r}\pnth{\e^{V}\sqrt{1 - \frac{2M}{r}}} + f_{r}\cp_{t} + \cf_{r}p_{t}\Bigg]\Bigg\}
    \end{align}
    From (\ref{divT3b}), we obtain
    \begin{align}
      \notag \rho_{t} &= \e^{V}\sqrt{1 - \frac{2M}{r}}\pnth{V_{r}(\mu + P) + P_{r} + \frac{2(P-Q)}{r}} - \frac{2\rho M_{t}}{r}\pnth{1 - \frac{2M}{r}}^{-1} \\
      \notag &=  - \frac{4\mu_{0}M_{t}}{r\Up^{2}}\Re(f_{r}\cp) + \e^{V}\sqrt{1 - \frac{2M}{r}}\Bigg[\frac{2\mu_{0}V_{r}}{\Up^{2}}\pnth{1 - \frac{2M}{r}}(\abs{f_{r}}^{2} + \abs{p}^{2}) \\
      \notag &\qquad + \mu_{0}\p_{r}\pnth{-\abs{f}^{2} + \pnth{1 - \frac{2M}{r}}\frac{\abs{f_{r}}^{2} + \abs{p}^{2}}{\Up^{2}}} + \frac{4\mu_{0}\abs{f_{r}}^{2}}{r\Up^{2}}\pnth{1 - \frac{2M}{r}}\Bigg] \displaybreak[0] \\
      \notag &= \frac{\mu_{0}\e^{V}}{\Up^{2}}\pnth{1 - \frac{2M}{r}}\Bigg\{2V_{r}\abs{f_{r}}^{2}\sqrt{1 - \frac{2M}{r}} + 2V_{r}\abs{p}^{2}\sqrt{1 - \frac{2M}{r}} \\
      \notag &\qquad - \Up^{2}(f_{r}\cf + \cf_{r}f)\pnth{1 - \frac{2M}{r}}^{-1/2} \displaybreak[0] \\
      \notag &\qquad + \pnth{1 - \frac{2M}{r}}^{-1/2}\p_{r}\pnth{1 - \frac{2M}{r}}\pnth{\abs{f_{r}}^{2} + \abs{p}^{2}} \\
      \notag &\qquad + \sqrt{1 - \frac{2M}{r}}\ \p_{r}\pnth{\abs{f_{r}}^{2} + \abs{p}^{2}} + \frac{4\abs{f_{r}}^{2}}{r}\sqrt{1 - \frac{2M}{r}}\Bigg\} - \frac{4\mu_{0}M_{t}}{r\Up^{2}}\Re(f_{r}\cp) \displaybreak[0] \\
      \notag &= \frac{\mu_{0}}{\Up^{2}}\pnth{1 - \frac{2M}{r}}\Bigg\{\cf_{r}\Bigg[f_{r}V_{r}\e^{V}\sqrt{1 - \frac{2M}{r}} \\
      \notag &\qquad - \e^{V}\Up^{2}f\pnth{1 - \frac{2M}{r}}^{-1/2} + \frac{2\e^{V} f_{r}}{r}\sqrt{1 - \frac{2M}{r}}\Bigg] \displaybreak[0]\\
      \notag &\qquad + f_{r}\brkt{\cf_{r}V_{r}\e^{V}\sqrt{1 - \frac{2M}{r}} - \e^{V}\Up^{2}\cf\pnth{1 - \frac{2M}{r}}^{-1/2} + \frac{2\e^{V} \cf_{r}}{r}\sqrt{1 - \frac{2M}{r}}} \displaybreak[0]\\
      \notag &\qquad + 2V_{r}\e^{V}\abs{p}^{2}\sqrt{1 - \frac{2M}{r}} + 2\abs{p}^{2}\e^{V}\p_{r}\pnth{\sqrt{1 - \frac{2M}{r}}} + \e^{V}\sqrt{1 - \frac{2M}{r}}\ \p_{r}(\abs{p}^{2}) \displaybreak[0]\\
      \notag &\qquad + 2\abs{f_{r}}^{2}\e^{V}\p_{r}\pnth{\sqrt{1 - \frac{2M}{r}}} + \e^{V}\sqrt{1 - \frac{2M}{r}} \pnth{f_{rr}\cf_{r} + f_{r}\cf_{rr}} \Bigg\} \\
      \notag &\qquad - \frac{4\mu_{0}M_{t}}{r\Up^{2}}\Re(f_{r}\cp) \displaybreak[0] \\
      \notag &= \frac{\mu_{0}}{\Up^{2}}\pnth{1 - \frac{2M}{r}}\Bigg\{\cf_{r}\Bigg[\p_{r}\pnth{f_{r}\e^{V}\sqrt{1 - \frac{2M}{r}}} \\
      \notag &\qquad + \e^{V} \pnth{- \Up^{2}f\pnth{1 - \frac{2M}{r}}^{-1/2} + \frac{2 f_{r}}{r}\sqrt{1 - \frac{2M}{r}}}\Bigg] \displaybreak[0]\\
      \notag &\qquad + f_{r}\brkt{\p_{r}\pnth{\cf_{r}\e^{V}\sqrt{1 - \frac{2M}{r}}} +\e^{V}\pnth{-\Up^{2}\cf\pnth{1 - \frac{2M}{r}}^{-1/2} + \frac{2 \cf_{r}}{r}\sqrt{1 - \frac{2M}{r}}}} \\
      \label{divTEKG2-bRHS} &\qquad + \abs{p}^{2}\p_{r}\pnth{\e^{V}\sqrt{1 - \frac{2M}{r}}} + \p_{r}\pnth{\abs{p}^{2}\e^{V}\sqrt{1 - \frac{2M}{r}}} \Bigg\} - \frac{4\mu_{0}M_{t}}{r\Up^{2}}\Re(f_{r}\cp)
    \end{align}
    Equating (\ref{divTEKG2-bLHS}) and (\ref{divTEKG2-bRHS}) yields
    \begin{align}
      \notag \frac{\mu_{0}}{\Up^{2}}\Bigg\{- & \frac{4M_{t}}{r}\Re(f_{r}\cp) + \pnth{1 - \frac{2M}{r}}\Bigg[\p_{r}\pnth{\abs{p}^{2}\e^{V}\sqrt{1 - \frac{2M}{r}}} \\
      \notag + \abs{p}^{2} & \p_{r}\pnth{\e^{V}\sqrt{1 - \frac{2M}{r}}} + f_{r}\cp_{t} + \cf_{r}p_{t}\Bigg]\Bigg\} \displaybreak[0] \\
      \notag &= \frac{\mu_{0}}{\Up^{2}}\pnth{1 - \frac{2M}{r}}\Bigg\{\cf_{r}\Bigg[\p_{r}\pnth{f_{r}\e^{V}\sqrt{1 - \frac{2M}{r}}} \\
      \notag &\qquad + \e^{V} \pnth{- \Up^{2}f\pnth{1 - \frac{2M}{r}}^{-1/2} + \frac{2 f_{r}}{r}\sqrt{1 - \frac{2M}{r}}}\Bigg] \displaybreak[0]\\
      \notag &\qquad + f_{r}\Bigg[\p_{r}\pnth{\cf_{r}\e^{V}\sqrt{1 - \frac{2M}{r}}} \displaybreak[0]\\
      \notag &\qquad + \e^{V}\pnth{-\Up^{2}\cf\pnth{1 - \frac{2M}{r}}^{-1/2} + \frac{2 \cf_{r}}{r}\sqrt{1 - \frac{2M}{r}}}\Bigg] \displaybreak[0]\\
      \notag &\qquad + \abs{p}^{2}\p_{r}\pnth{\e^{V}\sqrt{1 - \frac{2M}{r}}} + \p_{r}\pnth{\abs{p}^{2}\e^{V}\sqrt{1 - \frac{2M}{r}}} \Bigg\} \\
      \notag &\qquad - \frac{4\mu_{0}M_{t}}{r\Up^{2}}\Re(f_{r}\cp) \displaybreak[0] \\
      \notag 0 &= \cf_{r}\Bigg[- p_{t} + \p_{r}\pnth{f_{r}\e^{V}\sqrt{1 - \frac{2M}{r}}} \\
      \notag &\qquad + \e^{V}\pnth{-\Up^{2} f \pnth{1 - \frac{2M}{r}}^{-1/2} + \frac{2f_{r}}{r}\sqrt{1 - \frac{2M}{r}}} \Bigg] \displaybreak[0] \\
      \notag &\qquad + f_{r}\Bigg[- \cp_{t} + \p_{r}\pnth{\cf_{r}\e^{V}\sqrt{1 - \frac{2M}{r}}} \\
      \label{divTEKG2-bfin} &\qquad+ \e^{V}\pnth{-\Up^{2} \cf \pnth{1 - \frac{2M}{r}}^{-1/2} + \frac{2\cf_{r}}{r}\sqrt{1 - \frac{2M}{r}}}\Bigg]
    \end{align}
  \end{subequations}
  As before, since $M$ and $V$ are real valued, the second term above is the complex conjugate of the first term.  Then by equation (\ref{NCpde4}), equation (\ref{divTEKG2-bfin}) holds.  Then equations (\ref{NCpde4}) and (\ref{NCpde3}) imply equation (\ref{divT3}) holds, which completes the proof.
\end{proof}

\chapter{Spherically Symmetric Static States \\ of Wave Dark Matter}\label{ch3}

Now that we have described in detail spherically symmetric spacetimes and in particular presented some results about the spherically symmetric Einstein-Klein-Gordon equations, we can now begin to comment on the subject of wave dark matter.

Describing the Einstein-Klein-Gordon equations in spherical symmetry is an excellent place to start studying wave dark matter for two reasons.  First, it immensely simplifies the equations involved making them easier to solve.  Secondly, as mentioned in the previous chapter, there are some very interesting objects to be described that are approximately spherically symmetric.  In the case of wave dark matter, dwarf spheroidal galaxies are approximately spherically symmetric and almost entirely dark matter (\cite{Walker07,Mateo98}) implying that their kinematics are controlled by their dark matter component.  Thus determining what the wave dark matter model predicts in spherical symmetry would be important in showing the level of compatibility of wave dark matter with dwarf spheroidal galaxies.

The purpose of this chapter is to present a few results in this regard.  In particular, we describe an important class of spherically symmetric solutions of the Einstein-Klein-Gordon equations called static states and discuss some properties of these solutions.  We should also reference here others' results that, with the exception of the ground state, these static states are unstable under perturbations (\cite{Seidel90, Seidel98, MSBS, Lai07, Gleiser89, Hawley00}), which poses a problem in using the static states as a physical model.  However, in physical situations dark matter is always coupled with regular matter, which may have stabilizing effects on the dark matter.  Thus understanding the stability of these static states in the presence of regular matter is an important open problem.  It is also possible that dark matter may not exist as a single static state at all and as such, finding other kinds of stable configurations of wave dark matter is another important open problem.  These open problems are not addressed in this chapter, but are a goal of our future work and one of the reasons for creating a program like the one presented in Chapter \ref{ch5} to evolve the spherically symmetric Einstein-Klein-Gordon equations in time.

\section{The Spherically Symmetric Einstein-Klein-Gordon Equations}

In this section, we discuss the Einstein-Klein-Gordon equations in spherical symmetry on an asymptotically Schwarzschild spacetime.

To begin, let $N$ be a spherically symmetric spacetime with metric $g$ of the form in equation (\ref{metric}).  Additionally assume that $N$ is asymptotically Schwarzschild, that is, it is Schwarzschild as $r \to \infty$.  As in Chapter \ref{ch2}, define $f:N \to \C$ as a complex scalar field on $N$ such that $(g,f)$ satisfies the Einstein-Klein-Gordon equations (\ref{EKG-1}) and (\ref{EKG-2}).  By Chapter \ref{ch2}, the spherically symmetric Einstein-Klein-Gordon equations reduce to the system (\ref{NCpde}).  Recall that by Proposition \ref{NCpdeprop}, equations (\ref{NCceq1}) and (\ref{NCceq2}) are automatically satisfied by solving the system (\ref{NCpde}).

We note next the important behavior of the functions $f,p,V,$ and $M$ at the origin and as $r \to \infty$.

We use the previously mentioned fact that $M$ is the flat integral of the energy density to shed light on the initial behavior of the function $M$ near the origin, $r = 0$.  Using the construction in Chapter \ref{ch2}, the energy density of this system is defined as $\mu(t,r) = T(\nu_{t},\nu_{t})$, and the Einstein equation yields
\begin{equation}\label{intED}
    M(t,r) = \int_{E_{t,r}}\mu(t,s)\, dV = \int_{0}^{r} 4\pi s^{2}\mu(t,s)\, ds.
\end{equation}
Since the energy density is spherically symmetric and smooth, if the energy density is nonzero at the central value, $r=0$, at any time $t$, then the energy density must be an even function of $r$.  Moreover, reiterating a point made in Chapter \ref{ch2}, since $\mu(t,r)$ is also smooth at the origin, $\mu_{r}(t,0) = 0$ for all $t$.  Thus for small $r$, $\mu$ is approximately constant and nonnegative.  Then the above integral yields for small $r$ that
\begin{equation}\label{massCub}
    M(t,r) = \int_{0}^{r} 4\pi s^{2}\mu(t,s)\, ds \approx \int_{0}^{r} 4\pi s^{2}\mu(t,0)\, ds = \frac{4\pi \mu(t,0)}{3}r^{3}.
\end{equation}
Thus the initial behavior of $M$ near $r=0$ is that of a cubic power function (or in the case when $\mu(t,0)=0$, the zero function).  In particular, this implies that for all $t$
\begin{equation}\label{Mder}
    M(t,0) = 0, \qquad M_{r}(t,0) = 0, \qquad \text{and} \qquad M_{rr}(t,0) = 0.
\end{equation}
This is consistent with the fact that the metric functions, $\e^{2V}$ and $\pnth{1 - \dfrac{2M}{r}}$, are also smooth spherically symmetric functions that are nonzero at $r=0$ and hence even functions of $r$.  This implies that, since $r$ is an odd function, $M(t,r)$ must also act like an odd function near $r=0$.  Similarly, $V(t,r)$ must be an even function of $r$, implying that
\begin{equation}\label{Vder}
  V_{r}(t,0) = 0
\end{equation}
for all $t$.  Equation (\ref{Mder}) implies via \LHopital's rule that
\begin{equation}
  \lim_{r \to 0^{+}} \frac{M}{r} = \lim_{r \to 0^{+}} \frac{M}{r^{2}} = \lim_{r \to 0^{+}} \frac{M_{r}}{r} = \lim_{r \to 0^{+}} M_{rr} = 0.
\end{equation}
Next, since $f$ and $p$ are spherically symmetric and allowed to be nonzero at $r = 0$, we have that both $f$ and $p$ are even functions in $r$ as well, which, for regularity at $r=0$, implies that
\begin{equation}\label{sfder}
  f_{r}(t,0)=0 \qquad \text{and} \qquad p_{r}(t,0)=0.
\end{equation}
for all $t$.

Next we consider the behavior of the functions at the outer boundary.  Since $N$ is asymptotically Schwarzschild, there exist constants, $m\geq 0$, called the total mass of the system, and $\ka >0$, and a Schwarzschild metric $g_{S}$ given by
\begin{equation}\label{SW-metric}
  g_{S} = -\ka^{2}\pnth{1 - \frac{2m}{r}}\, dt^{2} + \pnth{1 - \frac{2m}{r}}^{-1}\, dr^{2} + r^{2}\, d\si^{2},
\end{equation}
such that $g$ approaches $g_{S}$ as $r \to \infty$.  This yields the following asymptotic boundary conditions.
\begin{align}
  \label{AB1a} \glap{g_{S}}f &\to \Upsilon^2 f \quad \text{and} \quad f \to 0 \quad \text{as } r \to \infty \\
  \label{AB2a} \e^{2V} &\to \ka^{2}\pnth{1 - \frac{2M}{r}} \quad \text{as } r \to \infty.
\end{align}
The first boundary condition (\ref{AB1a}) implies by equation (\ref{NCpde1a}) that $M_{r} \to 0$ as $r \to \infty$ and hence $M$ approaches a constant value as $r \to \infty$. Given equations (\ref{metric}), (\ref{SW-metric}), and the second boundary condition (\ref{AB2a}), this constant will be the parameter $m$ in (\ref{SW-metric}).

Now that we have described the Einstein-Klein-Gordon equations and its asymptotically Schwarzschild boundary conditions in spherical symmetry, we are ready to discuss the class of spherically symmetric solutions to the Einstein-Klein-Gordon equations that yield static metrics.

\section{Spherically Symmetric Static States}

In the remainder of this chapter, we will consider solutions of the spherically symmetric Einstein-Klein-Gordon equations where the scalar field $f$ is of the form
\begin{equation}\label{sfstaticstate}
  f(t,r) = \e^{i\om t}F(r)
\end{equation}
where $\om \in \RR$ and $F$ is complex-valued.  As $t$ increases, $f$ rotates the values of the function $F(r)$ through the complex plane with angular frequency $\om$ without changing their absolute value.  Thus, without loss of generality, we will assume that $\om \geq 0$, since if $\om < 0$, $F$ will simply rotate in the opposite direction.  With $f$ of this form, we have that
\begin{align}
  \label{fr-form} f_{r} &= \e^{i\om t} F'(r) \displaybreak[0]\\
  \notag p &= \e^{-V}\pnth{1 - \frac{2M}{r}}^{-1/2}f_{t} \\
  \label{ft-form} &= \e^{-V}\pnth{1 - \frac{2M}{r}}^{-1/2}\pnth{i\om \e^{i\om t}F(r)}.
\end{align}
With a scalar field of this form, the system (\ref{NCpde}) implies that the metric is static under certain conditions on the function $F$.  We collect this result in the following proposition.

\begin{prop}\label{staticprop}
  Let $(N,g)$ be a spherically symmetric asymptotically Schwarzschild spacetime that satisfies the Einstein-Klein-Gordon equations (\ref{EKG}) for a scalar field of the form in (\ref{sfstaticstate}).  Additionally, assume that $F(r) = h(r)\e^{i a(r)}$ for smooth real-valued functions $h$ and $a$, where $h$ has only isolated zeros, if any.  Then $(N,g)$ is static if and only if $a(r)$ is constant.
\end{prop}

\begin{proof}
  By definition, a spacetime metric is static if there exists a timelike Killing vector field and a spacelike hypersurface that is orthogonal to the Killing vector field.  For the metric in equation (\ref{metric}), if the metric components, $V$ and $M$, do not depend on $t$, then $\p_{t}$ is one such Killing vector field and it is already orthogonal to the $t~=~constant$ spacelike hypersurfaces.  If a spherically symmetric metric is static, then we can choose the $t$ coordinate to be in the direction of the timelike Killing vector field, making $\p_{t}$ the timelike Killing vector field in these coordinates, and choose the polar-areal coordinates on its orthogonal spacelike hypersurfaces, yielding a metric of the form in (\ref{metric}).  Then since the metric cannot change in the direction of $\p_{t}$, the metric components must be $t$-independent. It remains then to show that under and only under the given conditions on $F$, the metric components are $t$-independent.

  Note that by equation (\ref{NCceq1}), $M_{t}\equiv 0$ if and only if $\Re(f_{r}\conj{p}) \equiv 0$.  Using equation (\ref{fr-form}) and (\ref{ft-form}), we obtain
  \begin{align}
    \notag \Re(f_{r}\conj{p}) &= \Re\pnth{\e^{i\om t} F'(r)\e^{-V}\pnth{1 - \frac{2M}{r}}^{-1/2}\pnth{-i\om \e^{-i\om t}\overline{F(r)}}} \\
    \label{rp1} &= \e^{-V}\pnth{1 - \frac{2M}{r}}^{-1/2}\Re(-i\om F'(r)\overline{F(r)})
  \end{align}
  Thus $\Re(f_{r}\conj{p}) \equiv 0$ if and only if $\Re(-i\om F'(r)\overline{F(r)}) \equiv 0$, which is true if and only if $F'(r)\overline{F(r)}$ is real-valued.  By assumption, $F(r)$ can be written as
  \begin{equation}
    F(r) = h(r)\e^{i a(r)}
  \end{equation}
  for smooth real-valued functions $h$ and $a$.  If we write $F(r)$ this way, then $F'(r)$ is as follows.
  \begin{equation}
    F'(r) = h'(r) \e^{i a(r)} + i h(r) a'(r)\e^{i a(r)} = \e^{i a(r)}\pnth{h'(r) + i h(r) a'(r)}.
  \end{equation}
  Then we have that
  \begin{equation}
    F'(r)\overline{F(r)} = h'(r)h(r) + i h(r)^2 a'(r).
  \end{equation}
  Since $h$ and $a$ are both real-valued, we see that $F'(r)\overline{F(r)}$ is real if and only if $h(r)^{2}a'(r) \equiv 0$.  Since $h$ has only isolated zeros, $h(r)^{2}a'(r) \equiv 0$ if and only if $a'(r) \equiv 0$ or equivalently $a(r)$ is constant.

  It suffices from here to show that $M_{t} \equiv 0$ if and only if $g$ is $t$-independent.  Obviously, $g$ being $t$-independent implies $M_{t} \equiv 0$.  On the other hand, assume that $M_{t} \equiv 0$.  First note that, since $M_{t} \equiv 0$, $M_{rt} \equiv 0$ as well.  Moreover, since $\abs{f}^{2} = \abs{F}^{2}$ and $\abs{f_{r}}^{2} = \abs{F'}^{2}$ and $F$ has zero $t$-derivative, then $\abs{f}^{2}$ and $\abs{f_{r}}^{2}$ both have zero $t$-derivatives.  Differentiating (\ref{NCpde1a}) with respect to $t$ and using the fact that $\abs{f}^{2}$, $\abs{f_{r}}^{2}$, and $M$ all have zero $t$-derivatives, we have that
  \begin{align}
    \notag M_{rt} &= 4\pi r^{2}\mu_{0}\brkt{\p_{t}\pnth{\abs{f}^{2}} - \pnth{\frac{2M_{t}}{r}}\frac{\abs{f_{r}}^{2} + \abs{p}^{2}}{\Up^{2}} + \pnth{1 - \frac{2M}{r}}\frac{\p_{t}\pnth{\abs{f_{r}}^{2}} + \p_{t}\pnth{\abs{p}^{2}}}{\Up^{2}}} \\
    0 &= 4\pi r^{2}\mu_{0}\pnth{1 - \frac{2M}{r}}\frac{\p_{t}\pnth{\abs{p}^{2}}}{\Up^{2}}.
  \end{align}
  Thus since, $M$, $\Up$, and $\mu_{0}$ are all nonzero, $\p_{t}\pnth{\abs{p}^{2}} = 0$.

  Next, since (\ref{NCpde1a}) and (\ref{NCpde2a}) completely determine the Einstein equation, the function $V$ is determined at every value of $t$ by solving (\ref{NCpde2a}) at that value of $t$.  Since $\abs{f}^{2}$, $\abs{f_{r}}^{2}$, $\abs{p}^{2}$, and $M$ never change with $t$, $V_{r}$ never changes with $t$ and hence the solution, $V$, of (\ref{NCpde2a}), by uniqueness of the solution to an ODE, will never change with $t$ so long as $V(t,0) = constant$.  Since $V(t,0)$ is determined to be the value that makes $V$ satisfy (\ref{AB2a}) and since $M$ never changes with $t$, then the fact that $V_{r}$ is $t$-independent forces $V(t,0)=constant$.  Thus $V_{t} \equiv 0$ and the metric is t-independent.
\end{proof}

Since the value of $a$ simply adjusts the ``starting position'' of the values of $f$ before they rotate, we will, without loss of generality, set $a = 0$, which is the same assumption that $F(r)$ be real-valued.  This amounts to simply choosing the hypersurface that we denote as $t = 0$.  To summarize, we restrict our attention to static states of the form in (\ref{sfstaticstate}) with $\om \geq 0$ and $F$ real-valued.

\subsection{ODEs for Static States}

In this section, we input our ansatz (\ref{sfstaticstate}) with the requirement that $\om \geq 0$ and $F$ is real-valued into the Einstein-Klein-Gordon equations (\ref{NCpde1a})-(\ref{NCpde4a}).  Since the metric is $t$-independent, we summarize that
\begin{align}
  V&= V(r), & M &= M(r), & f(t,r) &= \e^{i\om t}F(r).
\end{align}
Also note that by (\ref{sfstaticstate})-(\ref{ft-form}),
  \begin{align}
    \abs{f} &= \abs{F}, \displaybreak[0]\\
    \abs{f_{r}} &= \abs{F'}, \displaybreak[0]\\
    \notag \abs{p} &= \abs{\e^{-V}\pnth{1 - \frac{2M}{r}}^{-1/2}\pnth{i\om \e^{i\om t}F}} \\
    &= \abs{F} \om\e^{-V}\pnth{1 - \frac{2M}{r}}^{-1/2}.
  \end{align}
Additionally, if we differentiate (\ref{ft-form}) with respect to $t$ and (\ref{fr-form}) with respect to $r$, we obtain,
\begin{align}
  \notag p_{t} &= \p_{t}\pnth{\e^{-V}\pnth{1 - \frac{2M}{r}}^{-1/2}\pnth{i\om \e^{i\om t}F}} \\
  &= -\om^{2}\e^{-V}\pnth{1 - \frac{2M}{r}}^{-1/2}\e^{i\om t}F \\
  f_{rr} &= \e^{i\om t}F''
\end{align}
Then equations (\ref{NCpde1a}) and (\ref{NCpde2a}) become
  \begin{align}
    \label{NCpde1b} M' &= 4\pi r^{2}\mu_{0}\brkt{\pnth{1 + \frac{\om^{2}}{\Up^{2}}\e^{-2V}}\abs{F}^{2} + \pnth{1 - \frac{2M}{r}}\frac{\abs{F'}^{2}}{\Up^{2}}} \\
    \label{NCpde2b} V' &= \pnth{1 - \frac{2M}{r}}^{-1}\brce{\frac{M}{r^{2}} - 4\pi r\mu_{0}\brkt{\pnth{1 - \frac{\om^{2}}{\Up^{2}}\e^{-2V}}\abs{F}^{2} - \pnth{1 - \frac{2M}{r}}\frac{\abs{F'}^{2}}{\Up^{2}}}}
  \end{align}
Equation (\ref{NCpde3a}) turns into (\ref{ft-form}), thereby becoming redundant.  The last equation, (\ref{NCpde4a}), becomes
  \begin{align}
    \notag -\om^{2}\e^{-V}\pnth{1 - \frac{2M}{r}}^{-1/2}\e^{i\om t}F &= \e^{V}\Bigg[-\Up^{2}\pnth{1 - \frac{2M}{r}}^{-1/2}\e^{i\om t}F \\
    \notag &\qquad + \frac{2\e^{i\om t}F'}{r}\sqrt{1 - \frac{2M}{r}}\Bigg] \displaybreak[0]\\
    \notag &\qquad + V'\e^{V}\e^{i\om t}F'\sqrt{1 - \frac{2M}{r}} + \e^{V}\e^{i\om t}F''\sqrt{1 - \frac{2M}{r}} \\
    \notag &\qquad + \e^{V}\e^{i\om t}F'\brkt{\frac{1}{2}\pnth{1 - \frac{2M}{r}}^{-1/2}\pnth{\frac{2M}{r^{2}} - \frac{2M'}{r}}} \displaybreak[0]\\
    \notag -\om^{2}\e^{-V}\pnth{1 - \frac{2M}{r}}^{-1/2}\e^{i\om t}F &= \e^{V}\Bigg\{\pnth{1 - \frac{2M}{r}}^{-1/2}\Bigg[2\e^{i\om t}F'\pnth{\frac{M}{r^{2}} - 4\pi r\mu_{0}\abs{F}^{2}} \\
    \notag &\qquad - \Up^{2}\e^{i\om t}F\Bigg]+ \sqrt{1 - \frac{2M}{r}}\pnth{\e^{i\om t}F'' + \frac{2}{r}\e^{i\om t}F'} \Bigg\} \displaybreak[0]\\
    \notag -\om^{2}F &= \e^{2V}\Bigg[2F'\pnth{\frac{M}{r^{2}} - 4\pi r \mu_{0}\abs{F}^{2}} \\
    \notag &\qquad - \Up^{2}F + \pnth{1 - \frac{2M}{r}}\pnth{F'' + \frac{2}{r}F'} \Bigg] \displaybreak[0]\\
    \notag \e^{2V}\pnth{1 - \frac{2M}{r}}F'' &= \e^{2V}\Bigg[\pnth{\Up^{2} - \frac{\om^{2}}{\e^{2V}}} F  \\
    \label{NCpde4b-derive} &\qquad - 2F'\pnth{\frac{1}{r} - \frac{M}{r^{2}} - 4\pi r\mu_{0}\abs{F}^{2}}\Bigg]
  \end{align}
which yields
  \begin{equation} \label{NCpde4b}
    F'' = \pnth{1 - \frac{2M}{r}}^{-1}\brkt{\pnth{\Up^{2} - \frac{\om^{2}}{\e^{2V}}} F + 2F'\pnth{\frac{M}{r^{2}} + 4\pi r\mu_{0}\abs{F}^{2} - \frac{1}{r}}}
  \end{equation}
To make the system first order, we will introduce a new function $H(r) = F'(r)$.  Then (\ref{NCpde1b}), (\ref{NCpde2b}), and (\ref{NCpde4b}) become the system
\begin{subequations} \label{NCpdec}
  \begin{align}
    \label{NCpde1c} M' &= 4\pi r^{2}\mu_{0}\brkt{\pnth{1 + \frac{\om^{2}}{\Up^{2}}\e^{-2V}}\abs{F}^{2} + \pnth{1 - \frac{2M}{r}}\frac{\abs{H}^{2}}{\Up^{2}}} \\
    \label{NCpde2c} V' &= \pnth{1 - \frac{2M}{r}}^{-1}\brce{\frac{M}{r^{2}} - 4\pi r\mu_{0}\brkt{\pnth{1 - \frac{\om^{2}}{\Up^{2}}\e^{-2V}}\abs{F}^{2} - \pnth{1 - \frac{2M}{r}}\frac{\abs{H}^{2}}{\Up^{2}}}} \\
    \label{NCpde3c} F' &= H \\
    \label{NCpde4c} H' &= \pnth{1 - \frac{2M}{r}}^{-1}\brkt{\pnth{\Up^{2} - \frac{\om^{2}}{\e^{2V}}} F + 2H\pnth{\frac{M}{r^{2}} + 4\pi r\mu_{0}\abs{F}^{2} - \frac{1}{r}}}
  \end{align}
\end{subequations}

\subsection{Boundary Conditions}

We will solve the system (\ref{NCpdec}) numerically, but in order to do so, we need to express how we will deal with our boundary conditions numerically.  Ideally, we would like to model the system in an infinite spacetime, but since we are computing these solutions numerically, we must introduce an artificial right hand boundary, say at $r=r_{max}$, to which we restrict our domain.  To simulate the fact that the spacetime is asymptotically Schwarzschild, which we detailed in (\ref{AB1a}) and (\ref{AB2a}), we will choose $r_{max}$ sufficiently large and impose the condition that the spacetime be approximately Schwarzschild at the boundary.  That is, we will impose (\ref{AB1a}) and (\ref{AB2a}) at the boundary $r=r_{max}$.

In this case, at $r = r_{max}$, equation (\ref{AB2a}) becomes
\begin{align}
  \notag \e^{2V(r_{max})} &\approx \ka^{2}\pnth{1 - \frac{2M(r_{max})}{r_{max}}} \\
  \label{AB2b} 0 &\approx V(r_{max}) - \frac{1}{2}\ln\pnth{1 - \frac{2M(r_{max})}{r_{max}}} - \ln \ka
\end{align}

For (\ref{AB1a}) at $r = r_{max}$, we require $f$ to approximately solve the Klein-Gordon equation (\ref{EKG-2}) in the Schwarzschild metric (\ref{SW-metric}).  Computing the Laplacian in the Schwarzschild metric, this equation becomes
\begin{equation}\label{lapS1}
  \pnth{1 - \frac{2m}{r}}^{2}F'' + \pnth{1 - \frac{2m}{r} + \pnth{1 - \frac{2m}{r}}^{2}}\frac{F'}{r} - \pnth{\Up^{2}\pnth{1 - \frac{2m}{r}} - \frac{\om^{2}}{\ka^{2}}}F = 0.
\end{equation}
For large $r$, this simplifies to
\begin{equation}\label{lapS2}
  F'' + \frac{2F'}{r} - \pnth{\Up^{2} - \frac{\om^{2}}{\ka^{2}}}F = 0.
\end{equation}
This ODE is routinely solved and has the general solution
\begin{equation}\label{Ssol}
  F = \frac{C_{1}}{r}\e^{r\sqrt{\Up^{2} - \frac{\om^{2}}{\ka^{2}}}} + \frac{C_{2}}{r}\e^{-r\sqrt{\Up^{2}-\frac{\om^{2}}{\ka^{2}}}}
\end{equation}
for some constants $C_{1},C_{2} \in \RR$.

We will show that this is consistent with the Wentzel-Kramers-Brillouin (WKB) approximation (\cite{Wentzel26, Kramers26, Brillouin26}) of the solution to equation (\ref{lapS1}) as $r \to \infty$, if we also assume that $m \ll 1$ (in mass units of years).

Following the procedure outlined in the book by Bender and Orszag (\cite{Bender78}), we have the following.  See also \cite{Carlini1817, Liouville1837, Green1838} for more references on this method.

Since $\pnth{1 - \dfrac{2m}{r}} > 0$ for all $r$, we can rewrite equation (\ref{lapS1}) as
\begin{equation}\label{lapS1-a}
  F'' + \pnth{\frac{1}{r} + \frac{1}{r}\pnth{1 - \frac{2m}{r}}^{-1}}F' - \pnth{\Up^{2}\pnth{1 - \frac{2m}{r}}^{-1} - \frac{\om^{2}}{\ka^{2}}\pnth{1 - \frac{2m}{r}}^{-2}}F = 0
\end{equation}
Define two new functions $\xi$ and $\zeta$ as follows.
\begin{align}
  \xi(r) &= \frac{1}{r} + \frac{1}{r}\pnth{1 - \frac{2m}{r}}^{-1} \\
  \zeta(r) &= \Up^{2}\pnth{1 - \frac{2m}{r}}^{-1} - \frac{\om^{2}}{\ka^{2}}\pnth{1 - \frac{2m}{r}}^{-2}
\end{align}
which allows us to rewrite (\ref{lapS1-a}) as
\begin{equation}\label{lapS1-b}
  F'' + \xi F' - \zeta F = 0
\end{equation}
To compute the leading term of the solution to (\ref{lapS1-b}) as $r \to \infty$, make the substitution
\begin{equation}
  F(r) = \e^{S(r)}, \qquad \text{where} \qquad S(r) = S_{0}(r) + S_{1}(r) + S_{2}(r) + \cdots
\end{equation}
with
\begin{equation}
  S_{0}(r) \gg S_{1}(r) \gg S_{2}(r) \gg \cdots
\end{equation}
as $r \to \infty$.  To compute the WKB approximation, we will compute the values of $S_{0}$ and $S_{1}$ in this expansion.  As is standard, we will also make the assumption that
\begin{equation}
  S'' \ll (S')^{2}.
\end{equation}
Making this substitution in (\ref{lapS1-b}) yields as $r \to \infty$
\begin{align}
  \notag F'' + \xi F' &\sim \zeta F \\
  \notag \e^{S}\pnth{S'' + (S')^2} + \xi \e^{S}S' &\sim \zeta \e^{S} \\
  \notag S'' + (S')^2 + \xi S' &\sim \zeta \\
  \notag (S')^2 + \xi S' &\sim \zeta \\
  \label{lapS1-c} S' &\sim -\frac{\xi}{2} \pm \frac{1}{2}\sqrt{\xi^{2} + 4\zeta}.
\end{align}
For this next step, we will compute the asymptotic expansion of the right hand side.  The asymptotic expansion of some function $B(r)$ is defined as
\begin{equation}
  \Taylor_{s = 0}\pnth{\hat{B}(s)}\bigg|_{s = \frac{1}{r}}
\end{equation}
where $\hat{B}(s) = B(1/s)$ and $\Taylor_{s=0}$ denotes the Taylor expansion centered at $s=0$.  If we let
\begin{equation}
  Q = \sqrt{\Up^{2} - \frac{\om^{2}}{\ka^{2}}}\pnth{ 1 - \frac{\om^{2}}{\Up^{2}\ka^{2} - \om^{2}}}
\end{equation}
Then the asymptotic expansion of the right hand side of (\ref{lapS1-c}), whose computation we have suppressed, is
\begin{equation}
  -\frac{\xi}{2} \pm \frac{1}{2}\sqrt{\xi^{2} + 4\zeta} \sim \pm\sqrt{\Up^{2} - \frac{\om^{2}}{\ka^{2}}} - \frac{1}{r} \pm \frac{mQ}{r} + O\pnth{\frac{1}{r^{2}}}.
\end{equation}
Thus as $r \to \infty$,
\begin{equation}\label{lapS1-d}
  S' \sim \pm\sqrt{\Up^{2} - \frac{\om^{2}}{\ka^{2}}} - \frac{1}{r} \pm \frac{mQ}{r} + O\pnth{\frac{1}{r^{2}}}
\end{equation}
which is consistent with our approximation that $S'' \ll (S')^2$.  Equation (\ref{lapS1-d}) yields
\begin{equation}\label{lapS1-e}
  S \sim \pm r\sqrt{\Up^{2} - \frac{\om^{2}}{\ka^{2}}} - \ln r \pm mQ\ln r  + \ln C + O\pnth{\frac{1}{r}}
\end{equation}
for some real constant $C>0$.  Thus we have
\begin{align}
  S_{0} &\sim \pm r\sqrt{\Up^{2} - \frac{\om^{2}}{\ka^{2}}} + \ln C \\
  S_{1} &\sim - \ln r \pm mQ\ln r
\end{align}
This yields that the leading term of $F$ is of the form
\begin{equation}\label{lapS1-f}
  \frac{Cr^{\pm m Q}}{r}\e^{\pm r\sqrt{\Up^{2} - \frac{\om^{2}}{\ka^{2}}}}.
\end{equation}
Thus as $r \to \infty$, the WKB approximation of the solution to (\ref{lapS1}) will be of the form
\begin{equation}\label{lapS1-g}
  F = \frac{C_{1}r^{m Q}}{r}\e^{r\sqrt{\Up^{2} - \frac{\om^{2}}{\ka^{2}}}} + \frac{C_{2}r^{-m Q}}{r}\e^{-r\sqrt{\Up^{2} - \frac{\om^{2}}{\ka^{2}}}}.
\end{equation}
However, in practice, the solutions we will be considering will usually have total mass values $m \ll 1$ (for a few examples of total mass values of solutions see Figure \ref{Ups50}).  Under these circumstances, the above WKB approximation can be further approximated by
\begin{equation}\label{lapS1-h}
  F = \frac{C_{1}}{r}\e^{r\sqrt{\Up^{2} - \frac{\om^{2}}{\ka^{2}}}} + \frac{C_{2}}{r}\e^{-r\sqrt{\Up^{2} - \frac{\om^{2}}{\ka^{2}}}}
\end{equation}
which is exactly what we have in (\ref{Ssol}).  It is important to note here, however, that if the mass value $m$ is not sufficiently small, requiring equation (\ref{lapS1-h}) would no longer maintain that the scalar field asymptotically satisfied the Klein-Gordon equation in the Schwarzschild metric.  We would instead need to impose the entire WKB approximation in (\ref{lapS1-g}) to get the appropriate asymptotics.  For our work here, though, we will use equation (\ref{lapS1-h}), or equivalently, equation (\ref{Ssol}).

Now we also require that $F \to 0$ as $r \to \infty$ so that $f \to 0$ as well.  Thus $C_{1} = 0$ and we relabel $C_{2}$ as simply $C$.  That is, at $r = r_{max}$, we require
\begin{equation}\label{pre-AB1}
  F = \frac{C}{r}\e^{-r\sqrt{\Up^{2} - \frac{\om^{2}}{\ka^{2}}}}.
\end{equation}
We have no way of directly determining the correct value of $C$ in equation (\ref{pre-AB1}) associated with a given static solution.  However, if we differentiate equation (\ref{pre-AB1}) with respect to $r$, we obtain
\begin{align}
  \notag F' &= -\frac{C}{r}\e^{-r\sqrt{\Up^{2} - \frac{\om^{2}}{\ka^{2}}}}\sqrt{\Up^{2} - \frac{\om^{2}}{\ka^{2}}} - \frac{C}{r^{2}}\e^{-r\sqrt{\Up^{2} - \frac{\om^{2}}{\ka^{2}}}} \\
  &= -\pnth{\sqrt{\Up^{2} - \frac{\om^{2}}{\ka^{2}}} + \frac{1}{r}}F
\end{align}
which does not depend on $C$.  Then the condition that at $r=r_{max}$, $f$ approximately satisfies the Klein-Gordon equation with the Schwarzschild background metric reduces to requiring that
\begin{equation}\label{AB1b}
  F'(r_{max}) + \pnth{\sqrt{\Up^{2} - \frac{\om^{2}}{\ka^{2}}} + \frac{1}{r_{max}}}F(r_{max}) \approx 0.
\end{equation}
This condition imposes that $F$ is decaying appropriately to $0$.  It also puts a restriction on the possible values of $\om$.  Since the left hand side of the above equation must be real, we have that $\Up^{2} - \frac{\om^{2}}{\ka^{2}} \geq 0$, or equivalently,
\begin{equation}
  \abs{\frac{\om}{\ka}} = \frac{\om}{\ka} \leq \Up.
\end{equation}
That is, $\om/\ka \in [0,\Up]$.  In our numerical calculations, we normalize this quantity and keep track of $\dfrac{\om}{\ka\Up} \in [0,1]$ instead.

Next we consider the boundary at the origin $r = 0$.  We have already noted above that $M(0) = 0$.  Moreover, since $f_{r}(t,0) = 0$ for all $t$,
\begin{equation}
  H(0) = F'(0) = \e^{-i\om t}f_{r}(t,0) = 0.
\end{equation}
Note that by the system (\ref{NCpdec}), $H(0)=0$ implies that if $F(0)=0$, then $F(r) = 0$ for all $r$, since $H'(r)$ and $F'(r)$ will never change.  We are interested in non-trivial solutions to these equations so we require that $F(0) \neq 0$.  However, for any constants $c$ and $\mu^{\ast}$, if $cf$ is a solution to the Einstein-Klein-Gordon equations, (\ref{EKG}), with $\mu_{0}=\mu^{\ast}$, then $f$ is a solution to the Einstein-Klein-Gordon equations with $\mu_{0} = c^{2}\mu^{\ast}$.  Thus, without loss of generality, we will set $F(0)=1$ absorbing any excess factors into $\mu_{0}$.

At this point, we have four remaining parameters to choose, namely, $V(0)$, $\om$, $\Up$, and $\mu_{0}$.  By requiring (\ref{AB2b}) and (\ref{AB1b}), two of these values are determined, leaving two remaining degrees of freedom.  The parameter $\Up$ is an as yet unknown fundamental constant, making it important in our computations to be able to freely set $\Up$ so that we can test different values and see how they match the data.  The parameter $\mu_{0}$ controls the magnitude of the energy density and so seems a natural choice as another parameter to freely choose.  However, this choice is not the only choice that could be made.  For example, one could instead freely choose $\Up$ and $\om$ and use the boundary conditions to pin down $V(0)$ and $\mu_{0}$ with equivalent results.

When freely choosing $\Up$ and $\mu_{0}$, to compute the other two parameters, $V(0)$ and $\om$, we solve a shooting problem to satisfy the desired boundary conditions.  We illustrate next in detail how we performed this shooting procedure.  However, first, we summarize the information about the boundary conditions.

For some choice of $\Up$ and $\mu_{0}$, we require at $r = 0$,
\begin{align}
  \label{cenval} F(0) &= 1, & H(0) &= 0, & M(0) &= 0, & V(0) &= V_{0},
\end{align}
and choose $\om$ and $V_{0}$ to satisfy
\begin{align}
  \label{AB1c} F'(r_{max}) + \pnth{\sqrt{\Up^{2} - \frac{\om^{2}}{\ka^{2}}} + \frac{1}{r_{max}}}F(r_{max}) &\approx 0, \\
  \label{AB2c} V(r_{max}) - \frac{1}{2}\ln\pnth{1 - \frac{2M(r_{max})}{r_{max}}} - \ln \ka &\approx 0.
\end{align}
For simplicity in our calculations, we set $\ka = 1$.  Then we use the standard forward Euler method to solve the system (\ref{NCpdec}) with these boundary conditions.

To understand the procedure we used to solve the shooting problems mentioned above, we first comment about what equation (\ref{NCpde4b}) tells us about the behavior of $F$ in our system.  We are solving these equations in the low field limit where the metric (\ref{metric}) is close to Minkowski.  That is, both $V$ and $M$ are approximately zero.  Recall that equation (\ref{NCpde4b}) came immediately from equation (\ref{NCpde4b-derive}), which results from substituting (\ref{sfstaticstate}) into (\ref{NCpde4a}).  Letting $V=M=0$ makes equation (\ref{NCpde4a}) and the first line of equation (\ref{NCpde4b-derive}) reduce to
\begin{equation}\label{prinpart}
  \lap_{r}F = \pnth{\Up^{2} - \om^{2}} F
\end{equation}
where $\lap_{r}$ is the Laplacian in $\RR^{3}$ in spherical coordinates.  The one dimensional analogue to the above equation is
\begin{equation}
  F_{xx} = k F.
\end{equation}
The solution of this differential equation depends on the sign of $k$.  If $k>0$, then the solutions either exponentially grow or decay.  If $k<0$, the solutions exhibit oscillatory behavior.  In equation (\ref{NCpde4b}), the analogous coefficient that will control whether the solutions exhibit oscillatory or exponential behavior is the following, which we will denote as $\la(r)$.
\begin{equation}
  \la(r) = \pnth{\Up^{2} - \frac{\om^{2}}{\e^{2V(r)}}}.
\end{equation}
The sign of $\la(r)$ depends on $r$.  While $\la(r) < 0$, the solution of (\ref{NCpde4b}) will exhibit oscillatory behavior.  On the other hand, while $\la(r) >0$, the solution of (\ref{NCpde4b}) will exhibit exponential growth or decay.

Since $\Up$ and $\om$ are constants, the sign of $\la(r)$ is completely controlled by the $\exp(2V)$ term.  If $V$ is strictly increasing and starts low enough, then for small $r$, $\exp(2V)$ will be small yielding that $\la(r) < 0$ and the initial part of the solution will oscillate.  Then as $r$ increases, $\exp(2V)$ will eventually be large enough that the $\Up^{2}$ term dominates $\la(r)$ making $\la(r) > 0$ and the end behavior will be exponential growth or decay.  Imposing boundary condition (\ref{AB1c}) will ensure decaying end behavior instead of growth.

The two parameters that control where this change from oscillation to exponential decay occurs are the initial value of $V$ and the parameter $\om$.  Larger values of $\om$ and lower values of $V(0)$ will increase the length of the oscillating region and push the point where it changes to exponential decay out to larger radii.  In fact, given a fixed $\Up$ and $\mu_{0}$, for each value of $V(0)$, there is a discrete infinite set of $\om$ values for which each $\om$ in the set corresponds to a solution $F$ that has a given number of zeros (caused by a lengthening of the oscillating region) and the appropriate end behavior.

Thus we perform our shooting problem as follows.  To find a solution with say $n$ zeros, first, fix a choice of $\Up$ and $\mu_{0}$.  Then pick a value of $V(0) = V_{0}$.  Since we have set $\ka = 1$, $\om < \Up$ in order to be able to satisfy equation (\ref{AB1c}).  Then we systematically pick different values of $\om$ in the interval $[0,\Up]$ until we obtain the appropriate number of zeros and satisfy (\ref{AB1c}).  Finally, we use a Newton's method approach to find the value of $V_{0}$ whose corresponding solution with $n$ zeros yields a potential function satisfying (\ref{AB2c}).

These solutions, now characterized by the number of zeros the resulting scalar field has, are referred to as spherically symmetric state states of wave dark matter.  A static state with no zeros is called a ground state; with one zero, it is called a first excited state; with two, it is called a second excited state, and so forth.  There is a considerable amount known about static states as they have been studied for decades, see \cite{MSBS, Seidel90, Seidel98, Gleiser89, Hawley00, Lai07} for just a few examples.  In the remaining sections, we will present some useful results about these static states.

\subsection{Plots of Static States}

\begin{figure}

  \begin{center}
    \includegraphics[height = 2.25 in, width = 2.85 in]{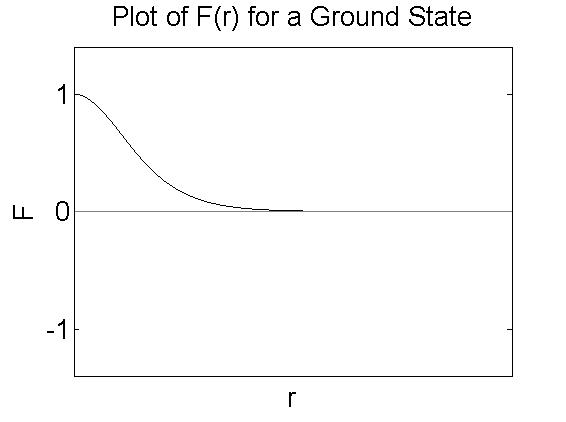}
    \includegraphics[height = 2.25 in, width = 2.85 in]{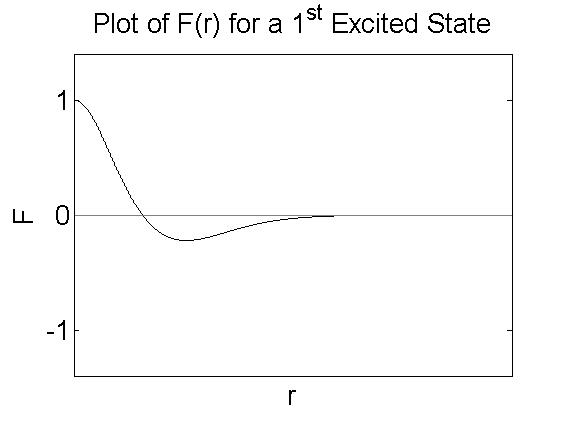}

    \includegraphics[height = 2.25 in, width = 2.85 in]{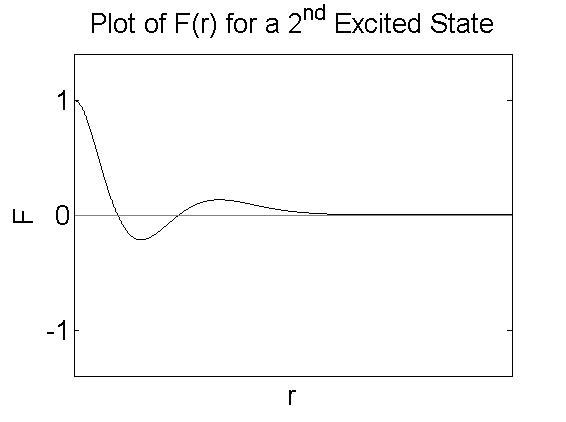}
    \includegraphics[height = 2.25 in, width = 2.85 in]{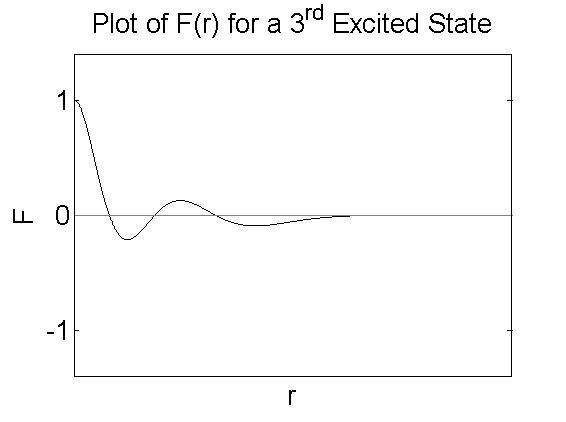}
  \end{center}

  \caption{Plots of static state scalar fields (specifically the function $F(r)$ in (\ref{sfstaticstate})) in the ground state and first, second, and third excited states.  Note the number of nodes (zeros) of each function.}

  \label{state_ex}

\end{figure}

\begin{figure}

    \begin{center}
        \includegraphics[height = 2.25 in, width = 2.85 in]{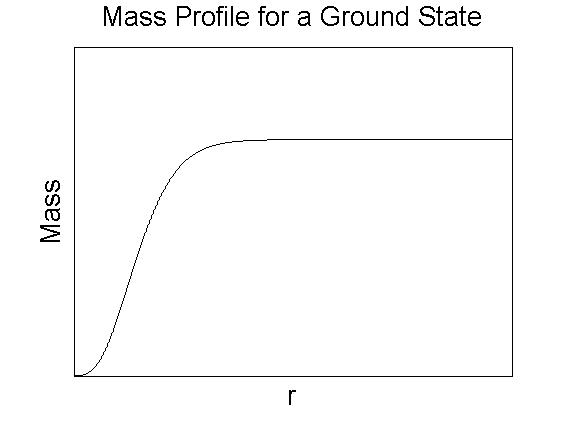}
        \includegraphics[height = 2.25 in, width = 2.85 in]{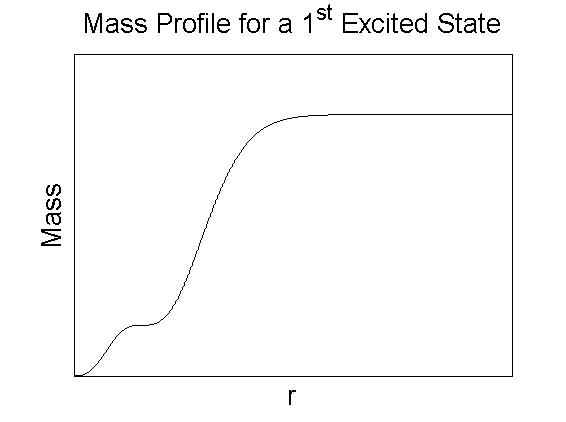}

        \includegraphics[height = 2.25 in, width = 2.85 in]{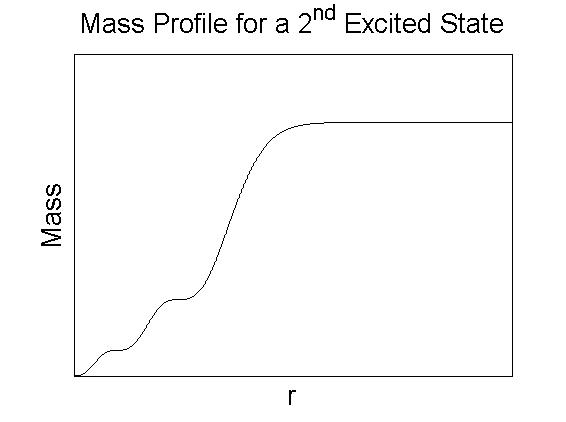}
        \includegraphics[height = 2.25 in, width = 2.85 in]{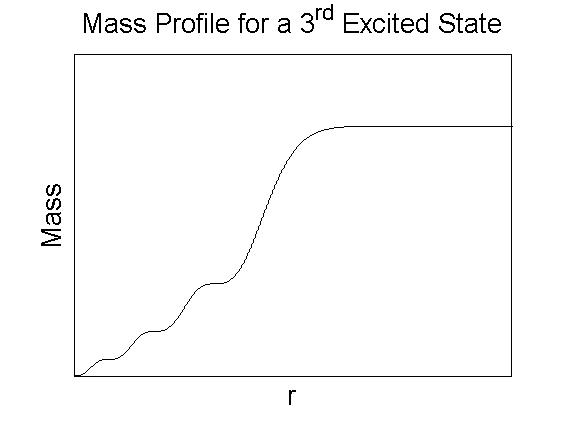}
    \end{center}

    \caption{Mass profiles for a static ground state and first, second, and third excited states of wave dark matter.}

    \label{state_m}

\end{figure}

\begin{figure}

    \begin{center}
        \includegraphics[height = 2.25 in, width = 2.85 in]{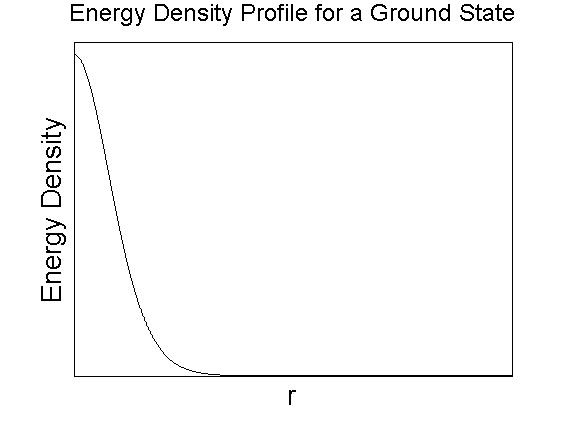}
        \includegraphics[height = 2.25 in, width = 2.85 in]{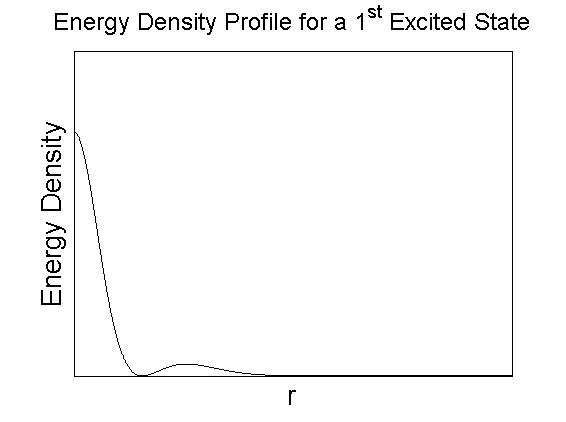}

        \includegraphics[height = 2.25 in, width = 2.85 in]{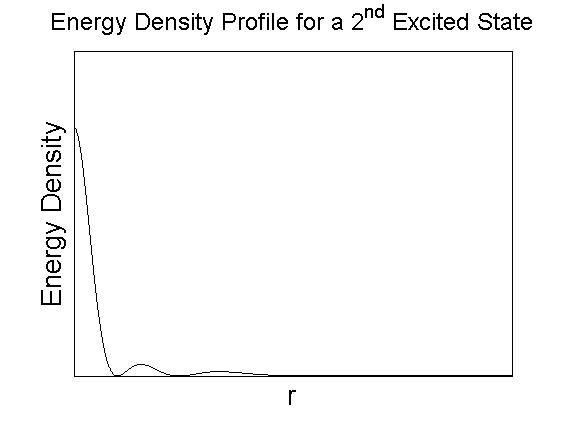}
        \includegraphics[height = 2.25 in, width = 2.85 in]{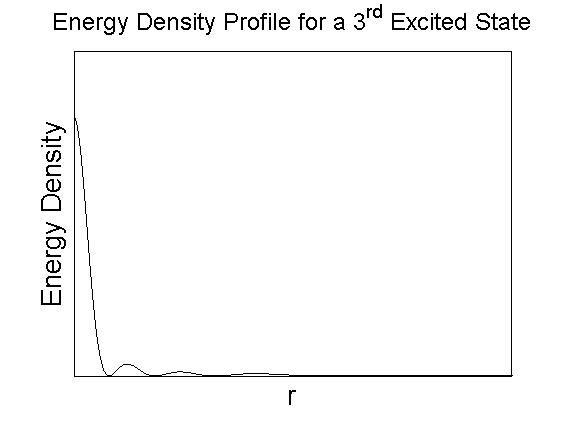}
    \end{center}

    \caption{Energy density profiles for a static ground state and first, second, and third excited states of wave dark matter.}

    \label{state_ED}

\end{figure}

\begin{figure}

    \begin{center}
        \includegraphics[height = 2.25 in, width = 2.85 in]{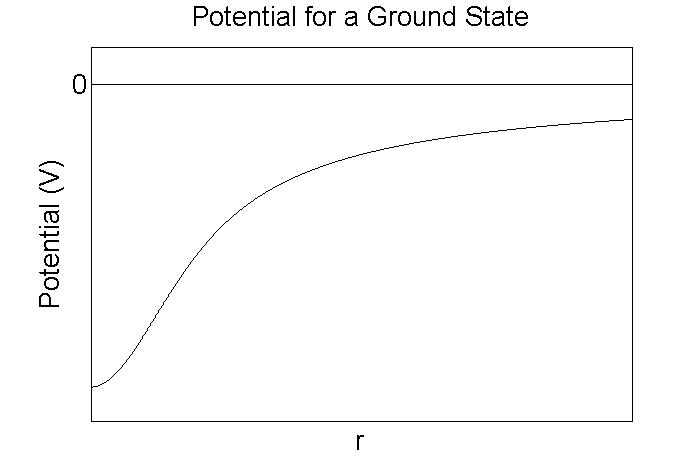}
        \includegraphics[height = 2.25 in, width = 2.85 in]{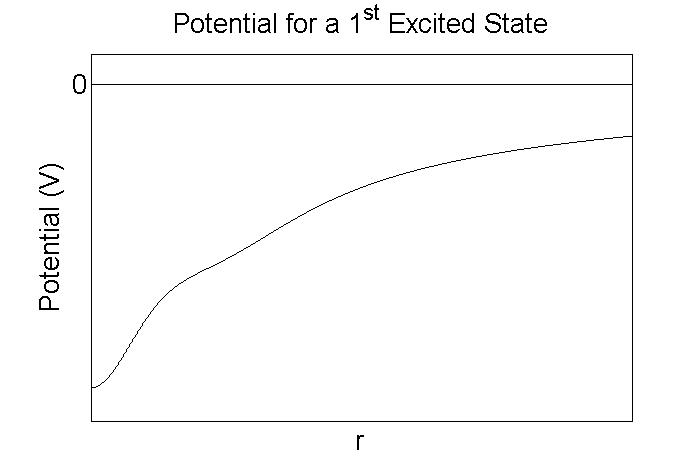}

        \includegraphics[height = 2.25 in, width = 2.85 in]{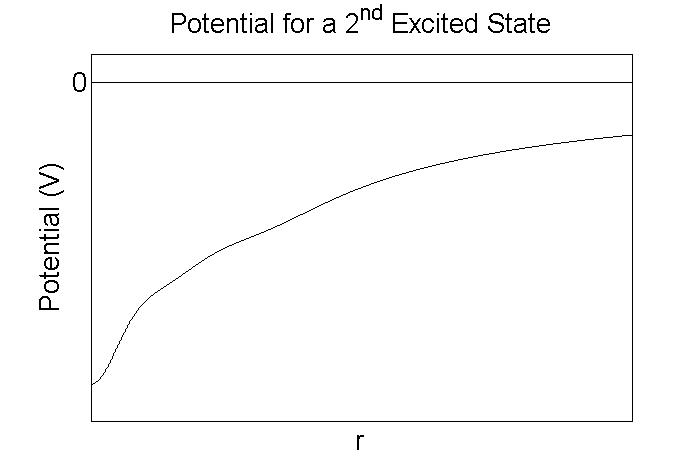}
        \includegraphics[height = 2.25 in, width = 2.85 in]{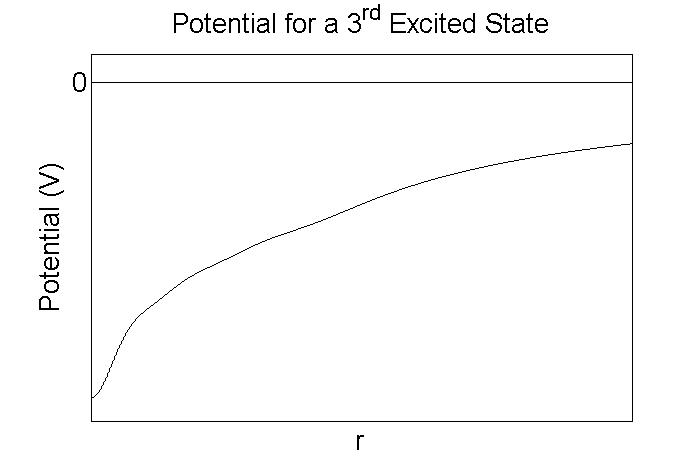}
    \end{center}

    \caption{Plots of the potential function, $V$, for a static ground state and first, second, and third excited states of wave dark matter.}

    \label{state_V}

\end{figure}

In Figures \ref{state_ex} to \ref{state_V}, we have plotted examples of the scalar field $F(r)$ (see Figure \ref{state_ex}), Mass $M(r)$ (see Figure \ref{state_m}), energy density $\mu(r)$ (see Figure \ref{state_ED}), and gravitational potential $V(r)$ (see Figure \ref{state_V}) for a generic ground state and first through third excited states.  We make here the following three observations.  First, in these plots, $V$ is strictly increasing, as expected, and $M$ approaches a constant value as $r \to \infty$, also as we expected.  Second, for each zero of the function $F$, there is a zero of the energy density plot $\mu$, a ripple in the mass profile $M$, and a slight but rapid change in the slope of the potential $V$.  And finally, the energy density appears to be approximately proportional to $\abs{F}^{2} = \abs{f}^{2}$.

\section{Families of Static States}

We explained above that four parameters control what static solution is generated by the equations.  However, since we require two conditions on the boundary, choosing only two of these parameters will define a static state.  As stated before, we choose to define $\Up$ and $\mu_{0}$ and solve the shooting problem for the parameters $V_{0}$ and $\om$.  For each $n$, this defines a function, $\S_{n}:\RR^{2}\to \RR^{2}$, which maps the pair $(\Up,\mu_{0})$ to the pair $(V_{0},\om)$ such that the choices $\Up,\mu_{0},V_{0},\om$ yields an $n^{\text{th}}$ excited state ($n=0$, of course, referring to the ground state).

\begin{figure}

    \begin{center}
        \includegraphics[height = 2.25 in, width = 2.85 in]{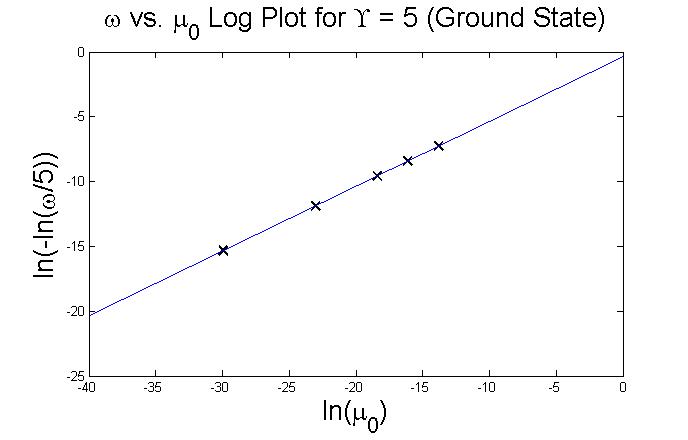}
        \includegraphics[height = 2.25 in, width = 2.85 in]{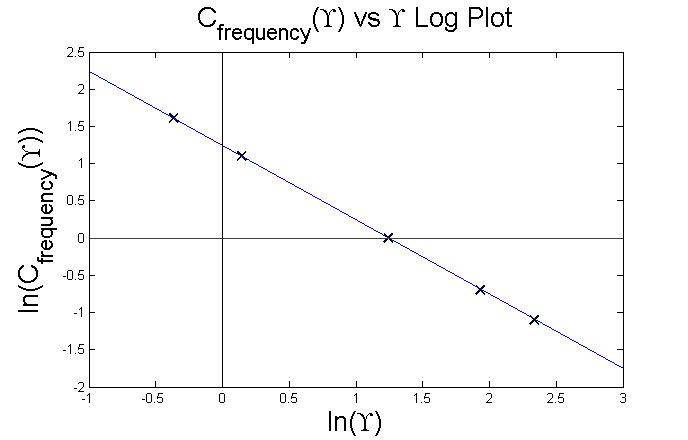}
    \end{center}

    \caption{Left: Log plot of the parameters $\om$ and $\mu_{0}$ for a ground state and constant value of $\Up = 5$.  The slope of this plot is almost exactly $1/2$.  We get the same slope for other values of $\Up$, thus $\om(\Up,\mu_{0}) = \Up \e^{C_{frequency}(\Up)\sqrt{\mu_{0}}}$.  Right: Log plot of the parameters $C_{frequency}(\Up)$ and $\Up$ for a ground state.  The slope of this plot is almost exactly $-1$.  Thus $C_{frequency}(\Up) = C_{frequency}/\Up$, where $C_{frequency}$ is a constant.  Similar plots exist for any $n^{\text{th}}$ excited state.}

    \label{logplot}

\end{figure}

A natural question to ask here would be ``Is there an expression for $\S_{n}$ for each $n$?''  The answer to this question is yes, at least in the low field limit.  In fact, we have also found expressions for the total mass $m$ of the system and the half-mass radius $r_{h}$, that is, the radius $r_{h}$ for which $M(r_{h}) = m/2$. These expressions were found by numerically computing the states for several different values of $\Up$ and $\mu_{0}$, all of which yield a state in the low field limit.  We analyzed the resulting values and found that certain log plots between the values were linear.  We have collected in Figure \ref{logplot} some of these plots for the ground state that led to this conclusion.

These log plots yielded the following expressions, which we emphasize are only expected to hold in the low field limit.  Let $\om^{n}$, $V_{0}^{n}$, $m^{n}$, $r_{h}^{n}$ be respectively the values of $\om$, $V(0)$, $m$, and $r_{h}$ for an $n^{\text{th}}$-excited state corresponding to a choice of $\Up$ and $\mu_{0}$.  Then we have that
\begin{subequations}\label{Scorr}
\begin{align}
  \label{Scorr1} \om^{n}(\Up,\mu_{0}) &\approx \Up \exp\pnth{C_{frequency}^{n}\frac{\sqrt{\mu_{0}}}{\Up}} \\
  \label{Scorr2} V_{0}^{n}(\Up,\mu_{0}) &\approx C_{potential}^{n}\frac{\sqrt{\mu_{0}}}{\Up} \\
  \label{Scorr3} m^{n}(\Up,\mu_{0}) &\approx C_{mass}^{n}\Up^{-3/2}\mu_{0}^{1/4} \\
  \label{Scorr4} r_{h}^{n}(\Up,\mu_{0}) &\approx C_{radius}^{n}\Up^{-1/2}\mu_{0}^{-1/4}
\end{align}
\end{subequations}
for some constants $C_{frequency}^{n}$, $C_{potential}^{n}$, $C_{mass}^{n}$ and $C_{radius}^{n}$ which depend only on $n$.  We have computed these constants for the ground through fifth excited states as well as for the tenth and twentieth excited states and have collected their values in Table \ref{ScorrVal}.  Note also that equations (\ref{Scorr1}) and (\ref{Scorr2}) constitute the $\S_{n}$ function mentioned above.

\begin{table}

  \caption{Values of the constants in the system (\ref{Scorr}) for the ground through fifth excited states as well as the tenth and twentieth excited states.  We have given them error ranges which encompass the interval we observed in our experiments.  However, it is possible that values outside our ranges here could be observed, though we don't expect them to be so by much if the discretization of $r$ used in solving the ODEs is sufficiently fine.  Note also that our values have less precision as we increase $n$.  This is because as $n$ increases, it becomes more difficult to compute the states with as much precision.}

  \begin{center}

  \begin{tabular}{c|cccc}
    $n$ & $C_{frequency}^{n}$ & $C_{potential}^{n}$ & $C_{mass}^{n}$ & $C_{radius}^{n}$ \\
    \hline & & & & \\
    $0$ & $-3.4638 \pm 0.010$ & $-6.7278 \pm 0.003$ & $4.567 \pm 0.05$ & $0.8462 \pm 0.004$ \\
    $1$ & $-3.2422 \pm 0.012$ & $-7.5411 \pm 0.007$ & $10.22 \pm 0.10$ & $2.2894 \pm 0.009$ \\
    $2$ & $-3.1566 \pm 0.014$ & $-7.9315 \pm 0.009$ & $15.81 \pm 0.16$ & $3.8253 \pm 0.014$ \\
    $3$ & $-3.1062 \pm 0.015$ & $-8.1823 \pm 0.010$ & $21.37 \pm 0.22$ & $5.3994 \pm 0.018$ \\
    $4$ & $-3.0714 \pm 0.015$ & $-8.3653 \pm 0.010$ & $26.91 \pm 0.27$ & $6.9860 \pm 0.022$ \\
    $5$ & $-3.0452 \pm 0.016$ & $-8.5086 \pm 0.011$ & $32.42 \pm 0.33$ & $8.5606 \pm 0.026$ \\
    $10$ & $-3.0076 \pm 0.052$ & $-9.0018 \pm 0.037$ & $60.32 \pm 1.18$ & $15.1357 \pm 0.039$ \\
    $20$ & $-2.9949 \pm 0.077$ & $-9.5061 \pm 0.074$ & $116.62 \pm 2.57$ & $29.6822 \pm 0.107$
  \end{tabular}
  \vspace{\baselineskip}

  \label{ScorrVal}

  \end{center}

\end{table}

In the wave dark matter model, $\Up$ is a fundamental constant that should be the same throughout the universe.  In the case of constant $\Up$, for each $n$, the equations in (\ref{Scorr}) define one-parameter families of the $n^{\text{th}}$ excited states, the parameter being the scaling constant $\mu_{0}$.  For constant $\Up$ then, we have that these static states only differ in size where, by equations (\ref{Scorr3}) and (\ref{Scorr4}), a larger $\mu_{0}$ corresponds to a more massive and more dense wave dark matter halo.

\subsection{Scalings of Static States}

Certain scalings exist for various quantities if we scale time and length in any coordinate system.  In particular, let us scale the time coordinate, $t$, and the standard spatial coordinates, $x^{i}$, so that
\begin{align}
  \label{Scal0}  \text{Time: } & \bar{t} = \be t & \text{Distance: } & \bar{x}^{i} = \al x^{i}
\end{align}
for some positive constants $\be,\al \in \RR$.  Then velocities in the $(t,x)$ system will scale to velocities in the $(\bar{t},\bar{x})$ system as follows
\begin{equation}
  \bar{v}^{i}  = \frac{d\bar{x}^{i}}{d\bar{t}} = \frac{\al}{\be}\frac{dx^{i}}{dt} = \frac{\al}{\be}v^{i}
\end{equation}
Similarly, other quantities scale as follows (velocity is included again for completeness).
\begin{align}
  \notag \text{Velocity: } & \bar{v} = \frac{\al}{\be}v & \text{Mass: } & \bar{m} = \frac{\al^{3}}{\be^{2}}m \\
  \notag \text{Energy Density: } & \bar{\mu} = \frac{1}{\be^{2}}\mu & \text{Gravitational Potential: } & \bar{V} = \frac{\al^{2}}{\be^{2}}V \\
  \label{Scal1} \text{Frequency: } & \bar{\la} = \frac{1}{\be}\la
\end{align}
These scalings are routine to derive and follow directly from how the units on each of these quantities scale with the scaling for mass being that which is required to keep the universal gravitational constant the same from one scaled coordinate system to the other.

From the system (\ref{Scorr}), the scalings in (\ref{Scal0}) and (\ref{Scal1}), and given that the constants $C_{\ast}^{n}$ are dimensionless, we infer how the parameters $\mu_{0}$ and $\Up$ would scale under these coordinate scalings in order to keep (\ref{Scorr}) invariant.  To do this, let $c_{1}$ and $c_{2}$ be such that under the coordinate scalings in (\ref{Scal0}),
\begin{equation} \label{Scal2}
  \bar{\mu}_{0} = c_{1}\mu_{0} \qquad \text{and} \qquad \bar{\Up} = c_{2}\Up.
\end{equation}
Since $m$ and $r_{h}$ are a mass and spatial quantity respectively, equations (\ref{Scorr3}) and (\ref{Scorr4}) yield
\begin{align}
  \notag \bar{m}^{n} &= \frac{\al^{3}}{\be^{2}}m^{n} \\
  \notag &\approx \frac{\al^{3}}{\be^{2}} C_{mass}^{n}\Up^{-3/2}\mu_{0}^{1/4} \\
  \notag &= \frac{\al^{3}}{\be^{2}} C_{mass}^{n}c_{2}^{3/2}\bar{\Up}^{-3/2} c_{1}^{-1/4}\bar{\mu}_{0}^{1/4} \\
  \label{invScorr3} &= \frac{\al^{3}c_{2}^{3/2}}{\be^{2}c_{1}^{1/4}} C_{mass}^{n}\bar{\Up}^{-3/2}\bar{\mu}_{0}^{1/4}
\end{align}
and
\begin{align}
  \notag \bar{r}_{h}^{n} &= \al r_{h}^{n} \\
  \notag &\approx \al C_{radius}^{n}\Up^{-1/2}\mu_{0}^{-1/4} \\
  \notag &= \al C_{radius}^{n}c_{2}^{1/2}\bar{\Up}^{-1/2}c_{1}^{1/4}\bar{\mu}_{0}^{-1/4} \\
  \label{invScorr4} &= \al c_{2}^{1/2} c_{1}^{1/4} C_{radius}^{n}\bar{\Up}^{-1/2}\bar{\mu}_{0}^{-1/4}
\end{align}
Then requiring these equations to be invariant under coordinate scalings is equivalent to
\begin{equation} \label{invScorr0}
  \frac{\al^{3}c_{2}^{3/2}}{\be^{2}c_{1}^{1/4}} = 1 \qquad \text{and} \qquad \al c_{2}^{1/2}c_{1}^{1/4} = 1
\end{equation}
Solving these two equations simultaneously for $c_{1}$ and $c_{2}$ yields
\begin{equation}\label{invScorr0a}
  c_{1} = \frac{1}{\be^{2}} \qquad \text{and} \qquad c_{2} = \frac{\be}{\al^{2}}.
\end{equation}
Thus equations (\ref{Scorr3}) and (\ref{Scorr4}) imply that $\mu_{0}$ and $\Up$ scale as follows
\begin{align}
  \label{Scal3} \text{Stress Energy Tensor Constant: }& \bar{\mu}_{0} = \frac{1}{\be^{2}}\mu_{0} & \text{Upsilon: }& \bar{\Up} = \frac{\be}{\al^{2}}\Up.
\end{align}
We observe here that $\mu_{0}$ scales as energy density, which is expected given that as we said before it controls the magnitude of the energy density function defined by the stress energy tensor.

Since $\Up$ is a fundamental constant in this system, it is appropriate to ask under what kinds of coordinate scalings of the form in (\ref{Scal0}) is $\Up$ invariant, that is, $\bar{\Up} = \Up$.  In light of (\ref{Scal3}), the answer to this question is readily apparent, those scalings where $\be = \al^{2}$.  Under this type of scaling, (\ref{Scal0}), (\ref{Scal1}), and (\ref{Scal3}) become
\begin{align}
  \notag \text{Time: } & \bar{t} = \al^{2} t & \text{Distance: } & \bar{x} = \al x \\
  \notag \text{Velocity: } & \bar{v} = \frac{1}{\al}v & \text{Mass: } & \bar{m} = \frac{1}{\al}m \\
  \notag \text{Energy Density: } & \bar{\mu} = \frac{1}{\al^{4}}\mu & \text{Gravitational Potential: } & \bar{V} = \frac{1}{\al^{2}}V \\
  \notag \text{Frequency: } & \bar{\la} = \frac{1}{\al^{2}}\la \\
  \label{Scal4} \text{Stress Energy Tensor Constant: }& \bar{\mu}_{0} = \frac{1}{\al^{4}}\mu_{0} & \text{Upsilon: }& \bar{\Up} = \Up.
\end{align}

The scalings in (\ref{Scal4}) are also consistent with keeping the remaining relations (\ref{Scorr1}) and (\ref{Scorr2}) invariant under coordinate scalings.  Showing this for (\ref{Scorr2}) is straightforward and follows from (\ref{Scal4}) and the fact that $V_{0}$ has gravitational potential units,
\begin{align}
  \notag \bar{V}_{0}^{n} &\approx C_{potential}^{n}\frac{\sqrt{\bar{\mu}_{0}}}{\bar{\Up}} \\
  \notag \frac{\al^{2}}{\be^{2}} V_{0}^{n} &\approx C_{potential}^{n}\frac{\sqrt{\be^{-2}\mu_{0}}}{\be\al^{-2}\Up} \displaybreak[0]\\
  \notag \frac{\al^{2}}{\be^{2}} V_{0}^{n} &\approx C_{potential}^{n}\frac{\al^{2}}{\be^{2}}\frac{\sqrt{\mu_{0}}}{\Up} \\
  V_{0}^{n} &\approx C_{potential}^{n}\frac{\sqrt{\mu_{0}}}{\Up}.
\end{align}
We should note that the above holds even if we do not assume that $\be = \al^{2}$.  However, the invariance of the next equation does rely on the fact that $\be = \al^{2}$.  In fact, it relies on the form the static Einstein-Klein-Gordon equations take in the low field limit.

Specifically, consider a static spacetime in the low field Newtonian limit and assume its static metric can be written in the form
\begin{equation}
  g = -\tV(\x)^{2}\ dt^{2} + \tV(\x)^{-2}\ d\x^{2}
\end{equation}
where $d\x^{2}$ is the flat metric on $\RR^{3}$ and $\tV \approx 1$.  Additionally, assume the spacetime is in the presence of a scalar field $f(t,\x)$ which can be written of the form
\begin{equation}
  f(t,\x) = \e^{i \Up t}\hF(t,\x)
\end{equation}
where $\hF$ has small time derivative and changes little on the scale of $\frac{1}{\Up}$.  Under these assumptions, the Einstein-Klein-Gordon equations reduce to the following system of equations
\begin{align}
  \lap_{\RR^{3}}\hV &= 4\pi \pnth{2\mu_{0}\big|\hF\big|^{2}} \\
  i\pa{\hF}{t} &= \frac{1}{2\Up}\lap_{\RR^{3}} \hF - \Up \hV\hF
\end{align}
where $\hV = \tV - 1$.  This system is called the Poisson-\Schrodinger\ equations.  In spherical symmetry, we would write this scalar field as $f(t,r) = \e^{i\Up t}\hF(t,r)$ and since the metric is static, the spherically symmetric static state would be contained in this class of solutions.  Since the static states are of the form $f(t,r) = \e^{i\om t}F(r)$, we have that
\begin{equation}
  \hF(t,r) = \e^{i\pnth{\om - \Up}t}F(r).
\end{equation}
Thus the angular frequency of the scalar field $\hF$ in the Poisson-\Schrodinger\ equations is $\om - \Up$.  Being a frequency in the low field limit, we would expect this to scale according to (\ref{Scal4}).

We now have what we need to show that (\ref{Scorr1}) is invariant under the scalings in (\ref{Scal4}).  Since (\ref{Scorr1}) is not a power function, we approximate the equation to first order using the Taylor expansion of $\e^{x}$ and show that the approximate equation is invariant which implies that the original equation is approximately invariant.  This is sufficient since all of the equations in (\ref{Scorr}) are only approximations anyway.
\begin{align}
  \notag \bar{\om}^{n} &\approx \bar{\Up}\exp\pnth{C_{frequency}^{n}\frac{\sqrt{\bar{\mu}_{0}}}{\bar{\Up}}} \\
  \notag \bar{\om}^{n} - \bar{\Up} &\approx \bar{\Up}\exp\pnth{C_{frequency}^{n}\frac{\sqrt{\bar{\mu}_{0}}}{\bar{\Up}}} - \bar{\Up} \displaybreak[0]\\
  \notag \frac{1}{\al^{2}}\pnth{\om^{n} - \Up} &\approx \bar{\Up} + \bar{\Up} C_{frequency}^{n}\frac{\sqrt{\bar{\mu}_{0}}}{\bar{\Up}} - \bar{\Up} \\
  \notag &= C_{frequency}^{n}\sqrt{\bar{\mu}_{0}}  \\
  \notag &= C_{frequency}^{n}\sqrt{\frac{\mu_{0}}{\al^{4}}}  \displaybreak[0]\\
  \notag \frac{1}{\al^{2}}\pnth{\om^{n} - \Up} &\approx \frac{1}{\al^{2}}C_{frequency}^{n}\sqrt{\mu_{0}}  \displaybreak[0]\\
  \notag \om^{n} - \Up &\approx C_{frequency}^{n}\sqrt{\mu_{0}} \displaybreak[0] \\
  \notag \om^{n} &\approx \Up + \Up C_{frequency}^{n}\frac{\sqrt{\mu_{0}}}{\Up} \\
  \om^{n} &\approx \Up \exp\pnth{C_{frequency}^{n}\frac{\sqrt{\mu_{0}}}{\Up}}.
\end{align}

\subsection{Properties of Static State Mass Profiles}

\begin{figure}

    \begin{center}
      \includegraphics[width=2.85 in]{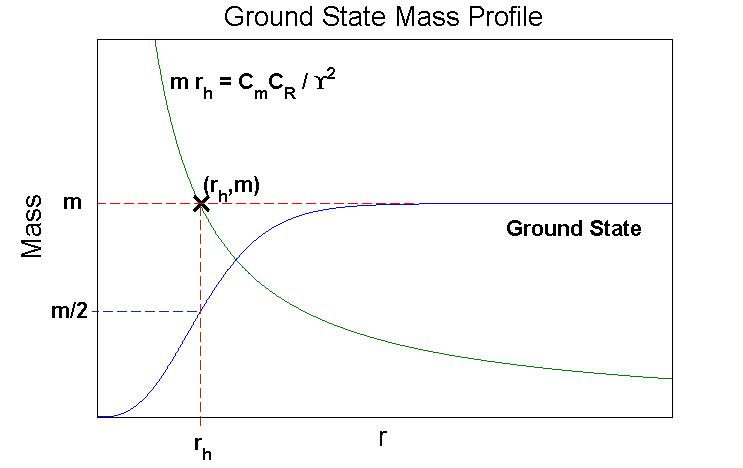}
      \includegraphics[width=2.85 in]{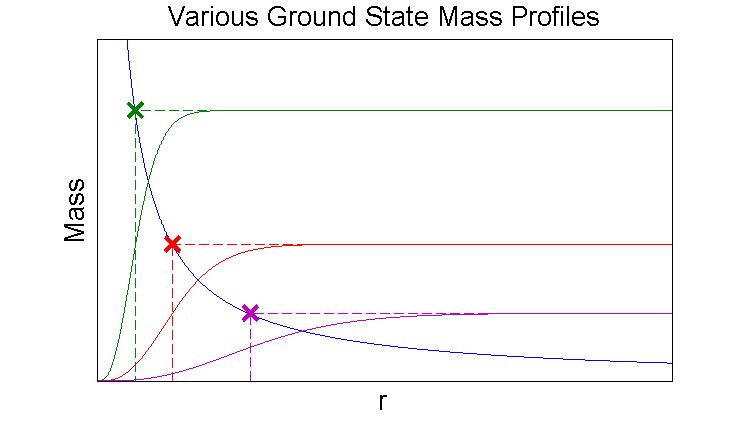}
    \end{center}

    \caption{Left: Plot of the mass profile of a ground state with its corresponding hyperbola of constant $\Up$ overlayed.  Any ground state mass profile that keeps the presented relationship with this hyperbola corresponds to the same value of $\Up$.  Right: Examples of different ground state mass profiles corresponding to the same value of $\Up$.  The corresponding hyperbola of constant $\Up$ is overlayed.  Notice that all three mass profiles have the same relationship with the hyperbola.}

    \label{mass-hyperbola-figure}

\end{figure}

In this last section, we discuss a few additional properties of static state mass profiles that are of particular interest to the study of wave dark matter.  The first is the relationship between the total mass $m$ and the half-mass radius $r_{h}$ for any $n^{\text{th}}$ excited state.  We observe from equations (\ref{Scorr3}) and (\ref{Scorr4}) that the product $mr_{h}$ does not depend on the parameter $\mu_{0}$.  Specifically,
\begin{equation}\label{mass-hyperbola}
    m r_{h} = \frac{C_{mass}C_{radius}}{\Up^{2}},
\end{equation}
where we have suppressed the notation of $n$.  If $\Up$ is constant, then, because both $C_{mass}$ and $C_{radius}$ are positive for all $n$ (see Table \ref{ScorrVal}), the right hand side of this equation is some positive constant, $k$, and we have that
\begin{equation}
  m r_{h} = k,
\end{equation}
which defines a hyperbola.  Thus, for a given $n^{\text{th}}$ excited state, all of the possible mass profiles for a constant value of $\Up$ lie along a hyperbola.  We illustrate this phenomenon in Figure \ref{mass-hyperbola-figure}.

\begin{figure}[t!]

    \begin{center}
      \includegraphics[width=2.85 in]{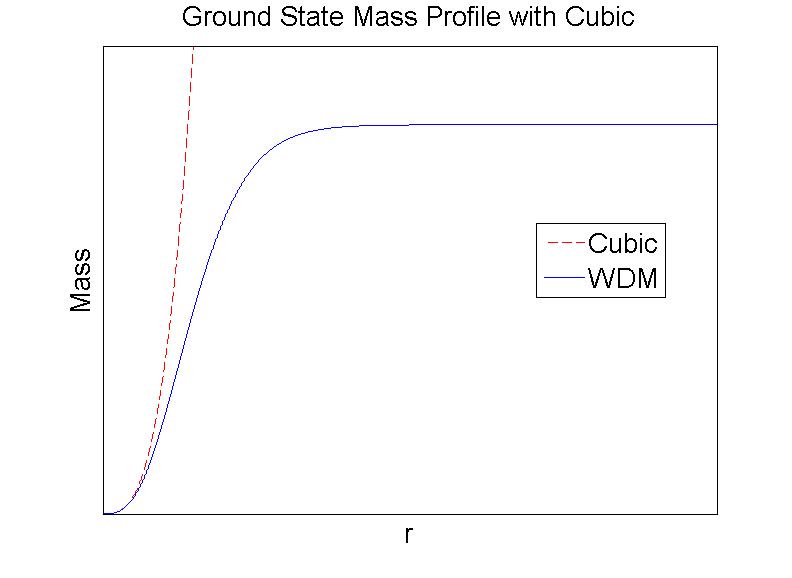}
      \includegraphics[width=2.85 in]{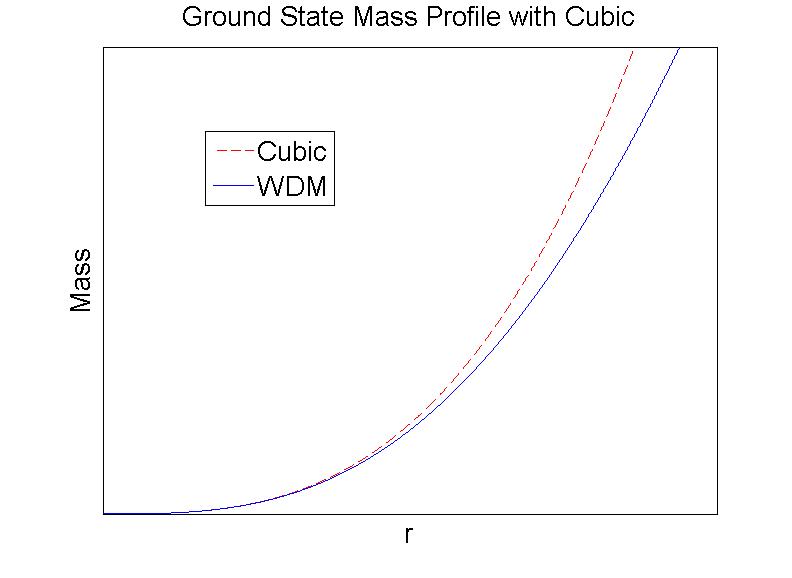}
    \end{center}

    \caption{Left: Plot of the mass profile of a ground state with its corresponding initial cubic function overlayed.  Right: Close up of the picture on the right in the region of small $r$.}

    \label{masswcubic-figure}

\end{figure}

Another property of interest is the initial behavior of a static state mass profile.  We explained previously that the mass function of any spherically symmetric solution to the Einstein-Klein-Gordon equations is initially cubic and approximately equal to the value in equation (\ref{massCub}).  It is routine to show that, for a static state, the value of $\mu(t,0) = \mu(0)$ (since the metric and stress energy tensor for a static state is independent of $t$) is
\begin{equation}
  \mu(0) = \mu_{0}\pnth{1 + \frac{\om^{2}}{\Up^{2}}\e^{-2V_{0}}}.
\end{equation}
Thus for an $n^{\text{th}}$ excited state and small $r$, we have that
\begin{equation}\label{initcubicreg}
  M(r) \approx \frac{4\pi r^{3} \mu_{0}}{3}\pnth{1 + \frac{\om^{2}}{\Up^{2}}\e^{-2V_{0}}}.
\end{equation}
We illustrate this initial behavior in Figure \ref{masswcubic-figure}.

We make one final note here about the stability of these static states.  The ground state is known to be stable under perturbations so long as its total mass is not too large (\cite{Seidel90, MSBS, Lai07, Gleiser89, Hawley00}).  In particular, the ground state is stable in the low field limit.  The higher excited states do not appear to be stable under perturbations, instead the agitated system either settles to a ground state or collapses into a black hole (\cite{Seidel98,MSBS}).  On the other hand, there has been some success in stabilizing higher excited states using interactions with the stable ground states (\cite{MSBS}).  However, as stated near the beginning of this chapter, in physical situations dark matter is always coupled with regular matter, which may have stabilizing effects on the dark matter.  Thus understanding the stability of these static states in the presence of regular matter is an important open problem.  It is also possible that dark matter may not exist as a single static state at all and as such, finding stable dynamical solutions of the Einstein-Klein-Gordon equations is another important open problem.


\section{Conclusion}

We summarize here the results of this chapter.  This chapter was concerned with spherically symmetric asymptotically Schwarzschild solutions of the Einstein-Klein-Gordon equations (\ref{EKG}) with a metric of the form
\begin{equation}
  g = -\e^{2V(t,r)}\, dt^{2} + \pnth{1 - \frac{2M(t,r)}{r}}^{-1}\, dr^{2} + r^{2}\, d\si^{2},
\end{equation}
where $V$ and $M$ are real-valued functions, and a scalar field of the form
\begin{equation}
  f(t,r) = \e^{i\om t}F(r),
\end{equation}
where $\om \geq 0$ and $F$ is complex-valued.  We proved the following proposition which is designated here with the same number as it appears earlier in the paper.

\begin{nprop}{\ref{staticprop}}
  Let $(N,g)$ be a spherically symmetric asymptotically Schwarzschild spacetime that satisfies the Einstein-Klein-Gordon equations (\ref{EKG}) for a scalar field of the form in (\ref{sfstaticstate}).  Additionally, assume that $F(r) = h(r)\e^{i a(r)}$ for smooth real-valued functions $h$ and $a$, where $h$ has only isolated zeros, if any.  Then $(N,g)$ is static if and only if $a(r)$ is constant.
\end{nprop}

Restricting our attention, without loss of generality, to solutions where $F$ was real-valued and hence the metric is static by the above proposition, we next showed that for chosen values of $(\Up,\mu_{0})$, there was a infinite number of solutions of this type that could be distinguished by the number of zeros, $n$, the scalar field $F$ contains.  These solutions are called $n^{\text{th}}$ excited states when $n > 0$ and ground states when $n = 0$.  Generic examples of these static states are displayed in Figures \ref{state_ex} - \ref{state_V}.

We also showed that the parameters defining such solutions, $\Up$, $\mu_{0}$, $\om$, and $V_{0}$ as well as the total mass, $m$, and the half-mass radius, $r_{h}$, are related via the following equations
\begin{subequations}
\begin{align}
  \om^{n}(\Up,\mu_{0}) &\approx \Up \exp\pnth{C_{frequency}^{n}\frac{\sqrt{\mu_{0}}}{\Up}} \\
  V_{0}^{n}(\Up,\mu_{0}) &\approx C_{potential}^{n}\frac{\sqrt{\mu_{0}}}{\Up} \\
  m^{n}(\Up,\mu_{0}) &\approx C_{mass}^{n}\Up^{-3/2}\mu_{0}^{1/4} \\
  r_{h}^{n}(\Up,\mu_{0}) &\approx C_{radius}^{n}\Up^{-1/2}\mu_{0}^{-1/4}
\end{align}
\end{subequations}
The values of the constants in these equations for various states are found in Table \ref{ScorrVal}.  We showed that these relations imply important scalings when we scale the coordinate functions as in (\ref{Scal0}) and restrict the types of scalings allowed if the parameter $\Up$ is to be invariant (see equation (\ref{Scal4})).

Finally, we showed that the last two of the above four relations imply that for a constant value of $\Up$, the mass profile of any $n^{\text{th}}$ excited state lies along a hyperbola (see Figure \ref{mass-hyperbola-figure}).  We also showed that the initial behavior of a static state mass profile was cubic (see Figure \ref{masswcubic-figure}).

All of these results are useful in understanding the implications of wave dark matter in the case where the spacetime is static and spherically symmetric.

\chapter{Modeling Wave Dark Matter \\ in Dwarf Spheroidal Galaxies}\label{ch4}

In the last two decades, there has been substantial progress on describing the distribution of dark matter on the galactic scale.  Navarro, Frenk, and White's popular model (\cite{NFW96}), which resulted from detailed $N$-body simulations, has been shown to agree well with observations outside the centers of galaxies (\cite{Humphrey06,Walker09}).  However, the Navarro-Frenk-White dark matter energy density profile also exhibits an infinite cusp at the origin, while observations favor a bounded value of the dark matter energy density at the centers of galaxies (\cite{Blok10,Gentile04}).  This has prompted many astrophysicists to employ a cored profile, such as the Burkert profile (\cite{Burkert95}), which ``cores'' out the infinite cusp, to model the dark matter energy density in a galaxy.  Resolving this ``core-cusp problem'' remains an important open problem in the study of dark matter.

Wave dark matter presents one possible solution to this problem and potentially other astrophysical problems.  In this chapter, we begin to test wave dark matter against observations at the galactic level.  In particular, we seek a working estimate of the fundamental constant in the wave dark matter model, $\Up$, to be used in future comparisons to data.  To do so, we will compare one of the simplest models defined by wave dark matter, namely the spherically symmetric static states described in the previous chapter, to models of dark matter that are already known to fit observations well.

\begin{figure}[t!]

    \begin{center}
        \includegraphics[height = 4.5 in]{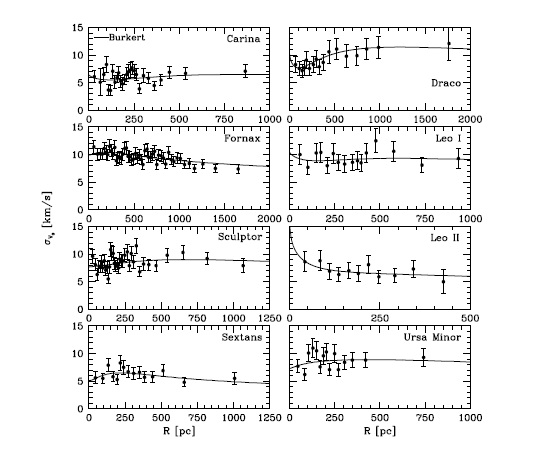}
    \end{center}

    \caption{Observed velocity dispersion profiles of the eight classical dwarf spheroidal galaxies are denoted by the points on each plot with its associated error bars.  The solid lines overlayed on these profiles are the best fit velocity dispersion profiles induced by the Burkert mass profile.  This figure is directly reproduced from the paper by Salucci et al.\ (\cite{Salucci11}) and the reader is referred to their paper for a complete description of how these models were computed.}

    \label{vel-disp}

\end{figure}

Salucci et al.\ recently used the Burkert profile to model the dark matter energy density profiles of the eight classical dwarf spheroidal galaxies orbiting the Milky Way.  They found excellent agreement between the observed velocity dispersion profiles of these galaxies and those velocity dispersion profiles induced by the Burkert profile (\cite{Salucci11}).  This can be seen in Figure \ref{vel-disp}, which we have reproduced exactly as it appears in the paper by Salucci et al.  In what follows, we will show that a value of $\Up = 50 \text{ yr}^{-1}$ produces wave dark matter mass models that are qualitatively similar to the Burkert mass models found by Salucci et al.  We will also show that under precise assumptions, comparisons to these Burkert profiles can be used to bound the value of $\Up$ above by $1000 \text{ yr}^{-1}$.

\section{Burkert Mass Profiles}

\begin{figure}[t!]

  \begin{center}
    \includegraphics[height = 2.25 in, width = 2.85 in]{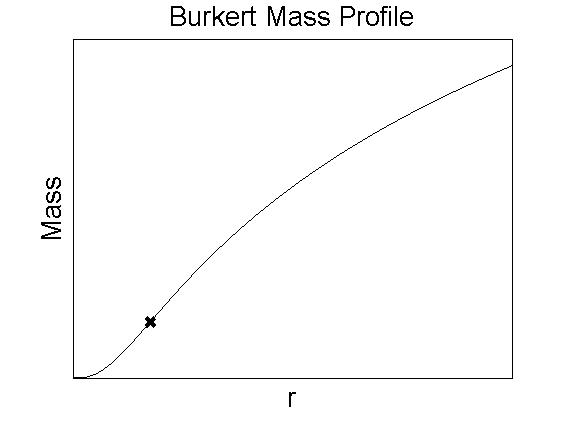}
  \end{center}

  \caption{Plot of a Burkert mass profile.  The inflection point is marked with an $\times$.}

  \label{BurMass}

\end{figure}

The Burkert energy density profile models the energy density of a spherically symmetric dark matter halo using the function
\begin{equation}\label{Burmod}
  \mu_{B}(r) = \frac{\rho_{0}r_{c}^{3}}{(r+r_{c})(r^{2}+r_{c}^{2})}
\end{equation}
where $\rho_{0}$ is the central density and $r_{c}$ is the core radius.  Integrating this function over the ball of radius $r$ centered at the origin, $B_{r}(0)$, with respect to the standard spherical volume form yields the Burkert mass profile as follows.
  \begin{align}
    \notag M_{B}(r) &= \int_{B_{r}(0)} \mu_{B}(s)\, dV_{\RR^{3}} \\
    \notag &= 4\pi \int_{0}^{r} s^{2}\mu_{B}(s)\, ds \displaybreak[0] \\
    \notag &= 4\pi \int_{0}^{r} \frac{s^{2}\rho_{0}r_{c}^{3}}{(s+r_{c})(s^{2}+r_{c}^{2})}\, ds \\
    \label{Bur-M} M_{B}(r) &= 2\pi \rho_{0}r_{c}^{3}\pnth{\ln\pnth{\frac{r+r_{c}}{r_{c}}} + \frac{1}{2}\ln\pnth{\frac{r^{2}+r_{c}^{2}}{r_{c}^{2}}} - \arctan\pnth{\frac{r}{r_{c}}}}
  \end{align}
A generic plot of a Burkert mass profile, $M_{B}(r)$, defined to be the dark matter mass in the ball of radius $r$, is shown in Figure \ref{BurMass}.  We make a few remarks about the behavior of this mass function.

Note that the behavior of the graph changes concavity at the inflection point $r = r_{ip}$, which we have marked on the plot in Figure \ref{BurMass} with an $\times$.  Recalling from equation (\ref{Bur-M}) the fact that $M_{B}(r)$ is the integral over the interval $[0,r]$ of the function $4\pi r^{2}\mu_{B}(r)$, we can compute this inflection point as follows.
\begin{equation}
  M_{B}'(r) = 4\pi r^{2}\mu_{B}(r) = \frac{4\pi\rho_{0}r_{c}^{3}r^{2}}{(r+r_{c})(r^{2}+r_{c}^{2})}
\end{equation}
Differentiating again yields
\begin{equation}
  M_{B}''(r) = \frac{-4\pi\rho_{0}r_{c}^{3}(r^{4} - r^{2}r_{c}^{2} - 2rr_{c}^{3})}{(r+r_{c})^{2}(r^{2}+r_{c}^{2})^{2}},
\end{equation}
which has two complex zeros and two real zeros.  The two real zeros are $r=0$ and
\begin{equation}\label{Bur_ip}
  r_{ip} = \pnth{\frac{3 + (27 + 3\sqrt{78})^{2/3}}{3(27 + 3\sqrt{78})^{1/3}}}r_{c} \approx 1.52r_{c},
\end{equation}
the latter being the inflection point of the mass model.

For $r \gg r_{ip}$, the plot grows logarithmically due to the fact that the $\arctan$ term in equation~(\ref{Bur-M}) approaches a constant value as $r \to \infty$.  To describe the behavior when $r \ll r_{ip}$, we note that the Taylor expansion of $M_{B}(r)$ centered at $r = 0$ is as follows,
\begin{equation}
  M_{B}(r) = \frac{4}{3}\pi \rho_{0}r^{3} + O(r^4).
\end{equation}
Thus for $r \ll r_{ip}$, $M_{B}(r)$ is dominated by an $r^{3}$ term making the initial behavior cubic.

\begin{figure}

  \begin{center}
    \includegraphics[height = 2.25 in, width = 2.85 in]{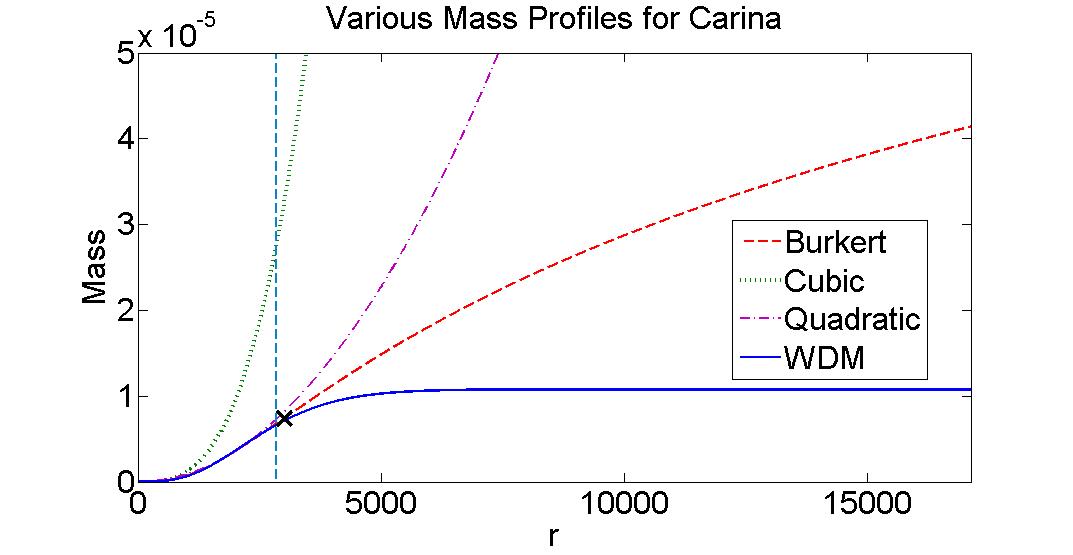}
    \includegraphics[height = 2.25 in, width = 2.85 in]{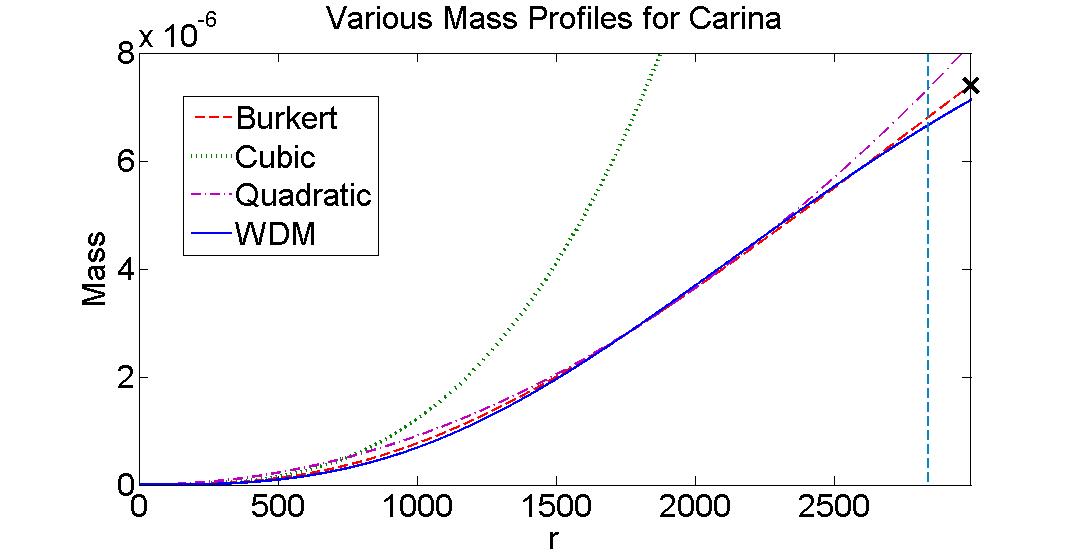}
  \end{center}

  \caption{Left: Plot of the Burkert mass profile for the Carina galaxy found by Salucci et al.\ (\cite{Salucci11}) along with a mass plot of a wave dark matter static ground state, the cubic function which is the leading term of the Taylor expansion of the Burkert mass profile, and the quadratic power function $\dfrac{M_{B}(r_{c})}{r_{c}^2}r^{2}$ where $r_{c}$ is the core radius of the Carina galaxy.  The $\mathbf{\times}$ marks the location of the inflection point of the Burkert mass profile, while the vertical line denotes the location of the outermost data point for the Carina galaxy and is presented for reference purposes only. Right: Closeup of the plot on the left over the $r$ interval $[0,r_{ip}]$.}

  \label{BurEKG_cmpr}

\end{figure}

In fact, several other models for dark matter mass profiles have similar initial behavior to the Burkert profile including a quadratic mass profile (which is not physical and is only included for the sake of comparisons) and wave dark matter mass profiles.  In Figure~\ref{BurEKG_cmpr}, we have collected several mass models that have similar behavior inside $r = r_{ip}$ to the Burkert mass profile computed by Salucci et al.\ for the Carina galaxy (\cite{Salucci11}).  While these models have similar behavior inside $r = r_{ip}$, they are very different outside $r = r_{ip}$.


\begin{figure}

  \begin{center}
    \includegraphics[height = 1.65 in, width = 1.9 in]{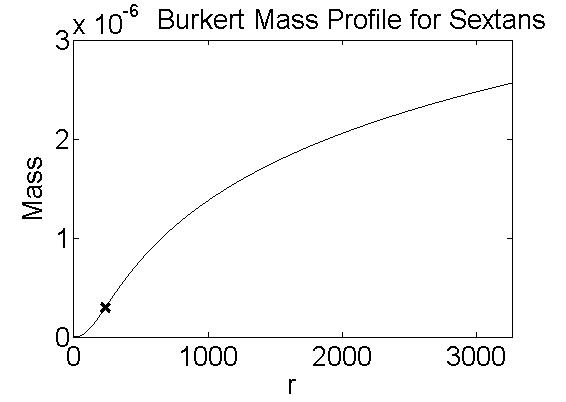}
    \includegraphics[height = 1.65 in, width = 1.9 in]{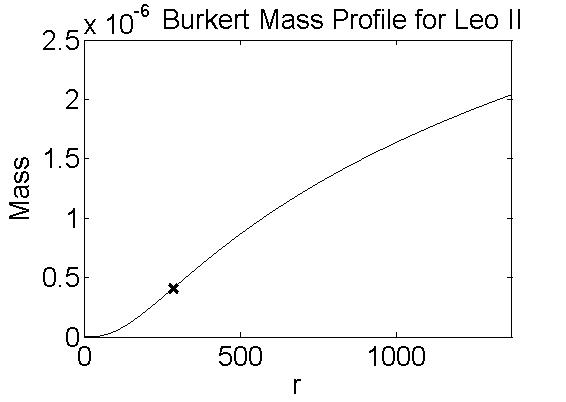}
    \includegraphics[height = 1.65 in, width = 1.9 in]{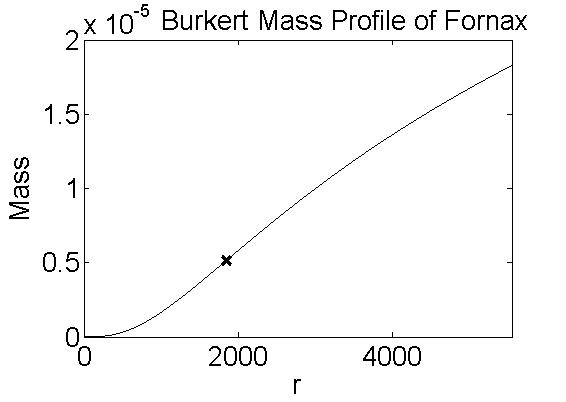}

    \includegraphics[height = 1.65 in, width = 1.9 in]{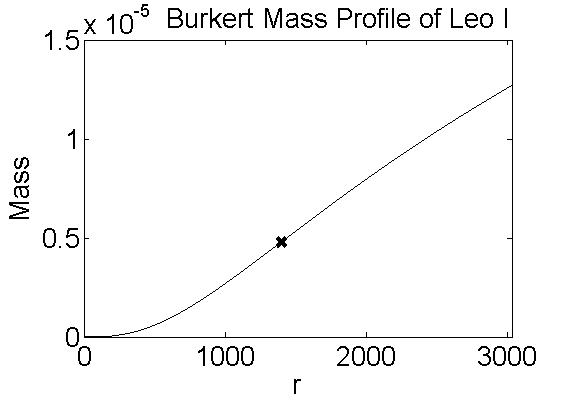}
    \includegraphics[height = 1.65 in, width = 1.9 in]{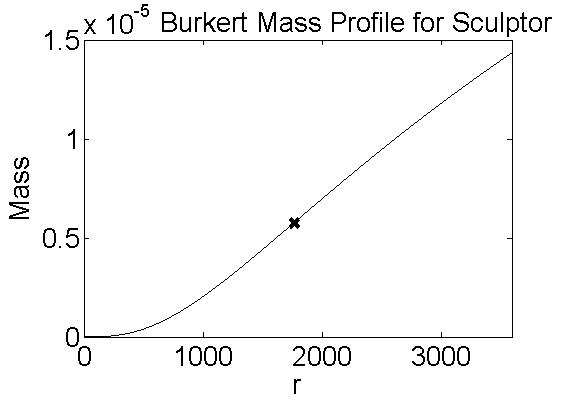}
    \includegraphics[height = 1.65 in, width = 1.9 in]{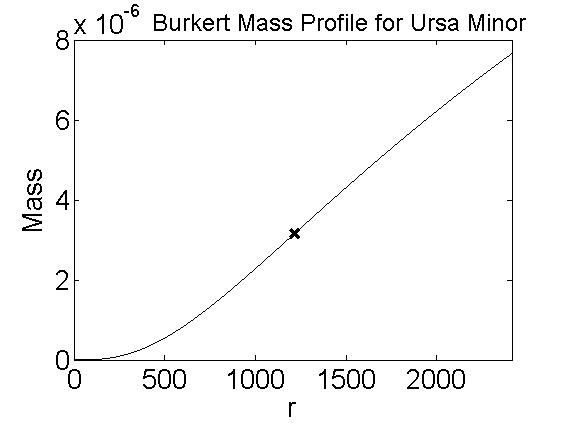}

    \includegraphics[height = 1.65 in, width = 1.9 in]{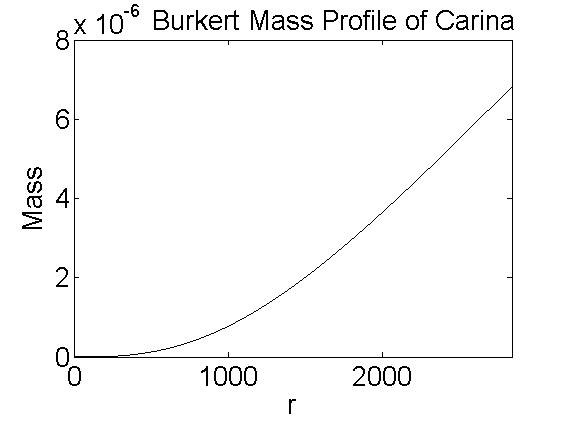}
    \includegraphics[height = 1.65 in, width = 1.9 in]{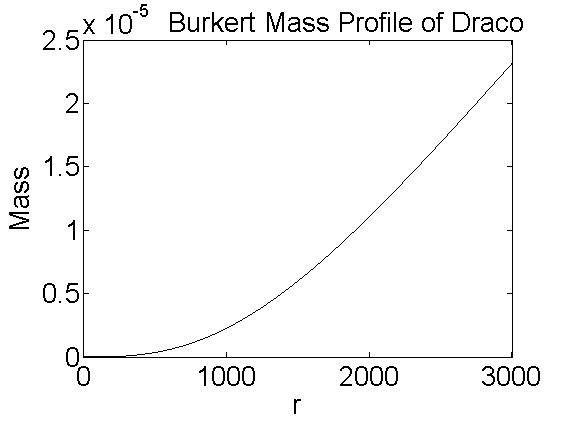}
  \end{center}

  \caption{Plots of the Burkert mass profiles computed by Salucci et al.\ of the eight classical dwarf spheroidal galaxies within the range of observable data.  The inflection point is marked on each plot by an $\mathbf{\times}$.  Carina and Draco have no inflection point marked because the inflection point for their Burkert mass profiles occurs outside the range of observable data.}

  \label{clsdSph_ip}

\end{figure}

We have computed the inflection points of each of the Burkert mass profiles computed by Salucci et al.\ for the eight classical dwarf spheroidal galaxies (\cite{Salucci11}) and have marked these points on a plot of each Burkert mass profile in Figure \ref{clsdSph_ip}.  We have constrained the viewing window of each plot to the range of data points collected.  That is, we plot the Burkert mass profiles on the interval $[0,r_{last}]$, where $r_{last}$ is the radius of the outermost data point given by Walker et al.\ (\cite{Walker09,Walker10}) for the observed velocity dispersion profiles.  We have presented them in order from greatest to least according to the ratio of $r_{last}/r_{ip}$.

In Table \ref{Bur-tab}, we have collected the defining parameters, $\rho_{0}$ and $r_{c}$, computed by Salucci et al.\ for the Burkert mass profiles which best predict the velocity dispersion profiles of each galaxy (\cite{Salucci11}).  We have also collected the outermost data point, $r_{last}$, of these velocity dispersion profiles (\cite{Walker09,Walker10}), as well as our computations of the inflection point, $r_{ip}$, and the ratio $r_{last}/r_{ip}$ for each of the classical dwarf spheroidal galaxies.  All quantities have been converted to geometrized units (the universal gravitational constant and the speed of light set to one) of (light)years for mass, length, and time.

\begin{table}

  \caption{Burkert mass profile data for the eight classical dwarf spheroidal galaxies converted to units of years for mass, length, and time.  The parameters $\rho_{0}$ and $r_{c}$ are those found by Salucci et al.\ for the best fit Burkert profiles (\cite{Salucci11}), and $r_{last}$ is the radius of the outermost data point given by Walker et al.\ (\cite{Walker09,Walker10}).  Also included is the value of the inflection point, $r_{ip}$, of the Burkert mass profile for each galaxy and the ratio of $r_{last}$ to $r_{ip}$.}

  \begin{center}

    \begin{tabular}{c|ccccc}
      Galaxy Name & $\rho_{0}$ ($\text{yr}^{-2}$) & $r_{c}$ (yr)         & $r_{last}$ (yr) & $r_{ip}$ (yr) & $r_{last}/r_{ip}$ \\
      \hline & & & \\
      Sextans     & $2.47 \times 10^{-14}$        & $1.53 \times 10^{2}$ & $3.26 \times 10^{3}$   & $2.32 \times 10^{2}$ & $14.05$ \\
      Leo II      & $1.83 \times 10^{-14}$        & $1.88 \times 10^{2}$ & $1.37 \times 10^{3}$   & $2.86 \times 10^{2}$ & $4.80$ \\
      Fornax      & $8.57 \times 10^{-16}$        & $1.21 \times 10^{3}$ & $5.54 \times 10^{3}$   & $1.84 \times 10^{3}$ & $3.01$ \\
      Leo I       & $1.83 \times 10^{-15}$        & $9.19 \times 10^{2}$ & $3.03 \times 10^{3}$   & $1.40 \times 10^{3}$ & $2.17$ \\
      Sculptor    & $1.10 \times 10^{-15}$        & $1.16 \times 10^{3}$ & $3.59 \times 10^{3}$   & $1.76 \times 10^{3}$ & $2.04$ \\
      Ursa Minor  & $1.83 \times 10^{-15}$        & $8.01 \times 10^{2}$ & $2.41 \times 10^{3}$   & $1.22 \times 10^{3}$ & $1.98$ \\
      Carina      & $2.90 \times 10^{-16}$        & $1.97 \times 10^{3}$ & $2.84 \times 10^{3}$   & $2.99 \times 10^{3}$ & $0.95$ \\
      Draco       & $8.19 \times 10^{-16}$        & $2.11 \times 10^{3}$ & $3.00 \times 10^{3}$   & $3.20 \times 10^{3}$ & $0.94$
    \end{tabular}

  \end{center}

  \label{Bur-tab}

\end{table}

\section{Static States of Wave Dark Matter}

To compare the wave dark matter model to dwarf spheroidal galaxies, we will use the spherically symmetric static states of wave dark matter defined in Chapter \ref{ch3}.  We will also take particular advantage of the property of these states that the possible mass profiles of an $n^{\text{th}}$ excited state of constant $\Up$ lie along a hyperbola.  This property will help us define a suitable definition for the $n^{\text{th}}$ excited state mass profile of constant $\Up$ that best fits a given Burkert mass profile.  We do exactly this in the next subsection.  In the subsections following, we present the resulting estimates of $\Up$ we can obtain.

\subsection{Fitting Burkert Mass Profiles}

In this subsection, we turn our attention to finding static state mass profiles that best fit the Burkert mass profiles computed by Salucci et al.\ (\cite{Salucci11}).  Given a Burkert mass profile, $M_{B}$, a value for $\Up$, and a specific state (i.e. value for $n$), we define for our purposes the best fit wave dark matter static state mass profile, $M_{W}$, as the one which minimizes the $L^{2}$ norm of the difference between these profiles, $E$, given by
\begin{equation}\label{L2Norm-exact}
    E = \norm{M_{B} - M_{W}}_{L^{2}}^{2} = \int_{0}^{r_{last}} (M_{B} - M_{W})^{2}\, dr.
\end{equation}
Of course, since we compute the static states numerically, we have to approximate this norm by an appropriate Riemann sum defined on a discretization of the interval $[0,r_{last}]$.

To find this minimum, we first note that since $\Up$ is fixed, we can write the total mass $m$ and the value of $\mu_{0}$ in terms of a choice of $r_{h}$ via equations (\ref{mass-hyperbola}) and (\ref{Scorr4}) respectively.  Thus, we parameterize the different mass profiles of constant $\Up$, and hence $E$ by $r_{h}$, that is, $E = E(r_{h})$.  Furthermore, since all of the static state mass profiles of constant $\Up$ lie on a hyperbola, there will be a value of $r_{h}$ that yields the minimum of $E(r_{h})$.

To make computing the best fits more uniform from galaxy to galaxy, we make the choice $r_{h} = b r_{c}$, where $b>0$ and $r_{c}$ is the core radius of the Burkert profile we wish to match, and vary the free parameter $b$.  To compute which value of $b$ produces a minimum value of $E(r_{h})$, we create a grid of $r_{h}$ values around an initial choice of $b$ of the form $[(b-step)r_{c}, br_{c}, (b+step)r_{c}]$ for some $step>0$.  Next we compute $E(r_{h})$ for each of the values of $r_{h}$ and shift the grid, if necessary, so that it is centered on the $r_{h}$ value which yielded the smallest value of $E(r_{h})$.  If the grid shifts, we recompute $E(r_{h})$ on the new grid and continue to shift, if necessary.  Once the minimum $E(r_{h})$ value occurs at the center of the grid, we keep that point as the center, but cut the step size in half.  We then run this shifting procedure again for this smaller grid until the minimum is at the center and then we shrink again.  We continue to shrink the step size until we get to a predetermined terminal value.  We generally would run the procedure until the step size was less than or equal to $2^{-10}$.

In the next subsection, we will use this matching procedure to show that there exists a value of $\Up$ which produces similar mass profiles to those computed by Salucci et al.\ (\cite{Salucci11}).  This best fitting procedure also provides a method of finding values of $\Up$, for $\Up$ sufficiently large, which produce untenable matches to the Burkert profiles of Salucci et al.\ (\cite{Salucci11}).  We will explore this in more detail in Section \ref{upbndsec}.

\subsection{Working Value of \texorpdfstring{$\Up$}{Upsilon}}

\begin{figure}

  \begin{center}
    \includegraphics[height = 1.6 in, width = 1.9 in]{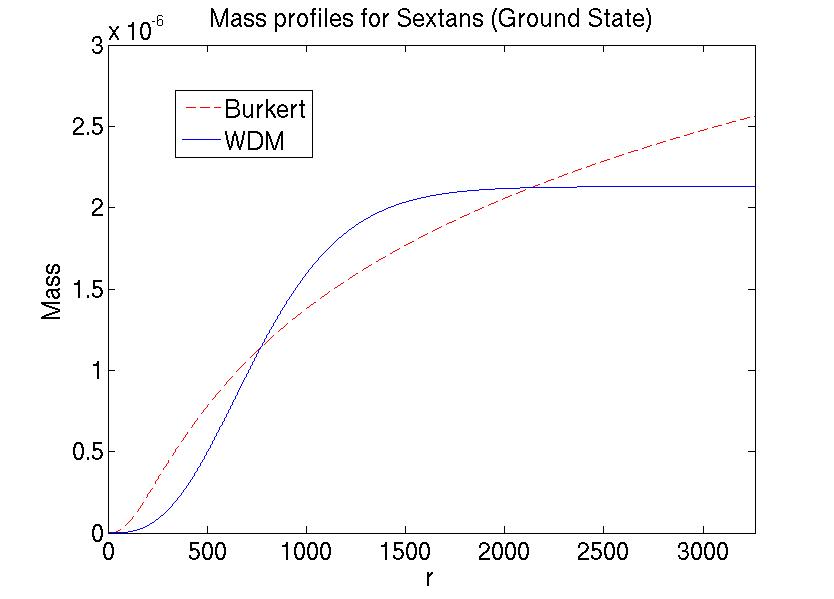}
    \includegraphics[height = 1.6 in, width = 1.9 in]{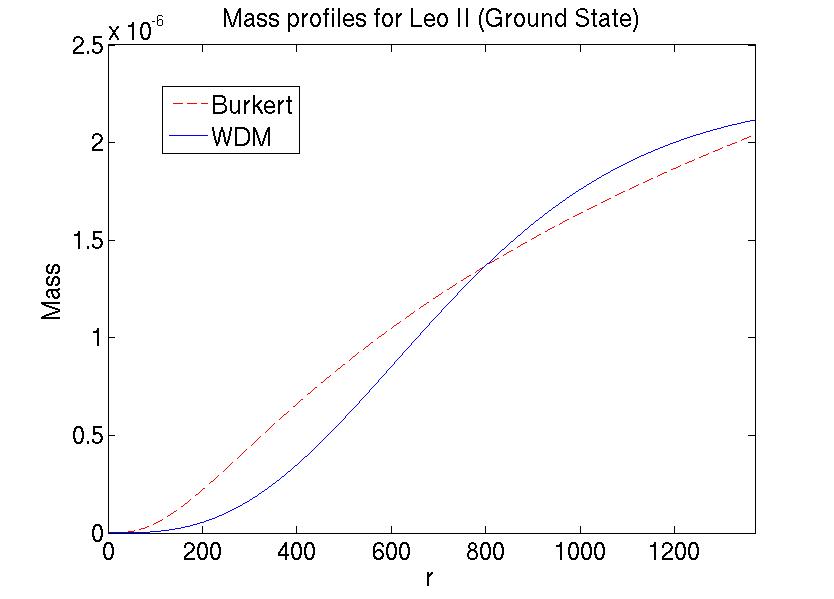}
    \includegraphics[height = 1.6 in, width = 1.9 in]{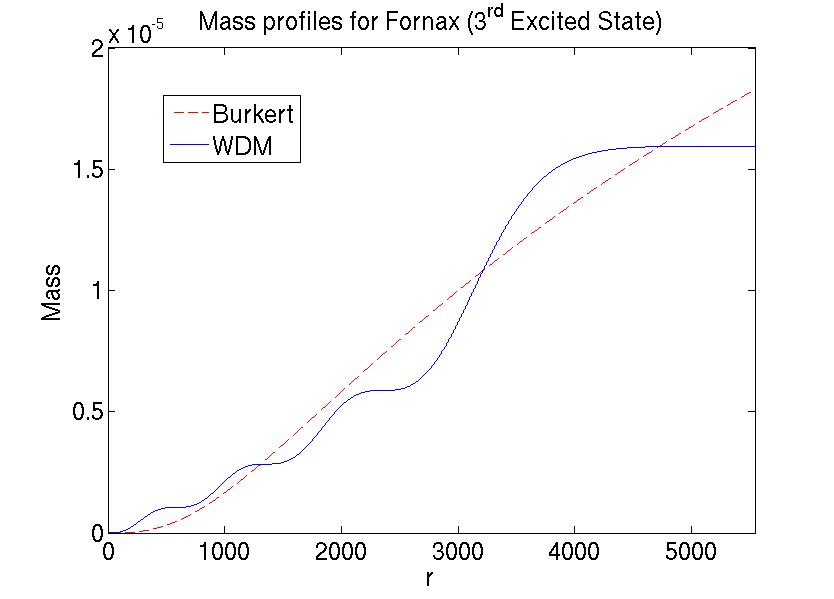}

    \includegraphics[height = 1.6 in, width = 1.9 in]{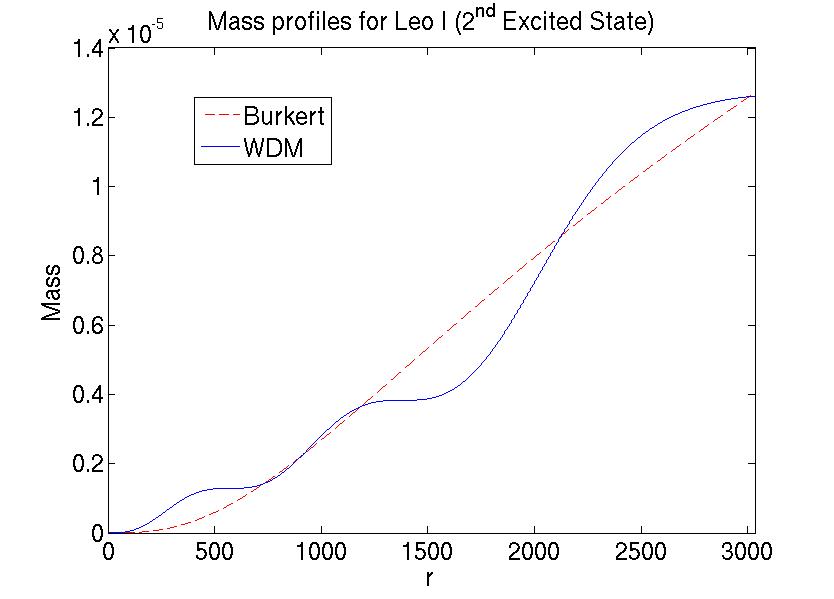}
    \includegraphics[height = 1.6 in, width = 1.9 in]{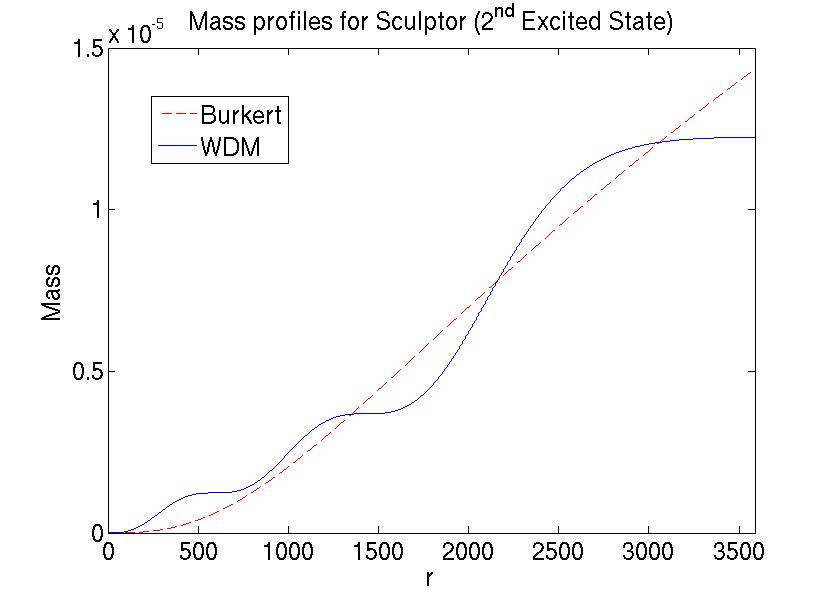}
    \includegraphics[height = 1.6 in, width = 1.9 in]{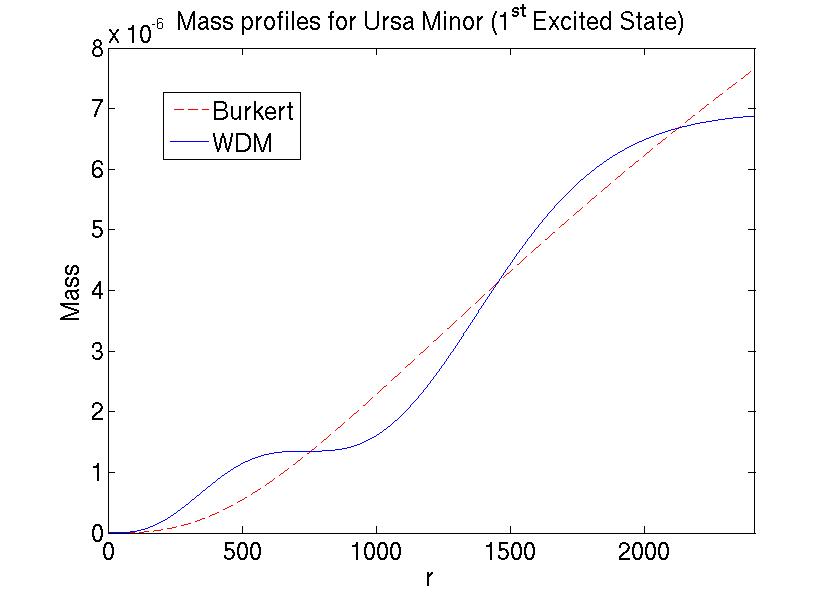}

    \includegraphics[height = 1.6 in, width = 1.9 in]{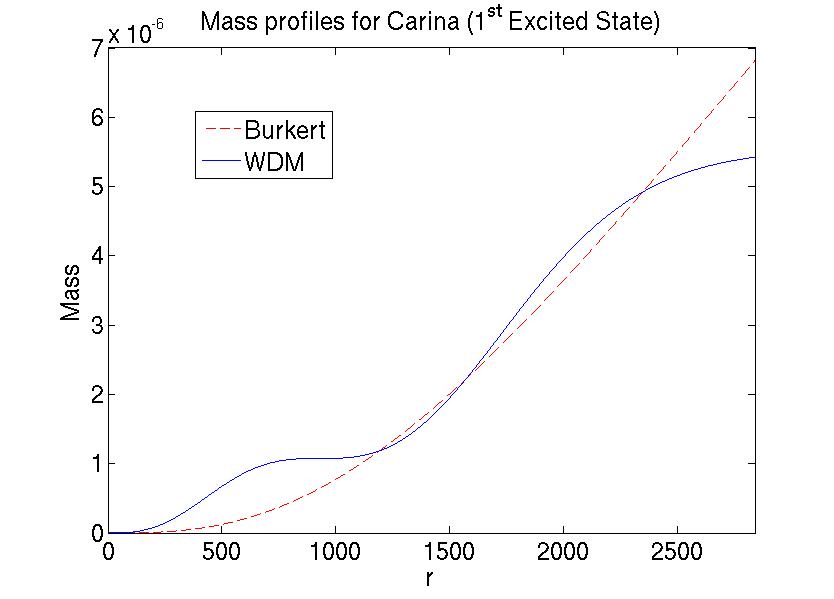}
    \includegraphics[height = 1.6 in, width = 1.9 in]{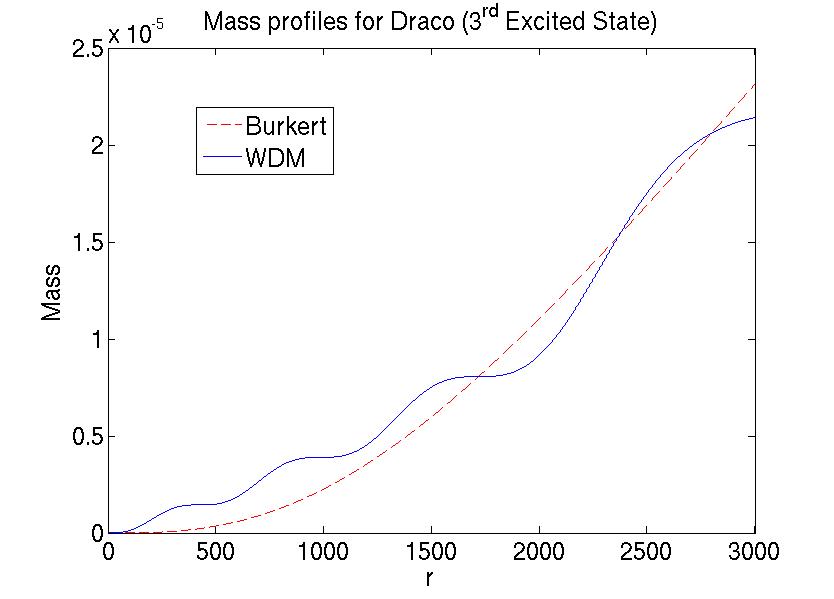}
  \end{center}

  \caption{Static state mass profiles for $\Up = 50$ which are each a best fit to the Burkert profiles of the corresponding  dwarf spheroidal galaxy.  For $\Up = 50$, we picked an $n^{\text{th}}$ excited state whose best fit profile matched the Burkert profile qualitatively well.  This shows that $\Up = 50$ is a reasonable working value of $\Up$.  However, it does not imply that the actual value of $\Up$ is $50$ or that these galaxies are correctly modeled by the presented $n^{\text{th}}$ excited state.  The units on $\Up$ are $\text{yr}^{-1}$.}

  \label{Ups50}

\end{figure}

As stated before, our overall goal is to find a value of $\Up$ that is compatible with the Burkert mass profiles that Salucci et al.\ (\cite{Salucci11}) computed to model the dark matter in the eight classical dwarf spheroidal galaxies.  For $\Up = 50 \text{ yr}^{-1}$, there is at least one wave dark matter static state mass profile that matches the Burkert mass profiles reasonably well.  We have plotted such matches in Figure \ref{Ups50}.  Thus we have chosen to use
\begin{equation}\label{Upwrkval}
  \Up = 50 \text{ yr}^{-1}
\end{equation}
as a working value of $\Up$ in our future work with wave dark matter until we have the capability to make a more accurate approximation or precise measurement of this value.

While we have chosen $\Up = 50 \text{ yr}^{-1}$ as a working value of $\Up$ since it corresponds to wave dark matter models compatible with other well fitting models, we note that the above does not constitute a precise measurement of the value of $\Up$.

In the remainder of this chapter, we will show that, under precise assumptions, the value of $\Up$ can be bounded above, a first step towards obtaining a direct measurement of $\Up$.

\subsection{Upper Bound for \texorpdfstring{$\Up$}{Upsilon}}\label{upbndsec}

To find an upper bound for $\Up$, we first need to explain how the static states change as $\Up$ gets large.  Equation (\ref{mass-hyperbola}) implies that for a given $n^{\text{th}}$ excited state, as $\Up$ increases, the product $mr_{h}$ decreases.  The hyperbolas corresponding to smaller values of $m r_{h}$ are those that lie closer to the mass and radius axes.

\begin{figure}

  \begin{center}
    \includegraphics[width=2.85 in]{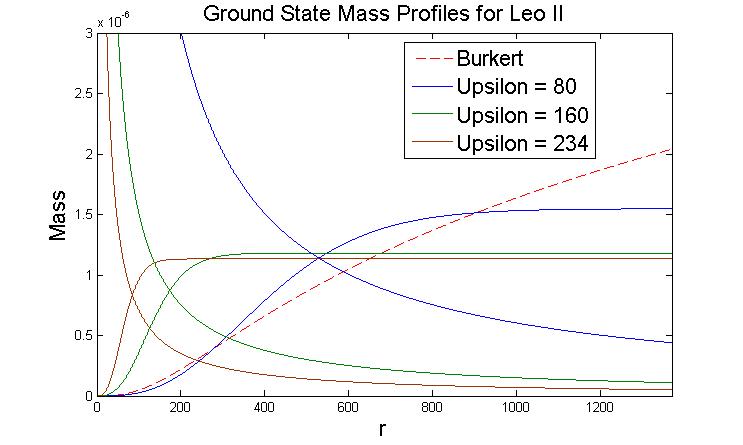}
    \includegraphics[width=2.85 in]{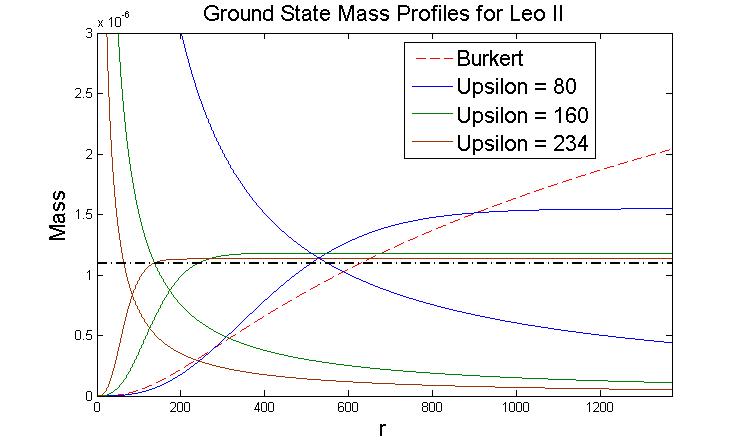}
  \end{center}

  \caption{Left: Ground state mass profiles of various values of $\Up$ that are best fits to the Burkert mass profile found by Salucci et al.\ (\cite{Salucci11}) for the Leo II galaxy.  The corresponding hyperbolas of constant $\Up$ on which these profiles lie are also plotted.  Ground states and their corresponding hyperbolas are drawn in the same color.  Right: The same plots as in the left frame, but with the constant function which best fits the Burkert profile also plotted.  Note that the best fit mass profiles approach this constant mass profile as $\Up$ increases.}

  \label{GroundStateMatchUpsUp}

\end{figure}

Now consider the $n^{\text{th}}$ excited state mass profile that is the best fit to a Burkert mass profile for a given $\Up$.  As $\Up$ increases, the hyperbola to which this static state mass profile corresponds will get closer to the mass and radius axes, but since the mass profile must also minimize $E(r_{h})$, the value of its total mass will not tend to $0$.  Since $m r_{h}$ tends to zero as $\Up \to \infty$, it must be instead that $r_{h} \to 0$ as $\Up \to \infty$.  This implies that, as $\Up$ increases, more of the constant portion of the best fitting $n^{\text{th}}$ excited state mass profile will be compared to the Burkert profile.  Thus, as $\Up \to \infty$, the best fitting $n^{\text{th}}$ excited state mass profile will limit to the constant function of $r$ that best fits the Burkert profile under the same fitting criteria used for the static states.  We illustrate this phenomenon in Figure \ref{GroundStateMatchUpsUp}.

As $\Up \to \infty$, the initial increasing region (i.e.\ the region before the constant portion) of the $n^{\text{th}}$ excited state mass profile that best fits a Burkert mass profile becomes more compressed.  This initial region is where all of the dark matter mass is located.  Thus as $\Up$ increases, the dark matter corresponding to the best fit $n^{\text{th}}$ excited state extends out to smaller radii.  However, observations suggests that dwarf spheroidal galaxies are dark matter dominated at all observable radii (\cite{Kleyna02}).  Thus the best fit $n^{\text{th}}$ excited state mass profiles for large $\Up$ do not represent observations well and can be rejected.  The question then is exactly when should we reject them.

Since every static state has the initial increasing region just described, the best fit $n^{\text{th}}$ excited state for any value of $\Up$ will be a better fit than the best fit constant function.  Moreover, since this initial region becomes more compressed as $\Up \to \infty$, for large $\Up$, the value of $E(r_{h})$ for the best fit $n^{\text{th}}$ excited state mass profile increases monotonically as $\Up \to \infty$ approaching the value of $E$ for the best fitting constant function.

This suggests a criteria for when to reject values of $\Up$.  We will reject a best fit $n^{\text{th}}$ excited state mass profile, and hence its corresponding value of $\Up$, as an untenable model of the dark matter mass if its value of $E(r_{h})$ is greater than some prescribed fraction of the value of $E$ for the best fitting constant function.  We choose to use $80\%$.  Explicitly, we use the following rejection criteria.

\begin{rej}\label{rejcrit_1}
  Given $\Up$, $n$, and a Burkert mass profile $M_{B}$, let $M_{W}$ be the spherically symmetric $n^{\text{th}}$ excited state mass profile corresponding to $\Up$ that best fits $M_{B}$, that is, that minimizes $E$ from equation (\ref{L2Norm-exact}) along the hyperbola defined by the value of $\Up$ and equation (\ref{mass-hyperbola}).  Let $E_{W}$ be the value of $E$ for this mass profile.  Furthermore, let $M_{C}$ be the constant function which best fits $M_{B}$, also by minimizing the corresponding function $E$, and let $E_{C}$ be the value of $E$ for the constant function.  Reject the given value $\Up$ as a tenable value for this fundamental constant if
    $$E_{W} \geq .8 E_{C}.$$
\end{rej}

In other words, any fit that is less than $20\%$ better than the best fitting constant function of $r$ is rejected as a bad fit.

\begin{figure}

  \begin{center}
    \includegraphics[height = 2.25 in]{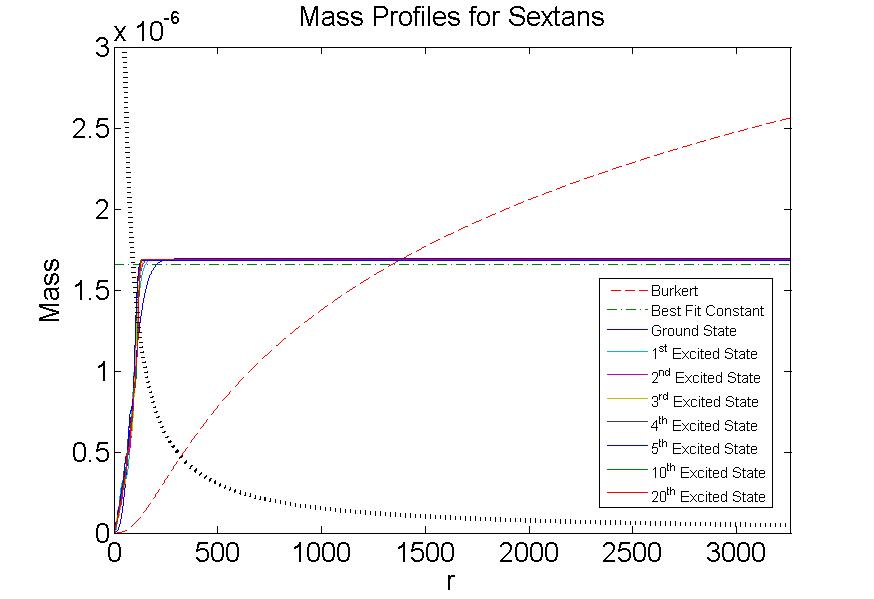}
  \end{center}

  \caption{The Burkert mass profile found by Salucci et al.\ (\cite{Salucci11}) for the Sextans galaxy.  The best fit static state mass profiles for a ground through fifth excited state, tenth excited state, and twentieth excited state all lying on the same hyperbola are overlayed on the plot.  The hyperbola here satisfies the rejection criteria for all of the different static states represented in the plot, thus all of these static states correspond to an upper bound on the value of $\Up$ for Sextans for their respective value of $n$ (i.e.\ the set of $n^{\text{th}}$ excited states).  Note how close together all of the states are.  This is due to the fact that the majority of their profiles which are being compared to the Burkert mass profile is the common and constant portion of the profiles.}

  \label{badfitSextans}

\end{figure}

For each of the eight dwarf spheroidal galaxies and $n \in \{0,1,2,3,4,5,10,20\}$, we computed values of $\Up$ that yielded $n^{\text{th}}$ excited state mass profiles that best fit that galaxy's Burkert profile which were rejected by the above criteria.  All of the values of $\Up$ above those computed are also rejected because they produce mass profiles even closer to the constant function.  In Table \ref{Upsilon_upbnd}, we have collected these upper bounds of $\Up$. In Figure \ref{badfitSextans}, for the galaxy Sextans, we present best fit static state mass profiles for the ground through fifth, tenth, and twentieth excited states for which $E(r_{h})$ is more than $80\%$ of the value of $E$ for the best fit constant function.

\begin{table}

  \caption{Upper bound values for $\Up$ corresponding to poor best fits of the Burkert mass profiles for each of the classic dwarf spheroidal galaxies.  The values in each column for each galaxy should be interpreted as an upper bound on the value of $\Up$, under the approximations explained in the paper, if that galaxy is best modeled by an $n^{\text{th}}$ excited state.  The units on $\Up$ are $\text{yr}^{-1}$.}

    \begin{center}
        \begin{tabular}{c|cccccccc}
            Galaxy $\backslash$ State & 0 & 1 & 2 & 3 \\
            \hline & & & & \\
            Sextans & $\Up < 160$ & $\Up < 394$ & $\Up < 633$ & $\Up < 875$ \\
            Leo II & $\Up < 234$ & $\Up < 576$ & $\Up < 926$ & $\Up < 1279$ \\
            Fornax & $\Up < 35$ & $\Up < 87$ & $\Up < 139$ & $\Up < 192$ \\
            Leo I & $\Up < 57$ & $\Up < 141$ & $\Up < 226$ & $\Up < 312$ \\
            Sculptor & $\Up < 49$ & $\Up < 121$ & $\Up < 194$ & $\Up < 268$ \\
            Ursa Minor & $\Up < 82$ & $\Up < 202$ & $\Up < 325$ & $\Up < 449$ \\
            Carina & $\Up < 84$ & $\Up < 207$ & $\Up < 333$ & $\Up < 459$ \\
            Draco & $\Up < 45$ & $\Up < 111$ & $\Up < 179$ & $\Up < 246$
        \end{tabular}

        \vspace{.2 in}

        \begin{tabular}{c|cccc}
           Galaxy $\backslash$ State & 4 & 5 & 10 & 20 \\
           \hline & & & & \\
           Sextans & $\Up < 1116$ & $\Up < 1356$ & $\Up < 2460$ & $\Up < 4789$ \\
           Leo II & $\Up < 1632$ & $\Up < 1983$ & $\Up < 3597$ & $\Up < 7003$ \\
           Fornax & $\Up < 245$ & $\Up < 297$ & $\Up < 538$ & $\Up < 1048$ \\
           Leo I & $\Up < 398$ & $\Up < 484$ & $\Up < 877$ & $\Up < 1706$ \\
           Sculptor & $\Up < 342$ & $\Up < 416$ & $\Up < 754$ & $\Up < 1467$ \\
           Ursa Minor & $\Up < 572$ & $\Up < 695$ & $\Up < 1261$ & $\Up < 2455$ \\
           Carina & $\Up < 586$ & $\Up < 712$ & $\Up < 1292$ & $\Up < 2514$ \\
           Draco & $\Up < 314$ & $\Up < 382$ & $\Up < 692$ & $\Up < 1347$
        \end{tabular}
    \end{center}

    \label{Upsilon_upbnd}

\end{table}

We observe from Table \ref{Upsilon_upbnd} that the upper bound values of $\Up$ increase as we increase the state we consider.  This is due to the following.  The rejected values of $\Up$ correspond to rejected hyperbolas of constant $\Up$, and hence constant $m r_{h}$.  Furthermore, the only qualitative difference between any two $n^{\text{th}}$ excited state mass profiles is the number of ripples in the initial increasing region of the profile.  For large $\Up$, the majority of a best fit $n^{\text{th}}$ excited state mass profile that is compared to the Burkert profile is the constant region which is shared by static state mass profiles for any $n$.  Thus the hyperbola corresponding to a rejected best fit ground state is close to the hyperbola corresponding to a rejected best fit $n^{\text{th}}$ excited state for any $n$.  In particular, there is a hyperbola of constant $m r_{h}$, for which the corresponding best fit $n^{\text{th}}$ excited state mass profiles for any $n$ are rejected by the above criteria.  Then, since $C_{mass}^{n}$ and $C_{radius}^{n}$ appear to monotonically increase as $n$ increases (see Table \ref{ScorrVal}), by equation (\ref{mass-hyperbola}), we would expect the same behavior for the value of $\Up$ in order for $m r_{h}$ to remain constant, which is what we observe in Table \ref{Upsilon_upbnd}.

Thus if a dwarf spheroidal galaxy is correctly modeled by a twentieth excited state or less, then an overall upper bound on the value of $\Up$ would be the upper bound corresponding to the twentieth excited state.  The least upper bound corresponding to the twentieth excited state over all eight galaxies is that value for the Fornax galaxy, which yields approximately that
\begin{equation}\label{Upupbnd}
  \Up < 1000 \text{ yr}^{-1}.
\end{equation}

\subsection{Lower Bound for \texorpdfstring{$\Up$}{Upsilon}}

\begin{figure}

  \begin{center}
    \includegraphics[height = 2.25 in]{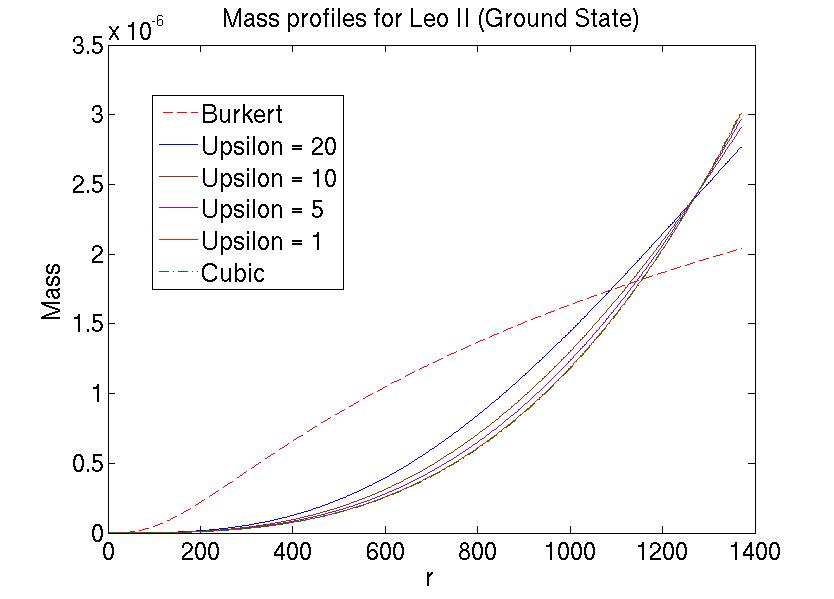}
  \end{center}

  \caption{The Burkert mass profile found by Salucci et al. \cite{Salucci11} for the Leo II galaxy.  The best fit ground state mass profiles for successively smaller values of $\Up$ are overlayed on the plot.  Also plotted is the best fit cubic power function, $ar^{3}$, which almost coincides with the mass plot for $\Up = 1$ and so is somewhat difficult to make out.  However, it is apparent that for successively smaller values of $\Up$, the best fit ground state approaches the best fit cubic function.}

  \label{Upsto0}

\end{figure}

The next most natural goal here would be to attempt to use similar methods to acquire a lower bound for $\Up$.  However, it appears that no such bound could be had without either new observations or imposing additional rather arbitrary assumptions.  This is due to the fact that as $\Up$ decreases, the hyperbolas on which the static state mass profiles reside would move arbitrarily far away from the axes.  Thus for small $\Up$, the initial region of the static state mass profile would dominate the part of the static state mass profile that gets compared to the Burkert mass profile in the observable range.  Hence, for any $n$, as $\Up \to 0$, the best fitting $n^{\text{th}}$ excited state mass profile would converge to the best fitting cubic profile since the initial region of any static state mass profile is cubic by equation (\ref{initcubicreg}).  Figure \ref{Upsto0} illustrates for a ground state this convergence to the best fitting cubic profile as $\Up \to 0$.

However, a cubic dark matter mass profile is not clearly rejectable like a constant dark matter mass profile was.  This is because a cubic profile is actually a decent fit in the observable range for some of the Burkert profiles of the classic dwarf spheroidal galaxies.  Thus there is no way to categorically reject best fitting mass profiles of small $\Up$ under our current set of assumptions.

That said, we should note that as $\Up \to 0$, these static state mass profiles have the additional property that they become arbitrarily massive and arbitrarily wide.  Thus if one could clearly bound the size of the mass or radius of the eight classical dwarf spheroidal galaxies, one could obtain a lower bound on $\Up$ by rejecting any value of $\Up$ which yields a dark matter halo that is too large.  Unfortunately, currently we can only detect the dark matter halo where we have stars to trace out its gravitational effects or strong evidence of gravitational lensing, which makes determining the precise overall size of the dark matter halo a very difficult undertaking.

Thus, short of making an arbitrary assumption on the size of the dark matter halos surrounding these dwarf spheroidal galaxies, we either must wait for bounds on the size of the halos to be observed or find another method of determining a lower bound.

\section{Utilized Approximations}

Now that we have presented our results, we list here the important approximations made in this chapter which led to these results and explain briefly why we make them.
\begin{description}
  \item[Approximation 1:] Dark matter is correctly described by the wave dark matter model.
  \item[Approximation 2:] Dark matter halos around dwarf spheroidal galaxies are spherically symmetric.
  \item[Approximation 3:] The dwarf spheroidal galaxies used in this chapter are in a state of dynamical equilibrium.
  \item[Approximation 4:] The spacetime metrics describing these dwarf spheroidal galaxies are static.
  \item[Approximation 5:] Wave dark matter predicts outcomes qualitatively similar to those of spherically symmetric static state solutions to the Einstein-Klein-Gordon equations.
  \item[Approximation 6:] The Burkert mass profiles computed by Salucci et al.\ (\cite{Salucci11}) fit the observational data very well.
  \item[Approximation 7:] The spacetime is in the low field limit, that is, $M \ll r$.
  \item[Approximation 8:] The spacetime is asymptotically Schwarzschild.
\end{description}

Approximation 1 is used because we are testing the wave dark matter model against observations.  Approximation 2 is a common approximation for dwarf spheroidal galaxies and is also necessary because we are comparing the wave dark matter model to the spherically symmetric Burkert mass profile.  Approximation 3 seems to be consistent with observations of dwarf spheroidal galaxies at least out to large radii (\cite{Salucci11,Cote99}). Approximation 6 is reasonable given Figure~\ref{vel-disp}.  Approximations 7 and 8 are standard when modeling galaxies.

Approximations 4 and 5 are used to simplify the types of solutions to the Einstein-Klein-Gordon equations we consider.  We note here, however, that there is a question of the stability of the spherically symmetric static state solutions.  It is known that, if the corresponding total mass is not too large, the ground state is stable under perturbations (\cite{Seidel90,Lai07}) but that, on their own, the excited states are not (\cite{Seidel98}) regardless of their mass.  However, it has also been shown that a coupling of an excited state with a ground state can produce a stable configuration (\cite{MSBS}).  We hypothesize that luminous matter distributions coupled with combinations of static states will produce a stabilizing effect allowing for more dynamically interesting systems to be physically plausible.

\section{Conclusions}

To summarize the results of this chapter, we have drawn effectively two conclusions, which we list here.

\begin{conc}
  Given Approximations 1 through 8, a value of $\Up$ which yields one or more spherically symmetric static state mass profiles which match well the best fit Burkert mass profiles computed by Salucci et al.\ (\cite{Salucci11}) for each of the eight classical dwarf spheroidal galaxies is $$\Up = 50 \text{ yr}^{-1}.$$
\end{conc}

\begin{conc}
  Given Approximations 1 through 8 and Rejection Criteria \ref{rejcrit_1}, if the dark matter halos of all of the eight classical dwarf spheroidal galaxies are correctly modeled by $20^{\text{th}}$ excited states or less, then $$\Up < 1000 \text{ yr}^{-1}.$$
\end{conc}

We note here that the main result of this chapter is more to describe a procedure of computing a working value and upper bound of $\Up$ rather than stating that the values that appear in the above conclusions are the best ones.  If one wished to alter the hypotheses of these conclusions, the corresponding values of $\Up$ might differ from what we presented here.  However, if one remains in the realm of using static states to model the dark matter, then the procedure presented in this paper for finding an upper bound would likely still apply and could be employed to get an upper bound under the new assumptions.  As such, the conclusions above should be taken as an example of the procedure applied to a set of assumptions we currently find reasonable or useful and not as final precise estimates.

For the interested reader, the Matlab code used for this dissertation to generate the spherically symmetric static states and to compute the best fits to a Burkert profile can be found on Bray's Wave Dark Matter Web Page at \url{http://www.math.duke.edu/\~bray/darkmatter/darkmatter.html}.

\chapter{A Numerical Scheme to Solve the \\ Einstein-Klein-Gordon Equations \\ in Spherical Symmetry} \label{ch5}

In this chapter, we present some of the progress we have made on the next part of our research into wave dark matter and dwarf spheroidal galaxies.  The next project we are interested in is numerically solving the Einstein-Klein-Gordon equations in spherical symmetry in order to determine more generic solutions to these equations and perhaps use these equations to obtain a better estimate of $\Up$.  Thus far, we have designed numerical code to evolve these equations, but there is still much to do in terms of analyzing the practical properties of the scheme and conducting experiments to generate generic solutions.

This chapter presents the numerical scheme we use to solve the spherically symmetric Einstein-Klein-Gordon equations.  We will discuss how to appropriately implement the boundary conditions into the scheme as well as some artificial dissipation.  We will also discuss the accuracy and stability of the scheme.  Finally, we will present some examples that show the scheme in action and illustrate how to use the code.

We wish to numerically evolve the spherically symmetric Einstein-Klein-Gordon equations found in equation (\ref{NCpde}).  The boundary conditions on these equations at $r = 0$ are found in equations (\ref{Mder}), (\ref{Vder}), and (\ref{sfder}).  These boundary conditions allow us to compute equation (\ref{NCpde}) at $r = 0$ by taking the limit as $r \to 0$, which yields
\begin{subequations} \label{NCpde_cen}
  \begin{align}
    M_{r}(t,0) &= 0 & V_{r}(t,0) &= 0 \\
    f_{t}(t,0) &= p(t,0)\e^{V(t,0)} & f_{r}(t,0) &= 0 \\
    p_{t}(t,0) &= \e^{V(t,0)}\pnth{3f_{rr}(t,0) - \Up^{2}f(t,0)} & p_{r}(t,0) &= 0.
  \end{align}
\end{subequations}
The above conditions on the central values are necessary for regularity at $r=0$ and so will always hold in the low field limit.  As in the case of the static states, even though theoretically we consider an infinite spacetime, to compute numerical solutions to these equations we must set an artificial right hand boundary point and impose boundary conditions.  If $r = r_{max}$ is the maximum $r$ value included in the computation, then we will require as a boundary condition that
\begin{equation} \label{Ref-Eq-2}
  p(t,r_{max}) + f_{r}(t,r_{max}) = \la (p(t,r_{max}) - f_{r}(t,r_{max})),
\end{equation}
for some $\la \in \C$ and $\abs{\la} \leq 1$.  This splits into three cases, $\la = 1$, $\la = -1$, and $\la \neq \pm 1$.  If $\la = 1$, then we have that
\begin{equation} \label{NBC}
  f_{r}(t,r_{max})=0,
\end{equation}
which is a Neumann boundary condition.  If $\la = -1$, then we have that
\begin{equation} \label{DBC}
    p(t,r_{max}) = 0 \text{ and hence } f(t,r_{max}) = f(0,r_{max})
\end{equation}
for all $t$.  This is a Dirichlet boundary condition.  If $\la$ is any other complex number, then we will have that
\begin{equation} \label{MBC}
    f_{r}(t,r_{max}) = \pnth{\frac{\la - 1}{\la + 1}}p(t,r_{max}).
\end{equation}

\begin{figure}

\begin{center}
  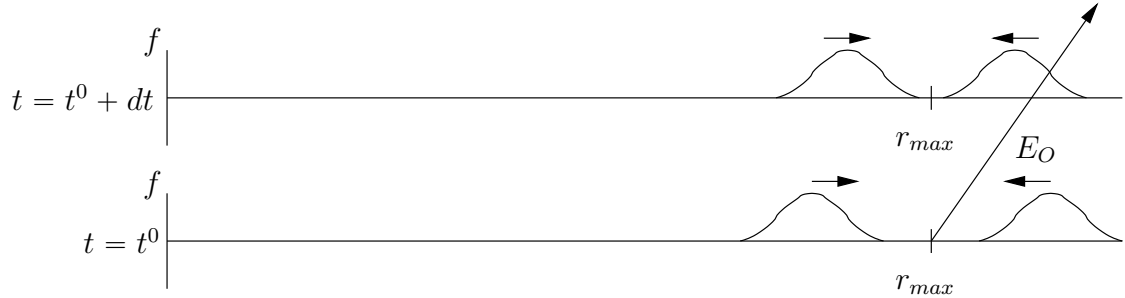
\end{center}

\caption{Image of an incoming and outgoing wave at the boundary $r_{max}$.  The future pointing outward null vector $E_{O}$ is also displayed.  Since no outgoing wave can travel faster than the speed of the null vector shown, then computing the change of $f$ along this null vector will only detect changes in $f$ due to incoming waves.}

\label{EO_rmax}

\end{figure}

\begin{figure}

\begin{center}
  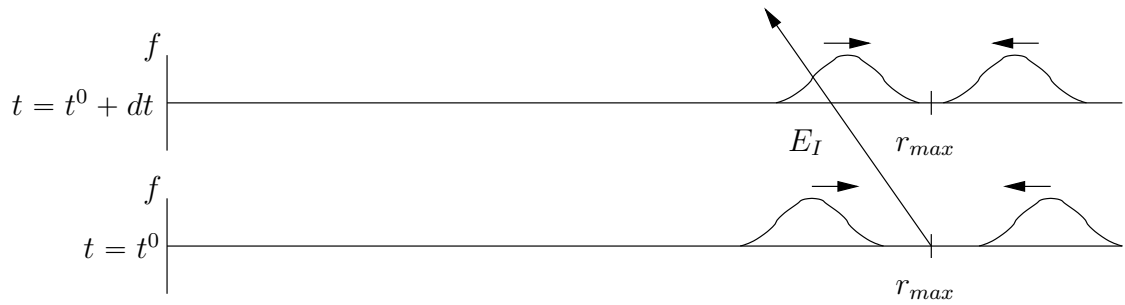
\end{center}

\caption{Image of an incoming and outgoing wave at the boundary $r_{max}$.  The future pointing inward null vector $E_{I}$ is also displayed.  Since no incoming wave can travel faster than the speed of the null vector shown, then computing the change of $f$ along this null vector will only detect changes in $f$ due to outgoing waves.}

\label{EI_rmax}

\end{figure}

This boundary condition has a physical interpretation.  Given the metric, the future-pointing null vectors in the radial direction are
\begin{align}
  \label{nulvecout} E_{O} &= \e^{-V}\, \p_{t} + \sqrt{1 - \frac{2M}{r}}\, \p_{r}, \\
  \label{nulvecin} E_{I} &= \e^{-V}\, \p_{t} - \sqrt{1 - \frac{2M}{r}}\, \p_{r}.
\end{align}
Then equation (\ref{Ref-Eq-2}) is a simplification of the following equation
\begin{align}
  \notag E_{O}(f) |_{r = r_{max}} &= \la E_{I}(f) |_{r = r_{max}} \\
  \notag \left. \e^{-V} f_{t} + \sqrt{1- \frac{2M}{r}} f_{r} \right|_{r = r_{max}} &= \left. \la\pnth{\e^{-V}f_{t} - \sqrt{1 - \frac{2M}{r}}f_{r}} \right|_{r = r_{max}} \\
  \notag \left. \e^{-V}\pnth{1 - \frac{2M}{r}}^{-1/2}f_{t} + f_{r} \right|_{r = r_{max}} &= \left. \la\pnth{\e^{-V}\pnth{1 - \frac{2M}{r}}^{-1/2}f_{t} - f_{r}} \right|_{r = r_{max}} \\
  p(t,r_{max}) + f_{r}(t,r_{max}) &= \la (p(t,r_{max}) - f_{r}(t,r_{max}))
\end{align}
Since $E_{O}(f)$ is the outward null derivative, it is the change of $f$ along a light ray leaving the system.  Since we are measuring the change along a light ray which moves faster than any amount of scalar field exiting the system, the only flux $E_{O}(f)$ can detect is the amount of scalar field moving into the system.  Similarly, $E_{I}(f)$ is the inward null derivative and by the same reasoning can only detect the amount of scalar field moving out of the system.  This idea is displayed pictorially in Figures \ref{EO_rmax} and \ref{EI_rmax}.  Thus the boundary condition in equation (\ref{Ref-Eq-2}) can be interpreted as the amount of matter that is flowing into the system is some multiple of the amount of matter that is flowing out. The parameter $\la$ controls how much matter is sent back in.  Since $\la$ can be any complex number with absolute value less than one, it can also phase shift the scalar field it sends back in.

\begin{figure}

\begin{center}
  \input{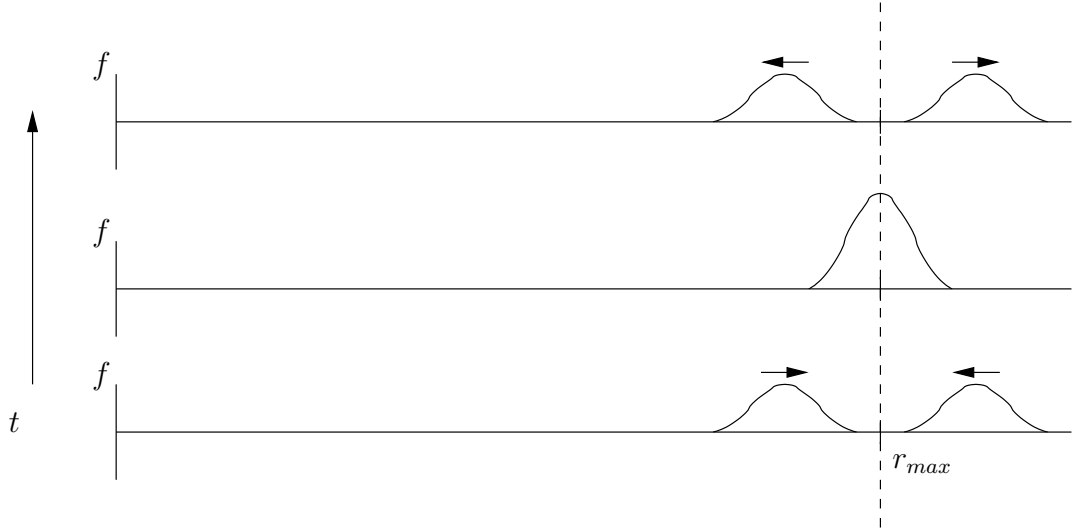}
\end{center}

\caption{A depiction of what is simulated by the boundary condition (\ref{Ref-Eq-2}) with $\la = 1$.  In this case, all outgoing waves are replaced by incoming waves of the same amplitude and phase.  If we restrict our attention to the left of the vertical dotted line representing $r = r_{max}$, we see that this is equivalent to the outgoing wave being immediately reflected by the boundary back into the system with the same amplitude and phase.  All values of $\la$ in the boundary condition (\ref{Ref-Eq-2}) can be interpreted as some amplifying or damping phase-shifting reflection.}

\label{lambda1}

\end{figure}

\begin{figure}

\begin{center}
  \input{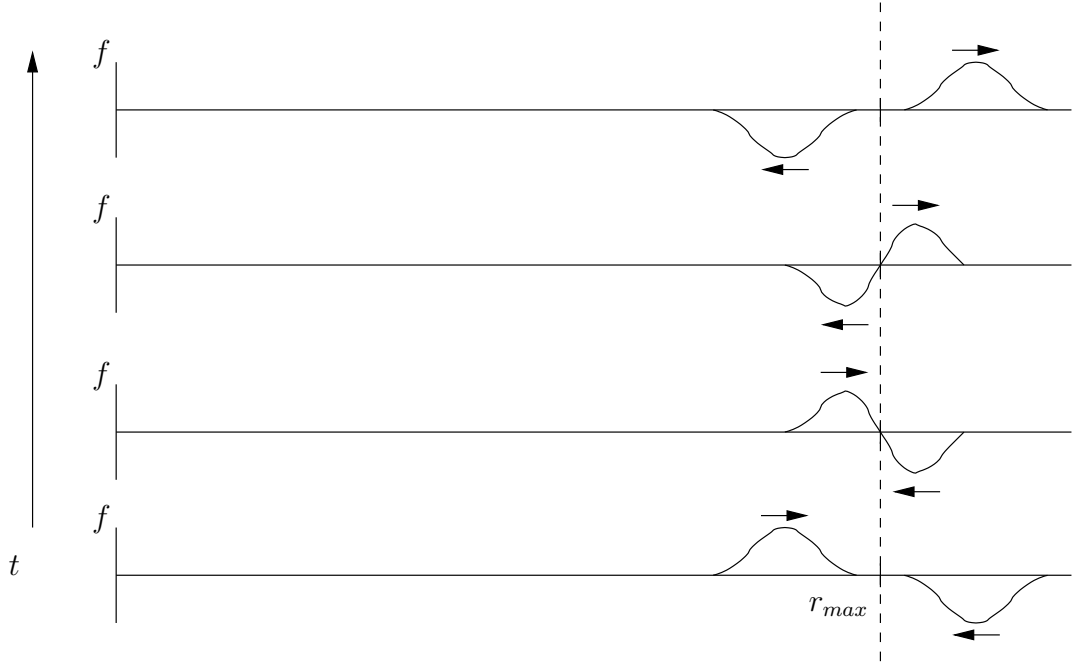}
\end{center}

\caption{A depiction of what is simulated by the boundary condition (\ref{Ref-Eq-2}) with $\la = -1$.  In this case, all outgoing waves are replaced by incoming waves of the same amplitude, but opposite phase (i.e. all peaks are valleys and vice versa).  If we restrict our attention to the left of the vertical dotted line representing $r = r_{max}$, we see that this is equivalent to the outgoing wave being immediately reflected by the boundary back into the system with the same amplitude, but opposite phase.  All values of $\la$ in the boundary condition (\ref{Ref-Eq-2}) can be interpreted as some amplifying or damping phase-shifting reflection.}

\label{lambdan1}

\end{figure}

An image that might help make this more clear, is that we have set up a sort of mirror on the outside boundary, where the level of reflectivity is controlled by $\la$, which we will call the reflectivity constant.  For example, for a value of $\la = 1$, all outgoing waves are immediately replaced by incoming waves of the same amplitude and phase.  Restricting the viewing window to the left of the $r=r_{max}$ boundary shows that this is equivalent to the outgoing wave reflecting back in exactly as it left with the same phase.  This is depicted in Figure \ref{lambda1}.  For a value of $\la = -1$, all outgoing waves are immediate replaced by incoming waves of the same amplitude but opposite phase (i.e. all of the peaks of the outgoing wave correspond to valleys of the incoming wave and vice versa).  Again, restricting the viewing window to the left of the $r = r_{max}$ boundary shows that this is equivalent to the outgoing wave being reflected back in with the opposite phase.  This is depicted in Figure \ref{lambdan1}.  Any value of $\la$ for which $\abs{\la} = 1$ will preserve the total mass at the right hand boundary.  Values of $\la$ with $\abs{\la} <1$ will allow the mass to decrease at the boundary but not increase, as it lets out more mass than it reflects back in.  Values of $\la$ will $\abs{\la} > 1$ will allow the mass to increase at the boundary but not decrease, as it ``reflects'' more mass in than it lets out.  A value of $\la = 0$ is the absence of any reflectivity, so things can only leave the system with nothing entering.  As is expected, we could use this boundary condition to simulate matter being artificially added to the system, but for the intents here, we only wish to allow the possibility that the amount reflected back in is bounded above by the amount that leaves the system.  Hence the condition mentioned before that $\abs{\la} \leq 1$.

Now that we have described the setup of the PDEs we wish to solve, we can discretize the system and apply a numerical scheme to solve it.

\section{The Numerical Scheme}

We will solve this system in two steps.  Given consistent initial conditions, we will first evolve equations (\ref{NCpde3a}) and (\ref{NCpde4a}) a single step in time and then compute the compatible metric components $M$ and $V$ using the system of ODEs (\ref{NCpde1a}) and (\ref{NCpde2a}).

Since the system (\ref{NCpde3a}) and (\ref{NCpde4a}) is first order in time, we will solve it using the method of lines.  This entails discretizing every spatial derivative and then solving the resulting system of ODEs in time using a standard ODE technique.  In this case, we will discretize the spatial derivatives using the standard second order centered finite difference and then use the third order Runge-Kutta method (\cite{Gott01,Alic07}), which is stable so long as the usual CFL condition is satisfied, to evolve in time.  For purposes of smoothing and to dissipate any high oscillation noise that may occur, we will also add in artificial Kreiss-Oliger dissipation (\cite{KO73,Lai04}) after each full Runge-Kutta time step.  While how these are implemented in our problem will all be described below in detail, for convenience, we will write down the forward Euler discretization of a first derivative in time, forward Euler of a first derivative in space, and second order centered finite difference of both a first and second derivative in space on a mesh $(t^{n},r_{k})$ for a function $f$, where $f^{n}_{k} = f(t^{n},r_{k})$.  We have that
\begin{align*}
  \text{Forward Euler in Time:} \ \ (f_{t})^{n}_{k} &\approx \frac{f^{n+1}_{k} - f^{n}_{k}}{\Del t} \displaybreak[0]\\
  \text{Forward Euler in Space:} \ \  (f_{r})^{n}_{k} &\approx \frac{f^{n}_{k+1} - f^{n}_{k}}{\Del r} \displaybreak[0]\\
  2^{\text{nd}} \text{ Order Center FD in space:} \ \ (f_{r})^{n}_{k} &\approx \frac{f^{n}_{k+1} - f^{n}_{k-1}}{2\Del r}, \\
  (f_{rr})^{n}_{k} &\approx \frac{f^{n}_{k+1} - 2f^{n}_{k} + f^{n}_{k-1}}{(\Del r)^{2}}
\end{align*}

After each complete partial time step, we will recompute the corresponding metric using a modified forward Euler scheme to solve the spatial ODEs (\ref{NCpde1a}) and (\ref{NCpde2a}) at the current time.  This is the method we will follow to update the metric components at each step.

Now we will setup the discretization.  We will compute everything in terms of $r$ and $t$.  Let the computational domain be $[0,r_{max}]$.  Then we will discretize this into $N$ subintervals of equal length, $\Del r$, with endpoints $r_{k}$ as follows.
\begin{equation}
  0 = r_{1} < r_{2} < \cdots < r_{N} < r_{N+1} = r_{max}
\end{equation}
where $r_{k} = (k-1)\Del r$.  We use $r_{1}$ as the initial point as opposed to $r_{0}$ in order to remain consistent with our Matlab code.  To choose $\Del t$, we must first compute the CFL condition.  Because the system in question comes from the Einstein equation, we know that no signal can travel faster than light in the system.  Thus the maximum wave speed for transmitting any information is the speed of light.  Equations (\ref{nulvecout}) and (\ref{nulvecin}) are the future-pointing null vectors in the radial direction.  If we move infinitesimally along either direction from the point $(t,r)$ to the point $(t+\del t, r + \del r)$, then, given the metric (\ref{metric}), the length of the infinitesimal line we traveled, being in a null direction will be 0.  Coupled with the fact that the movement in space is entirely radial, this yields that
\begin{align}
  \notag 0 &= -\e^{2V}(\del t)^{2} + \pnth{1 - \frac{2M}{r}}^{-1}(\del r)^{2} \\
  \notag \pnth{\frac{\del r}{\del t}}^{2} &= \e^{2V}\pnth{1 - \frac{2M}{r}} \\
  \frac{\del r}{\del t} &= \pm \e^{V}\sqrt{1 - \frac{2M}{r}}
\end{align}
letting $\del t$ tend to zero (and hence $\del r$ because we are moving in the direction of either the vector $E_{O}$ or $E_{I}$), yields that the speed of movement in a radial null direction, that is, the speed of light in this metric in the radial direction is given by
\begin{equation} \label{spoli}
  \pnth{\da{r}{t}}_{\text{light}} = \pm\e^{V}\sqrt{1 - \frac{2M}{r}}.
\end{equation}
The positive value corresponds to the speed of light in the outward radial direction, while the negative value corresponds to the speed of light in the inward radial direction.  Because this is the fastest that information can travel in this metric, it must be the maximum wave speed of the equations above, which model the metric.  Thus in order for our discretization to not lose information and remain stable, we must update before information from one mesh point can travel to either neighboring mesh point on the next time step.  In other words, the mesh speed, $\Del r/\Del t$, must be greater than the absolute value of the maximum wave speed, $\abs{(dr/dt)_{\text{light}}}$.  This is known as the CFL condition and is given in this case by
\begin{align} \label{CFL}
  \notag \frac{\Del r}{\Del t} &\geq \max_{r,t}\abs{\pnth{\da{r}{t}}_{\text{light}}} = \max_{r,t} \abs{\e^{V}\sqrt{1 - \frac{2M}{r}}} \\
  \Del t &\leq \min_{r,t} \abs{\e^{-V}\pnth{1 - \frac{2M}{r}}^{-1/2}} \Del r.
\end{align}
Note that since we have to pick a $\Del t$ in order to compute $V$ and $M$ and at any given time, we can't choose a $\Del t$ that is guaranteed to always be smaller than the CFL condition.  To deal with this, we will make a significantly smaller choice than the CFL condition on the initial conditions and keep the option open that if stability issues arise, we can make an even smaller choice.  Explicitly, we will choose a $c \geq 1$ and define
\begin{equation} \label{CFL-2}
  \Del t = \pnth{\frac{1}{c}}\min_{r} \abs{\e^{-V}\pnth{1 - \frac{2M}{r}}^{-1/2}} \Del r
\end{equation}
where the $V$ and $M$ here are the initial ($t=0$) values of $V$ and $M$.  Then we discretize $t$ as follows,
\begin{equation}
  0 = t^{0} < t^{1} < \cdots < t^{n} < \cdots
\end{equation}
where $t^{n} = n \Del t$.  We can solve the equations out as far out as we desire, say $t_{max}$, but since we have made the choice (\ref{CFL-2}), we will actually stop at the smallest $t_{k}$ value that is greater than or equal to $t_{max}$.

Now that we have defined the mesh, we can now discretize the system.  First, note that for any function, $f$, on the mesh, we will define $f^{n}_{k}$ as the approximation of the point $f(t^{n},r_{k})$.  We will first apply the above mentioned schemes to $f$ and $p$, however, since the third order Runge-Kutta is an iteration using forward Euler steps, we will first compute only a single forward Euler step on $f$.  Using forward Euler for the time derivatives and second order centered finite differences for the space derivatives, away from the boundaries, equation (\ref{NCpde3a}) becomes
\begin{align} \label{dis_f_m}
  \notag \frac{f^{n+1}_{k} - f^{n}_{k}}{\Del t} &= p^{n}_{k}\e^{V^{n}_{k}}\sqrt{1 - \frac{2M^{n}_{k}}{r_{k}}} \\
  f^{n+1}_{k} &= f^{n}_{k} + \Del t p^{n}_{k}\e^{V^{n}_{k}}\sqrt{1 - \frac{2M^{n}_{k}}{r_{k}}}.
\end{align}
To discretize (\ref{NCpde4a}), we first note that by (\ref{NCpde1a}) and (\ref{NCpde2a}), we have
\begin{align}
  \notag \p_{r}\pnth{\e^{V}f_{r}\sqrt{1 - \frac{2M}{r}}} &= V_{r}\e^{V}f_{r}\sqrt{1 - \frac{2M}{r}} + \e^{V}f_{rr}\sqrt{1 - \frac{2M}{r}} \\
  \notag &\qquad + \e^{V}f_{r}\pnth{1 - \frac{2M}{r}}^{-1/2}\pnth{\frac{M}{r^{2}} - \frac{M_{r}}{r}} \displaybreak[0]\\
  \notag &= \e^{V}\Bigg[f_{rr}\sqrt{1 - \frac{2M}{r}} \\
  &\qquad + 2f_{r}\pnth{1 - \frac{2M}{r}}^{-1/2}\pnth{\frac{M}{r^{2}} - 4\pi r\mu_{0}\abs{f}^{2}}\Bigg]
\end{align}
This makes (\ref{NCpde4a}) become
\begin{align}
  \notag p_{t} &= \e^{V}\Bigg[\pnth{1 - \frac{2M}{r}}^{-1/2}\pnth{2f_{r}\pnth{\frac{M}{r^{2}} - 4\pi r\mu_{0}\abs{f}^{2}} - \Up^{2}f} \\
  &\qquad + \sqrt{1 - \frac{2M}{r}}\pnth{f_{rr} + \frac{2f_{r}}{r}}\Bigg]
\end{align}
which, away from the boundaries, discretizes as
\begin{align}\label{dis_p_m}
  \notag \frac{p^{n+1}_{k} - p^{n}_{k}}{\Del t} &= \e^{V^{n}_{k}}\Bigg[ \sqrt{1 - \frac{2M^{n}_{k}}{r_{k}}}\pnth{\frac{f^{n}_{k+1} - 2f^{n}_{k} + f^{n}_{k-1}}{(\Del r)^{2}} + \frac{f^{n}_{k+1} - f^{n}_{k-1}}{r_{k}\Del r}} \\
  \notag &\qquad + \pnth{1 - \frac{2M^{n}_{k}}{r_{k}}}^{-1/2}\Bigg(2\pnth{\frac{M^{n}_{k}}{r_{k}^{2}} - 4\pi r_{k}\, \mu_{0}\abs{f^{n}_{k}}^{2}}\pnth{\frac{f^{n}_{k+1} - f^{n}_{k-1}}{2\Del r}} \\
  \notag &\qquad - \Up^{2}f^{n}_{k}\Bigg) \Bigg] \displaybreak[0]\\
  \notag p^{n+1}_{k} &= p^{n}_{k} + \Del t \e^{V^{n}_{k}}\Bigg[\sqrt{1 - \frac{2M^{n}_{k}}{r_{k}}}\pnth{\frac{f^{n}_{k+1} - 2f^{n}_{k} + f^{n}_{k-1}}{(\Del r)^{2}} + \frac{f^{n}_{k+1} - f^{n}_{k-1}}{r_{k}\Del r}} \\
  \notag &\qquad  + \pnth{1 - \frac{2M^{n}_{k}}{r_{k}}}^{-1/2}\Bigg(2\pnth{\frac{M^{n}_{k}}{r_{k}^{2}} - 4\pi r_{k}\, \mu_{0}\abs{f^{n}_{k}}^{2}}\pnth{\frac{f^{n}_{k+1} - f^{n}_{k-1}}{2\Del r}} \\
  &\qquad - \Up^{2}f^{n}_{k}\Bigg)\Bigg]
\end{align}

To deal with updating the central ($r=0$) values of these functions, we use the system (\ref{NCpde_cen}) and (\ref{sfder}).  Then using our finite difference formulas at $r=0$, we have that
\begin{equation}
  \frac{f^{n}_{2} - f^{n}_{-1}}{2\Del r} = 0
\end{equation}
where $f^{n}_{-1}$ is the approximation of $f(t^{n},-\Del r)$ (recall that $f^{n}_{1} = f(t^{n},0)$ and there is no $f^{n}_{0}$).  This implies that $f^{n}_{-1} = f^{n}_{2}$.  Using this fact we have that at $r=r_{1}=0$,
\begin{equation}
  (f_{rr})^{n}_{1} = \frac{f^{n}_{2} - 2f^{n}_{1} + f^{n}_{-1}}{(\Del r)^{2}} = \frac{2\pnth{f^{n}_{2} - f^{n}_{1}}}{(\Del r)^{2}}.
\end{equation}
Then by (\ref{NCpde_cen}), we get that $f$ and $p$ update as follows
\begin{align}\label{dis_f_c}
  \notag \frac{f^{n+1}_{1} - f^{n}_{1}}{\Del t} &= p^{n}_{1}\e^{V^{n}_{1}} \\
  f^{n+1}_{1} &= f^{n}_{1} + \Del t p^{n}_{1}\e^{V^{n}_{1}}
\end{align}
and
\begin{align}\label{dis_p_c}
  \notag \frac{p^{n+1}_{1} - p^{n}_{1}}{\Del t} &= \e^{V^{n}_{1}}\pnth{\frac{6\pnth{f^{n}_{2} - f^{n}_{1}}}{(\Del r)^{2}} - \Up^{2}f^{n}_{1}} \\
  p^{n+1}_{1} &= p^{n}_{1} + \Del t \e^{V^{n}_{1}}\pnth{\frac{6\pnth{f^{n}_{2} - f^{n}_{1}}}{(\Del r)^{2}} - \Up^{2}f^{n}_{1}}
\end{align}

At the right hand boundary, we have to impose one of the boundary conditions, which one depends on the choice of $\la$.  We will treat each one individually.  First, if $\la = 1$, equation (\ref{NBC}) holds and discretizes as
\begin{equation}
  \frac{f_{N+2} - f_{N}}{2\Del r} = 0
\end{equation}
or that $f_{N+2} = f_{N}$.  Applying this to the discretizations for $f$ and $p$, (\ref{dis_f_m}) and (\ref{dis_p_m}) respectively, yields at $r=r_{N+1}=r_{max}$ that
\begin{equation} \label{dis_f_rNBC}
  f^{n+1}_{N+1} = f^{n}_{N+1} + \Del t\, p^{n}_{N+1}\, \e^{V^{n}_{N+1}}\sqrt{1 - \frac{2M^{n}_{N+1}}{r_{N+1}}}
\end{equation}
and
\begin{align}
  \notag p^{n+1}_{N+1} &= p^{n}_{N+1} + \Del t\,  \e^{V^{n}_{N+1}} \, \Bigg[- \Up^{2}f^{n}_{N+1}\pnth{1 - \frac{2M^{n}_{N+1}}{r_{N+1}}}^{-1/2} \\
  \label{dis_p_rNBC} &\qquad \qquad \qquad \qquad \qquad + 2\sqrt{1 - \frac{2M^{n}_{N+1}}{r_{N+1}}}\pnth{\frac{f^{n}_{N} - f^{n}_{N+1}}{(\Del r)^{2}}}\Bigg]
\end{align}
Next, if $\la = -1$, then equation (\ref{DBC}) holds and we don't have to evolve anything at the boundary.  Instead, we have that for all $n$,
\begin{align}
  \label{dis_f_rDBC} f^{n+1}_{N+1} &= f^{n}_{N+1} \\
  \label{dis_p_rDBC} p^{n+1}_{N+1} &= 0
\end{align}
Finally, if $\la \neq \pm 1$, we have equation (\ref{MBC}), which yields
\begin{align}
  \notag \frac{f^{n}_{N+2} - f^{n}_{N}}{2\Del r} &= \pnth{\frac{\la - 1}{\la + 1}}p^{n}_{N+1} \\
  f^{n}_{N+2} &= f^{n}_{N} + 2\Del r \, p^{n}_{N+1}\pnth{\frac{\la - 1}{\la + 1}}.
\end{align}
Applying this to the discretizations for $f$ and $p$, (\ref{dis_f_m}) and (\ref{dis_p_m}) respectively, yields at $r = r_{N+1} = r_{max}$ that
\begin{equation} \label{dis_f_rMBC}
  f^{n+1}_{N+1} = f^{n}_{N+1} + \Del t \, p^{n}_{N+1}\, \e^{V^{n}_{N+1}}\sqrt{1 - \frac{2M^{n}_{N+1}}{r_{N+1}}}
\end{equation}
and
\begin{align}
  \notag p^{n+1}_{N+1} &= p^{n}_{N+1} + \Del t\,  \e^{V^{n}_{N+1}}\, \Bigg[\pnth{1 - \frac{2M^{n}_{N+1}}{r_{N+1}}}^{-1/2}\Bigg(2\bigg(\frac{M^{n}_{N+1}}{r_{N+1}^{2}} \\
  \notag &\qquad - 4\pi r_{N+1}\, \mu_{0}\abs{f^{n}_{N+1}}^{2}\bigg)\pnth{\frac{\la - 1}{\la + 1}}p^{n}_{N+1} - \Up^{2}f^{n}_{N+1} \Bigg) \\ \label{dis_p_rMBC} &\qquad + 2\sqrt{1 - \frac{2M^{n}_{N+1}}{r_{N+1}}}\pnth{\frac{f^{n}_{N} - f^{n}_{N+1}}{(\Del r)^{2}} + p^{n}_{N+1}\pnth{\frac{1}{r_{N+1}} + \frac{1}{\Del r}}\pnth{\frac{\la - 1}{\la + 1}}}\Bigg]
\end{align}

To summarize, given that we know $f$, $p$, $V$, and $M$ on the time step $t=t^{n}$, we evolve $f$ and $p$ a single forward Euler step in time as follows.  At $r = r_{1} = 0$, we have from (\ref{dis_f_c}) and (\ref{dis_p_c})
\begin{align}
  f^{n+1}_{1} &= f^{n}_{1} + \Del t p^{n}_{1}\e^{V^{n}_{1}} \\
  p^{n+1}_{1} &= p^{n}_{1} + \Del t \e^{V^{n}_{1}}\pnth{\frac{6\pnth{f^{n}_{2} - f^{n}_{1}}}{(\Del r)^{2}} - \Up^{2}f^{n}_{1}}.
\end{align}
For $k = 2, \ldots, N$, we have from (\ref{dis_f_m}) and (\ref{dis_p_m})
\begin{align}
  f^{n+1}_{k} &= f^{n}_{k} + \Del t p^{n}_{k}\e^{V^{n}_{k}}\sqrt{1 - \frac{2M^{n}_{k}}{r_{k}}} \displaybreak[0] \\
  \notag p^{n+1}_{k} &= p^{n}_{k} + \Del t \e^{V^{n}_{k}}\Bigg[\pnth{1 - \frac{2M^{n}_{k}}{r_{k}}}^{-1/2}\Bigg(2\bigg(\frac{M^{n}_{k}}{r_{k}^{2}} \\
  \notag &\qquad - 4\pi r_{k}\, \mu_{0}\abs{f^{n}_{k}}^{2}\bigg)\pnth{\frac{f^{n}_{k+1} - f^{n}_{k-1}}{2\Del r}} - \Up^{2}f^{n}_{k}\Bigg) \\
  &\qquad + \sqrt{1 - \frac{2M^{n}_{k}}{r_{k}}}\pnth{\frac{f^{n}_{k+1} - 2f^{n}_{k} + f^{n}_{k-1}}{(\Del r)^{2}} + \frac{f^{n}_{k+1} - f^{n}_{k-1}}{r_{k}\Del r}}\Bigg].
\end{align}
And then finally, at $r = r_{N+1} = r_{max}$, we have the following possible boundary conditions determined by the value of $\la$.  If $\la = 1$, then we use (\ref{dis_f_rNBC}) and (\ref{dis_p_rNBC}) as follows
\begin{align}
  f^{n+1}_{N+1} &= f^{n}_{N+1} + \Del t\, p^{n}_{N+1}\, \e^{V^{n}_{N+1}}\sqrt{1 - \frac{2M^{n}_{N+1}}{r_{N+1}}} \\
  \notag p^{n+1}_{N+1} &= p^{n}_{N+1} + \Del t\,  \e^{V^{n}_{N+1}} \, \Bigg[- \Up^{2}f^{n}_{N+1}\pnth{1 - \frac{2M^{n}_{N+1}}{r_{N+1}}}^{-1/2} \\
  &\qquad \qquad \qquad \qquad \qquad + 2\sqrt{1 - \frac{2M^{n}_{N+1}}{r_{N+1}}}\pnth{\frac{f^{n}_{N} - f^{n}_{N+1}}{(\Del r)^{2}}}\Bigg].
\end{align}
If $\la = -1$, we use (\ref{dis_f_rDBC}) and (\ref{dis_p_rDBC}) and obtain
\begin{align}
  f^{n+1}_{N+1} &= f^{n}_{N+1} \\
  p^{n+1}_{N+1} &= 0.
\end{align}
Finally, if $\la \neq \pm 1$, we use (\ref{dis_f_rMBC}) and (\ref{dis_p_rMBC}) and obtain
\begin{align}
  f^{n+1}_{N+1} &= f^{n}_{N+1} + \Del t \, p^{n}_{N+1}\, \e^{V^{n}_{N+1}}\sqrt{1 - \frac{2M^{n}_{N+1}}{r_{N+1}}} \\
  \notag p^{n+1}_{N+1} &= p^{n}_{N+1} + \Del t\,  \e^{V^{n}_{N+1}}\, \Bigg[\pnth{1 - \frac{2M^{n}_{N+1}}{r_{N+1}}}^{-1/2}\Bigg(2\bigg(\frac{M^{n}_{N+1}}{r_{N+1}^{2}} \\
  \notag &\qquad - 4\pi r_{N+1}\, \mu_{0}\abs{f^{n}_{N+1}}^{2}\bigg)\pnth{\frac{\la - 1}{\la + 1}}p^{n}_{N+1} - \Up^{2}f^{n}_{N+1} \Bigg) \\
  &\qquad + 2\sqrt{1 - \frac{2M^{n}_{N+1}}{r_{N+1}}}\pnth{\frac{f^{n}_{N} - f^{n}_{N+1}}{(\Del r)^{2}} + p^{n}_{N+1}\pnth{\frac{1}{r_{N+1}} + \frac{1}{\Del r}}\pnth{\frac{\la - 1}{\la + 1}}}\Bigg]
\end{align}
To perform this evolution step, we will need to know all four functions at $t^{n}$.  Define the array $U^{n}$ as $U^{n} = [f^{n}, p^{n}, V^{n}, M^{n}]$.  Then we will denote the above update of $f$ and $p$ as
\begin{equation} \label{E1}
  [f^{n+1}, p^{n+1}] = E1(U^{n},\Del t).
\end{equation}

After performing that evolution step, we will need to compute the corresponding $V$ and $M$.  We will write here the discretizations of equations (\ref{NCpde1a}) and (\ref{NCpde2a}).  Unfortunately, due to the fact that $V$ doesn't appear on the right-hand side of (\ref{NCpde2a}), using the second order centered finite difference approximation for $V_{r}$ produces a decoupling of the even and odd discretizations points, sometimes called a ``checkerboard instability''.  In order to remedy this, we will used a modified forward Euler scheme that uses both $V_{r}$ and $V_{rr}$ to compute the next point of $V$.  To do this, we will simply use the Taylor series of $V$ truncated at the second derivative to compute $V_{k+1}$ from points at $r_{k}$ as follows,
\begin{equation} \label{2ndEul}
  V_{k+1} = V_{k} + (V_{r})_{k} \, \Del r + (V_{rr})_{k}\, \frac{\Del r^{2}}{2}.
\end{equation}
To be consistent, we will compute the points of $M$ using the same scheme as above, but we will continue to use the centered finite differences to approximate the space derivatives of $f$ and $p$.  This should keep the overall scheme roughly second order accurate in space.

In order to follow this overall scheme, we will have to compute formulas for $M_{rr}$ and $V_{rr}$ and compute their values at $r=0$.  To that end, we simply compute the derivatives of equations (\ref{NCpde1a}) and (\ref{NCpde2a}) and obtain
\begin{align}
  \notag M_{rr} &= \frac{2M_{r}}{r}\brkt{1 - 4\pi r^{2}\mu_{0}\pnth{ \frac{\abs{f_{r}}^{2} + \abs{p}^{2}}{\Up^{2}}}} \\
  \notag &\qquad + 4\pi r^{2}\mu_{0}\Bigg\{ 2 \Re(f_{r}\conj{f}) + \frac{2M}{r^{2}}\pnth{\frac{\abs{f_{r}}^{2} + \abs{p}^{2}}{\Up^{2}}} \\
  \label{NCpde1a2nd} &\qquad + 2\pnth{1 - \frac{2M}{r}}\frac{\Re(f_{rr}\conj{f}_{r} + p_{r} \conj{p})}{\Up^{2}} \Bigg\} \displaybreak[0] \\
  \notag V_{rr} &= \pnth{1 - \frac{2M}{r}}^{-1}\Bigg\{\frac{M_{r}}{r}\brkt{\frac{1}{r} + 2V_{r} - 8\pi r\mu_{0}\pnth{\frac{\abs{f_{r}}^{2} + \abs{p}^{2}}{\Up^{2}}}} \\
  \notag &\qquad + \frac{V_{r}}{r}\pnth{1 - \frac{4M}{r}} - \frac{3M}{r^{3}} - 4\pi r\mu_{0}\Bigg[2 \Re(f_{r}\conj{f}) \\
  \label{NCpde2a2nd} &\qquad - 2\pnth{1 - \frac{2M}{r}}\frac{\Re(f_{rr}\conj{f}_{r} + p_{r}\conj{p})}{\Up^{2}} - \frac{2M}{r^{2}}\pnth{\frac{\abs{f_{r}}^{2} + \abs{p}^{2}}{\Up^{2}}}\Bigg] \Bigg\}
\end{align}
There is no need, in general, to substitute in the values for $M_{r}$ and $V_{r}$ since we will need to compute the values of each at every point and so can just use those values here.  Now, at $r=0$, we already know that
\begin{equation}
  M = 0, \qquad M_{r} = 0, \qquad M_{rr} = 0, \qquad \text{and} \qquad V_{r} = 0.
\end{equation}
So we already have enough information to compute $M_{2} = M(\Del r)$, and we only need to know $V$ and $V_{rr}$ at the origin in order to have the information necessary to compute the point $V_{2} = V(\Del r)$.

We want $V$ and $M$ to define a metric that is asymptotically Schwarzschild as with the static states.  We will impose the asymptotically Schwarzschild condition at $r = r_{max}$.  Thus $V(0)$ must be chosen to satisfy (\ref{AB2a}) on the boundary.  Note also that given the system (\ref{NCpde}), for a given $f$ and $p$, if $M$ and $V$ satisfy (\ref{NCpde1}) and (\ref{NCpde2}), then so do $M$ and $V+b$ for some constant $b \in \RR$.  This allows us to set $V(0) = 0$ to initially solve for $V$, followed by setting $b$ as follows
\begin{equation} \label{Vmax}
  b = \frac{1}{2}\ln\pnth{1 - \frac{2M(t,r_{max})}{r_{max}}} + \ln \ka - V(t,r_{max}) = \frac{1}{2}\ln\pnth{1 - \frac{2M^{n}_{N+1}}{r_{N+1}}} + \ln \ka - V^{n}_{N+1},
\end{equation}
where as before, for convenience, we set $\ka = 1$.  We will then relabel $V + b$ as simply $V$.  Thus for the intents and purposes of computing $V$ at $r = 0$, we will set $V(0) = 0$, imposing the right hand boundary condition after computing $V$ and $M$ completely.

For $V_{rr}$, if we attempt to take the limit of (\ref{NCpde2a2nd}) as $r\to 0$, by \LHopital's Rule, we will acquire no new information.  However, if we rewrite the following
\begin{align}
  \notag \frac{V_{r}}{r}\pnth{1 - \frac{4M}{r}} &= \frac{V_{r}}{r}\pnth{1 - \frac{2M}{r}} - \frac{2V_{r}M}{r} \\
  &= \frac{M}{r^{3}} - 4\pi \mu_{0}\pnth{\abs{f}^{2} - \pnth{1 - \frac{2M}{r}}\frac{\abs{f_{r}}^{2} + \abs{p}^{2}}{\Up^{2}}} - \frac{2V_{r}M}{r},
\end{align}
then we get the equation
\begin{align}
  \notag V_{rr} &= \pnth{1 - \frac{2M}{r}}^{-1}\Bigg\{\frac{M_{r}}{r}\brkt{\frac{1}{r} + 2V_{r} - 8\pi r\mu_{0}\pnth{\frac{\abs{f_{r}}^{2} + \abs{p}^{2}}{\Up^{2}}}} \\
  \notag &\qquad - \frac{2V_{r}M}{r} - \frac{2M}{r^{3}}  - 4\pi\mu_{0}\pnth{\abs{f}^{2} - \frac{\abs{f_{r}}^{2} + \abs{p}^{2}}{\Up^{2}}} \\
  \label{NCpde2a2ndA} &\qquad - 4\pi r\mu_{0}\brkt{2 \Re(f_{r}\conj{f}) - 2\pnth{1 - \frac{2M}{r}}\frac{\Re(f_{rr}\conj{f}_{r} + p_{r}\conj{p})}{\Up^{2}}} \Bigg\}.
\end{align}
In this case, we get something useful from the limit at $r \to 0$.  By \LHopital's rule, we obtain
\begin{align}
  \notag \lim_{r \to 0} V_{rr} &= \lim_{r\to 0} \pnth{1 - \frac{2M}{r}}^{-1}\Bigg\{\frac{M_{r}}{r}\brkt{\frac{1}{r} + 2V_{r} - 8\pi r\mu_{0}\pnth{\frac{\abs{f_{r}}^{2} + \abs{p}^{2}}{\Up^{2}}}} \\
  \notag &\qquad - \frac{2V_{r}M}{r} - \frac{2M}{r^{3}} - 4\pi\mu_{0}\pnth{\abs{f}^{2} - \frac{\abs{f_{r}}^{2} + \abs{p}^{2}}{\Up^{2}}} \\
  \notag &\qquad - 4\pi r\mu_{0}\brkt{2 \Re(f_{r}\conj{f}) - 2\pnth{1 - \frac{2M}{r}}\frac{\Re(f_{rr}\conj{f}_{r} + p_{r}\conj{p})}{\Up^{2}}} \Bigg\} \displaybreak[0]\\
  \notag &= \lim_{r \to 0} \frac{M_{r}}{r^{2}} - \frac{2M}{r^{3}} - 4\pi \mu_{0}\pnth{\abs{f}^{2} - \frac{\abs{p}^{2}}{\Up^{2}}} \displaybreak[0]\\
  \notag &= \lim_{r\to 0} \frac{M_{rr}}{2r} - \frac{2M_{r}}{3r^{2}} - 4\pi \mu_{0}\pnth{\abs{f}^{2} - \frac{\abs{p}^{2}}{\Up^{2}}} \displaybreak[0]\\
  \notag &= \lim_{r \to 0} \frac{M_{rrr}}{2} - \frac{2M_{rr}}{6r} - 4\pi \mu_{0}\pnth{\abs{f}^{2} - \frac{\abs{p}^{2}}{\Up^{2}}} \displaybreak[0]\\
  \notag &= \lim_{r \to 0} \frac{M_{rrr}}{2} - \frac{M_{rrr}}{3} - 4\pi \mu_{0}\pnth{\abs{f}^{2} - \frac{\abs{p}^{2}}{\Up^{2}}} \\
  &= \left. \frac{M_{rrr}}{6} - 4\pi \mu_{0}\pnth{\abs{f}^{2} - \frac{\abs{p}^{2}}{\Up^{2}}} \right|_{r=0}
\end{align}
Now, while we don't have enough information to pin down $M_{rrr}$ at the central value exactly, due to the fact that (\ref{NCpde1a}) and (\ref{NCpde1a2nd}) do not depend directly on the values of $V$, we will first compute $M_{2}=M(\Del r)$ and $M_{3} = M(2\Del r)$ and then use a second order centered finite difference formula to approximate $M_{rrr}$ at $r=0$.  Then, because $M$ is an odd function, we obtain
\begin{align}
  \notag M_{rrr}(0) &\approx \frac{-M(-2\Del r) + 2M(-\Del r) - 2M(\Del r) + M(2 \Del r)}{2(\Del r)^{3}} \\
  \notag &= \frac{2M(2\Del r) - 4M(\Del r)}{2(\Del r)^{3}} \displaybreak[0]\\
  \notag &= \frac{M(2\Del r) - 2M(\Del r)}{(\Del r)^{3}} \\
  &= \frac{M_{3} - 2M_{2}}{(\Del r)^{3}}
\end{align}
And since
\begin{equation}
  M_{2} = M_{1} + (M_{r})_{1}\Del r + (M_{rr})_{1}\frac{(\Del r)^{2}}{2} = M_{1} = 0
\end{equation}
we have that
\begin{equation}
  M_{rrr}(0) = \dfrac{M_{3}}{(\Del r)^{3}}.
\end{equation}
and hence
\begin{equation} \label{Vrr0}
  V_{rr}(0) = (V_{rr})_{1} = \dfrac{M_{3}}{6(\Del r)^{3}} - 4\pi\mu_{0}\pnth{\abs{f_{1}}^{2} - \frac{\abs{p_{1}}^{2}}{\Up^{2}}}
\end{equation}
This gives us enough information to compute $M$ and $V$ near the left hand boundary $r=0$, which in turn, because the scheme is explicit, gives us the necessary information to continue the approximation out to our right hand boundary $r=r_{max}$.  Then we only need to ensure that we have the appropriate asymptotic behavior, which we accomplish by selecting the appropriate value $b$ to add to the metric $V$, where $b$ is as defined in equation (\ref{Vmax}).

We now solve for $V$ and $M$ as follows; note that since all of the values will be defined at the same $t^{n}$ point, we will omit the superscripts $n$ from these equations.  At $r=r_{1}=0$, we have
\begin{align}
  \label{dis_M_c1} M_{1} &= 0 \\
  \label{dis_V_c1} V_{1} &= 0.
\end{align}
Next, since $M_{r}(0) = M_{rr}(0)=0$, we have that
\begin{equation}
  M_{2} = M_{1} = 0,
\end{equation}
but, as stated before, we first need to compute $M_{3}$ before we can compute $V_{2}$.  To compute $M_{3}$, we have that
\begin{align}
  (M_{r})_{2} &= 4\pi r_{2}^{2}\, \mu_{0}\brkt{\abs{f_{2}}^{2} + \pnth{1 - \frac{2M_{2}}{r_{2}}}\pnth{\frac{\abs{f_{3} - f_{1}}^{2}}{4\Up^{2}(\Del r)^{2}} + \frac{\abs{p_{2}}^{2}}{\Up^{2}}}} \\
  \notag (M_{rr})_{2} &= \frac{2(M_{r})_{2}}{r_{2}}\brkt{1 - 4\pi r_{2}^{2}\, \mu_{0}\pnth{\frac{\abs{f_{3} - f_{1}}^{2}}{4\Up^{2}(\Del r)^{2}} + \frac{\abs{p_{2}}^{2}}{\Up^{2}}}} \\
  \notag &\qquad + 4\pi r_{2}^{2}\, \mu_{0}\Bigg\{2\Re\brkt{\pnth{\frac{f_{3} - f_{1}}{2\Del r}}\conj{f}_{2}} + \frac{2M_{2}}{r_{2}^{2}}\pnth{\frac{\abs{f_{3} - f_{1}}^{2}}{4\Up^{2}(\Del r)^{2}} + \frac{\abs{p_{2}}^{2}}{\Up^{2}}} \\
  &\qquad + \frac{2}{\Up^{2}}\pnth{1 - \frac{2M_{2}}{r_{2}}}\Re\brkt{\pnth{\frac{f_{3} - 2f_{2} + f_{1}}{(\Del r)^{2}}}\pnth{\frac{\conjL{f_{3} - f_{1}}}{2\Del r}} + \pnth{\frac{p_{3} - p_{1}}{2\Del r}}\conj{p}_{2}} \Bigg\}
\end{align}
Then we compute $M_{3}$ as follows
\begin{equation}
  M_{3} = M_{2} + (M_{r})_{2} \, \Del r + (M_{rr})_{2}\, \frac{(\Del r)^{2}}{2}.
\end{equation}
Then we compute $V_{2}$ as follows.
\begin{align}
  (V_{r})_{1} &= 0 \\
  (V_{rr})_{1} &= \dfrac{M_{3}}{6(\Del r)^{3}} - 4\pi\mu_{0}\pnth{\abs{f_{1}}^{2} - \frac{\abs{p_{1}}^{2}}{\Up^{2}}}
\end{align}
which yields
\begin{equation}
  V_{2} = V_{1} + (V_{r})_{1}\, \Del r + (V_{rr})_{1}\, \frac{(\Del r)^{2}}{2}.
\end{equation}
To get $V_{3}$, we have
\begin{align}
  \notag (V_{r})_{2} &= \pnth{1 - \frac{2M_{2}}{r_{2}}}^{-1}\Bigg[\frac{M_{2}}{r_{2}^{2}} - 4\pi r_{2}\, \mu_{0}\Bigg( \abs{f_{2}}^{2} \\
  &\qquad \qquad \qquad \qquad \qquad \qquad \qquad - \pnth{1 - \frac{2M_{2}}{r_{2}}} \pnth{\frac{\abs{f_{3} - f_{1}}^{2}}{4\Up^{2}(\Del r)^{2}} + \frac{\abs{p_{2}}^{2}}{\Up^{2}}} \Bigg) \Bigg] \displaybreak[0] \\
  \notag (V_{rr})_{2} &= \pnth{1 - \frac{2M_{2}}{r_{2}}}^{-1}\Bigg\{\frac{(M_{r})_{2}}{r_{2}}\brkt{\frac{1}{r_{2}} + 2(V_{r})_{2} - 8\pi r_{2}\, \mu_{0}\pnth{\frac{\abs{f_{3} - f_{1}}^{2}}{4\Up^{2}(\Del r)^{2}} + \frac{\abs{p_{2}}^{2}}{\Up^{2}}}} \\
  \notag &\qquad + \frac{(V_{r})_{2}}{r_{2}}\pnth{1 - \frac{4M_{2}}{r_{2}}} - \frac{3M_{2}}{r_{2}^{3}} - 4\pi r_{2}\, \mu_{0}\Bigg[2 \Re\pnth{\pnth{\frac{f_{3} - f_{1}}{2\Del r}}\conj{f_{2}}} \displaybreak[0] \\
  \notag &\qquad - \frac{2}{\Up^{2}}\pnth{1 - \frac{2M_{2}}{r_{2}}}\Re\pnth{\pnth{\frac{f_{3} - 2f_{2} + f_{1}}{(\Del r)^{2}}}\pnth{\frac{\conjL{f_{3} - f_{1}}}{2\Del r}} + \pnth{\frac{p_{3} - p_{1}}{2\Del r}}\conj{p}_{2}} \\
  &\qquad - \frac{2M_{2}}{r_{2}^{2}}\pnth{\frac{\abs{f_{3} - f_{1}}^{2}}{4\Up^{2}(\Del r)^{2}} + \frac{\abs{p_{2}}^{2}}{\Up^{2}}} \Bigg] \Bigg\}
\end{align}
which yields
\begin{equation}
  V_{3} = V_{2} + (V_{r})_{2}\, \Del r + (V_{rr})_{2}\, \frac{(\Del r)^{2}}{2}.
\end{equation}
Then for $k = 3, \ldots, N$, we compute the following
\begin{align}
  (M_{r})_{k} &= 4\pi r_{k}^{2}\, \mu_{0} \pnth{\abs{f_{k}}^{2} + \pnth{1 - \frac{2M_{k}}{r_{k}}}\pnth{\frac{\abs{f_{k+1} - f_{k-1}}^{2}}{4 \Up^{2} (\Del r)^{2}} + \frac{\abs{p_{k}}^{2}}{\Up^{2}}}} \displaybreak[0]\\
  \notag (M_{rr})_{k} &= \frac{2(M_{r})_{k}}{r_{k}}\brkt{1 - 4\pi r_{k}^{2}\, \mu_{0}\pnth{\frac{\abs{f_{k+1} - f_{k-1}}^{2}}{4\Up^{2}(\Del r)^{2}} + \frac{\abs{p_{k}}^{2}}{\Up^{2}}}} \\
  \notag &\qquad + 4\pi r_{k}^{2}\, \mu_{0}\Bigg\{2\Re\brkt{\pnth{\frac{f_{k+1} - f_{k-1}}{2\Del r}}\conj{f}_{k}} + \frac{2M_{k}}{r_{k}^{2}}\pnth{\frac{\abs{f_{k+1} - f_{k-1}}^{2}}{4\Up^{2}(\Del r)^{2}} + \frac{\abs{p_{k}}^{2}}{\Up^{2}}} \displaybreak[0]\\
  \notag &\qquad + \frac{2}{\Up^{2}}\pnth{1 - \frac{2M_{k}}{r_{k}}}\Re\Bigg[\pnth{\frac{f_{k+1} - 2f_{k} + f_{k-1}}{(\Del r)^{2}}}\pnth{\frac{\conjL{f_{k+1} - f_{k-1}}}{2\Del r}} \\
  &\qquad + \pnth{\frac{p_{k+1} - p_{k-1}}{2\Del r}}\conj{p}_{k}\Bigg] \Bigg\} \displaybreak[0] \\
  \notag (V_{r})_{k} &= \pnth{1 - \frac{2M_{k}}{r_{k}}}^{-1}\Bigg(\frac{M_{k}}{r_{k}^{2}} \\
  &\qquad - 4\pi r_{k}\, \mu_{0}\pnth{\abs{f_{k}}^{2} - \pnth{1 - \frac{2M_{k}}{r_{k}}}\pnth{\frac{\abs{f_{k+1} - f_{k-1}}^{2}}{4 \Up^{2} (\Del r)^{2}} + \frac{\abs{p_{k}}^{2}}{\Up^{2}}}}\Bigg) \displaybreak[0]\\
  \notag (V_{rr})_{k} &= \pnth{1 - \frac{2M_{k}}{r_{k}}}^{-1}\Bigg\{\frac{(M_{r})_{k}}{r_{k}}\brkt{\frac{1}{r_{k}} + 2(V_{r})_{k} - 8\pi r_{k}\, \mu_{0}\pnth{\frac{\abs{f_{k+1} - f_{k-1}}^{2}}{4\Up^{2}(\Del r)^{2}} + \frac{\abs{p_{k}}^{2}}{\Up^{2}}}} \\
  \notag &\qquad + \frac{(V_{r})_{k}}{r_{k}}\pnth{1 - \frac{4M_{k}}{r_{k}}} - \frac{3M_{k}}{r_{k}^{3}} - 4\pi r_{k}\, \mu_{0}\Bigg[2 \Re\pnth{\pnth{\frac{f_{k+1} - f_{k-1}}{2\Del r}}\conj{f_{k}}} \displaybreak[0]\\
  \notag &\qquad - \frac{2}{\Up^{2}}\pnth{1 - \frac{2M_{k}}{r_{k}}}\Re\Bigg(\pnth{\frac{f_{k+1} - 2f_{k} + f_{k-1}}{(\Del r)^{2}}}\pnth{\frac{\conjL{f_{k+1} - f_{k-1}}}{2\Del r}} \\
  &\qquad + \pnth{\frac{p_{k+1} - p_{k-1}}{2\Del r}}\conj{p}_{k}\Bigg) - \frac{2M_{k}}{r_{k}^{2}}\pnth{\frac{\abs{f_{k+1} - f_{k-1}}^{2}}{4\Up^{2}(\Del r)^{2}} + \frac{\abs{p_{k}}^{2}}{\Up^{2}}} \Bigg] \Bigg\}
\end{align}
and obtain
\begin{align}
  M_{k+1} &= M_{k} + (M_{r})_{k}\, \Del r + (M_{rr})_{k}\, \frac{(\Del r)^{2}}{2} \\
  V_{k+1} &= V_{k} + (V_{r})_{k}\, \Del r + (V_{rr})_{k}\, \frac{(\Del r)^{2}}{2}
\end{align}
Finally, we compute $b$ according to equation (\ref{Vmax}) and redefine the entire array $V$ as $V+b$.

To perform the above update of $M$ and $V$, we will only have to know the updated values of $f$ and $p$ computed in $E1$ (from (\ref{E1})).  Since it only depends on the $f$ and $p$ at any given time, we will denote this process as
\begin{equation} \label{E2}
  [f^{n},p^{n},V^{n},M^{n}] = E2(f^{n},p^{n})
\end{equation}
Composing the two processes, $E1$ and $E2$ (from (\ref{E1}) and (\ref{E2}) respectively), together we obtain a process of updating all four functions a single forward Euler step in time, which we will call simply $E(U^{n},\Del t)$.  Explicitly, we write
\begin{equation} \label{E}
  U^{n+1} = [f^{n+1},p^{n+1},V^{n+1},M^{n+1}] = E(U^{n},\Del t) = E2(E1(U^{n},\Del t))
\end{equation}

Now we have enough to describe in simple terms the entire process of evolving the four functions $f,p,V,M$.  We start with given initial conditions $f^{0}$ and $p^{0}$ which can be freely defined.  Then we perform the following process.
\begin{enumerate}
  \item\label{step1} Compute the compatible initial values of the metric, $V^{0}$ and $M^{0}$, using $E2$.  That is,
      \begin{equation}
        U^{0} = E2(f^{0},p^{0}).
      \end{equation}
  \item\label{step2} Use the third order Runge-Kutta method (\cite{Gott01,Alic07}) as follows to evolve the four functions, $f,p,V,M$ a single time step from $t^{n}$ to $t^{n+1}$.  This is a three step iterative process of successive weighted averages of the four input functions and a forward Euler step.  Explicitly, we compute $U^{n+1}$ as follows.
      \begin{subequations}
        \begin{align}
          U^{\ast} &= E(U^{n},\Del t) \\
          U^{\ast\ast} &= \frac{3}{4}U^{n} + \frac{1}{4}E(U^{\ast},\Del t) \\
          U^{n+1} &= \frac{1}{3}U^{n} + \frac{2}{3}E(U^{\ast\ast},\Del t).
        \end{align}
      \end{subequations}
  \item\label{step3} Apply Kreiss-Oliger dissipation (\cite{KO73,Lai04}) to the updated functions $f$ and $p$ as follows.  This adds a smoothing factor that is dependent on the previous time step which removes some of the high frequency oscillation that can occur due to numerical errors.  We only add this in for $k = 3,\ldots,N-1$, the other points we leave unchanged.
      \begin{subequations}\label{KO-diss}
        \begin{align}
          f^{n+1}_{k} &= f^{n+1}_{k} - \frac{\vep}{16}\pnth{f^{n}_{k-2} - 4f^{n}_{k-1} + 6f^{n}_{k} - 4f^{n}_{k+1} + f^{n}_{k+2}} \\
          p^{n+1}_{k} &= p^{n+1}_{k} - \frac{\vep}{16}\pnth{p^{n}_{k-2} - 4p^{n}_{k-1} + 6p^{n}_{k} - 4p^{n}_{k+1} + p^{n}_{k+2}}
        \end{align}
      \end{subequations}
      where, for stability, $0 \leq \vep < 1$ (\cite{KO73,Lai04}) ($\vep = 0$ corresponds to no dissipation).  In our case, we choose $\vep = 1/2$.  We choose that value for two reasons.  First, this value is the typical one used (\cite{Lai04}).  And second, in a few trial simulations we performed of a simpler PDE where the solution was known, this value of $\vep$ had the desired smoothing effect but the lowest mean error when compared to other attempted values of $\vep$ including $\vep = 0$.
  \item\label{step4}  Since we have once again changed the values of $f$ and $p$, we need to once again compute the compatible metric components $V$ and $M$. Thus we apply $E2$ again and obtain
      \begin{equation}
        U^{n+1} = E2(f^{n+1},p^{n+1})
      \end{equation}
  \item\label{step5} Repeat steps \ref{step2} - \ref{step4} until the desired time value $t_{max}$ is reached.
\end{enumerate}

\section{Stability and Accuracy of the Scheme}

In this section, we make a few comments about the stability and accuracy of the numerical scheme presented above.

\subsection{Stability of the Scheme}

We have already noted that the third order Runge-Kutta scheme is stable so long as the CFL condition in equation (\ref{CFL}) is satisfied.  Moreover, the implementation of Kreiss-Oliger dissipation is stable so long as the $\vep$ in equation (\ref{KO-diss}) is positive and less than one.  We used $\vep = 1/2$ and so we should expect stability from this aspect of the scheme.  These stability results are consistent with our initial runs of the scheme which show stable behavior for hundreds or even thousands of periods of static states in time.

However, we did find experimentally an additional stability concern due to the right hand boundary condition involving the reflectivity constant $\la$. Due to the last term in equation (\ref{dis_p_rMBC}), which is used whenever $\la \neq \pm 1$, we have that, since all of the other values are generally very close to zero due to the asymptotically Schwarzschild condition, the values of $p_{N+1}^{n+1}$ will progressively get larger whenever
\begin{equation}
  \abs{2p_{N+1}^{n}\pnth{\e^{V_{N+1}^{n}}\sqrt{1 - \frac{2M_{N+1}^{n}}{r_{N+1}}}}\frac{\Del t}{\Del r}\pnth{\frac{\la-1}{\la + 1}}} = \abs{2(f_{t})_{N+1}^{n}\frac{\Del t}{\Del r}\pnth{\frac{\la - 1}{\la +1}}} \geq \abs{(f_{t})_{N+1}^{n}}
\end{equation}
This depends on the speed of the mesh $\Del r /\Del t$ and hence is mesh-dependent and not an effect inherent in the pdes.  Due to the growing values of $p_{N+1}^{n+1}$ and hence $f_{N+1}^{n+1}$, if the above condition is not satisfied, the numerical scheme will be unstable at the right hand boundary which will quickly propagate to the rest of the system.  Thus the scheme is only stable so long as the above condition with the opposite inequality is satisfied.  That is, when
\begin{equation}
  \abs{\frac{\la - 1}{\la +1}} < \frac{\Del r}{2\Del t}
\end{equation}
The term $\Del r/ \Del t$ occurs in the CFL condition.  The CFL condition we use in practice is found in equation (\ref{CFL}) and implies that the above equation makes the above inequality become
\begin{equation}
  \abs{\frac{\la - 1}{\la + 1}} < \frac{c}{2}\max_{r,t}\abs{\e^{V}\sqrt{1 - \frac{2M}{r}}}
\end{equation}
This is only a problem as $\la$ approaches $-1$.  In order to deal with this issue in practice, if $\la$ is close enough to $-1$ to cause problems with our chosen mesh, we will either increase the value of $c$, which makes our mesh finer in the $t$ direction, or we simply consider $\la$ close enough to $-1$ and use the boundary condition corresponding to $\la = -1$.

\subsection{Accuracy of the Scheme}

This overall scheme uses second order accurate finite difference approximations in space and a third order accurate finite difference approximations in time.  While these are the formal orders of accuracy, every scheme has a practical or experimental order of accuracy.  This practical order of accuracy can be determined by evolving the same initial conditions on different meshes and comparing their differences to either the known solution with those initial conditions, or the evolution of the initial conditions on an extremely fine mesh.

One of the first things we will do now that we know the scheme functions is to conduct an experiment to determine the practical order of accuracy.

\section{Explanation of the Matlab Code}

We have encoded this numerical scheme into a package of Matlab programs.  These programs are set up with a parent program, NCEKG.m, and several subroutines that perform the various steps of the numerical integration process as well as graph the solutions.  The output generated is a .avi movie file or files that displays plots the evolving functions.  The frame rate used can be set internally in the program.

The call sequence for this program is as follows
\begin{align*}
  NCEKG(f,&p,c,rmax,tmin,tmax,tmaxult,Upsilon,mu0,\\
  &lambda,vidtype,rt,st,PlotMax,U0,Mloss,tlist)
\end{align*}
In addition to the output .avi movie files, this matlab file also generates several output variables, specifically the list
\begin{equation*}
  [U,rt,st,PlotMax,U0,Mloss,tlist]
\end{equation*}
Each of these variables are defined as follows.  The NCEKG function evolves the Einstein-Klein-Gordon scalar field $f$ and its ``time derivative'' $p$, as defined in this dissertation, through time from a specified initial $t$ value, $tmin$, to a specified $t$ value, $tmax$. It plots the scalar field and metric functions, as well as different types of energy density plots.  Superimposed on the plot of the potential function $V$ is a vertical line that represents a point particle placed in the gravitating system.  It approximately follows a radial geodesic and so simulates the gravitational effects of the system.  Additionally, superimposed on all of the plots and depicted as a red dotted line is their initial value.

The parameters $f$ and $p$ are defined at equidistant radial points from 0 to $rmax$, that is, on the discretized mesh we defined in this dissertation.  The parameter $c$ is the CFL ratio parameter defined in this dissertation.  The number $lambda$ is the reflectivity constant, $Upsilon$ is the fundamental constant of the Einstein-Klein-Gordon equations, and $mu0$ is the scaling parameter in the Einstein-Klein-Gordon equations.  The parameters $rt$ and $st$ are optional arguments that specify the initial location and radial velocity of the tracked point particle moving along a geodesic.  $PlotMax$ is an optional argument and must be a list of five integers that define the order of magnitude of the vertical window sizes for the plots.  The inputs $U0$, $Mloss$, and $tlist$ are defined identically as their output counterparts and keep track of the original initial conditions, the change in mass at the right hand boundary, and the $t$-mesh currently computed upon.  These functions are used if the program is employed in succession to evolve the Einstein-Klein-Gordon equations for a long period of time.  The parameter $tmaxult$ is the largest value of $t$ that we expect to evolve to.  Finally, the parameter $vidtype$ tells the program what kind of plot is desired.  A $vidtype$ value of 1 will plot individual .avi files for each of the plots of the functions, while a $vidtype$ value of 2 will plot the individual function .avi files and the .avi file with all the plots.  Any other vidtype value will plot only the .avi file with all the plots.

The output $U$ is an array whose rows are the values of the scalar field $f$, $p$, the metric components $V$, $M$, and the energy density $muV$ respectively at the final $t$ value computed by the program.  The parameter $U0$ has the same makeup, but consists of the initial values of these functions.  The parameters $PlotMax$, $rt$, and $st$ are outputs so that one can continue a simulation with the geodesic in the same position and the plot windows the same size.  This facilitates a more seamless transition between the movies.  The parameters $Mloss$ and $tlist$ also facilitate continuing simulations.  $Mloss$ defines the total change in mass at the right hand boundary point defined on the $t$-values in $tlist$.

\subsection{Examples}

We present two examples here to give the reader a feel for the output plots.  The first is a complex static ground state.  Note that after $t=5000$, it still looks like a ground state.  Moreover, as can be seen in Figure \ref{evolve_ground}, the output plot plotted eight functions, including individual plots of $\Re(f)$, $\Im(f)$, and $\abs{f}$.  This is how the program automatically treats all complex solutions.

The picture generated in Figure \ref{evolve_ground} was produced by first computing a ground state using the method described in Chapter \ref{ch3}.  The program that generates the static states is called NCEKG\_state.m.  For reference, the Matlab command list that generated the output in Figure \ref{evolve_ground} is the following (you must, of course, have both the NCEKG\_state and NCEKG software packages open in your current folder).

\vspace{\baselineskip}

  \noindent $N = 10000;$ $rmax = 150;$ $r = 0:rmax/N:rmax;$

  \noindent $[lambda,U] = NCEKG\_state(1,N,rmax,1,10e-6,0,1);$

  \noindent $f = U(1,1:N+1);$ $p = U(2,1:N+1);$

  \noindent $[dr,N1,dt] = NCEKG\_mesh(f,p,10,1,150,0,5000,1,10e-6,lambda)$

  \noindent $Ntru = 333;$

  \noindent $rmaxtru = 149.85;$

  \noindent $fini = zeros(1,Ntru+1);$ $pini = zeros(1,Ntru+1);$ $rtru = zeros(1,Ntru+1);$

  \noindent $fini(1)=f(1);$ $pini(1)=p(1);$

  \noindent $for \ k = 2:Ntru+1$

  \noindent \ \  $fini(k) = f(30*(k-1)+1);$

  \noindent \ \  $pini(k) = p(30*(k-1)+1);$

  \noindent \ \  $rtru(k) = r(30*(k-1)+1);$

  \noindent $end$

  \noindent $f0 = fini;$ $p0 = pini;$

  \noindent $[U1,rt1,st1,PlotMax1,U0,Mloss1,tlist1]$

  \noindent \ \ $= NCEKG(f0, p0, 10, rmaxtru, 0, 1000, 5000, 1, 10e-6, lambda, 0, 40, 0);$

  \noindent $f1 = U1(1,1:Ntru+1);$ $p1 = U1(2,1:Ntru+1);$

  \noindent $[U2,rt2,st2,PlotMax2,U0,Mloss2,tlist2]$

  \noindent \ \ $= NCEKG(f1, p1, 10, rmaxtru, tlist1(964), 2000, 5000, 1, 10e-6, lambda, 0,$

  \noindent \ \ $rt1, st1, PlotMax1, U0, Mloss1, tlist1);$

  \noindent $f2 = U2(1,1:Ntru+1);$ $p2 = U2(2,1:Ntru+1);$

  \noindent $[U3,rt3,st3,PlotMax3,U0,Mloss3,tlist3]$

  \noindent \ \ $= NCEKG(f2, p2, 10, rmaxtru, tlist2(1927), 3000, 5000, 1, 10e-6, lambda, 0,$

  \noindent \ \ $rt2, st2, PlotMax2, U0, Mloss2, tlist2);$

  \noindent $f3 = U3(1,1:Ntru+1);$ $p3 = U3(2,1:Ntru+1);$

  \noindent $[U4,rt4,st4,PlotMax4,U0,Mloss4,tlist4]$

  \noindent \ \ $= NCEKG(f3, p3, 10, rmaxtru, tlist3(2890), 4000, 5000, 1, 10e-6, lambda, 0,$

  \noindent \ \ $rt3, st3, PlotMax3, U0, Mloss3, tlist3);$

  \noindent $f4 = U4(1,1:Ntru+1);$ $p4 = U4(2,1:Ntru+1);$

  \noindent $[U5,rt5,st5,PlotMax5,U0,Mloss5,tlist5]$

  \noindent \ \ $= NCEKG(f4, p4, 10, rmaxtru, tlist4(3853), 5000, 5000, 1, 10e-6, lambda, 0,$

  \noindent \ \ $rt4, st4, PlotMax4, U0, Mloss4, tlist4);$

  \noindent $f5 = U5(1,1:Ntru+1);$ $p5 = U5(2,1:Ntru+1);$

\vspace{\baselineskip}

On the other hand, we plot in Figure \ref{evolve_real1st} the real version of a first excited state evolved for a certain amount of time.  Since there are no complex components in this case, as the Einstein-Klein-Gordon equations do not inherently generate complex values, we will plot simply $f$, instead of the three functions we plot in the complex case.  Thus there are only six plots plotted in this version.  This is how the program automatically treats all real solutions.

By choosing a $vidtype$ of $1$ or $2$, the program will generate individual plots of each of these functions.  This is useful if one wished to make a more detailed analysis of the evolution of just one of these equations.

The program set can be found on Hubert Bray's wave dark matter web page found at \url{http://www.math.duke.edu/~bray/darkmatter/darkmatter.html} and is called NCEKG, an acronym for Newtonian Compatible Einstein-Klein-Gordon.  All of the functions are well annotated to explain what each function does and how it does it.

\begin{landscape}

\begin{figure}

    \begin{center}
        \includegraphics[width = 8 in, height = 4 in]{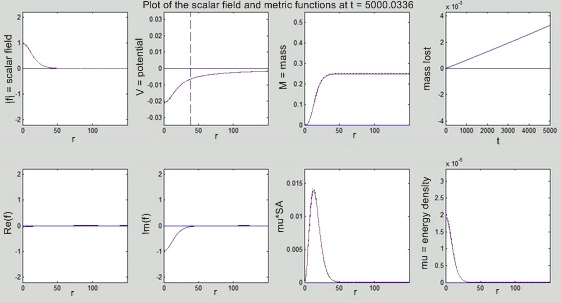}
    \end{center}

    \caption{Output plot generated by the NCEKG program for a complex scalar field.  The upper left panel depicts the absolute value of the scalar field, while the first two panels on the bottom row are plots of the real and imaginary parts of the scalar field respectively.  The second and third upper panels are plots of the potential and mass functions respectively.  The vertical dotted line in the plot of the potential traces the position of a particle in free fall in the system.  The last two plots on the bottom row are plots of the energy density function $\mu$ and $\mu$ times the surface area of a metric sphere of radius $r$.  The upper right panel plots the difference between the current value of $M$ at $r=r_{max}$ the initial value of $M$ at $r=r_{max}$ as a function of $t$.  This ``mass lost'' plot can be a visual measure of the error if the plot in question is not supposed to lose mass, or it can serve as a visual representation of the mass lost due to escaping waves.}

    \label{evolve_ground}

\end{figure}

\newpage

\begin{figure}

    \begin{center}
        \includegraphics[width = 8 in, height = 4 in]{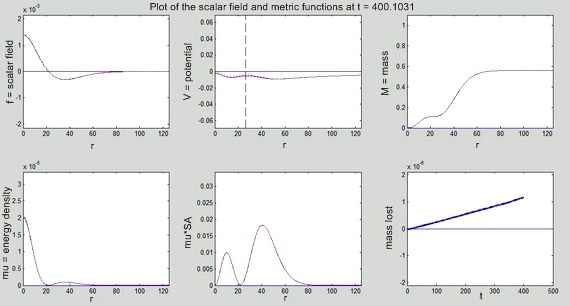}
    \end{center}

    \caption{Output plot generated by the NCEKG program for a real scalar field.  The upper left panel plots the scalar field.  The second and third upper panels are plots of the potential and mass functions respectively.  The vertical dotted line in the plot of the potential traces the position of a particle in free fall in the system.  The first two plots on the bottom row are plots of the energy density function $\mu$ and $\mu$ times the surface area of a metric sphere of radius $r$.  The lower right panel plots the difference between the current value of $M$ at $r=r_{max}$ the initial value of $M$ at $r=r_{max}$ as a function of $t$.  This ``mass lost'' plot can be a visual measure of the error if the plot in question is not supposed to lose mass, or it can serve as a visual representation of the mass lost due to escaping waves.}

    \label{evolve_real1st}

\end{figure}

\end{landscape}

\chapter{Conclusions}\label{ch6}

In this short concluding chapter, we summarize the main results of this dissertation.  In this dissertation, we have considered a possible description of dark matter called wave dark matter, which uses a scalar field to model the dark matter.  Specifically, we have considered a spherically symmetric spacetime with a scalar field and metric that satisfies the Einstein-Klein-Gordon equations.  We have compared the properties of such a spacetime with observations of dwarf spheroidal galaxies with the hopes of constraining the fundamental constant of the Einstein-Klein-Gordon equations, $\Up$.

To do this, we have surveyed the result that the Einstein equation under a general spherically symmetric metric of the form in (\ref{metric}) is remarkably simple.  In particular, it defines a first order system of partial differential equations.  The Klein-Gordon equation can be used to complete this system to evolve the metric corresponding to a complex scalar field.  Moreover, for a scalar field of the form
\begin{equation}
  f(t,r) = \e^{i\om t} F(r),
\end{equation}
with $\om \in \RR$, we have shown that the metric is static if and only if $F = \e^{ia}h(r)$ for a real constant $a$ and a real valued function $h$.  Without loss of generality, we consider these static solutions where $a = 0$.  These solutions can be distinguished by the number, $n$, of zeros they contain and are called static ground states when $n = 0$ and static $n^{\text{th}}$ excited states otherwise.

We compared these static states to the Burkert dark matter mass profiles computed by Salucci et al. (\cite{Salucci11}) in order to constrain $\Up$.  We found that under precise assumptions, a value of
\begin{equation}
  \Up = 50 \text{ yr}^{-1}
\end{equation}
produces wave dark matter mass models that are qualitatively similar to the Burkert mass models found by Salucci et al.  We also showed that comparisons to these Burkert profiles can be used to bound the value of $\Up$ by
\begin{equation}
  \Up \leq 1000 \text{ yr}^{-1}.
\end{equation}

Our future work will continue in this vein, using the numerical scheme described in Chapter \ref{ch5} to numerically evolve the spherically symmetric Einstein-Klein-Gordon equations.  With this tool, we can remove the assumption that the metric is static and consider more generic solutions of the spherically symmetric Einstein-Klein-Gordon equations.  We plan to directly compare these to observations of dwarf spheroidal galaxies in the hopes of obtaining a better estimate of the constant $\Up$.  These results will help us in our ultimate quest of determining if dark matter in the universe can be described by this wave dark matter theory. 

\appendix
\chapter{The Derivation of the \\ Einstein-Klein-Gordon Equations}\label{app1}

In this appendix, we derive the Einstein-Klein-Gordon system of equations for a complex scalar field by computing the Euler-Lagrange equations of its corresponding action on the manifold.

Let $(N,g)$ be a (3+1)-dimensional Lorentzian spacetime without boundary, where $g$ is the metric.  Let $f:N \to \C$ be a complex scalar field on the manifold.  Let $U \subset N$ be any open set for which a coordinate chart is defined.  Consider the action on $U \subset N$ given by
  \begin{equation} \label{action} \F(g,f) =  \int_{U} R - 2\La - 16\pi \mu_{0}\pnth{\frac{\abs{df}^{2}}{\Up^{2}} + \abs{f}^{2}} \, dV \end{equation}
with $\Up \in \RR$.  This action arises in a natural way from the axioms of general relativity as described by Bray (\cite{Bray10}).  We seek a pair $(g,f)$ that is a critical point of $\F$ with respect to any variation of $(g,f)$ that is compactly supported in $U$ for any $U$.  We compute the critical point via computing the Euler-Lagrange equations.

To compute the Euler-Lagrange equations of this action, we will vary $g$ and $f$ independently and compute the conditions under which $(g,f)$ would be a critical point of $\F$.  We will vary the metric $g$ first.

\section{Varying the Metric}

Let $g_{s}$ be a variation of $g$, compactly supported in $U$, with $g_{0}=g$ and $\dtg = h$, where the dot will always denote the variational derivative $\da{}{s}$ at $s=0$.  We want to find the conditions under which $(g,f)$ is a critical point of $\F$ for all variations of $g$.  That is, we wish to solve the equation
  \begin{equation} \left. \da{}{s}\F(g_{s},f) \right|_{s=0} = 0 \end{equation}
for $g$.  To this end, we have that
\begin{align}
  \notag 0 &= \left. \da{}{s}\F(g_{s},f) \right|_{s=0} \\
  \notag &= \left. \da{}{s}\int_{U} R -2\La - 16\pi \mu_{0}\pnth{\frac{\abs{df}^{2}}{\Up^{2}} + \abs{f}^{2}}\, dV \right|_{s=0} \\
  \notag &= \int_{U} \dot{R} - \frac{16\pi\mu_{0}}{\Up^{2}} \left.\da{}{s}g_{s}(df,d\cf) \right|_{s=0}\, dV  \\
  \label{dFvarg-1} &\qquad + \int_{U} R - 2\La - 16\pi\mu_{0}\pnth{\frac{\abs{df}^{2}}{\Up^{2}} + \abs{f}^{2}} \, \dot{dV}
\end{align}
It is well known that under a variation of the metric, the following are true
\begin{align}
  \label{Rdot} \dot{R} &= -\ip{h,\Ric}_{g} + \div(\div h) - \lap(\tr_{g} h), \\
  \label{dVdot} \dot{dV} &= \frac{1}{2}\tr_{g}h \, dV, \\
  \label{ginvdot} \dot{g^{\mu\nu}} &= -h^{\mu\nu}.
\end{align}
Now write $f = f^{R} + i f^{C}$, where $f^{R} = \Re(f)$ and $f^{C} = \Im(f)$.  Since $f$ is independent of the variation, then by (\ref{ginvdot}), we have that
\begin{align}
  \notag \left. \da{}{s} g_{s}(df, d\cf) \right|_{s=0} &= \left. \da{}{s} g_{s}^{\mu\nu}(f^{R}_{\mu} + i f^{C}_{\mu})(f^{R}_{\nu} - i f^{C}_{\nu}) \right|_{s=0} \\
  \notag &= -h^{\mu\nu}(f^{R}_{\mu} f^{R}_{\nu} + f^{C}_{\mu}f^{C}_{\nu} - if^{R}_{\mu}f^{C}_{\nu} + i f^{C}_{\mu}f^{R}_{\nu}) \displaybreak[0]\\
  \notag &= -h^{\mu\nu}(f^{R}_{\mu}f^{R}_{\nu} + f^{C}_{\mu}f^{C}_{\nu}) \\
  \label{dfdot-1} &= -\ip{h,df^{R} \otimes df^{R} + df^{C}\otimes df^{C}}_{g}
\end{align}
where the subscripts on $f$ denote partial differentiation and the second to last line is obtained from the previous one by relabeling of the indices on the purely imaginary terms.  Now on the other hand,
\begin{align}
  \notag \frac{1}{2}(df \otimes d\conj{f} + d\conj{f}\otimes df) &= \frac{1}{2}\bigg( \pnth{df^{R} + i df^{C}} \otimes \pnth{df^{R} - idf^{C}} \\
  \notag &\qquad + \pnth{df^{R} - idf^{C}} \otimes \pnth{df^{R} + idf^{C}}\bigg) \displaybreak[0]\\
  \notag &= \frac{1}{2}\big(2df^{R} \otimes df^{R} + 2df^{C} \otimes df^{C} - idf^{R}\otimes df^{C} \\
  \notag & \qquad \qquad + idf^{C} \otimes df^{R} + i df^{R} \otimes df^{C} - idf^{C} \otimes df^{R}\big) \\
  \label{dfdfbar} &= df^{R} \otimes df^{R} + df^{C} \otimes df^{C}.
\end{align}
This makes (\ref{dfdot-1}) become
  \begin{equation} \label{dfdot-2} \left. \da{}{s} g_{s}(df, d\cf) \right|_{s=0} = - \ip{h,\frac{1}{2}\pnth{df \otimes d\cf + d\cf \otimes df}}_{g}. \end{equation}
Next, equations (\ref{Rdot}), (\ref{dVdot}), and (\ref{dfdot-1}), and the fact that $\tr_{g}h = \ip{g,h}_{g}$ make (\ref{dFvarg-1}) become
\begin{align}
  \notag 0 &= \int_{U} -\ip{h,\Ric}_{g} + \div(\div h) - \lap(\tr_{g}h) \\
  \notag &\qquad + \frac{16\pi\mu_{0}}{\Up^{2}} \ip{h,\frac{1}{2}\pnth{df \otimes d\cf + d\cf \otimes df}}_{g}\, dV \\
  \notag & \qquad + \int_{U} \brkt{R - 2\La - 16 \pi \mu_{0}\pnth{\frac{\abs{df}^{2}}{\Up^{2}} + \abs{f}^{2}}}\pnth{\frac{1}{2}\ip{h,g}_{g}} \, dV \displaybreak[0]\\
  \notag &= \int_{U} -\ip{h,\Ric}_{g} + \frac{8\pi \mu_{0}}{\Up^{2}}\ip{h,df \otimes d\cf + d\cf \otimes df}_{g} \\
  \notag &\qquad + \ip{h,\brkt{\frac{1}{2}R - \La - 8\pi \mu_{0}\pnth{\frac{\abs{df}^{2}}{\Up^{2}} + \abs{f}^{2}}}g}_{g}\, dV \\
  \notag &\qquad + \int_{U} \div(\div h)\, dV - \int_{U} \lap(\tr_{g}h)\, dV. \displaybreak[0]\\
  \notag &= \int_{U} \ip{h,-G - \La g + 8\pi \mu_{0}\pnth{\frac{df \otimes d\cf + d\cf \otimes df}{\Up^{2}} - \pnth{\frac{\abs{df}^{2}}{\Up^{2}} + \abs{f}^{2}}g}}_{g}\, dV \\
  \notag & \qquad + \int_{U} \div(\div h)\, dV - \int_{U} \div \cov(\tr_{g}h) \, dV. \displaybreak[0] \\
  \notag 0 &= \int_{U} \ip{h,-G - \La g + 8\pi \mu_{0}\pnth{\frac{df \otimes d\cf + d\cf \otimes df}{\Up^{2}} - \pnth{\frac{\abs{df}^{2}}{\Up^{2}} + \abs{f}^{2}}g}}_{g}\, dV \\
  \label{dFvarg-2} &\qquad + \int_{\p U} \ip{\div h, \eta}_{g} \, dA - \int_{\p U} \ip{\cov(\tr_{g}h), \eta}_{g} \, dA
\end{align}
where $G = \Ric - \frac{1}{2}Rg$ is the Einstein curvature tensor, $\eta$ is an outward unit normal vector to $\p U$, and $dA$ is the induced volume form on $\p U$.  The last line uses the divergence theorem.  Since the variation $g_{s}$ is compactly supported in $U$, making $h = 0$ on $\p U$, the last two terms here are zero.  Since we want (\ref{dFvarg-2}) to be true for all compactly supported variations on any coordinate chart, we must require that
\begin{equation} \label{EL-1}
  G + \La g = 8\pi \mu_{0}\pnth{\frac{df \otimes d\cf + d\cf \otimes df}{\Up^{2}} - \pnth{\frac{\abs{df}^{2}}{\Up^{2}} + \abs{f}^{2}}g}
\end{equation}
on all of $N$.  This equation is the Einstein equation in the presence of a scalar field.

\section{Varying the Scalar Field}

Next, consider a variation of $f$, $f_{s}$, compactly supported in $U$ with $f_{0} = f$ and $\dtf = q$.  In this case, we want a condition under which $(g,f)$ is a critical point of $\F$ for all variations of $f$.  This time, since $g$ is independent of the variation, we have that
\begin{align}
  \notag 0 = \left. \da{}{s} \F(g,f_{s}) \right|_{s=0} &= \left. \da{}{s} \int_{U} R - 2\La - 16\pi\mu_{0}\pnth{\frac{\abs{df_{s}}^{2}}{\Up^{2}} + \abs{f_{s}}^{2}} \, dV \right|_{s=0} \\
  \label{dFvarf-1} &= \int_{U} \left. -16 \pi \mu_{0} \da{}{s}\pnth{\frac{\abs{df_{s}}^{2}}{\Up^{2}} + \abs{f_{s}}^{2}} \right|_{s=0}\, dV
\end{align}
Now,
  \begin{equation} \label{f2dot}
    \left. \da{}{s} \abs{f_{s}}^{2} \right|_{s=0} = \left. \da{}{s} f_{s}\cf_{s} \right|_{s=0} = q\cf + f\cq
  \end{equation}
and
  \begin{equation} \label{gradfdot}
    \left. \da{}{s} g\pnth{df_{s},d\cf_{s}} \right|_{s=0} = \left. \da{}{s} g\pnth{\cov f_{s}, \cov \conj{f_{s}}} \right|_{s=0} = g\pnth{\cov f, \cov \cq} + g\pnth{\cov q, \cov \cf}.
  \end{equation}
Then (\ref{dFvarf-1}) becomes
  \begin{equation} \label{dFvarf-2}
    0 = -16\pi\mu_{0}\int_{U} \frac{1}{\Up^{2}}\pnth{g\pnth{\cov f, \cov \cq} + g\pnth{\cov q, \cov \cf}} + q\cf + f\cq \, dV.
  \end{equation}
By the divergence theorem and the fact that the variation is compactly supported in $U$ so that $q = 0$ on $\p U$,
\begin{align}
  \label{div1} 0 &= \int_{U} \div\pnth{\cq\cov f}\, dV = \int_{U} g\pnth{\cov \cq,\cov f} + \cq\glap{g} f \, dV \\
  \label{div2} 0 &= \int_{U} \div\pnth{q\cov \cf}\, dV = \int_{U} g\pnth{\cov q, \cov \cf} + q\glap{g} \cf \, dV
\end{align}
Equations (\ref{dFvarf-2}) - (\ref{div2}) imply that
\begin{align}
  \notag 0 &= -16 \pi \mu_{0} \int_{U} \frac{1}{\Up^{2}}\pnth{-\cq \glap{g}f - q \glap{g}\cf} + q\cf + \cq f \, dV \\
  \label{dFvarf-3} &= -\frac{16\pi \mu_{0}}{\Up^{2}}\int_{U} \cq\pnth{\Up^{2}f - \glap{g} f} + q\pnth{\Up^{2}\cf - \glap{g}\cf} \, dV.
\end{align}
Equation (\ref{dFvarf-3}) holds for all compactly supported variations of $f$ in any coordinate chart.  This implies that
  \begin{equation} \label{EL-2} \glap{g} f = \Up^{2}f \end{equation}
on all of $N$.  This equation is called the Klein-Gordon equation.

\section{The Einstein-Klein-Gordon Equations}

We have shown that the Euler-Lagrange equations for the action (\ref{action}) are (\ref{EL-1}) and (\ref{EL-2}), which are rewritten below for convenience.
\begin{subequations} \label{EKG-agn}
  \begin{align}
    \label{EKG-1-2} G + \La g &= 8\pi \mu_{0}\pnth{\frac{df \otimes d\cf + d\cf \otimes df}{\Up^{2}} - \pnth{\frac{\abs{df}^{2}}{\Up^{2}} + \abs{f}^{2}}g} \\
    \label{EKG-2-2} \glap{g} f &= \Up^{2}f.
  \end{align}
\end{subequations}
These paired equations are called the Einstein-Klein-Gordon equations.



\bibliographystyle{jasa} 
\cleardoublepage
\normalbaselines 
\addcontentsline{toc}{chapter}{Bibliography} 
\bibliography{References} 


\end{document}